\RequirePackage{fix-cm}
\documentclass[11pt,a4paper,oneside]{book}

\usepackage{graphicx,xcolor,textpos}
\usepackage{Packages}
\usepackage{Bibliography}
\usepackage{emptypage}
\pdfoutput=1

\makeglossaries

\newcommand*\cleartoleftpage{%
  \clearpage
  \ifodd\value{page}\hbox{}\newpage\fi
}

\definecolor{green}{RGB}{172,196,0}
\definecolor{bluetitle}{RGB}{29,141,176}
\definecolor{blueaff}{RGB}{0,0,128}
\definecolor{blueline}{RGB}{82,189,236}

\newglossaryentry{pi}
{
  name={\ensuremath{\pi}},
  description={ratio of circumference of circle to its
               diameter},
  sort=pi
}

\newglossaryentry{alpha}
{
  name={\ensuremath{\alpha}},
  description={a random greek letter},
  sort=alpha
}
\newglossaryentry{QMC}{
name={QMC},
description={Quantum Monte Carlo}
}
\newglossaryentry{PIMC}{
name={PIMC},
description={Path Integral Monte Carlo}
}
\newglossaryentry{WAPIMC}{
name={WA-PIMC},
description={Worms Algorithm Path Integral Monte Carlo}
}
\newglossaryentry{PIGS}{
name={PIGS},
description={Path Integral Ground State Monte Carlo}
}
\newglossaryentry{WA}{
name={WA},
description={Worms Algorithm}
}
\newglossaryentry{pdf}{
name={PDF},
description={Probability Density Function}
}
\newglossaryentry{ZRP}{
name={ZRP},
description={Zero-Range Potential}
}
\newglossaryentry{NP}{
name={NP},
description={nondeterministic polynomial}
}
\newglossaryentry{VMC}{
name={VMC},
description={Variational Monte Carlo}
}
\newglossaryentry{GFMC}{
name={GFMC},
description={Green's-function Monte Carlo}
}
\newglossaryentry{DMC}{
name={DMC},
description={Diffusion Monte Carlo}
}
\newglossaryentry{BCH}{
name={BCH},
description={Baker-Campbell-Hausdorff}
}
\newglossaryentry{AMO}{
name={AMO},
description={Atomic, Molecular and Optical Physics}
}
\newglossaryentry{MI}{
name={MI},
description={Mott Insulator}
}
\newglossaryentry{SRQ}{
name={SRQ},
description={Superfluid Ring Qubit}
}
\newglossaryentry{SRL}{
name={SRL},
description={Superfluid Ring Lattices}
}
\newglossaryentry{SQUID}{
name={SQUID},
description={Superconducting QUantum interference Devices Ring Qubit}
}
\newglossaryentry{ILM}{
name={ILM},
description={Integer Lattice method}
}
\newglossaryentry{SFQ}{
name={SFQ},
description={Superconducting Flux Qubits}
}
\newglossaryentry{SLM}{
name={SLM},
description={Spatial Light Modulator}
}

\begin{document}
\fontfamily{qcr}
\frontmatter

\thispagestyle{empty}
\newcommand{\form}[1]{\scalebox{1.087}{\boldmath{#1}}}
\sffamily
\begin{textblock}{191}(-24,-11)
\colorbox{green}{\hspace{123mm}\ \parbox[c][18truemm]{68mm}{\textcolor{white}{FACULTY OF SCIENCES}}}
\end{textblock}
\begin{textblock}{70}(-18,-19)
\textblockcolour{}
\includegraphics*[height=19.8truemm]{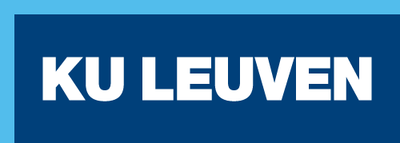}
\end{textblock}
\begin{textblock}{160}(-6,63)
\textblockcolour{}
\vspace{-\parskip}
\flushleft
\fontsize{31}{32}\selectfont \textcolor{bluetitle}{Path Integral Monte Carlo Simulation of Superfluid Ring Lattices}\\[1.5mm]
\fontsize{20}{22}\selectfont Building Quantum Devices with Optical Landscapes 
\end{textblock}

\begin{textblock}{79}(20,102)
\textblockcolour{}
\vspace{-\parskip}
\flushleft
\includegraphics[height=7cm]{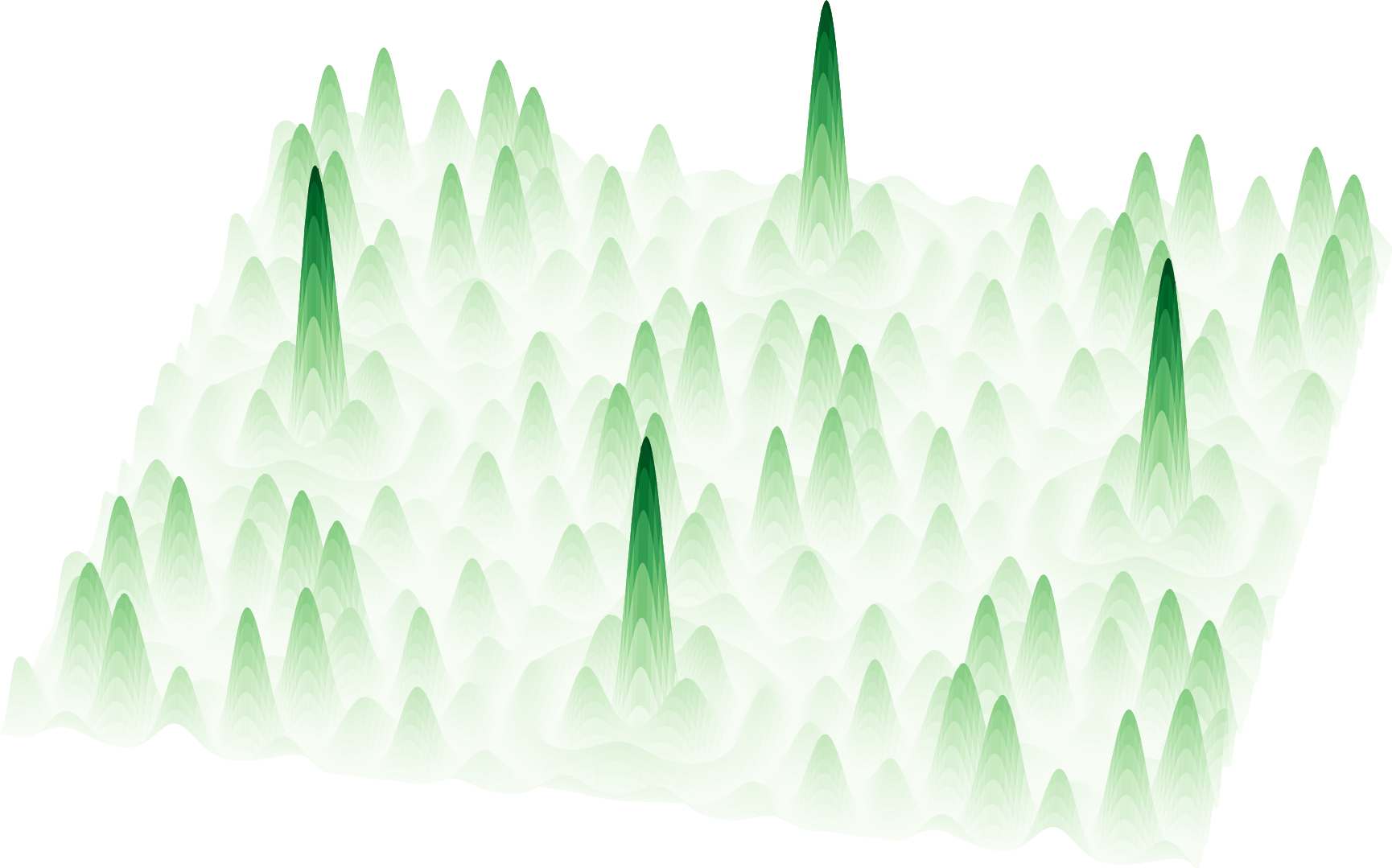}
\end{textblock}

\begin{textblock}{160}(8,173)
\textblockcolour{}
\vspace{-\parskip}
\flushright
\fontsize{14}{16}\selectfont \textbf{Orjan AMEYE}
\end{textblock}
\begin{textblock}{70}(-6,191)
\textblockcolour{}
\vspace{-\parskip}
\flushleft
Supervisor: {Prof. P. Van Dorpe}\\[-2pt]
\textcolor{blueaff}{Imec $\vert$ KU Leuven }\\[5pt]
Co-supervisor: {N. Verellen}\\[-2pt]
\textcolor{blueaff}{Imec $\vert$ KU Leuven }\\[5pt]
Mentor: {D. Kouznetsov}\\[-2pt]
\textcolor{blueaff}{Imec $\vert$ KU Leuven }\\
\end{textblock}
\begin{textblock}{160}(8,191)
\textblockcolour{}
\vspace{-\parskip}
\flushright
Master thesis submitted in fulfillment  \\[4.5pt]
 of the requirements for the degree in \\[4.5pt]
Master of Science in Physics\\
\end{textblock}
\begin{textblock}{160}(8,232)
\textblockcolour{}
\vspace{-\parskip}
\flushright
Academic year 2021-2022
\end{textblock}
\begin{textblock}{191}(-24,248)
{\color{blueline}\rule{550pt}{5.5pt}}
\end{textblock}
\vfill
 \cleardoublepage
\thispagestyle{empty}
\vspace*{\fill}
\cleardoublepage
\rmfamily
\setcounter{page}{0}
\pagenumbering{roman}

\chapter*{Scientific summary}
\addcontentsline{toc}{chapter}{Scientific summary}%


In modern physics, computer simulations have become an essential part of performing research. However, many quantum mechanical systems are difficult to solve computationally as the dimensionality of Hilbert space becomes exceedingly large for only moderately complex systems. This results in unfeasible computation times when using naive algorithms. To solve this problem, several simulation techniques have been proposed --- almost all relying on the Monte Carlo method or so-called quantum Monte Carlo method --- to tackle the extensiveness of the configuration space. In particular, Path Integral Monte Carlo (PIMC) provides exact ab-initio results in thermodynamic equilibrium by solving the path integral formulation in quantum statistical mechanics. The technique is often complemented with the Worm Algorithm (\gls{WA}) to sample the right particle statistics at low temperatures.

Although \gls{WAPIMC} was originally developed for studying bulk 4-helium gases, in recent years the technique has been extensively exploited in the field of Atomic, Molecular, and Optical (AMO) physics to investigate systems of ultracold atoms in optical lattices due to its ability to simulate Bose-Einstein condensation and superfluidity. Recently, the platform that it provides has been investigated to implement quantum engineered devices --- which led to the novel field of atomtronics. In particular, the combined work of \citet{Edwards2013, Amico2014, ryu_quantum_2020} resulted in the experimental realization of controllable two-state quantum systems or qubits, coined the Superfluid Ring Qubit (\gls{SRQ}). The designs are analogous to Josephson flux qubits but enjoy a higher decoherence time due to the use of neutral bosons instead of charged current carriers. The challenge remains to design a scalable, versatile multi-qubit system to construct qubit gates and eventually circuits.

The goal of this work is to lay the groundwork to construct and characterize a quantum device --- which we refer to as a superfluid ring lattice --- that could serve as a multi-qubit system in the future. Accordingly, a mathematical framework, called the Integer Lattice Method (\gls{ILM}), is exploited to construct a two-dimensional optical landscape which could facilitate a superfluid ring qubit. The Integer Lattice Method allows one to design and explore both periodic and quasi-periodic structured coherent wave interference patterns. Furthermore, the formalism allows for a direct link to experimental realization. The lattices obtained from \gls{ILM} can be investigated using \gls{WAPIMC}. In particular, the spatial and superfluid density at equilibrium are observables of interest to obtain our objective.

In this thesis, the ab-initio Path Integral Monte Carlo Worm algorithm is implemented with the future aim of investigating the construction of superfluid ring lattices. It is written in independent and interchangeable modules in the novel programming language Julia. To ensure scalability, a custom nearest-neighbour algorithm is added. The simulation software is benchmarked by comparing it with exact analytical results from the non-interacting Bose gas model. Emerging issues with our implementation of the Worm algorithm are identified and thoroughly investigated. Finally, some preliminary results on the equilibrium density distribution of the found \gls{ILM} lattices are presented and discussed.

\cleardoublepage
\chapter*{Summary for a general audience}
\addcontentsline{toc}{chapter}{Summary for a general audience}
Over the past 100 years, quantum mechanics --- the physical theory that describes the behaviour of matter at the atomic scale --- has helped our understanding of physical systems immensely. Lately, it has also begun to influence our everyday technology. For example, today you can purchase a quantum dot TV screen that uses artificial atoms, i.e., quantum dots, that emit light with a specific colour when you apply energy. One of the most popular devices we would like to see realized is a quantum computer. This is a computer where the processor uses the principles of quantum mechanics. There are many physical systems that could implement a device like this.

One of the potential candidates would be ultracold atoms in optical lattices. These are very cold gases the are subjected to a strong concentration of light beams, also called laser beams. The low of temperature make the atoms in the gas very manipulable and controllable. Thus, the movement of the atoms can be limited to a two-dimensional landscapes or lattices. These landscapes can be created in many different shapes. \citet{Edwards2013, Amico2014, ryu_quantum_2020} have introduced a ring-shaped structure into the landscape in which the atoms are fixed. By applying a magnetic flux they can control the clockwise and counter-clockwise movement of the atoms, leading to a computing unit, called Superfluid Ring Qubit (SRQ). Due to the low temperature, the flow of the atoms has the property of a fluid with no viscosity, called a superfluid.

To build a computer, we would need multiple computing units in the system. Using a mathematical method, called the Integer Lattice Method (ILM), to find new lattices, we found an optical landscape that contains multiple ring structures. In this work, we aim to lay the groundwork to investigate whether optical surfaces are good candidates to host superfluid ring qubits. We do this by implementing a computational technique that can simulate the equilibrium state of atoms for any temperature, called Path Integral Monte Carlo (PIMC). This technique is then extended with the Worm Algorithm (WA), which gives the ability to simulate the superfluid phase of the atomic gas.


 \cleardoublepage

{
\hypersetup{linkcolor=black}
\addcontentsline{toc}{chapter}{Table of contents}
\tableofcontents \cleardoublepage
\listoffigures \cleardoublepage
\listoftables \cleardoublepage
\printglossary[title={List of Abbreviations}] \cleardoublepage 
}

\newpage
\mainmatter
\rmfamily
\setcounter{page}{0}
\pagenumbering{arabic}

\chapter{Introduction}
\label{ch: Introduction}

Quantum engineering has matured immensely during the 21st century. From the deployment of quantum dots in industry to the development of true modular single-photon transmitters on which many future technologies rely \cite{Aharonovich2016}. One of the most sought-after devices is the realization of quantum circuits, because of their potential to solve complex computing problems like polymer folding, logistics optimization, etc. In parallel to classical circuits, they consist of a collection of gates formed by qubits --- a two-state system which can be in a coherent superposition of both. Over the years, many physical systems have been proposed to allow for such a binary quantum-mechanical system. One of the more promising system is the superconducting flux qubit, characterized by a superconducting loop interrupted by several Josephson junctions \cite{Dmitriev2021}. Nevertheless, superconducting electronic circuits are known to have a fast decoherence time, which makes it hard to scale to a multi-qubit system \cite{Kjaergaard2020}.

Recently, a similar system has been proposed which uses ultracold neutral atoms as current carriers in a ring-shaped optical lattice --- instead of electrons in a superconducting loop --- to form Superfluid Ring Qubit (\gls{SRQ}), or atomtronic flux qubits. By the same principles as the Superconducting Flux Qubit (\gls{SFQ}), it has been demonstrated experimentally that a \gls{SRQ} defines an effective quantum two-level system \cite{Amico2014, aghamalyan2015}. A remaining challenge is the establishment of scalable communication between the neighbouring counterparts to create quantum gates or circuits \cite{amico2021}.

A way to overcome this difficulty lies in creating an optical landscape that enables such interactions and can arrange ultracold atom qubits on very small scales. A potential technique is the Integer Lattice method (\gls{ILM}) developed by \citet{DmitryILM2020}. \gls{ILM} is a mathematical framework which provides a way to design coherent wave interference patterns with a well-behaved structure. The method lends itself to concrete experimental realization of the \mbox{(quasi-)periodic} optical landscapes with the possibility to tune the periodicity between the intensity maxima \cite{DmitryBEC2022}. Using \gls{ILM}, a collection of optical lattices is found which could potentially host quantum circuits of superfluid ring qubits or Superfluid Ring Lattices (\gls{SRL}s).

In this thesis, a numerical tool called Path Integral Monte Carlo (PIMC) is constructed with the future aim of assessing the viability of the proposed optical lattices for hosting parallel superfluid ring qubit circuits \cite{Ceperley1984}. \gls{PIMC} is an ab-initio quantum Monte Carlo approach for numerically solving quantum mechanical problems using the path integral formulation. It has been successfully applied in many condensed matter applications \cite{Dornheim2018, Ceperley1987, DelMaestro2011}, including ultracold atoms in optical lattices \cite{Yao2020LiebLiniger, Gauter2021}. The simulation software is written in the programming language Julia from the ground-up with a focus on two-dimensional bosons systems. Furthermore, we pursue an extension of the method called the Worm-Algorithm (\gls{WA}) \cite{Boninsegn2006PRL} to assist the simulation at temperatures near absolute zero. \pagebreak

In the following, ultracold atoms in optical lattices are introduced. A preface on how this technology is used to engineer qubit systems, is presented. Lastly, a more in-depth sketch of the potential solution using \gls{ILM} is given.

\section{Tinkering with nature}

``We are currently in the midst of a \emph{second quantum revolution}.'' This is how \citet{Dowling2003} started their 2003 review article on quantum technology\footnote{The term did only take off when it was used by Alain Aspect in his introduction to the renowned book by John Bell \cite{Bell2004}}. They argue the first scientific and technological period of quantum mechanics is wrapped up, and a second has started. The discovery of quantum mechanics at the beginning of the 20th century initiated the \emph{first quantum revolution}, which led to the foundational technologies of contemporary society. More concretely, the framework of quantum mechanics resulted in the understanding of atomic physics (e.g. periodic table), chemical and nuclear interactions, which underpin energy production, but also the behaviour of electrons, which led to electronic semiconductor physics and resulted in the ``information age'' we live in today. The first revolution is marked by understanding the quantum properties of what already exists. In contrast, in the last 30 years, we have started to develop techniques to manipulate or engineer quantum systems for our own purposes. Instead of describing the atoms in the periodic table, we engineer artificial atoms, e.g., quantum dots, which have the properties of our own liking. We are in the midst of a second quantum revolution.

There are many fields which lead this ongoing revolution, from quantum computation to quantum communication, metrology, sensing, and imaging \cite{Antonio2018}. A major contribution was provided in the field of Atomic, Molecular and Optical (\gls{AMO}) physics by Steven Chu, Clause Cohen-Tannoudji and William D.Phillips. This was for their ``pioneering contributions to the development of methods to cool and trap atoms with laser light'' as stated by the Nobel Committee \cite{Nobelprize1997}. These Nobel Laureates developed methods using laser light to cool gases to the \unit{\micro K} temperature range. In particular, they have paved the way for a better understanding of the quantum physical properties of gases at low temperatures and the emergent new physics that arises. Furthermore, at low temperatures, gases are easier to control, thereby opening the door to manipulating these \emph{ultracold atoms} to construct quantum engineered systems.


\section{Ultracold atoms in optical lattices}
Ultracold gases are dilute neutral atomic gases, often made of alkali atoms, held at a temperature of practically absolute zero. The creation of such gases had an immediate impact on our understanding of physics as it led to the experimental confirmation of \textit{Bose-Einstein condensation} \cite{Bradley1995, Davis1995, Anderson1995} (1995) and not that long after \textit{Fermi degeneracy} \cite{Schreck2001, DeMarco1999, Truscott2001} (1999-2001); two phenomena described by a coherent macroscopic matter wave function where the many-body problem can be represented as a system of non-interacting quasiparticles. Although these weakly-interacting gases have been studied extensively in the 1950s and 1960s, the experimental realization of such condensates in dilute gases was a major milestone in quantum engineering.

\begin{figure}[htbp]
    \centering
    \includegraphics[width=0.9\linewidth]{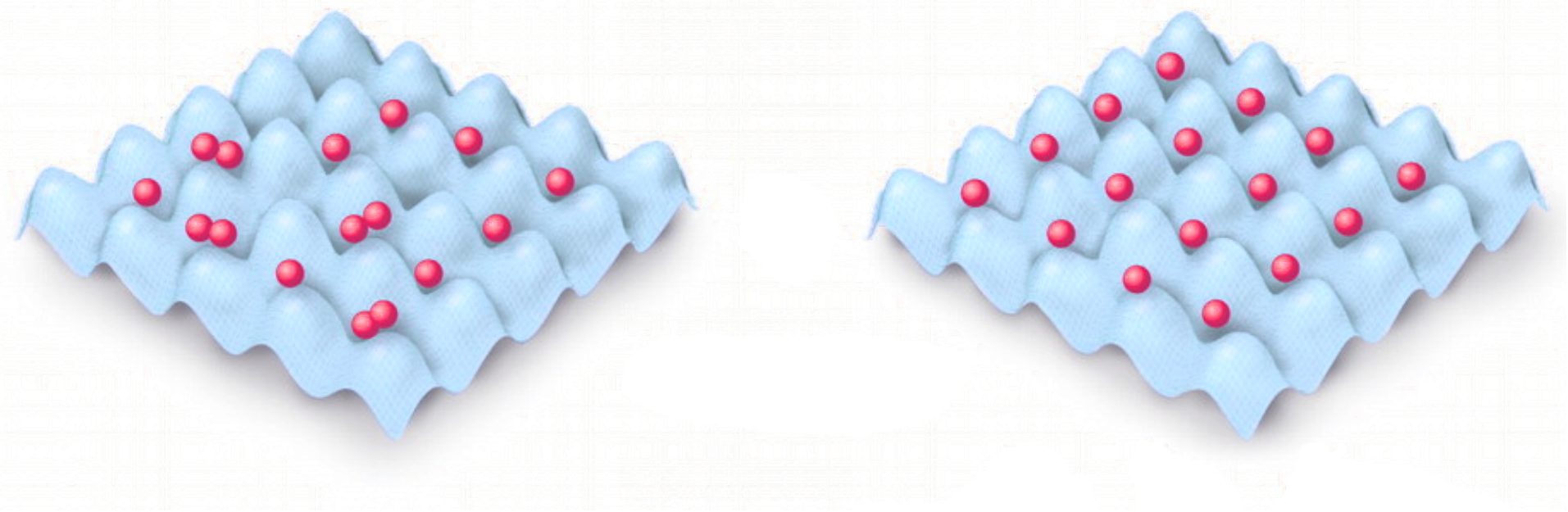}
    \caption{Three-dimensional representation of ultracold atoms loaded into a two-dimensional square optical lattice. On the left, the atoms are free to roam the potential landscape, whereas on the right the atoms are localized --- for example, due to a high potential depth. Taken and amended from \citet{Buluta2009} under licence of The American Association for the Advancement of Science (AAAS).}
    \label{fig: UAOL}
\end{figure}

In the last 20 years, great contributions have been made to the manipulation of these ultracold gases. For example, the dynamics of the atoms can be controlled by generating coherent electromagnetic wave interference patterns which results in a system called \textit{ultracold atoms in optical lattices}. First independently realized by \citet{Dalibard1989,Ungar1989}, ultracold atoms can be loaded into a one-, two-, or three-dimensional optical potential landscape made from interfering electromagnetic laser beams --- an example of which can be seen in figure \ref{fig: UAOL}. They play the role of an effective potential, i.e., it has the effect of an external potential term in the Hamiltonian. Just as the sign in the term in the Hamiltonian can be reversed, the peaks in the optical potential can be made attractive or repulsive. Optical lattices provide a great deal of versatility: not only the dimensionality, symmetry, and depth of the optical landscape can be modified, but also the overall pattern can be tuned. Furthermore, more recently, potential landscapes that carry effective magnetic fields have been successfully engineered, which makes it a platform to create \textit{topological quantum devices} \cite{Goldman2016}.

The system od ultracold atoms loaded in an optical lattice is known for being a \textit{quantum simulator}. A quantum simulator is ``an experimental system that mimics a simple model of condensed matter or other physics fields'' \cite{Lewenstein2012}. The physical outcome of the experiment can then be extended to the original model, as they theoretically should both respect the same physical equations. This approach can be beneficial when a system is experimentally unattainable as the parameter scales are not accessible. A recent example is the experimental realization of the Haldane model by \citet{Jotzu2014} with ultracold atoms in optical lattices. It illustrates how the quantum Hall effect might arise as an intrinsic characteristic in two-dimensional semiconductors instead of being induced by an external magnetic field. The model has had a large influence on research in topological insulators and superconductors. However, it has never been possible to realize the phenomenon in two-dimensional semiconductors \cite{Simon2014}. Nevertheless, ultracold atoms in an optical lattice were the ideal platform to confirm the model.


\section{Superfluid Ring Qubit}
Ultracold atoms in optical lattices are not only helpful as quantum simulators. They are also a toolbox to create quantum engineered systems, such as in the field of \emph{atomtronics}. Atomtronics utilizes the techniques used in the field of ultracold atoms research to create matter wave circuits as an analogue to canonical electrical circuits. Using atoms instead of electrons has a couple of advantages. First, there is a lot more freedom in the properties of the current carriers. They can both adhere to Bose or Fermi statistics; the strength of inter-particle interaction is tunable with Feshbach resonance \cite{Chin2010}; the interaction can be long or short range, attractive or repulsive, etc. Furthermore, because the optical landscape can be tuned, one can create time-dependent circuits whose layout can be adjusted during operation. Examples of analogue electrical devices created using ultracold atoms are transistors, capacitors, diodes, atomic Superconducting QUantum Interference Devices or \gls{SQUID}'s, etc. However, the new environment, and the flexibility of it, gives the possibility to create novel devices with other topologies and functionality.
\begin{figure}[htbp]
    \centering
    \includegraphics[width=0.8\linewidth]{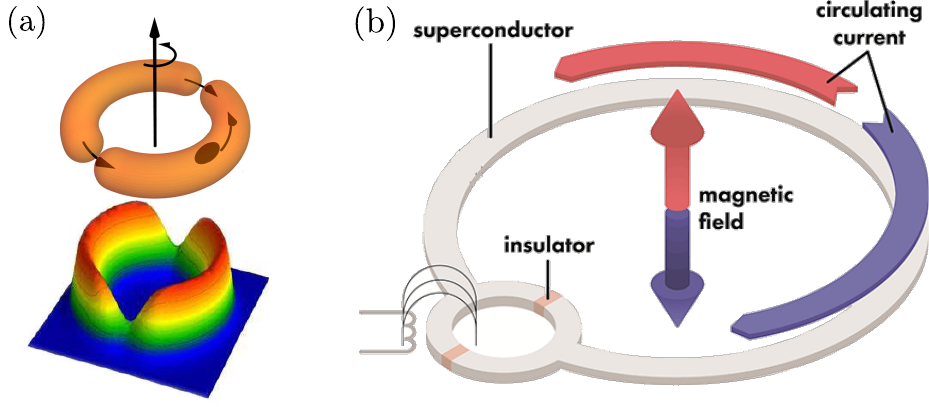}
    \caption{Experimental realization of a superfluid ring qubit. \textcolor{red}{(a)} A realization of a two-state quantum system with ultracold atoms in an optical lattice with two insulator components. \textcolor{red}{(b)} A schematic eluding the principles of a canonical superconductive flux qubit made in electronic circuits. The two figures are taken and amended under the Creative Commons Attribution 4.0 International Licence from \citet{ryu_quantum_2020} and \citet{Grant2020}, respectively.}
    \label{fig: ring qubit}
\end{figure}

An example of a state-of-the-art quantum engineered system in atomtronics is the \textit{superfluid ring qubit} (\gls{SRQ}) \cite{Amico2014} --- seen in figure \ref{fig: ring qubit}\textcolor{red}{(a)}. The architecture follows the same concepts as the more well-known \emph{superconductive flux qubits} (\gls{SFQ}) --- seen in figure \ref{fig: ring qubit}\textcolor{red}{(b)}. As with \gls{SFQ}s, \gls{SRQ} facilitates an effective two-state quantum mechanical system resulting from the interplay between conducting and insulating components. This is realized by a conducting ring optical potential containing insulating gaps (see  figure \ref{fig: ring qubit}\textcolor{red}{(a)}). Analogously, a \gls{SFQ} is constructed out of superconducting material closed with an insulating dielectric. The two states are formalized by the symmetric and anti-symmetric superposition of clockwise and anti-clockwise current flows of the ultracold atoms (\gls{SRQ}) or electrons (\gls{SFQ}) on the circuits. The state of the device, or qubit, is then controlled by changing the phase of the loop generated by artificial magnetic flux pierced through the ring structure.

\gls{SRQ}s also enjoy some differences from \gls{SFQ}s. One of the main problems with the latter is the small coherence time, i.e. the time after which the initial state become uncorrelated from the final state. Although decoherence is a complex process, it has two main sources in \gls{SFQ}s: the influence of magnetic flux fluctuations on charged current carriers and defects in the material \cite{Martinis2014, Clarke2008}. As \gls{SRQ}s use neutral atoms as current carriers and the great control over the optical landscape, both issues are addressed. This leads to a decoherence time in the range of tens of seconds with a fidelity greater than 99\% \cite{Moulder2012, amico2021}. 

Naturally, superfluid ring qubits still have to face some challenges. In particular, allowing multiple superfluid ring qubits to communicate in a controllable manner to generate quantum gates and circuits. An effort was made by \citet{Amico2014}, who stacked the qubits in a one-dimensional array by using the Spatial Light Modulator (\gls{SLM}) technique. Nevertheless, having a system in which the qubits only interact in a one dimensional manner has a pitfall. To protect the processed quantum information from yielding errors, such as phase flips, one needs to implement quantum error correction algorithms in the qubits circuits \cite{Fowler2012}. However, to implement the widely studied surface code --- a quantum phase flip error correction algorithm --- one requires a two-dimensional lattice \cite{Corcoles2015}. To date, no two-dimensional lattice configuration multi-qubit system in atomtronics has been experimentally achieved.

The problem lies in constructing a coherent optical landscape with as little infrastructure as possible --- for example, lasers, mirrors, liquid crystals, etc. Furthermore, the constructed qubit gates and circuits favourably must be compact, thereby making the devices more easily scalable and eventually producible on-chip. However, often the optical lattices are realized with the help of \gls{SLM}s which are inherently three-dimensional setups. This prevents \gls{SLM} apparatus from being integrated into a fully two-dimensional system, and thereby limits the scalability. Therefore, strictly, two-dimensional techniques like \emph{multibeam interference} --- an optical lattice formed by the interference of counter-propagating laser beams in a horizontal plane --- are favoured. In figure \ref{fig: Harmonic_trap} a sketch of such an experimental setup is given.

\section{Integer Lattice Method}

\begin{figure}[htbp]
    \centering
    \includegraphics[width=\linewidth]{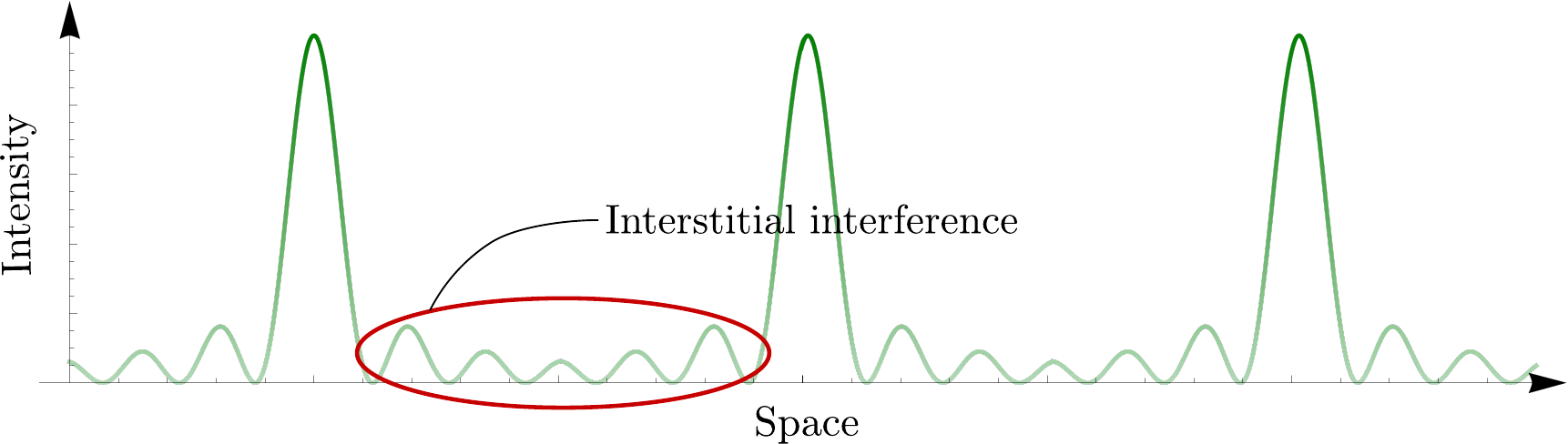}
    \caption{Three subsequent one-dimensional Bessel spots with unwanted interstitial interference between them. The pattern can be created by focussing three subsequent laser spots next to each other.}
    \label{fig: bessel}
\end{figure}

In the previous section, we established we need a two-dimensional optical lattice that can ideally be integrated in a two-dimensional system to construct multi-\gls{SRQ} system or, from now on, coined Superfluid Ring Lattices (\gls{SRL}). As such, we would like to create a landscape with well-defined circular or ring structure. However, this is not a trivial task as even a simple Bessel spot has sidelobes that might cause additional/auxiliary interstitial interference patterns, as shown in figure \ref{fig: bessel}. These often unwanted stencils could adversely influence the performance of the device.

\citet{DmitryILM2020} developed an algebraic technique to design structured coherent wave interference patterns that produce optical lattices with the theoretically most sideband suppression, i.e., the spots in the unit cell show limited interstitial interference.
It uses moir\'e theory in the complex plane to compute the orientations of laser beams for the generation of optical lattices with variable symmetry. Furthermore, it has the ability to tune the periodicity of the optical lattices --- which contributes to the dynamics of the ultracold atoms --- without changing the wavelength of the laser \cite{DmitryBEC2022}.
\begin{figure}[htbp]
    \centering
    \begin{subfigure}[]{.45\linewidth}
        \begin{overpic}[width=\linewidth]{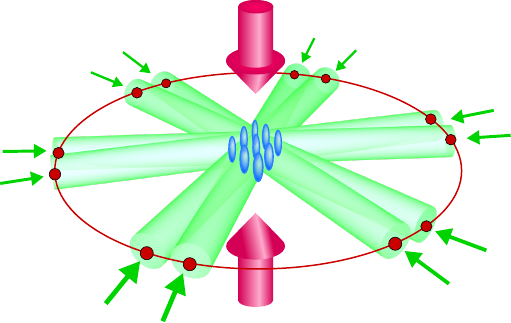}
        \put(0,60){(a)}
        \end{overpic}
        \phantomsubcaption\label{fig: Harmonic_trap}
    \end{subfigure}
    \hspace*{1cm}
    \begin{subfigure}[]{.45\linewidth}
        \begin{overpic}[width=\linewidth]{Preamble/Titlepage/optlat.png}
        \put(0,60){(b)}
        \put(37,-1){$x[\lambda]$}
        \put(96,23){$y[\lambda]$}
        \end{overpic}
        \phantomsubcaption \label{fig: optlat}
    \end{subfigure}%
    \caption{Creation of an optical lattice using the integer lattice method. \subref{fig: Harmonic_trap} A planar laser beam configuration resulting in a multibeam interference setup to generate an optical landscape. The red dots on the circle correspond to the placement of a laser in the direction of the centre region of the circle where the beams overlap and the lattice is formed. The red arrows represent the harmonic confinement of the ultracold atoms to a two-dimensional plane. Taken from \cite{DmitryBEC2022} with permission from the author. \subref{fig: optlat} The resulting optical lattice generated using the integer lattice method configuration corresponding to the ring $\mathbb{Z}\left[\zeta_{m}\right]=\left\{a+b \zeta_{m} \mid a, b \in \mathbb{Z}\right\}$ with $\zeta_{m} = e^{2 \pi i / m}$ with $m=4$ and field norm $n=65$ (see chapter \ref{ch: ILM}). The length between the interstitials is determined by the wavelength $\lambda$ of the monochromatic lasers beams.}
\end{figure}
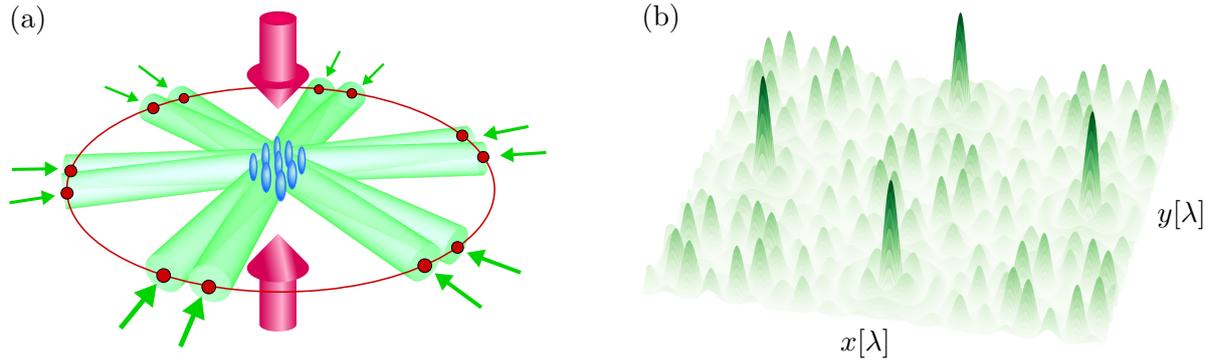

\pagebreak
In this thesis, \gls{ILM} is used to find optical landscapes which could facilitate multi-qubit systems. In particular, there is searched for ring-shaped structures such that the design of a superfluid ring qubit can be implemented. An example of such an optical landscape produced with ILM is seen in figure \ref{fig: optlat}. 

\section{Path Integral Monte Carlo}
To investigate if the optical landscapes provided by \gls{ILM} could host superfluid ring qubits, we would like to use numerical techniques to simulate the properties and phases of the ultracold atoms loaded into the selected optical lattice. For this purpose, Path Integral Monte Carlo or \gls{PIMC} was chosen. It is an ab-initio quantum Monte Carlo algorithm which can simulate systems in continuous space in every statistical ensemble. \gls{PIMC} solves perfectly the thermodynamic equilibrium properties of ultracold bosonic systems in various inhomogeneous external potentials, at any regime of interactions and temperature \cite{Gauter2021}. Furthermore, it is renowned for correctly representing Feynman’s theory of superfluidity \cite{Feynman1953}, which is necessary for studying the phases of the ultracold atom system. However, the classical \gls{PIMC} implementation suffers from a critical slowing down phenomenon in the sampling of the correct particle statistics at very low temperatures. Therefore, it is extended with the Worm Algorithm (\gls{WA}), an additional sampling scheme which addresses the hurdle.

This thesis focuses on implementing a Path Integral Monte Carlo Worm algorithm (\gls{WAPIMC}) in the programming language Julia. The implementation concentrates on simulating ultracold bosons in two-dimensional optical lattices, although an effort is made to make the code as general as possible so that it can be used for other use cases. Furthermore, special attention was given to separating the functionality of the program into independent and interchangeable modules such that each contains everything necessary to execute only one aspect of the desired code. In addition, to ensure scalability to systems with many particles, a custom nearest-neighbour scheme is implemented. The algorithm is benchmarked using various test, and it is compared against exact analytical results. Finally, some preliminary results on the \gls{ILM} optical lattices --- such as the spatial  --- are discussed. \gls{WAPIMC} will allow us to lay the groundwork to eventually construct superfluid ring lattices.

\section{Outline}
In what follows, a brief introduction to ultracold atoms in optical lattices is presented in chapter \ref{ch: Ultracold Atoms In Optical Lattices}. Both how the ultracold atoms interact with each other and with the optical lattice is discussed. In addition, a primed on superfluid ring qubits is given. Next, in chapter \ref{ch: ILM} the mathematical description of Integer Lattice Method (\gls{ILM}) is outlined and the prospective \gls{ILM} optical lattices are discussed. Afterwards, in chapter \ref{ch: PIMC}, the theory surrounding Path Integral Monte Carlo is given. Finally, various aspect related to the implementation of the simulation software in Julia are discussed in chapter \ref{ch: Implementation}. Comparisons to analytical systems are used to verify the correctness of the created code. Finally, the equilibrium density distribution of the obtained \gls{ILM} optical lattices are presented and analysed.



\chapter{From Bose-Einstein condensates to Superfluid Ring Qubits}
\label{ch: Ultracold Atoms In Optical Lattices}

The notion of \emph{Bose-Einstein condensation} takes us almost 100 years back tp 1925, where, on the basis of a publication by the Indian physicist S. N. Bose, A. Einstein suggested the occurrence of a phase transition in a gas of non-interacting atoms at low temperature \cite{Stone2013}. The phenomenon is characterized by a description of coherent macroscopic matter due to collective occupation of the lowest quantum state. This depiction led to the suspicion of a connection between to observation of \emph{superfluidity} in helium ${}^4$He, also formulated by a coherent macroscopic wavefunction by \citet{Ginzburg1950}. Nevertheless, condensed helium exhibits strong interaction between the atoms which is not described by the original theory of Bose-Einstein condensation.

In 1995, the realization of ultracold atomic gases due to lased-based techniques to cool the atoms resulted in the observation of a pure Bose-Einstein condensate \cite{Bradley1995, Davis1995, Anderson1995}, by virtue of --- in contrast to liquid helium ${}^4$He, --- the weak interactions in the dilute gases. The many-body aspect of a Bose-Einstein condensate can be captured with an effective single particle description given by a nonlinear Schr\"odinger equation, i.e., the Gross-Pitaevski equation. A weakly interacting Bose gas, such as ultracold atoms, is covered by Bogoliubov theory which added small fluctuation to this zero-order description. The result being that the many-body physics solved entirely from the perspective of a collection of non-interacting quasiparticles.

As presented in the introduction, the 21st century gave us the possibility to experimentally diverge from the non-interacting quasiparticles to a strongly interacting system. For example, established by \citet{Inouye1998,Courteille1998}, \citet{Cornish2000} used \emph{Feshbach resonance} to change the interaction strength between the atoms thereby reaching the strong coupling regime. To be precise, the technique allows for the scattering length $a$ to be extended above the average inter-particle spacing. Unfortunately, This approach is best suited for fermions, as for bosons the lifetime is limited due to three-body effects \cite{Bloch2008}. Another way to accomplish a strongly coupled regime would be to restrict the dynamics via optical lattices. These effective potential landscape generated from interfering electromagnetic laser beams can be used to construct neighbouring traps. In a sufficient high lattice depth, tunnelling energy plays the role of the kinetic energy. Further, increasing the lattice depth, manifests a strongly correlated regime, as observed by \citet{greiner2002}. This phase transition by tuning the potential depth, i.e., laser beam intensity, is called the \emph{superfluid to Mott insulator transitions}. In this work, our focus will lie on the latter system (optical lattices) and parameter (potential depth).


Here, a primer on ultracold atoms in optical lattices is given. First, we discuss the scattering of ultracold atoms in section \ref{sec: scattering problem}. As we will focus on two-dimensional optical lattices the transition of scattering to such regime is made. The obtained knowledge in this section will be employed to implement interaction in our simulation in chapter \ref{ch: PIMC}. Afterwards, in \ref{sec: ucol}, we discuss the creation of optical lattices and how they interact with the ultracold atomic gasses. At last, we explain how Superfluid Ring Qubits (\gls{SRQ}s) are engineered with ultracold atoms in optical lattices.

\section{Scattering of ultracold atoms}\label{sec: scattering problem}
\subsection{Two-body weak scattering problem}\label{sec: two-body scattering problem}
In general, when describing a dilute Bose gas with weak interactions, one can make the so called \emph{pair-production approximation} where interactions involving three or more bosons are neglected \cite{Bloch2008}. The general quantum mechanical problem of two particles of mass $m$ colliding is covered by the theory of elastic scattering.
\begin{figure}[hbtp]
    \centering
    \includegraphics[width=0.6\textwidth]{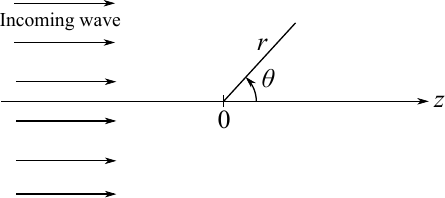}
    \caption{Sketch of an incoming harmonic wave along the $z$-direction which scatters into a spherical wave whose amplitude depends on the angle $\theta$ with the $z$-axis. The figure is based on the drawings from \citet{piatecki2005}.}
    \label{fig: scatteringsetup}
\end{figure}

Considering that the interaction potential $U(r)$ is spherical symmetric and depends only on the relative distance between the two particles, the motion of the centre-of-mass may be separated from the relative motion of the two particles. The former is just a free particle of mass $2m$, whereas the latter corresponds to a particle of mass $m^*=m/2$ subjected to the potential $U(r)$ --- described by the Schr\"odinger equation
\begin{align}
\left(-\lambda^*\nabla^{2}+U(r)\right) \psi(\mathbf{r})=E \psi(\mathbf{r}),
\end{align}
where $E=\lambda^*k^2$ with $k$ the wave vector and $\lambda^*=\frac{\hbar^{2}}{2m^*}$ for convenience. We consider only the \emph{scattered states}, i.e., with energy $E>0$ because $\lim_{r\to\infty}U(r)=0$. From the Lippmann-Schwinger equation one can derive that, at long distances, the basic shape of the relative wave-function is the superposition of an entering plane wave $e^{i k r}$ and an outgoing spherical wave $e^{i k r}/r$ whose amplitude depends on the angle $\theta$ (see figure \ref{fig: scatteringsetup}), i.e, \cite{sakurai2017,Dalibard2021}
\begin{align}\label{eq: Lippmann-Schwinger equation}
\psi_k(r, \theta) \sim e^{i k z}+f(k, \theta) \frac{e^{i k r}}{r}
\end{align}
where $f(k, \theta)$ is called the \emph{scattering amplitude}. Naturally, it still depends on the scattering potential\footnote{It is related to the expectation value of the interaction potential, see \citet{Dalibard2021}}. To find an expression for the scattering amplitude $f(k, \theta)$ one abuses the spherical symmetry to decompose the wavefunction in the angular momentum basis
\begin{align}\label{eq: eigenvalue decomposition}
\psi_k(r, \theta)=\sum_{l=0}^{\infty} P_{l}(\cos \theta) R_{k l}(r),
\end{align}
with $P_{l}$ the Legendre polynomial of order $l$ and $l\in\mathbb{Z}$ the quantum number. The resulting radial Schr\"odinger equation can than be solved for $r\to\infty$ to give the traditional result $R_{k l}(r)\propto\sin \left(k r-l \pi / 2+\delta_{l}(k)\right)/r$ with a phase shift $\delta_{l}(k)$. The phase shifts restores the interaction in the region where the potential $U(r)$ cannot be neglected \cite{Dalibard2021}. It is found by solving the radial equation and is given by $\delta_{l}(k)=0$, if no interaction is present. Substituting this result in eigenvalue decomposition \eqref{eq: eigenvalue decomposition} and also expanding the plane waves into the angular momentum eigenbasis in equation \eqref{eq: Lippmann-Schwinger equation}, we can compare both depictions. As a result we can find an expression for the scattering amplitude in terms of phase shifts $\delta_{l}(k)$ given by
\begin{align}\label{eq: scattering amplitude}
f(k,\theta)=\frac{1}{2ik}\sum_{l=0}^{\infty}(2 l+1)\left(e^{2 i \delta_{l}(k)}-1\right)  P_{l}(\cos \theta)=\sum_{l=0}^{\infty}(2 l+1)f_l(k)P_{l}(\cos \theta),
\end{align}
where \emph{partial scattering amplitudes} can be defined as
\begin{align}\label{eq: partial scattering amplitude}
     f_l(k)\equiv\frac{e^{2 i \delta_{l}}-1}{2ik} = \frac{1}{k \cot{\delta_0(k)}-ik}.
\end{align}
Notice that the partial scattering amplitudes $f_l(k)$ of angular momenta $l$ is zero if and only if the interactions are negligible at the angular momenta $l$, i.e., $\delta_{l}(k)=0$.

\subsection{The s-wave approximation}\label{sec: s-wave approximation}
In the previous section, we treated the formalism of two particles weakly scattering by taking advantage of the rotation invariance of the interaction potential to simplify the problem. The result being that the interaction depends on a phase shift $\delta_{l}(k)$ for each angular momenta $l$. Now, we consider the case of a low-energy scattering, a good description for gases in the sub-milikelvin regime. By analysing the radial Schr\"odinger equation, one can show that the following scaling law holds at low energies \cite{landau2013quantum, Dalibard2021}
\begin{align}\label{eq: phase shift scaling law}
    \tan \left[\delta_{l}(k)\right] \propto k^{2 l+1},
\end{align}
It shows that scattering primarily contributes at low angular momentum, beginning with the s-wave, $l = 0$. Hence, the \emph{s-wave approximation} consist of making the assumption that $\delta_{l}(k)=0$ for $l \geq 1$. This amounts to taking the scattering amplitude in equation \eqref{eq: scattering amplitude} to be $f(k, \theta)=f_{0}(k)$. The Pauli exclusion principle infers that this approximation does not hold for fermions, where the lowest angular momenta channel is the p-wave. Hence, from here on, we restrict our analyses to bosons.

From equation \eqref{eq: phase shift scaling law}, it is clear that $\tan \left[\delta_{0}(k)\right] \propto k$ for $k\to0$. Hence, we one can define a \emph{scattering length}\footnote{Note that the proportionality coefficient has the dimension of length.} $ a_{3 \mathrm{D}}$, so that for very low temperatures the \emph{low energy s-wave scattering amplitude} becomes\footnote{One could also add correction term. However, it turns out these are negligible \cite{Dalibard2021}.}
\begin{align}\label{eq: low energy s-wave scattering amplitude}
f(k) =\frac{-1}{1/a_{3 \mathrm{D}}+i k },
\qq{with} a_{3 \mathrm{D}}\equiv-\lim _{k \rightarrow 0} \frac{\tan \left[\delta_{0}(k)\right]}{k}.
\end{align}
Hence, for s-wave scattering the interaction is completely determined by one parameter, i.e., the s-wave scattering length, which has to be quantified in experiment.

For very low temperatures the scattering amplitude is $f(k) \approx -a_{3 \mathrm{D}}$. As a result, the s-wave scattering length may be thought of as the size at which the incoming plane wave in figure \ref{fig: scatteringsetup} is affected, i.e., the effective distance on which particles interact. With this in mind, we can deduce a criterion for the weakly-interacting regime in three-dimensions $na_{3 \mathrm{D}}^3 \ll 1$, which asserts that the weakly-interacting regime is attained at low density. $na_{3 \mathrm{D}}^3$ is often called the \emph{diluteness parameter}.

Because the s-wave scattering length is the sole significant parameter, it is permissible to substitute the real potential with a model potential that yields the same results as above. One can show that indeed the contact potential
\begin{align}
\hat{U}[\psi(\boldsymbol{r})]=\left. g \ \delta(\vb{r}) \frac{\partial}{\partial r}[r \psi(\boldsymbol{r})]\right|_{\mathrm{r}=0} \qq{with} g = 4\pi \lambda^* a_{3 \mathrm{D}}
\end{align}
leads the same scattering amplitude as equation \eqref{eq: low energy s-wave scattering amplitude}. For the derivation the reader is referred to the lecture notes of \citet{Dalibard2021}. $g$ is called the coupling constant of the contact interaction.

\subsection{Scattering for a two-dimensional Bose gas}\label{sec: quasi two-dimensional Bose gas}
The following approach in describing two-dimensional Bose interaction is based on the supplementary material of \citet{Gauter2021}. It describes the two-dimensional interactions based on confining a three-dimensional gas harmonically, which is more realistic compared to treating the system purely two-dimensional.

As concluded in section \ref{sec: s-wave approximation}, the interaction of an ultracold atomic gas is solely characterized by the three-dimensional scattering length $a_{3D}$. However, as stated earlier we want to load the atoms in two-dimensional optical landscape. Hence, we want to detain or freeze one of the spatial degrees of freedom. Generally, one traps the bosons to a quasi two-dimensional geometry, by applying a tight harmonic trap with an angular frequency $\omega_{\perp}$ in the direction perpendicular to a desired two-dimensional plane. This problem of two bosons scattering in this two-dimensional plane was addressed by Petrov et al. \cite{Petrov2000, Petrov2001}. They demonstrated that in such a system, the two-dimensional scattering length $a_{2 \mathrm{D}}$ can be related to the trap's characteristic length $l_{\perp} = \sqrt{\hbar/m\omega_{\perp}}$ and the three-dimensional scattering length $a_{3D}$, given by\footnote{The numerical constant comes from taking $B=0.915$ in equation (21) of \cite{Petrov2001} such that $2.092 \approx \sqrt{\pi/B}/q$ as done in \cite{Pricoupenko2007}.}
\begin{align}\label{eq: a_2D}
a_{2 \mathrm{D}} \simeq 2.092 \ l_{\perp} \exp \left(-\sqrt{\frac{\pi}{2}} \frac{l_{\perp}}{a_{3 \mathrm{D}}}\right).
\end{align}
The equation holds assuming that both the chemical potential $\mu$ and thermal energy $k_\mathrm{B}T$ is much smaller than the trap harmonic energy $\hbar\omega_{\perp}$, i.e,  $\mu\ll\hbar\omega_{\perp}$ and $k_\mathrm{B}T\ll\hbar\omega_{\perp}$. In a homogeneous atomic Bose gas the energy per particle can be written as $E / N=\tilde{g} \times\left(\hbar^2n/2m\right)=\tilde{g} \times\left(\lambda n \right)$ where $n$ is the 2D density, $\lambda=\hbar^2/2m$ and $\tilde{g}$ a dimensionless mean field coupling constant. In perturbation theory, the coupling constant is equal to the vertex of inter-particle interaction in zero order and hence, at low energies, corresponds with the amplitude of scattering $f(k)$ for energy $\hbar k/2m$. Therefore, noticing that the motion of two bosons in a harmonic tight confinement can still be divided into a relative and centre-of-mass portion, such as is section \ref{sec: two-body scattering problem}, \citet{Petrov2000} computes that
\begin{align}\label{eq: coupling constant for a_3D}
\tilde{g}=f(k)=\int\dd{\vb{r}} \psi(\mathbf{r})_k U(r) \psi^{*}(\mathbf{r})_k
\simeq \frac{2 \sqrt{2 \pi}}{l_{\perp} / a_{3 \mathrm{D}}+1 / \sqrt{2 \pi} \ln \left(\lambda / \pi \mu l_{\perp}^{2}\right)},
\end{align}
where $U(r)$ is the interaction potential, $\mathbf{r}$ the relative position vector between the two bosons, and $\mu$ the chemical potential of the Bose gas. Besides the spherical symmetry, the only assumption made for the interaction potential $U(r)$ is their exist a certain distance $R$ for which $U(R)=0$ if $r>R$.

It would be convenient to rewrite the mean field coupling constant in terms of the 2D scattering length $a_{2 \mathrm{D}}$. Hence, using \eqref{eq: a_2D} to write $l_{\perp} / a_{3 \mathrm{D}}=\sqrt{2 / \pi} \ln \left(2.092 \ l_{\perp} / a_{2 \mathrm{D}}\right)$ and substituting this in \eqref{eq: coupling constant for a_3D} gives
\begin{align}\label{eq: intermediate coupling constant}
\tilde{g} 
&=\frac{4\pi}{2\ln \left(2.092 \ l_{\perp} / a_{2 \mathrm{D}}\right)+ \ln \left(\lambda / \pi\mu \, l_{\perp}^{2}\right)},
\end{align}
where the quasi-momentum was written in terms of the chemical potential $\mu$. If one now defines a quantity $a \equiv 2.092 \ l_{\perp}$ we can write
\begin{align}\label{eq: coupling constant for a_2D}
\tilde{g} \simeq \frac{4 \pi}{2 \ln \left(a / a_{2 \mathrm{D}}\right)+\ln \left(\Lambda E_{\mathrm{r}} / \mu\right)},
\end{align}
where we defined the recoil energy to be $E_{\mathrm{r}}=\pi^{2} \hbar^{2} / 2 m a^{2}$ and the numerical constant $\Lambda=(2.092^2/\pi^3)$. Above representation yields a suitable form for our purposes as all the degrees of freedom in the transverse direction have been removed. The length $a$ can be seen as an arbitrary length scale which cancels from equation \eqref{eq: coupling constant for a_2D} if we use the logarithmic identities. It is left such that it can be used as the units of length of the system $a$. Notice, that the interaction in two dimensions does not only depend on the scattering length $a_{2\mathrm{D}}$ any more, but also on the chemical potential $\mu$.  Moreover, equation \eqref{eq: intermediate coupling constant} provides a coupling constant with a separated contribution from the chemical potential $\mu$ and the 2D scattering length $a_{2\mathrm{D}}$, i.e. the relevant interaction strength 
\begin{align} \label{eq: interaction strength}
    \tilde{g}_{0}\equiv\frac{2 \pi}{\ln \left(a / a_{2 \mathrm{D}}\right)}.
\end{align}
The latter interaction strength will be the one employed in our work, e.g., see section \ref{sec: interaction implementation} and appendix \ref{app: rel int prop}.

\section{Ultracold atom in optical lattices}\label{sec: ucol}
Now that we understand how the atoms in the ultracold gas interact with each other, we have to understand how the lights beams or optical lattice interacts with the gas. First, we give a short review how the light manifest itself as an effective potential on the atoms. It will set constraints on the frequency and intensity of the lasers which determine major effects, e.g., attractive or repulsive forces, of the atoms in the lattice. Afterwards, we show how an optical lattice is constructed. A more in-depth reference is found in \citet{Grimm2000}.

\subsection{Atom-light interactions}
Take a neutral atom positioned in a monochromatic light field of frequency $\omega$, wavelength $\lambda$ or wave vector $2\pi/\lambda$. The electric field is given by
\begin{align}
\vb{E}=\frac{\vb{\mathcal{E}}(\vb{r})}{2} e^{-i \omega t+i \varphi(\vb{r})}+\text { c.c. },
\end{align}
with $\varphi$ the phase, $\mathcal{E}$ the real amplitude and c.c. meaning the complex conjugate of the first term. The electric field induced an non-zero electric dipole on the neutral atoms yielding the vector $\vb{D}=\tilde{\alpha}\left(\omega\right) \vb{E}$. The electric polarisability of the atom $\bar{\alpha}(\omega)=\alpha(\omega)+i \alpha^{\prime}(\omega)$ has a real and imaginary part. The former contributes to the time-averaged dipole potential, given by\footnote{The bar represents the time-averaging over an optical period and the factor $\frac{1}{2}$ comes the induced Stark shift \cite{Gerbier2018}.}
\begin{align}
V_{\mathrm{dip}}=-\frac{1}{2} \overline{\vb{D} \cdot \vb{E}}=-\frac{1}{4} \alpha\left(\omega \right)|\vb{\mathcal{E}}(\vb{r})|^{2}.
\end{align}
The time averaging is valid when the timescale of the motion of the atom is neglectiable compared to the inverse light beam frequency. The imaginary part of the polarisability contributes to the energy radiated by the dipole
\begin{align}
E_\mathrm{rad}=-\frac{\omega\alpha^{\prime}(\omega)}{2}\abs{\vb{\mathcal{E}}(\vb{r})}^{2}=-\hbar \omega \Gamma_{\mathrm{sp}},
\end{align}
where we made the connection to photons by equating the expression to number of spontaneously emitted photons per unit time $\Gamma_{\mathrm{sp}}$.

To compute the dipole potential one needs to find an expression for the electric polarisability of the atom. To explain the essence, we shall restrict our selves to the simple case of a two-level atom with a ground $\ket{g}$ and excited state $\ket{e}$ at frequency $\omega_0$ close the laser frequency $\omega$. The optical Maxwell-Bloch equations describe the interaction between the atom and the electromagnetic mode. Solving these one finds that \cite{tannoudji1992atom, Gerbier2018}
\begin{align}
\tilde{\alpha}\left(\omega \right)=\frac{d^{2} / \hbar}{-\Delta+i \Gamma / 2}
\approx
\frac{-d^{2} / \hbar}{1 / \Delta+i \Gamma / 2 \Delta^{2}},
\end{align}
where $\Delta = \omega-\omega_0$ is the detuning of the laser from the atomic resonance and $\Gamma$ inverse emission time of the transition. The electric dipole matrix element $d=|\bra{e}{\hat{d}}\ket{g}|$ with the $\hat{d}$ the dipole operator indicates coupling strength of the transition. Taking the limit to far-off resonance, i.e. a large detuning $\abs{\Delta}\gg\Gamma$, we find that the dipole potential and spontaneous emission rate become
\begin{align}
V_{\mathrm{dip}}
=\frac{d^{2}}{4 \hbar}\frac{|\vb{\mathcal{E}}|^{2}}{ \Delta}
=\frac{\hbar }{4}\frac{\Omega^{2}}{\Delta}
\qq{and} 
\Gamma_{\mathrm{sp}}
=\frac{\Gamma d^{2}}{8\hbar}\frac{|\vb{\mathcal{E}}|^{2}}{\Delta^2}
=\frac{\hbar\Gamma}{8}\frac{\Omega^{2}}{\Delta^2}
=\frac{\Gamma}{2\left|\Delta\right|}\left|V_{\mathrm{dip}}\right|,
\end{align}
where we introduced the Rabi frequency $\Omega=d|\mathcal{E}|/\hbar$. We make two observations from this result. First, the atoms are attracted  or repulsed from the nodes or anti-nodes of laser intensity if the laser is red-detuned ($\omega <\omega_0$) or blue-detuned ($\omega_0 <\omega$)
respectively. Second, we have different scaling of the detuning for the dipole potential $V_{\mathrm{dip}}\propto 1/\Delta$ and spontaneous emission rate $\Gamma_{\mathrm{sp}}\propto 1/\Delta^2$. For a given dipole potential $V_{\mathrm{dip}}$ one can adjust the absolute value of the detuning to be arbitrary large on condition that one also increasing the laser intensity $\abs{\mathcal{E}}$. Hence, the dipole force, is seen as conservative in practice.
\begin{table}[htpb]
    \caption{Typical parameters associated with optical lattices for ${ }^{87} \mathrm{Rb}$ atoms. Taken from \citet{Gerbier2018}.}
    \label{tab: parameters OL}
    \centering
\begin{tabular}{llllll}
    \toprule
 $\boldsymbol{\lambda}$ & \textbf{Power} & \textbf{Beam waist} & $\boldsymbol{\Omega /2 \pi}$ & \textbf{potential depth} ($\boldsymbol{V_{0} / k_{B}}$) & $\boldsymbol{\Gamma_{\text{sp }}}$ \\
 \midrule[\heavyrulewidth] $850  \mathrm{~nm}$ & $100 \mathrm{~mW}$ & $100~\mu \mathrm{m}$ & $1 \mathrm{~GHz}$ & $500 \mathrm{~nK}$ & $0.01 \mathrm{~s}^{-1}$ \\
\bottomrule
\end{tabular}
\end{table}

We conclude that choice of the atomic gas fixes the freqeuncy range of the lasers we can use. Furthermore, the power or intensity of laser beams must be high enough such that the force is conservative. The typical scales of the parameters for a optical lattice experiment with ${ }^{87} \mathrm{Rb}$ atoms can be observed in table \ref{tab: parameters OL}.

\subsection{Construction of the optical potential}
In general, optical lattices are constructed by orienting light wavefronts, e.g, laser beams, at relative angles to each other. The lasers emit a beam of radiation with a single frequency, i.e., monochromatic light, which has the property of being coherent. In other words, there is a stationary relation between the phase of the individual photons and their frequency. The interference between the laser result in a stationary interference pattern.

\begin{figure}[htbp]
    \centering
    \includegraphics{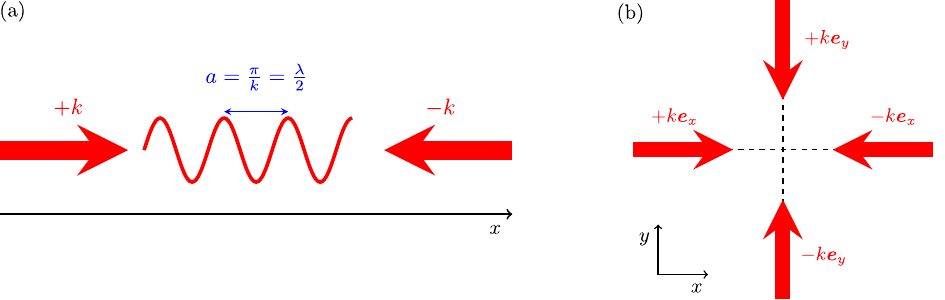}
    \caption{Illustrations of the laser configuration used to generate a one or two-dimensional lattice. \figa~Counter-propagating lasers with the same wavelength $\lambda$ generate a one-dimensional standing wave with a spatial period half the wavelength, i.e., $a=\lambda/2$. \figb~Duplicating the setup and placing it perpendicular on the x-axis gives two-dimensional optical stationary interference pattern.}
    \label{fig: OL_lasers}
\end{figure}

So in principle, we can take an arbitrary number of light fields $\vb{E}_n$ with real amplitudes $\mathcal{E}_{n}$, wave vector $\vb{k}_n$, and phases $\phi_{n}$ interfering with each other. Let us first discuss a one dimensional example seen in figure \ref{fig: OL_lasers}\figa. Choosing two counter-propagating laser beams with wave vectors $\mathbf{k}_{1}=k  \mathbf{e}_{x}$ and $\mathbf{k}_{2}=-\mathbf{k}_{1}$, parallel polarizations and equal amplitude we get that intensity yields $|\mathcal{E}|^{2}=E_{0}^{2} \cos^{2} \left(k_{L} x+\phi_{21} / 2\right)$ with $E_{0}$ the amplitude of the interfering beams. This leads to a effective potential of the form \cite{Gerbier2018}
\begin{align}
V_{1 \mathrm{D}}=\operatorname{sgn}(\Delta) V_{0} \cos^2 \left(k  x+\phi_{21}/2\right)
\end{align}
where $\phi_{21} = \phi_2-\phi_1$ is the relative phase between the two laser beams and the potential depth $V_{0}\propto d^{2} {E}_{0}^{2} /\left|\delta_{L}\right|$. The resulting period or lattice spacing $a$ of the potential is $\lambda/2$. Moreover, as shown in the previous subsection, the resulting potential will be attractive or repulsive depending on the sign of the detuning $\Delta$. Last, notice that the minima and maxima of the interference pattern or lattice point move when the relative phase $\phi_{21}$ changes.

For two dimensions the result does not change much. For example, consider the setup with four beams with the same frequency, two counter-propagating along the $x$-axis and two along $y$-axis as drawn in figure \ref{fig: OL_lasers}\figb. The two beams on the two different orthogonal direction have polarizations $\boldsymbol{\epsilon}_{x}$ and $\boldsymbol{\epsilon}_{y}$. The intensity of the interference pattern can be found to give \cite{Gerbier2018}
\begin{align}
|\vb{\mathcal{E}}(\vb{r})|^{2}=\left|E_{0} \cos \left(k  x+\phi_{x}\right) \boldsymbol{\epsilon}_{x}+E_{0} \cos \left(k  y+\phi_{y}\right) e^{i \phi_{x y}} \boldsymbol{\epsilon}_{y}\right|^{2}
\end{align}
where $\phi_{x}$ and $\phi_{y}$ are the relative phase between the beams on the $x$ and $y$-axis, respectively. The main difference lies in relative phase $\phi_{xy}$ between $x$ and $y$. The phases  $\phi_{x}$ and $\phi_{y}$ only move the interference pattern in the $xy$-plane, whereas the phase $\phi_{xy}$ can change the geometry of the optical lattice. The problem is resolved by taking orthogonal polarization vectors ($\boldsymbol{\epsilon}_{x}\perp\boldsymbol{\epsilon}_{y}$) and using slightly different wavelengths. 
The result being that any multi-dimensional factor oscillates. Nevertheless, this happens much faster than the time scales of atomic dynamics such only the time-averaged potential has effect on the atoms \cite{Gerbier2018}. Again the effective potential will have the form
\begin{align}
V_{2 \mathrm{D}}=\operatorname{sgn}(\Delta) V_{0}\left( \cos ^{2}\left(k x+\phi_{x} / 2\right) + \cos ^{2}\left(k y+\phi_{y} / 2\right) \right).
\end{align}
The same analyses can be peformed for three dimensional optical lattices.

As summarised by \citet{Betzig2005}, \citet{Petsas1994} made some key observations regarding the optical lattice created by the electromagnetic plane waves. First, the lattice spatial features, such as symmetry, primitive cell form, and periodicity, are predominantly determined by the plane wave wave vectors $\vb{k}_n$. Second, a $d$-dimensional lattice ($d=2$ or $3$) requires a minimum of $d+1$ wave vectors. Indeed, two wave vectors $\vb{k}_{0}, \vb{k}_{1}$ define a one-dimensional lattice with intensity $\vb{e}(\vb{x}, t)=e(x, t)$, where $\hat{\vb{e}}_{x} \|\left(\vb{k}_{1}-\vb{k}_{0}\right)$, and three wavevectors $\vb{k}_{0}, \vb{k}_{1}, \vb{k}_{2}$ define a two-dimensional lattice $\vb{e}(\vb{x}, t)=\vb{e}(x, y, t)$, where $\hat{\vb{e}}_{z} \|\left(\vb{k}_{1} \times \vb{k}_{2}+\vb{k}_{2} \times \vb{k}_{3}+\vb{k}_{3}\times \vb{k}_{1}\right)$.




\section{Superfluid Ring Qubits} \label{sec: Superfluid Ring Qubits}
Here, we discuss the realization of a atomic SQUID or superfluid ring qubit. The theoretical description is explained by comparing to the Flux qubit. The experimental realization will only be briefly covered. For a more in-depth coverage, the reader is referred to \citet{amico2021, Amico2014, aghamalyan2015} with their corresponding supplementary materials.

\subsection{The analogy with the Flux qubit}
A qubit device is a two-state quantum system that may be controlled, connected to its peers, and measured in a coherent manner. Several such physical systems have been explored in the past couple of decades, each with its own set of benefits and drawbacks at different aspects. Flux qubits or persistent current qubits are superconducting metal rings that are separated by Josephson junctions, i.e., insulator components or weak links between two superconducting currents. In practice, one implemented multiple Josephson junction, but we shall concentrate on one to portray the idea.
\begin{figure}[htb]
    \centering
    \includegraphics[width=\textwidth]{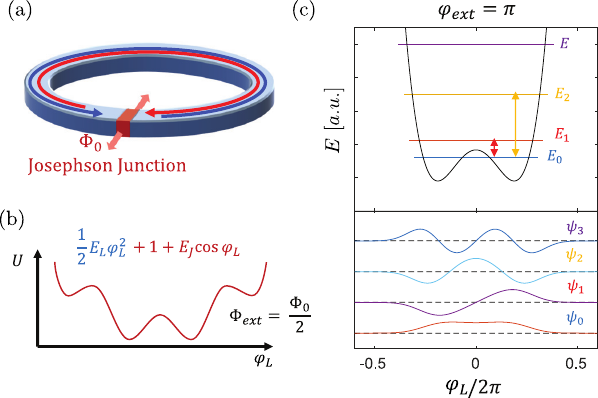}
    \caption{A flux qubit. \figa~Illustration of a flux with a superconding ring and one josephson junction. \figb~Potential energy of the flux qubit with the external flux $\Phi_{\text {ext }}=\Phi_{0} / 2$. \figc~Energy eigenvalues and eigenvectors of the Hamiltonian of a flux qubit at $\Phi_{\text {ext }}=\Phi_{0} / 2$. All three figures are taken under the a license with AIP Publishing from \citet{Dmitriev2021}.}
    \label{fig: Flux_qubit}
\end{figure}

Figure \ref{fig: Flux_qubit} alludes the basic principles. When an external magnetic flux $\Phi_{\mathrm{ext}}$ is supplied, the flux qubit characteristics are designed so that a persistent current flows through the superconducter (inductance $L$) and the Josephson junction ($J$) continuously. The external flux wants to be cancelled, hence, the following relation is satisfied $\varphi_{L} + \varphi_{J} = - \varphi_{\mathrm{ext}}$ \cite{Dmitriev2021}. The external phase can be written as $\varphi_{\mathrm{ext}}=2 \pi \Phi_{\mathrm{ext}} / \Phi_{0}$ where $\Phi_{0}=2e/h$ is called the flux quantum and $e$ is the electron charge. Some insight can be gained by conidering the energy associated with the superconductor and Josephson junction which are given by $U_{L}=\frac{1}{2} E_{L} \varphi^{2}$ and $U_{J}=E_{J}\left[1-\cos \left(\varphi_{L}+\varphi_{\mathrm{ext}}\right)\right]$, respectively. Here is $E_J$ the Josephson energy and $E_L$ the magnetic energy. The resulting potential energy for the system is than written as
\begin{align}
U(\varphi_{L})=U_{L}+U_{J}=\frac{1}{2} E_{L} \varphi_{L}^{2}+E_{J}\left[1-\cos \left(\varphi_L+\varphi_{\mathrm{ext}}\right)\right].
\end{align}
Applying a magnetic field of half flux quanta, i.e. $\Phi_{\text {ext }}=\Phi_{0} / 2$, we find that the potential energy has the form of a double-well, shown in figure \ref{fig: Flux_qubit}\figb. It displays two-fold degeneracy on the condition that Josephson energy $E_J$ is large enough compared to $E_L$. Hence, the situation where $\Phi_{\text {ext }}=\Phi_{0} / 2$ is called the \emph{degenaracy point}. The two wells corresponds to the clockwise and anti-clockwise supercurrent flow as the phase $\varphi_{L}$ is smaller or larger than $\Phi_{0} / 2$. 

The complete system can be described by particle confined in the potential $U$ with Cooper-pair charging energy $E_{C}=(2 e)^{2} / 2 C$ where $C$ the Josephson junction capacitance \cite{Dmitriev2021}. The resulting Hamiltonian yields $H=U(\varphi)-E_{C} \partial^{2} / \partial \varphi^{2}$. Taking $\Phi_{\text {ext }}=\Phi_{0} / 2$, the lowest energy eigenvalues are plotted with there corresponding eigenstates in figure \ref{fig: Flux_qubit}\figc. The two lowest energy eigenstates are a quantum symmetric or anti-symmetric superposition of the two states of the currents or ring flux. The sole difference between the two lowest energy eigenstates is the relative quantum phase of the composed current-direction states. Because the potential is anharmonic the higher energy eigenstates are energetically separated from the lowest two. This ``qubit non-linearity'' criterion, allows for a two-level quantum system. Pulsing the qubit with an energy comparable to the gap between the energy of the two basis states can than be used to conduct computational operations. The complete bloch sphere is than traversed by also changing the external flux $\Phi_{\text {ext }}$.

The advantages of using flux qubits in quantum computations is their fast gate operations and the ability to build them with the techniques of the current silicon industry \cite{Dmitriev2021}. Nevertheless, they suffer from low coherence times due to the fluctuations in the applied field \cite{amico2021}. These interact with the charged carrier of the supercurrent, i.e. electrons. A solutions to this issue would be to use neutral carriers, e.g. bosonic atoms.

\subsection{A two-level system in optical lattices}
\begin{figure}[htbp]
    \centering
    \includegraphics[width=\textwidth]{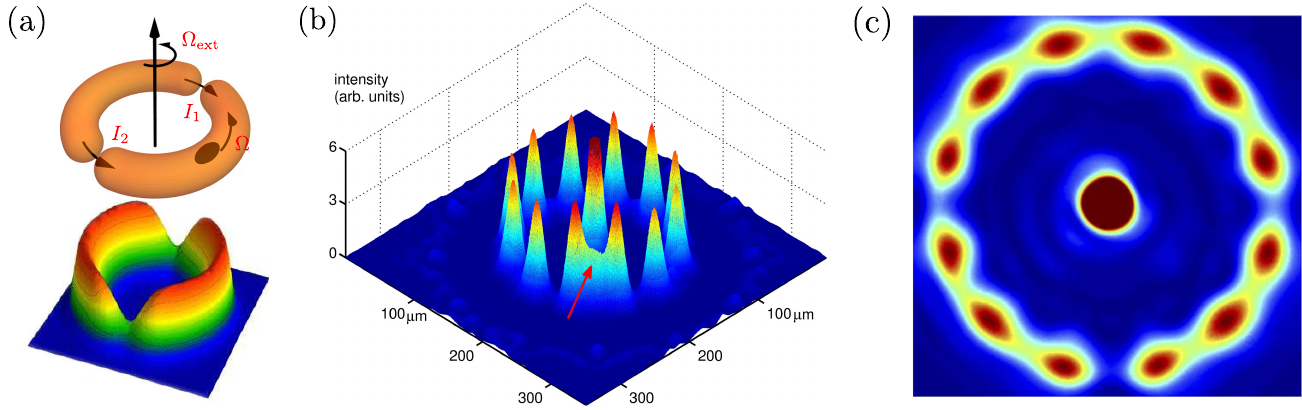}
    \caption{Three different realisation of a qubit using neutral cold atoms in optical lattices with one \figb, two \figa~or three \figc~insulator components. All three figures are taken and amended under the Creative Commons Attribution 4.0 International License from \citet{ryu_quantum_2020}, \citet{amicoSuperfluidQubitSystems2015}, and \citet{aghamalyanAtomtronicFluxQubit2016} respectively.}
    \label{fig: Qubits}
\end{figure}

Superfluid Ring qubits are designed to combine the advantages of superconducting flux qubit with that of ultracold atoms in optical lattices. The idea is much the same. We want a closed circuit where a persistent current of ultracold atoms can rotate in clockwise and anti-clockwise direction with respect to the two-dimensional plane. In optical lattices this is facilitated by a circular (ring-like) interfere pattern, seen in figure \ref{fig: Qubits} where the intensity maxima are made attractive with respect to the ultracold atom species. As with the flux qubit a symmetry of the ring or the flow of the atoms must be broken. One breaks the symmetry by applying a weak link between two wells such the tunneling rate is different between them. This can be done by implementing a gap in ring as in \ref{fig: Qubits}\figa~and \ref{fig: Qubits}\figc. Conversely, one can create a lattice with a hopping sites in a circular pattern such as in figure \ref{fig: Qubits}\figb. As with flux qubits, the number of gaps or weak links may vary. Lastly, the ring must be pierced by an magnetic flux\footnote{This can be achieved in many ways, e.g., a synthetic gauge flux by applying a phase directly to the atoms via suitably designed laser fields \cite{Goldman2016, Dalibard2011} or a Coriolis flux by revolving the lattice around the origin of ring at a constant velocity \cite{Fetter2009}.}.

\begin{figure}[htbp]
    \centering
    \includegraphics[width=\textwidth]{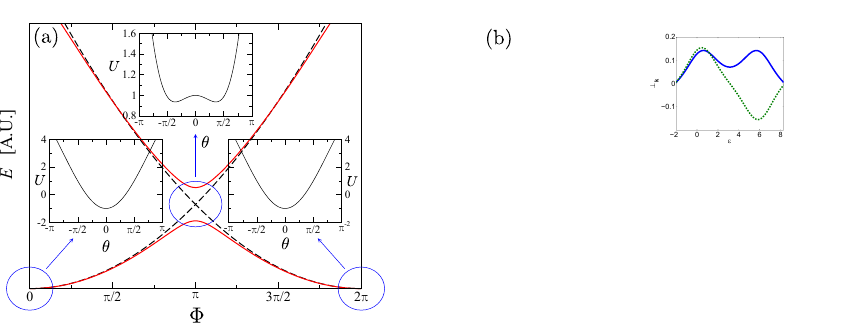}
    \caption{Potential and energy eigenvalues for the superfluid ring qubit seen in figure \ref{fig: Qubits}\figb. \figa~The first two eigenvalues depending on the external flux $\Phi$. The black dashed line indicates the energy eigenvalues in the absence of the barrier. For $Phi=0,\pi,2\pi$ the potentials in function of the phase $\theta$ is given in the insets. \figb~The potentials in function of the phase $\theta$ for $\Phi=\pi$ with the two eigenstates of the corresponding two-lowest energy eigenvalues shown in the inset. Both figures are taken and amended under the Creative Commons Attribution 3.0 International License from \citet{Aghamalyan_2015}.}
    \label{fig: Two_level_system}
\end{figure}

For our analyses, we restrict ourselves to an implementation based on figure \ref{fig: Qubits}\figb. In \citet{Amico2014}, the effective Hamiltonian of the system can be obtained by looking at the one-dimensional periodic Bose-Hubbard model. Here the Bosons dynamics are described by the hopping between lattice sites where the hopping amplitude at weak link differs from the rest. The system dynamics is solely described by the phases of the superfluid at each site $\phi_i$, if the density of the superfluid is large enough to ignore the number fluctuations on each site. Accordingly, each lattice site has a Josephson coupling $J_i$ associated with it. The resulting effective Hamiltonian can be computed to be\footnote{The original notation from \citet{Amico2014} and \cite{Aghamalyan_2015} has been respected.} \cite{aghamalyan2015}
\begin{align} \label{eq: ham SRQ}
     H=\frac{1}{2 \mu}\pdv[2]{\theta}+\frac{J}{N}\theta^{2}-J^{\prime}\cos( \theta-\Phi)
\end{align}
where $\theta=\phi_{N}-\phi_1$ is the the phase difference associated with the slip or weak link, $\mu$ the effective mass of the system, $J$ a parameter associated with Josephson energy, $J^\prime$ the Josephson coupling of the weak link \cite{Aghamalyan_2015} and $N$ the number of sites. Comparing equation \eqref{eq: ham SRQ} to that of the flux qubit, we can conclude that the superfluid ring qubits adhere similar dynamics as the flux qubit. Indeed, how the first two energy levels and the potential depend on the external flux can be seen in figure \ref{fig: Two_level_system}\figa. The ground-state energy in the absence of the weak link is represented by black dashed lines. Adding a barrier with $NJ^\prime/J$ larg eneough, makes the potential anharmonic, such that energy levels have unequal spacing and a gap opens. Again a degeneracy point at $\Phi=\pi$ is observed. Hence, we can define a two levels system. Moreover, at degeneracy point the two first energy levels are represented by a symmetric and anti-symmetric superposition of the two minima --- as seen in figure \ref{fig: Two_level_system}\figb. These minima correspend to the clockswise and anti-clockwise flow of the current carriers. The Bloch sphere is again traversed changing the flux $\Phi$.

\subsection{Towards a multi-qubit setup}
We have shown in that an \gls{SRQ} indeed provides a Bloch sphere where the states one the sphere can be controlled. However, to do quantum computations one has show one can create quantum gates wich can be readout. In \citet{Amico2014}, they showed theoretically that indeed a set of universal quantum gates can be realised. Furthermore, they discussed the gates can be readout by using time-of-flight measurement which was effentually demonstrated by \cite{Haug2018}.

Nevertheless, a experimental realisation of a quantum computation in the field of atomtronics is missing. The central bottleneck is the fulfilment of comunication between different circuital parts or gates \cite{amico2021}. Moreover, a concrete strategy for scalability is lacking. The challenge is to build a coherent optical environment using as minimal equipment as possible, such as lasers, mirrors, liquid crystals, and so on. The designed qubit gates and circuits must be compact, making the devices more easily scaled and, eventually, on-chip producible. However, \gls{SRQ}s are frequently created using Spatial Light Modulators (\gls{SLM})s. \gls{SLM}s are often employed due to their flexibility in creating light interference patterns of varying spatial shapes \cite{amico2021}. However. they are essentially three-dimensional systems. This precludes \gls{SLM} equipment from being fully incorporated into a two-dimensional system, limiting scalability. As a result, two-dimensional approaches such as multibeam interference are preferred. Nevertheless, as will be shown in chapter \ref{ch: ILM}, it does have a large parameter space to construct lattices, which, to date, has not been fully maped out. \citet{DmitryILM2020} developed a method to roam the parameter space of the multibeam interference setup in which optical lattices with ring-like structures can be pursued. This will be done in the next chapter.


\chapter{Designing Superfluid Ring Lattices using the Integer Lattice Method}
\label{ch: ILM}
In short, the integer lattice method uses \emph{algebraic number theory} to aid with the design of structured coherent wave interference patterns. It provides a list of potentially useful optical landscapes that can physically be achieved and gives you the recipe --- i.e., the number of input lasers, relative angles between the lasers, the phase and the polarization of the lasers beams --- to realize them. In this chapter a brief overview of the integer lattice method is given as introduced by Kouznetsov et al. \cite{DmitryILM2020,DmitryBEC2022}.

First, in section \ref{sec: Multibeam Equidistant Interference}, we will outline the setup to create coherent interference patterns using the integer lattice method. We restrict the discussion of the formulation of the integer lattice method to only two-dimensional periodic lattices, although the method could be applied to any number of spatial dimensions. A formulation for two-dimensional quasi-period lattices can be found in \citet{DmitryILM2020}. Next, in section \ref{sec: Algebraic Number Theory}, the number theoretic formalism is discussed. How the analytical framework is linked to generation of exotic optical lattice potentials for trapping of ultracold atoms is covered in section \ref{sec: Ol and ILM}. An analysis of the different structures obtained can be found in section \ref{sec: Moire superlattices} and \ref{sec: matter-wave}. Finally, the lattices under investigation to build a potential qubit system are discussed in section \ref{sec: SRQ}.

\section{Multibeam Interference}\label{sec: Multibeam Equidistant Interference}
\begin{figure}[htbp]
    \centering
    \includegraphics[width=\textwidth]{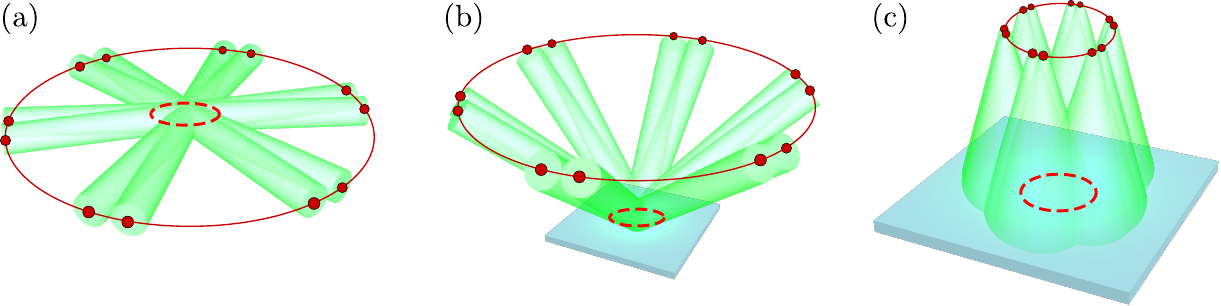}
    \caption{Three different configuration --- planar \figa, tilted \figb, perpendicular \figc ~--- orientations to construct two-dimensional optical coherent lattices. Taken from \citet{DmitryILM2020} under permission from the author.}
    \label{fig: different-setups}
\end{figure}

In chapter \ref{ch: Ultracold Atoms In Optical Lattices}, we discussed how the interference of coherent plane waves can lead to the emergence of an optical lattice. Here, we restrict the configuration to designing the landscapes with the integer lattice method. First, 
we place the incoming light beams at different angles from the centre where the optical lattice is created. This way, they form a circle as seen in \ref{fig: different-setups}. The setup can still be in different arrangement in comparison with the optical lattice. This can range from planar \figa, needed in two-dimensional confined systems like in photonic integrated circuits \cite{Dmitry22}, to tilted \figb, like in conventual laser microscopy, and diffraction cones \figc ~used in novel application such as lattice light-sheet microscopy \cite{Betzig2014}.

There are still some parameters to fix before the lattice can be created. As discussed in chapter \ref{ch: Ultracold Atoms In Optical Lattices}, the frequencies $\omega_j$ or wavelengths $\lambda_j$ of the laser beams are taken to be uniform, i.e., $\lambda_j=\lambda$. It determines the distance between the intensity maxima of the optical lattices. Also, the light beams are taken to be in phase. Changing the common phase of the laser beams will have no effect on the coherent pattern. Only the difference between the phases has an impact. Hence, for simplicity, we take the phase $\varphi_i=0$. How different phase can change the optical lattice has been investigated by Dmitry Kouznetsov in reference \cite{Dmitrythesis}. Moreover, the radius of the circle must be smaller than the length where the laser beam cannot be approximated by a plane due to beam divergence. Last but not least, as discussed in chapter \ref{ch: Ultracold Atoms In Optical Lattices}, polarization has an immediate impact on the coherent interference pattern. For our analysis, we assume all in-going light beams are parallel polarized with respect to each other. In short, for our analyses we use \emph{linear polarized coherent light beams}.

Neglecting the linear polarization for a moment, the laser beams can be seen as coherent plane waves. Consider a complex plane $\mathbb{C}$ spanned out by the $x$- and $y$-axis such that an arbitrary point on the plane is given by $z=x+iy\in\mathbb{C}$. A plane wave propagating along the y-axis with the wavefront parallel to the x-axis is written as
\begin{align}
f\colon\mathbb{C}\to\mathbb{C} \colon f(z)=e^{2 \pi i \operatorname{Im}(z)},
\end{align}
where it is expressed in units of wavelength $\lambda$. This depiction diverges from chapter \ref{ch: Ultracold Atoms In Optical Lattices} where we make use of the complex plane to write the plane waves as scalar waves. This is done to make an easier connection to the mathematical formalism introduced later. The resulting spatial pattern in the complex plane is a result of the interference or normalized total superposition of the $N$ number of plane waves \cite{DmitryILM2020}, i.e.
\begin{align}
\hat{f}(z)=\frac{1}{N} \sum_{g \in G} f(g \cdot z).
\end{align}
The finite set $G$ of size $N$ contains the transformation of the plane waves in the complex plane which moves them on the circle. Hence, they are pure rotations and only affect the orientation or the wave vector $k_j$ of each plane wave. The plane waves are kept directed at the centre of the circular configuration. The resulting intensity of the interference pattern is obtained from the complex amplitude by multiplying it with its complex conjugate \cite{DmitryILM2020, fowles1989introduction}
\begin{align}
I(z)=\hat{f}(z) \hat{f}^{*}(z).
\end{align}
This formalization in scalar coherent plane waves may also apply to ``acoustic, interfacial, and mono-energetic matter waves'' \cite{Betzig2005}. Hence, the resulting structures will be called \emph{coherent lattices}, whereas we call it an \textit{optical lattice} if it is made from electromagnetic waves.

What remains are two free parameters: the number of laser beams and the relative angles between the laser beams. Both have a big influence on the resulting spatial pattern. These two degrees of freedom is a large parameter space to cover --- in the sense that it is unclear what the result will be, before calculating the pattern. There seems to be no order.


\section{A formulation in Algebraic Number Theory} \label{sec: Algebraic Number Theory}
\begin{figure}[htbp]
    \centering
    \includegraphics[width=\textwidth]{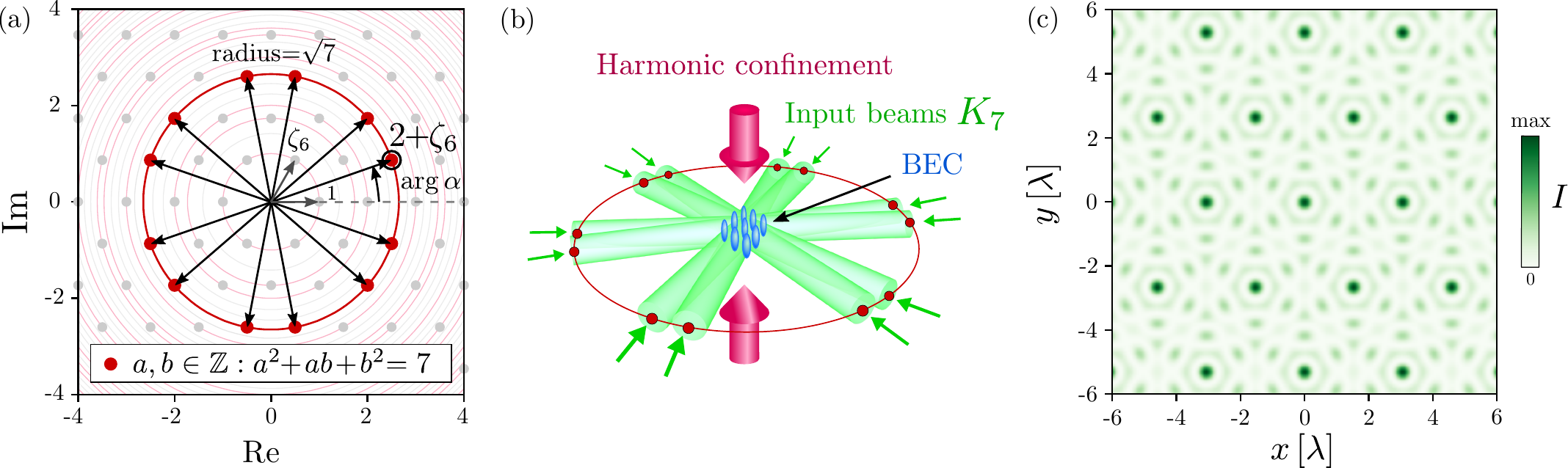}
    \caption{The connection between the number theoretic framework of the integer lattice method and the physical creation of coherent optical lattices. The concyclic integer lattice sites corresponding to the field norm $n=7$ in the integer lattice $\mathbb{Z}[\zeta_6]$ in \figa, yield the input beams for the planar setup in \figb, resulting in the total field intensity $I$ interference pattern observed in \figc. Taken and amended from Kouznetsov et al. \cite{Dmitry22} under permission from the author.}
    \label{fig: mbi}
\end{figure}

Integer lattice method is an algorithmic approach to traversing the chaos. To explain how it works, we have to introduce an algebraic number theoretical framework. Consider an integer lattice described in the complex plane $\mathbb{C}$ by the ring
\begin{align}
\mathbb{C}\supset\mathbb{Z}[\zeta_m] = \{ \alpha=a+b\zeta_m \mid a, b \in \mathbb{Z} \}\cong\mathbb{Z}^2,
\qq{with}
\zeta_m = \mathrm{e}^{2\pi i/m}.
\end{align}
The ring elements are generated by the unit vectors or complex numbers $\zeta_m$ and $1$ with the result being that the integer lattice is cyclotomic \cite{washington1997introduction}. Together the points form a lattice structure as seen by the grey dots in figure \ref{fig: mbi}\figa ~for $m=6$. We shall restrict our self to only periodic lattices. Hence, by the \emph{crystallographic restriction theorem}, only the triangular $(m=6)$ or rectangular $(m=4)$ lattice are allowed \cite{senechal2009quasicrystals}. Quasi-periodic optical lattices are easily constructed with the Integer Lattice method by allowing other integer values of $m$, see for example Kouznetsov et al. \cite{DmitryILM2020, DmitryBEC2022}. Although, also for $m=4,6$ quasi-periodic interference lattices are attainable.

An interesting number theoretic structure to look at, are the circles with origin $(0,0)$ where a lattice site $\alpha=(a, b)\in\mathbb{Z}[\zeta_m]$ can sit on. Considering that the distance of the lattice point $\alpha$ to the origin is equal to $\sqrt{\alpha\bar{\alpha}}=\sqrt{a^2 +a b + b^2}$ with $a$ and $b$ integer numbers, the circles must have a radius of the square root of an integer number $n\in\mathbb{N}$. Hence, the lattice points on a circle with radius $\sqrt{n}$ is given by the set\footnote{Both the asterisk ${}^*$ and the bar over symbol $\bar{ \ }$ are used to indicate the complex conjugate respecting the notation of the relevant mathematical field.}
\begin{align}\label{eq: concyclic lattice sites}
P(n)=\left\{N(\alpha)=n \mid \alpha \in \mathbb{Z}\left[\zeta_{m}\right]\right\},
\qq{with}
N(\alpha)=\alpha \bar{\alpha}= \begin{cases}a^{2}+b^{2} & \text { for } m=4, \\ a^{2}+a b+b^{2} & \text { for } m=6,\end{cases}
\qq{ \ }
\end{align}
where $N(\alpha)$ is called the field norm of the lattice $\mathbb{Z}\left[\zeta_{m}\right]$. It is said that the lattice points $\alpha \in P(n)$ for a $n\in\mathbb{N}$ in the lattice $\mathbb{Z}[\zeta_m]$ are concyclic to each other, i.e., they lie on the same circle with radius $\sqrt{n}$. In figure \ref{fig: mbi}\figa~the circles for every $n\in\mathbb{N}$ is drawn on the triangular lattice $\mathbb{Z}[\zeta_6]$. The circles for which $P(n)$ is empty or not, are, coloured in opaque grey and red, respectively. Also, the circle corresponding to field norm $n=7$ with the lattice points $P(n=7)$ is indicated in solid red.

The concyclic lattice point $P(n)$ has some interesting properties. First, as seen from the definition \eqref{eq: concyclic lattice sites}, the concyclic lattice points $\alpha \in P(n)$ correspond to a representation of $n$ as the product of $\alpha$ and the complex conjugate \cite{washington1997introduction}. This has the consequence that if $\alpha$ is in the set $P(n)$, also $\bar{\alpha}$ is in $P(n)$. For example, in the triangular lattice $\mathbb{Z}[\zeta_6]$, the integer $n=7$ can be factorised as $(3-\zeta_6)(3-\bar{\zeta}_6)$. Furthermore, the number of elements in the set $P(n)$, i.e., number of point on the circle with radius $\sqrt{n}$, is equal to the number of combination of complex factors in the prime factorization\footnote{The integer lattices $\mathbb{Z}\left[\zeta_{4}\right]$ and $\mathbb{Z}\left[\zeta_{6}\right]$ are euclidean and have therefore a unique factorization domain \cite{hardy2008introduction}.} of $n$ \cite{hardy2008introduction}. In addition, it can be proven that this is always in a multiple of $m$ \cite{cox2011primes}. How these properties are exploited is shown in section \ref{sec: Moire superlattices}.

\section{Optical Lattices generated using Integer Lattice Method}\label{sec: Ol and ILM}
This mathematical framework can be connected to the interference of coherent plane waves. In section \ref{sec: Multibeam Equidistant Interference}, it was concluded that we have two free parameters in our optical setup: the number of laser beams (the number of elements in $G$) and relative angles between the laser beams (the elements in $G$). In the integer lattice method, we link these unknowns to the lattice points in the set $P(n)$. In particular, we identify the relation
\begin{align}\label{eq: G=P(n)}
     G=n^{-1/2}P(n),
\end{align}
such that the light beams are placed in the configuration provided by the set $P(n)$. An example, in figure \ref{fig: mbi}, can be seen for $n=7$ in the triangular integer lattice $\mathbb{Z}[\zeta_6]$. The relative angles between the red lattice points in \figa~are used in the planar optical setup in \figb~with the result of the emergence of the optical setup seen in \figc. The normalisation in equation \eqref{eq: G=P(n)} guarantees that the lattice points have unit magnitude such that the resulting transformations in $G$ are pure rotations.

Stepping away from the general formulation of coherent waves in section \ref{sec: Multibeam Equidistant Interference} by using electromagnetic waves, we can link the relation \eqref{eq: G=P(n)} with the description of the general optical potential in a two-dimensional plane $\mathbb{R}^2$ given by \cite{DmitryBEC2022}
\begin{align}\label{eq: theoretical potential}
V_{\mathrm{latt }}(\vb{r})=V_{0}\left|\sum_{j} \mathcal{E}_{j} \boldsymbol{\epsilon}_{j} e^{-i\left(\vb{k}_{j} \cdot \mathbf{r}+\varphi_{j}\right)}\right|^{2},
\end{align}
where $V_{0}$ is the overall strength of the potential, $\mathcal{E}_{j} \in[0,1]$ is the relative intensity of the light beam $j$, $\boldsymbol{\epsilon}_{j}$ is the polarization, $\vb{k}_{j}$is the wave vector, and $\varphi_{j}$ is the phase. Here, we restored the polarization and phase for completeness. The relation \eqref{eq: G=P(n)} can then be rewritten in terms of the wave vectors $\vb{k}_{j}$ by the set
\begin{align}\label{eq: def K_n}
K_{n} \equiv \frac{2 \pi}{\lambda} \operatorname{vec}\left(\frac{P(n)}{\sqrt{n}}\right),
\qq{with}
\operatorname{vec} \colon x+yi \mapsto (x, y),
\end{align}
where the function $\operatorname{vec}$ converts the complex number to a spatial vector in $\mathbb{R}^2$. The concyclic points $P(n)$ are scaled by the wavelength $\lambda$ of the laser beams after being normalized to have unit length.

\begin{figure}[htbp]
    \centering
    \includegraphics[width=\textwidth]{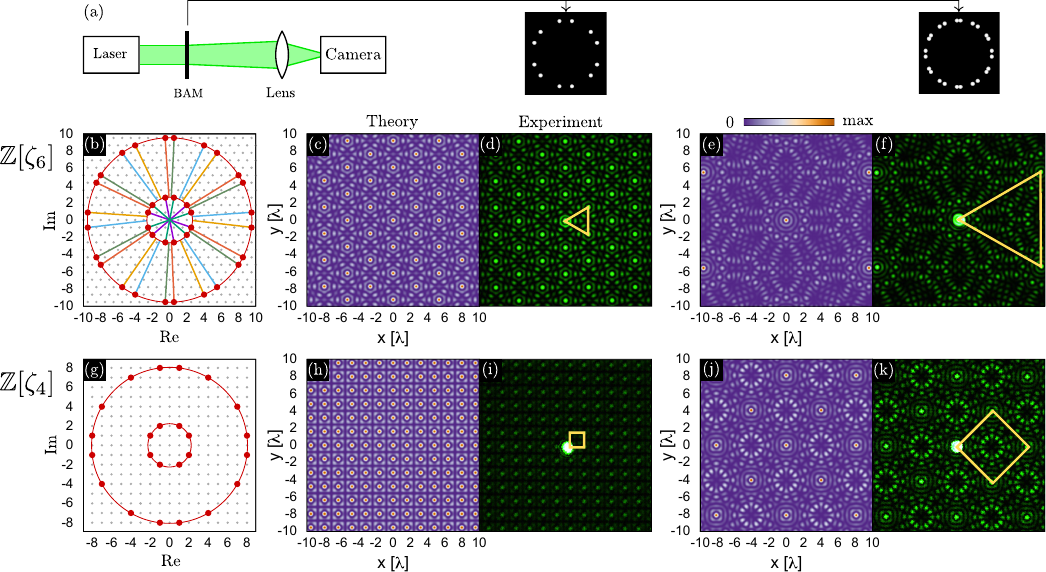}
    \caption{Experimental and theoretical comparison of the coherent optical latices for $m=6$ \textcolor{red}{(b-f)} and $m=4$ \textcolor{red}{(g-k)} by passing a laser beam through binary amplitude mask and measuring the resulting diffraction pattern \figa. Taken and amended from \citet{DmitryILM2020} under permission from the author.}
    \label{fig: ILM_examples_alt}
\end{figure}

In short, the integer lattice method links the wave vectors of the laser beams of the optical setup (see figure \ref{fig: mbi}\figb) to sets of concyclic lattice points of the integer lattices $\mathbb{Z}[\zeta_m]$. In that way that the resulting optical lattices are classified by two symmetry classes: $m=4$ and $m=6$. Each class consists of a countable infinite optical landscape represented by the field norm $n$. In figure \ref{fig: ILM_examples_alt} you can see an experimental validation of the integer lattice method by using the~diffraction setup --- as previously shown in figure \ref{fig: different-setups}\figc. More concretely, as seen in figure \ref{fig: ILM_examples_alt}\figa~the diffraction cones are made by passing linearly polarised coherent light through apertures in a binary amplitude mask, determined by the elements in $P(n)$. The measured optical lattices show a good agreement with the plotted theoretical potential \eqref{eq: theoretical potential}.

\section{Sieve of Eratosthenes with Moir\'e superlattices} \label{sec: Moire superlattices}
So far, the selection of the number of the light beams and their wave vectors, i.e., choosing the field norm $n$, seems random. One could iterate through all possibilities until an optical lattice of interest becomes apparent. However, a more clever approach is to exploit the properties of the sets $P(n)$ --- listed in section \ref{sec: Algebraic Number Theory}.

\begin{figure}[htb]
    \centering
    \includegraphics[width=\textwidth]{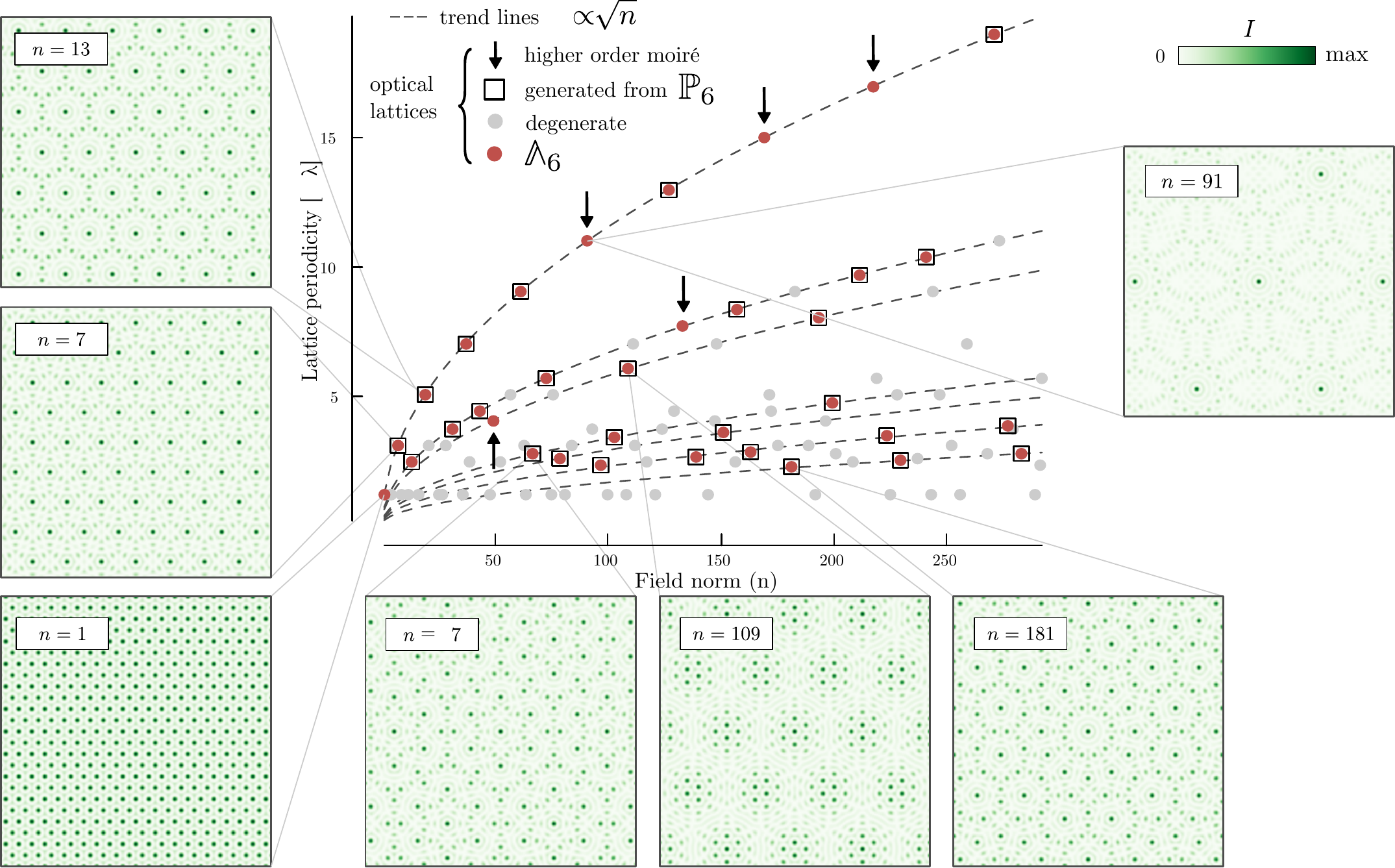}
    \caption{
    Classification of optical lattices using the integer lattice technique. An optical lattice formed from an integer $n$ corresponds to each point in the diagram. The classification between lattices is based on lattice periodicity, with the first appearing lattices $\mathbb{\Lambda}_6$ being highlighted in red and identical lattices being highlighted in gray. The periodicity's trend lines (dashed lines) are provided as a visual help. Notice, the non-inert prime numbers $\mathbb{P}_{6}$ (black squares) are always $\mathbb{\Lambda}_6$ as they are first-order moir\'e lattices. Some of the lattices corresponding total field intensity $I$ is shown. intensity Prime number products are used to create higher-order moir\'e optical lattices (black arrows).
    Taken and amended from Kouznetsov et al. \cite{DmitryBEC2022} under permission from the author.
    }
    \label{fig: eratosthenes}
\end{figure}

A first observation that can be made by cycling through all possible circle raddii or field norms $n$, is that some optical lattices tend to reoccur with an identical periodicity. But first, to quantify the periodicity of the optical lattice, we can make a connection to \emph{moir\'e theory}. Here, it is known that the smallest frequency space components define the dominant spatial characteristics \cite{amidror2009theory}. With this fact in mind, Kouznetsov et al. \cite{DmitryBEC2022, DmitryILM2020} defined the periodicity $d$ of an optical lattice generated with the integer lattice method to be
\begin{align}\label{eq: def periodicity}
d=\min \left|\mathbf{k}_{i}-\mathbf{k}_{j}\right|^{-1}, \qq{with} i \neq j,
\end{align}
with $\mathbf{k}_{i}\in K_n$ defined in equation \eqref{eq: def K_n}. The quantity can be used as a distinguishing factor between optical lattices. For example, in figure \ref{fig: eratosthenes} the periodicity $d$ is plotted against the field norm $n$ for the triangular lattice $\mathbb{Z}\left[\zeta_{6}\right]$. One can see that some optical lattices follow a trend line $propto \sqrt{n}$. This corresponds to the normalization factor in equation \eqref{eq: def K_n}.

A deeper understanding can be gained by looking at the set of prime numbers $\mathbb{P} = \{p \in \mathbb{N} \mid \text{p is prime}\}$. By the \emph{unique factorisation theorem}, an integer number can be written as a product of prime numbers. However, not all primes remain prime in an integer lattice $\mathbb{Z}[\zeta_m]$. For example, $7$ can be factorised in $\left( 3 - \zeta_{6} \right) \left( 3 - \bar{\zeta}_{6} \right)$ over $\mathbb{Z} \left[ \zeta_{6} \right]$. The primes who do remain prime in $\mathbb{Z}\left[\zeta_{m}\right]$ are called inert primes \cite{washington1997introduction}. However, if a field norm $n$ is inert prime, then, it is clear from the definition \eqref{eq: concyclic lattice sites} that there will be no concyclic point, i.e, $P(n)=\emptyset$. Identifying the set of all inert primes in $\mathbb{Z} \left[ \zeta_{m} \right]$ as $\mathbb{I}_m$, we may select the field norms which will generate an optical lattice by $\mathbb{P}_m \equiv \mathbb{P} \setminus \mathbb{I}_m$. 

The field norm $n\in\mathbb{P}_m$ can be seen as the \emph{first order moir\'e superlattice}. To see this, consider the units $\mathbb{Z}[\zeta_m]^\times$ of the an integer lattice $\mathbb{Z}[\zeta_m]$. It is defined by the lattice points with $N(\alpha)=1$ or $P(1)$ which are given by
\begin{align}
    \mathbb{Z}\left[\zeta_{m}\right]^{\times}= \begin{cases}\{\pm 1, \pm i\} & \text { for } m=4, \\ \left\{\pm 1, \pm \zeta_{6}, \pm \zeta_{6}^{2}\right\} & \text { for } m=6\end{cases}.
\end{align}
Using the units of integer lattice, it can be proven that if $n\in\mathbb{P}_m$ than the corresponding concyclic points $P(n)$ can be written as \cite{conway1998book}
\begin{align}
     P(n) = \mathbb{Z}[\zeta_m]^\times \cdot \alpha \  \cup \  \mathbb{Z}[\zeta_m]^\times \cdot \bar{\alpha},
\end{align}
where $\alpha \in P(n)$. For example, taking again the factorisation the field norm $n=7=\left( 3 - \zeta_{6} \right) \left( 3 - \bar{\zeta}_{6} \right)$ in the triangular lattice $\mathbb{Z}\left[\zeta_{6}\right]$, we get that
\begin{align}
     P(7) = \mathbb{Z}[\zeta_6]^\times \cdot \left( 3 - \zeta_{6} \right) \  \cup \  \mathbb{Z}[\zeta_6]^\times \cdot \left( 3 - \bar{\zeta}_{6} \right).
\end{align}
In this perspective, using the orientations in $\operatorname{arg}[P(7)]$ in the optical setup is equivalent to rotating multiple copies of the units configuration. Hence, the units  $\mathbb{Z}[\zeta_m]^\times$ can be seen to establish a ``base pattern'' and lattices generated using the integer lattice method are moir\'e rotations from it. As an example, the points on the inner circle in figure \ref{fig: ILM_examples_alt}\figb correspond to the set $P(7)$ and are colour coded green and purple according to multiples of the units.

Also related \emph{higher-order moir\'e superlattices} can easily be found using the integer lattice method. Indeed, take $n=91$, the moir\'e superpattern is exposed by looking at the prime factorisation, $91=7 \cdot 13=\left(3-\zeta_{6}\right)\left(3-\bar{\zeta}_{6}\right)\left(4-\zeta_{6}\right)\left(4-\bar{\zeta}_{6}\right)$, such that:
\begin{align}
P(91)=&
 \ \mathbb{Z}[\zeta_6]^\times \cdot\left(3-\zeta_{6}\right)\left(4-\zeta_{6}\right)
\cup \mathbb{Z}[\zeta_6]^\times \cdot\left(3-\zeta_{6}\right)\left(4-\bar{\zeta}_{6}\right)\\
&\cup \mathbb{Z}[\zeta_6]^\times \cdot\left(3-\bar{\zeta}_{6}\right)\left(4-\zeta_{6}\right)
\cup \mathbb{Z}[\zeta_6]^\times \cdot\left(3-\bar{\zeta}_{6}\right)\left(4-\bar{\zeta}_{6}\right)
\end{align}
Again, in \ref{fig: ILM_examples_alt}\figb the concyclic points are colour coded according to multiples of the units. The corresponding lattice can be seen in figure \ref{fig: different-setups}\fige. It has very similar spatial symmetry to $n=7$, but a larger periodicity. This is due to the $\sqrt{n}$ dependence as seen in equation \eqref{eq: def periodicity}. The symmetry contribution of $n=13$ can also be observed by comparing the two lattices in figure \ref{fig: eratosthenes}. 

With the above in mind, we can find all optical lattices that occur for the first time, i.e., with the lowest field norm $n$ for a particular periodicity $d$ and collect them in the set $\mathbb{\Lambda}_m$. Just like finding primes with the \emph{sieve of Erathones} --- ``an algorithm in which all multiples of a number are marked iteratively such that only all primes remain'' \cite{Baker1997}. In figure \ref{fig: eratosthenes} everything comes together. Each dot corresponds to the optical lattice generated with the integer lattice method and has therefore a nonempty set of concyclic points $P(n)$ in the integer lattice $\mathbb{Z}[\zeta_6]$. The optical lattices that appear for the first time, i.e., $n\in\mathbb{\Lambda}_m$, are indicated in red. The other lattices with the same periodicity are observed to give the same optical lattices\footnote{This is an observation and no concrete prove exist yet.} and is therefore degenerate. They are marked in grey. For example, the field norms $n=1$ and $n=4$, lead to the same set of wavevectors $K_1=K_4$. Non-inert field norms, i.e., $n\in\mathbb{P}$, are indicted with a black square around the dot. Notice, these are always in $\mathbb{\Lambda}_m$. The lattices are the first order moir\'e lattices in the integer lattice method. However, not every $n\in\mathbb{\Lambda}_m$ is non-inert. These are the higher moir\'e superlattices --- indicated with a black arrow --- where the field norm can be factorised into a product of prime numbers. They are constructed as a superposition of previous field norms.

\section{Matter-wave interference}\label{sec: matter-wave}
\begin{figure}[htbp]
    \centering
    \includegraphics[width=\textwidth]{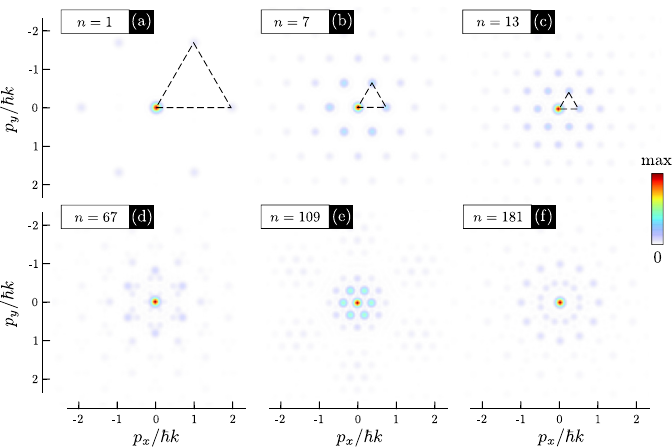}
    \caption{Computed matter-wave interference patterns corresponding to the momentum distribution of the ultracold in the optical lattices constructed using the integer lattice method. The dashed triangles from \figa~to \figc~represent a reduction in the first Brillouin zone, which indicates to denser lattice sites. Taken and amended from Kouznetsov et al. \cite{DmitryBEC2022} under permission from the author.}
    \label{fig: matter-wave}
\end{figure}

In figure \ref{fig: eratosthenes} some of the lattices corresponding to the fields norm in $\mathbb{Z}[\zeta_6]$ are shown. To investigate these lattices, one can look at their effect on ultracold bosons forming a Bose-Einstein condensate. In particular, the momentum distribution of a BEC can provide key information of the system. This is frequently observed experimentally in time-of-flight experiments by monitoring matter-wave diffraction. In Kouznetsov et al. \cite{DmitryBEC2022}, the time-dependent Gross-Pitaevskii equation ---  thought to describe the ground state of a quantum system of identical bosons --- was solved numerically. This yielded the momentum distribution or matter-wave interference of the Bose-Einstein condensate seen in figure \ref{fig: matter-wave}.  As explained in chapter \ref{ch: Ultracold Atoms In Optical Lattices}, the lattice depth has a big influence on the phase of the ultracold atom in the lattice. Tuning the depth such that the atoms only tunnel/hop between the intensity maxima yields a discrete momentum distribution (collection of Bragg peaks) depending on the geometry of the lattice. 

The first three lattices in the upper trend line of figure \ref{fig: eratosthenes}, i.e., $n=1,7,13$, are shown in figure \ref{fig: matter-wave}. We clearly observe a decrease in Bragg peaks in the momentum distribution as the periodicity defined in equation \eqref{eq: def periodicity} increases. More complex matter-wave interference's are shown in figure \ref{fig: matter-wave}\textcolor{red}{(d-f)}. For instance, in $n=67$, auxiliary dynamics of the Bose-Einstein condensate in the lattice are revealed, seen by secondary interference peaks inside the first brouillon zone. New exotic dynamics is seen in $n=109$ corresponding to a first-order moir\'e lattice. Last, quasiperiodic lattices are observed using the Integer lattice Method as can be seen for $n=181$ where the momentum space reveals the twelvefold rotational symmetry. The lattices are of great interest as they create hybrid crystalline/amorphous dynamics for the ultracold atoms.

\section{Superfluid Ring lattices}\label{sec: SRQ}
We want to use the integer lattice method to create optical lattices which can facilitate superfluid ring Qubits. For this we are looking for interference pattern which contain ring structures. It is best to concentrate on the first-order moir\'e lattices as they determine the interference pattern for higher orders. Once we find such an optical lattice, it can be used to construct their higher-order moire lattices with the hope it contains the same ring structures.

\begin{figure}[htbp]
    \centering
    \includegraphics[width=0.8\textwidth]{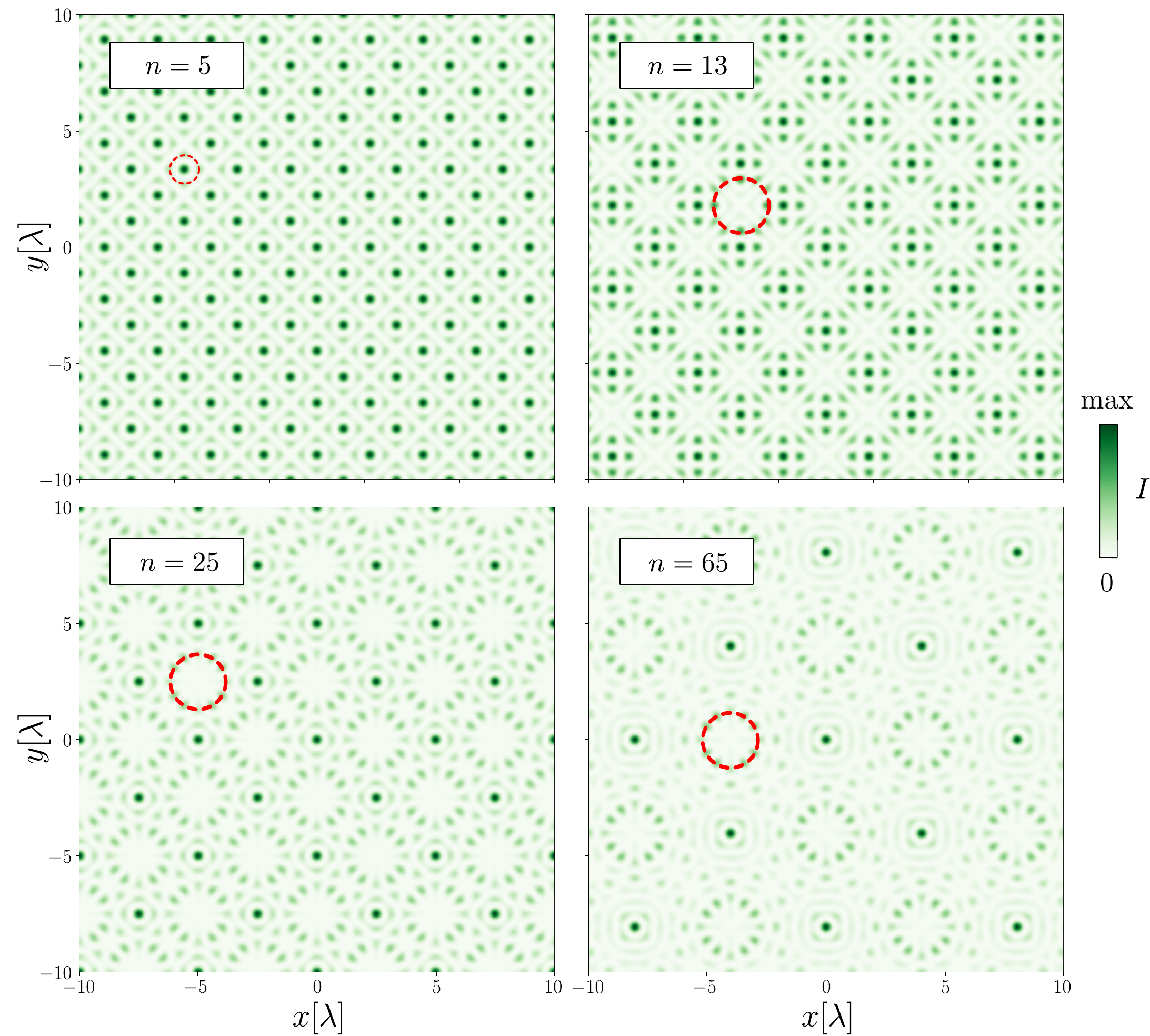}
    \caption{Optical lattices constructed from the integer lattice $\mathbb{Z}[\zeta_4]$ with field norms $n=5,13,25$ and $65$ with the goal to find circular structures which can host superfluid ring qubits. The field norms $n=5,13$ are first order moir\'e optical lattices, whereas $n=25,65$ correspond to higher orders. The red dashed line indicate the ring-shaped interference pattern observed.}
    \label{fig: SRCfromto}
\end{figure}

We looked both at the square integer lattice $m=4$ and the triangular integer lattice $m=6$. For the integer lattice with $m=6$ no circular ring structure could be found. Nevertheless, the square integer lattice $\mathbb{Z}[\zeta_4]$ with field norms $n=5$ and $n=13$ both contains circular  interferencestructures, as can be seen in figure \ref{fig: SRCfromto}. Using these first-order moir\'e lattices we can construct higher order interference patterns. In particular $n=5 \cdot 5 = 25$ and $n= 5 \cdot 13 = 65$ both yield well-defined rings. For both lattices the ring are constructed out of hopping sites. Hence, comparing with the qubit design covered in section \ref{sec: Superfluid Ring Qubits}, both seem able to facilitate superfluid ring qubits.

To prove that indeed $n= 25$ and/or $n= 65$ can host a qubit system, a first step would be to demonstrate that the lattices can hold a superfluid ring and therefore can host a superfluid current of neutral atomic bosons. As explained in chapter \ref{ch: Introduction}, this can be done with Path Integral Monte Carlo Worm algorithm, renowned for correctly simulating the behaviour of superfluids.


\chapter{Path Integral Monte Carlo for bosons in continuous space}
\label{ch: PIMC}
Monte Carlo methods have become a popular approach to tackling physics problems. Although the statistics community was already promoting the technique in the early 1900s, it was \citet{Metropolis1949} who, 50 years later\footnote{Of course, it became more practical with the rise of computers.}, first published the approach in a physical context to compute the ground-state energy of the Schr\"odinger equation \cite{Reynolds1990}\footnote{Actually, Stanislaw Ulam and John von Neumann used the Markov Chain Monte Carlo method when working on neutron diffusion for nuclear weapons in the Manhattan Project at the Los Alamos National Laboratory \cite{Metropolis1987, Benov2018}.}. The idea was inspired by Fermi's remarks given at a symposium \cite{Fermi1957} where he noticed the resemblance of the time-dependent Schr\"odinger equation to a diffusion equation (see later) and hence can be treated as a stochastic process. Diffusion Monte Carlo was discovered.

Nowadays, Monte Carlo methods are used in all fields of physics, from molecular modelling in biophysics \cite{Carlon2002} to designing detectors in particle physics \cite{Agostini2018} or understanding galaxy evolution in astrophysics \cite{MacGillivray1982}. In statistical physics, the overall purpose of using the Monte Carlo method is to compute a multivariate integral over phase space to compute the expectation value of some macroscopic variable. The motivation being that the error converges as $\sim 1/\sqrt{n}$, where $n$ represents the total number of independent samples\footnote{In practice, the samples are correlated (see section \ref{sec: error}).}, regardless of the dimension of the phase space. This property is especially advantageous for quantum systems where the Hilbert space expands exponentially with the number of particles. Other methods, such as exact diagonalization, thereby scale very unfavourably with system size due to memory requirements. By contrast, \emph{Quantum Monte Carlo} (\gls{QMC}) algorithms are computational methods developed to produce solutions to quantum many-body systems by handling the multivariate integrals that emerge in various formulations of the many-body problem.

The key disadvantage of the \gls{QMC} approach is that it is almost always only effective for bosonic systems\footnote{There are some exceptions to this rule, e.g., bold diagrammatic Monte Carlo \cite{Prokofev2007}.}. The notorious numerical \emph{sign problem}, in which near-cancellation of the positive and negative contributions to the integral, prevents general fermionic systems from being sampled in this way \cite{Pollet2012}. This does not prohibit the Monte Carlo approach from being used, but it does restore the exponential scaling of computational resources with system size. Solving the fermion-sign problem is seen to be impossible, grounded on the fact that it has been proven to be non-deterministic polynomial (NP)-hard \cite{Troyer2005}.

The \gls{QMC} algorithm we will be using is called Path Integral (Quantum) Monte Carlo, or \gls{PIMC}. The method was developed by Pollock and Ceperley \cite{Ceperley1984, Ceperley1987, Ceperley1995} to understand the $4$He superfluid transition. Nowadays, it is used to understand a wide variety of condensed matter systems, e.g., dipolar systems \cite{Cinti2017}, ultracold gases \cite{Cinti2014,Pascual2021,Carleo2013} and quantum fluids and solids \cite{Boninsegni2009, Corboz2008}. By developing the method, Pollock and Ceperley were able to turn Feynman's theory of superfluidity \cite{Feynman1953} into a quantitative theory that could be validated using experimental evidence. However, the method was still limited by the number of particles in the system, as the sampling of Bose statistics becomes inefficient at low temperatures. Nevertheless, the hurdle was overcome by using an alternative sampling approach, known as the worm algorithm (\gls{WA}) \cite{Prokofev1998, Boninsegn2006, Boninsegn2006PRL}. The combination of both methods is often called Worm Algorithm Path integral Monte Carlo (\gls{WAPIMC}). In the case of ultracold atoms in optical lattices, \gls{WAPIMC} has been shown to be in good agreement with experiment \cite{Trotzky2010}. \gls{WAPIMC} allows for a lot of thermodynamic observables of interest at temperatures greater than zero, including structural, e.g., pair correlation functions, the structure factor, and compressible, as well as quantities such as superfluid density and condensate fraction.

In this chapter, we will explain the \gls{WAPIMC} method in continuous space, starting with the basics of the \gls{PIMC} algorithm (\ref{sec: pimc}) and thereby also Markov chain Monte Carlo (\ref{sec: mcmc}). Subsequently, in section \ref{sec: stp}, a discussion of the small-time propagator can be found. After, a depiction of the sampling updates (\ref{sec: update}) is given, with a special focus on the Worm Algorithm (\ref{sec: Worms}). Finally, a brief overview of the computation of physical observables (\ref{sec: observable}) is given, complemented by an explanation of Monte Carlo data error analysis (\ref{sec: error}), including autocorrelation effects.

\section{Path Integral Monte Carlo} \label{sec: pimc}
 In the following subsections, \gls{WAPIMC} is presented and follows the approach used by \citet{Yao2020, Ceperley1995, Boninsegn2006}. Some other useful didactic sources considered are \citet{Pollet2012, YanBlume2017, dornheim2014}.

\subsection{Feynman's path integral} \label{sec: feynman path integral}
Consider a system of $N$ identical \emph{bosons} in a volume $V$ of $d$ dimensions at a finite temperature $T$, which is represented in the canonical ensemble by a Hamiltonian $\ham$. Quantum statistical mechanics \cite{feynman1965, feynman1972} dictates that the expectation value of an observable $\hat{O}$ at thermodynamic equilibrium is given by
\begin{align} \label{eq: def expval}
    \expval*{\hat{O}} = \frac{1}{Z}\Tr[\hat{\rho} \ \oper] \qq{with} Z=\Tr[\hat{\rho}] \qq{and} \hat{\rho} = e^{\beta \ham},
\end{align}
where $Z$ is the canonical partition function, $\hat{\rho}$ the matrix density operator, and $\beta= (k_B T)^{-1}$ the inverse temperature, with $k_B$ the Boltzmann constant. In order to evaluate the trace over the configurations, one has to choose a convenient basis. Let us take the spatial locations of the N particles $\ket{\mathbf{R}} \equiv \ket{\mathbf{r}_1, \cdots, \mathbf{r}_N}$, so that the trace for an operator $\hat{X}$ can be written as
\begin{align} \label{eq: def trace}
     \operatorname{Tr}[\hat{X}]=\int \dd{\mathbf{R}} \expval{\hat{X} \hat{S}}{\mathbf{R}} \qq{where}  \hat{S}=\frac{1}{N !} \sum_{\sigma \in \mathcal{P}}\ket{\sigma \cdot \mathbf{R}}\bra{\mathbf{R}}.
\end{align}
The symmetrization operator $\hat{S}$ must be included in the trace to account for the indistinguishability of the identical bosons and, thereby, a crucial part of the physics of the system. The canonical partition function $Z$ is then written by
\begin{align}
     Z = \frac{1}{N !} \sum_{\sigma \in \Pi} \int \dd{\mathbf{R}}  \mel{\mathbf{R}}{ \ \hat{\rho} \ }{\sigma \cdot \mathbf{R}} = \frac{1}{N !} \sum_{\sigma \in \mathcal{P}} \int \dd{\mathbf{R}}  \rho(\mathbf{R},\sigma \cdot \mathbf{R}; \beta),
\end{align}
where $\rho(\mathbf{R},\mathbf{R}^\prime; \beta)$ is called the \emph{density matrix} at inverse temperature $\beta$. If the Hamiltonian $\ham$ is hermitian, the density matrix is symmetric in its first two arguments.

Often, the potential energy\footnote{My apologies for the unconventional notation. Nevertheless, this part of the contribution to the density matrix is often referred to as the action (see section \ref{sec: stp}). In the Hamiltonian, the potential for the optical lattice and the interaction will be noted as V and U, respectively.} part $\hat{\mathcal{S}}$ of the Hamiltonian $\ham = \hat{\mathcal{T}} + \hat{\mathcal{S}}$ is diagonal with respect to the position eigenbasis. Hence, it would be preferable to decompose the density matrix into a product containing the kinetic $\hat{\mathcal{T}}$ and potential $\hat{\mathcal{S}}$ contributions to the Hamiltonian. Nevertheless, in general, we have that these contributions do not commute. One could remedy this problem by using the Baker-Campbell-Hausdorff (\gls{BCH}) formula \cite{Campbell1897} to construct an approximation given by
\begin{align} \label{eq: Baker-Campbell-Hausdorff}
    e^{-\beta \ham}= e^{-\beta\hat{\mathcal{T}}}e^{-\beta\hat{\mathcal{S}}}e^{-\frac{\beta^2}{2}\comm{\hat{\mathcal{T}}}{\hat{\mathcal{S}}} \ + \ \cdots} = e^{-\beta\hat{\mathcal{T}}}e^{-\beta\hat{\mathcal{S}}} + \mathcal{O}(\beta^2).
\end{align}
Splitting the kinetic and potential contribution with the help of \gls{BCH} formula is often called the \emph{primitive approximation}. Notice that, the error to our approximation diverges in the low temperature (quantum) limit $\beta\gg 1$, the limit one is often interested in. To remedy this problem, one can split the density matrix into $J\in\mathbb{Z}$ parts of step size $\tau \equiv \beta/J$ by using the normalization condition $\hat{I} = \int \dd{\mathbf{R}} \ket{\mathbf{R}}\bra{\mathbf{R}}$. Doing this yields
\begin{align} \label{eq: modified expval}
    Z=\frac{1}{N !} \sum_{\sigma \in \Pi} \int \dd{\mathbf{R}_{J-1}} \mydots \dd{\mathbf{R}_{0}}\mel{ \mathbf{R}_{0}}{e^{-\tau \hat{H}}}{\mathbf{R}_{1}} \mel{\mathbf{R}_{1}}{e^{-\tau \hat{H}}}{\mathbf{R}_{2}}\mydots \mel{\mathbf{R}_{J-1}}{e^{-\tau \hat{H}}}{\sigma \cdot \mathbf{R}_{0}},
\end{align}
where we introduce the notation\footnote{In general, I would like to reserve the indices $i$ and $j$ for enumerating the number of particles and time slices respectively. The index $k$ can be used as a spare index.} $\ket{\mathbf{R}_j} \equiv \ket{\mathbf{r}_{1,j}, \cdots, \mathbf{r}_{N,j}}$ and $\sigma$ is a permutation operator of the group of permutations of the $N$ elements $\Pi$. Notice that for the above expression, we used that the symmetrization operator $\hat{S}$ commutes with the Hamiltonian $\ham$. Furthermore, we define $\ket{\mathbf{R}_J} \equiv \ket{\sigma\cdot\mathbf{R}_0}$ such that we can write the partition function in a more compact form:
\begin{align}\label{eq: partition function compact form}
    Z = \frac{1}{N !} \sum_{\sigma \in \Pi}\prod_{j=0}^{J-1} \int \dd{\mathbf{R}_j}  \rho(\mathbf{R}_{j},\mathbf{R}_{j+1}; \tau).
\end{align}
The expression for the partition function is still exact for every $J\geq1$. Although, we modified the problem from integrating one density matrix at temperature $\beta$ to integrating $J$ density matrices at $\tau$. Now it will be easier to find an expression for the density matrix. In practice, one takes $J$ very large such that $\tau \ll 1$ and an accurate expression of $\rho(\mathbf{R}_{j},\mathbf{R}_{j+1}; \tau)$ can be found using \eqref{eq: Baker-Campbell-Hausdorff} and higher-orders (see section \ref{sec: stp}). A broader discussion is given in section \ref{sec: stp}.

A key insight into the problem can be made if one identifies $\beta=it/\hbar$ such that the operator $\operatorname{exp} (-\beta \ham)$ turns into the unitary time evolution operator. Consequently, the system can be seen as N particles evolving through imaginary time, i.e. the density matrix becomes a \emph{time evolution propagator}\footnote{The same substitution can be made in the time-dependent Schrödinger equation which results in a diffusion equation with a diffusion constant $D = \tfrac{\hbar^2}{2m}$.}. This mapping of a $d$-dimensional quantum system to a $d+1$ classical system is known as the quantum-classical mapping of Feynman path integral formulation of quantum statistical mechanics. From now on, we shall use both representations interchangeably.
\begin{figure}[ht]
    \centering
    \includegraphics[width=\textwidth]{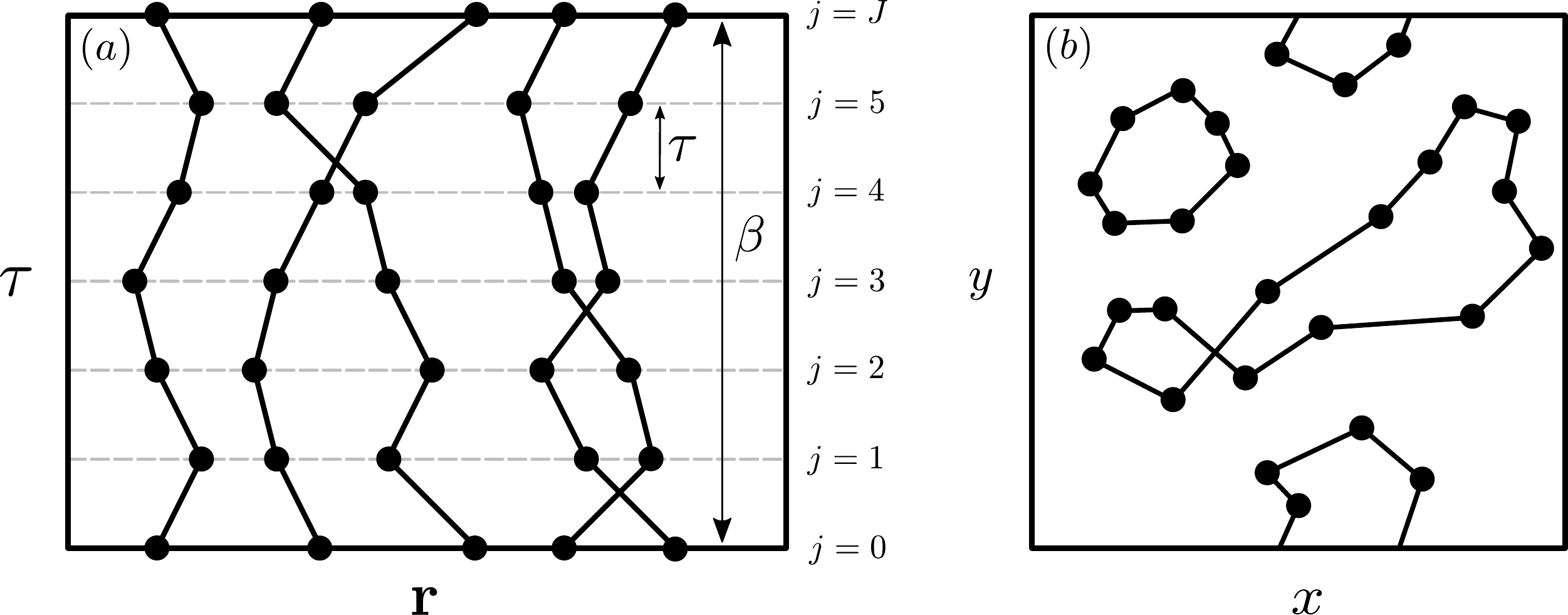}
    \caption{Two representations of the worldline configurations. The schematic are inspired by and adapted from \citet{Yao2020, dornheim2014}. \textcolor{red}{(a)} A collection of beads connected by links plotted in function of imaginary time. \textcolor{red}{(b)} Two-dimensional spatial position of the beads connected by their links to form a ring or polymer.}
    \label{fig: Configuration}
\end{figure}

A configuration $\config$ is defined as the collection $(\mathbf{R}_{1}, \cdots, \mathbf{R}_{J})$ which represents $N$ discrete worldlines in imaginary time from 0 to $\beta$. An example of a typical configuration in one spatial dimension can be found in figure \ref{fig: Configuration}\textcolor{red}{(a)}. The position of the i-th particle at the j-th time slice, i.e. $r_{i,j}$ is called a \emph{bead} and the propagator $\rho\left(\mathbf{r}_{i, j}, \mathbf{r}_{i, j+1} ;  \tau\right)$ who connects two beads of consecutive time slices is called a \emph{link}. In table \ref{tab: jargon}, a list of all the jargon of a \gls{PIMC} simulation van be found. As can be seen in \eqref{eq: modified expval}, due to the imaginary time boundary condition $\ket{\mathbf{R}_J} = \ket{\sigma\cdot\mathbf{R}_0}$, the worldlines of are $n\beta$ periodic with $1\leq n\leq N$. In the case $n>1$, the particles are interchanged in the configuration, which accounts for the indistinguishability of the particles.

\citet{ChandlerWolynes1981} introduced another perspective. In the classical representation, one could view the worldlines as ring polymers where they form a polyatomic fluid, see figure \ref{fig: Configuration}\textcolor{red}{(b)}. As the particle permutes, the polymers grow with a maximum of $NJ$ beads. It has been shown that the polymer spatial area is proportional to the de Broglie wavelength \cite{Ceperley1995}
\begin{align}\label{eq: de Broglie wavelength}
    \lambda_{\beta} = \frac{2 \pi \hbar}{\sqrt{2 \pi m k_{\mathrm{B}} T}} =\sqrt{4\pi\lambda\beta},
\end{align}
with $\lambda=\hbar^2/2m$. Hence, the quantum statistics become important when the thermal de Broglie wavelength of the particles is of the same order as the inter-particle distance of the particles and, thereby, the polymers can permute, i.e. the particles become indistinguishable. Accordingly, if the temperature is raised, i.e., $\beta\to 0$, the polymer shrinks and becomes distinguishable again, analogously to the single-particle quantum wavefunction.

With the procedure of constructing \eqref{eq: modified expval} in mind, equation \eqref{eq: def expval} can be written in an insightful form. Namely, one can define the normalized \emph{probability density function} (\gls{pdf}) of a configuration $\config$ by
\begin{align} \label{eq: def weight}
     \pi(\config)\equiv\frac{\mathcal{W}(\config)}{Z}=\frac{1}{Z}\frac{1}{N!}\mel{ \mathbf{R}_{J}}{e^{-\tau \hat{H}}}{\mathbf{R}_{J-1}} \ldots \mel{\mathbf{R}_{1}}{e^{-\tau \hat{H}}}{\mathbf{R}_{0}},
\end{align}
where it satisfies the normalization condition $\int\dd{\config}\pi(\config)=1$ and $\mathcal{W}(\config)$ is called the weight of a configuration $\config$. Hence, we write the expectation value of the observable $\oper$ as
\begin{align} \label{eq: expval mc integral}
     \expval*{\hat{O}}=\int \dd{\config} \pi(\config) \mathcal{Q}(\config) \qq{with}\mathcal{Q}(\config) =\frac{\mel{\mathbf{R}_{1}}{e^{-\tau \hat{H}}\oper}{\mathbf{R}_{0}}}{\mel{\mathbf{R}_{1}}{e^{-\tau \hat{H}}}{\mathbf{R}_{0}}},
\end{align}
where $\mathcal{Q}(\config)$ is called the \emph{estimator} for observable $\oper$ associated with the configuration $\config$. The \gls{pdf} $\pi(\config)$ and estimator $\mathcal{Q}(\config)$ are worked out in sections \ref{sec: stp} and \ref{sec: observable}, respectively. Nevertheless, evaluating these identities for all configurations would be unfeasible. This can be solved using Monte Carlo integration.

\subsection{Monte Carlo integration} \label{sec: mc}
Consider a function $f$ from $\mathbb{R}^d$ to $\mathbb{R}$ where $d\in \mathbb{Z}$ and the integral
\begin{align}
I=\int_{\mathbb{R}^{d}} f(\vb{x}) \dd{\vb{x}}.
\end{align}
The integral is often difficult to compute in high-dimensional integration spaces. A common way is to divide the integration space into small cells, each with its own evaluation point $\vb{x}_i$, and then sum the value of the function at these locations multiplied by thee volume of the cells $\dd{\vb{x}_i}$, i.e. $I \approx \sum_i f(\vb{x}_i) \dd{\vb{x}_i}$. Nevertheless, the computational cost scales exponentially with the dimension $d$. For equation \ref{eq: expval mc integral} the dimension of integration space, i.e., the configuration space, $D=dNJ$ becomes very large if one wants to find an accurate small-time propagator $\rho(\mathbf{R}, \mathbf{R}^\prime; \tau)$ and thus taking $J\gg 1$.

One of the methods to prevail the so-called curse of dimensionality is Monte Carlo integration, which samples the configuration stochastically. The key reason for the above method's failure is that it treats every part of the configuration on the same footing. Conversely, Monte Carlo integration uses importance sampling, where the configuration with a large contribution has a higher chance of being generated than the samples with a low contribution. More concretely, suppose the stochastic variable $X$ which lies in $\mathbb{R^D}$ distributed by a \gls{pdf} $\pi$, one writes
\begin{align} \label{eq: importance sampling}
    I=\int_{\mathbb{R}^{D}} \frac{f(x)}{\pi(x)} \pi(x) \dd{x}=\langle\mathcal{I}(X)\rangle_{\pi} \approx\frac{\mathcal{I}\left(X_{1}\right)+\mathcal{I}\left(X_{2}\right)+\ldots+\mathcal{I}\left(X_{n}\right)}{n},
\end{align}
where $\mathcal{I}(X)= f(X)/\pi(X)$. The outcome of the integral is thus the expectation value of $\mathcal{I}(x)$ under the \gls{pdf} $\pi$. Consequently, because of the law of large numbers, the integral can be approximated by a large set of independent samples $X_1, X_2,\ldots, X_n$ as shown in \eqref{eq: importance sampling}. The central limit theorem tells us that the error will be, $\varepsilon_I=\sigma_I/\sqrt{n}$ where $\sigma_I$ is the standard deviation of $\mathcal{I}(x)$. In section \ref{sec: error} a more in-depth analysis of the error of \gls{PIMC} will be given.

If one makes the identification with equation \ref{eq: expval mc integral} obtained in the previous section, we find that the expectation value can be approximated by
\begin{align} \label{eq: estimator average}
\expval*{\oper}\approx \frac{\mathcal{Q}\left(\config_{1}\right)+\ldots+\mathcal{Q}\left(\config_{n}\right)}{n},
\end{align}
where $\config_{1}, \ldots,\config_n$ are sampled out of \gls{pdf} $\pi$ given by \eqref{eq: def weight}. However, the direct sampling of the configuration $\config$ out of a \gls{pdf} as complex as $\pi$ can be a hard task. The solution is sampled with the help of a Markov chain, which will be discussed in the next section.

\subsection{Markov Chain Monte Carlo} \label{sec: mcmc}
A \emph{Markov chain} is a discrete stochastic sequence, in our case, of configurations that starts from an initial configuration $\config$ and generates new configurations with the help of a set of updates, each with their own transition probability $\mathcal{T}(\config\to\config^\prime)$. The result is a chain of configurations
\begin{align}
\config_{1} \rightarrow \config_{2} \rightarrow \ldots \rightarrow \config_{k} \rightarrow \ldots\ .
\end{align}
Each transition is independent of the history of the chain, i.e., they satisfy \emph{Markov's property}, and the normalization condition $\int \dd{\config^\prime} \mathcal{T}(\config\to\config^\prime)=1$. The goal is to generate a Markov chain where the \emph{equilibrium distribution} of the configurations in the chain is the \gls{pdf} $\pi$ given by equation \ref{eq: def weight}. To achieve this, the Markov chain performs a random walk in the configuration space that is attracted to configurations with high contributions to the integral \eqref{eq: expval mc integral}.

For the walker to produce a stationary distribution, one usually requires two conditions: \emph{detailed balance} and \emph{ergodicity}. The former states that the probability of being in a configuration $\config$ times the transition probability of $\config\to\config^\prime$ is equal to the probability of being in a configuration $\config^\prime$ times the transition probability of $\config^\prime\to\config$, i.e.,
\begin{align}\label{eq: detailed balance equation}
\pi(\config) \mathcal{T}\left(\config \rightarrow \config^{\prime}\right)=\pi\left(\config^{\prime}\right) \mathcal{T}\left(\config^{\prime} \rightarrow \config\right).
\end{align}
In other words, each transition is reversible. The latter condition states that given two arbitrary configurations $\config$ and $\config^\prime$ there always exists a path or sequence of transitions with a combined non-zero transition probability connecting them. It ensures that the whole configuration space can be sampled. Note that detailed balance and ergodicity together are sufficient for the Markov chain to generate a stationary distribution, but not necessary. There are also walkers which do not satisfy detail balance and produce stationary distributions that correspond with the desired distribution $\pi$. A practical example is given in \citet{Bernard2009}.

The walker of interest for our purposes is the \emph{Metropolis-Hastings algorithm}. It is constructed in such a way that it satisfies the above two conditions, and follows two steps:
\begin{enumerate}[(i)]
    \item \label{itm: propose new el MC}
    First, propose a move or update which modifies the last element of the Markov chain such that a new configuration $\config^\prime$ is offered with \emph{proposal probability} $\mathcal{P}\left(\config_i \rightarrow \config^{\prime}\right)$. How this update is constructed is discussed in section \ref{sec: update}.

    \item \label{itm: eval new el MC}
    Afterwards, the new configuration $\config^\prime$ is evaluated to check whether we want to accept or reject the move, determined by the \emph{acceptance probability}
    \begin{equation}
        \mathcal{A}\left(\config_i \rightarrow \config^{\prime}\right)=\min \left(1, \frac{\pi\left(\config^{\prime}\right) \mathcal{P}\left(\config^{\prime} \rightarrow \config_i\right)}{\pi(\config_i) \mathcal{P}\left(\config_i \rightarrow \config^{\prime}\right)}\right),
        \label{eq: acceptance probability}
    \end{equation}
    such that the \emph{transition probability} $\mathcal{T}(\config_t\to\config^\prime)=\mathcal{P}\left(\config_i \rightarrow \config^{\prime}\right)\mathcal{A}\left(\config_i \rightarrow \config^{\prime}\right)$. If it is accepted, one sets $\config_{i+1}=\config^\prime$, otherwise, $\config_{i+1}=\config_i$.
\end{enumerate}
Together, the steps \ref{itm: propose new el MC} and \ref{itm: eval new el MC} lead to a new configuration. The acceptance probability \eqref{eq: acceptance probability} can easily be checked to satisfy the detailed balance equation by substituting it into equation \ref{eq: detailed balance equation} and considering the two cases $\pi\left(\config^{\prime}\right) \mathcal{P}\left(\config^{\prime} \rightarrow \config_i\right) < \pi(\config_i) \mathcal{P}\left(\config_i \rightarrow \config^{\prime}\right)$ and $\pi(\config_i) \mathcal{P}\left(\config_i \rightarrow \config^{\prime}\right) < \pi\left(\config^{\prime}\right) \mathcal{P}\left(\config^{\prime} \rightarrow \config_i\right)$\footnote{Note that the partition function $Z$ is the same for both configuration $\config^{\prime}$ and $\config_i$, and hence, thereby cancels.}.

\subsection{PIMC Algorithm}
In summary, a common \gls{PIMC} simulation is formulated as follows.
\begin{enumerate}[(i)]
    \item By using the detailed balance equation \eqref{eq: detailed balance equation} an ergodic set of updates can be derived, each with its own proposal probability $\mathcal{P}$ (see section \ref{sec: update}). Note that the ergodicity of the set is system dependent.
    \item Compute an (approximated) expression for the density matrix with the help of the Hamiltonian $\ham$ which yields an exact result in the limit $\tau\to0$ (see section \ref{sec: stp}).
    \item Start from an arbitrary initial configuration $\config_k$ with $k=0$ that respect the system parameters set by the Hamiltonian $\ham$.
    \item  Create a new configuration $\config^\prime$ with a random update from the set constructed in (i) and evaluate the acceptance probability $\mathcal{A}\left(\config_k \rightarrow \config^{\prime}\right)$ given by equation \ref{eq: acceptance probability}.
    \item If the acceptance probability $\mathcal{A}\left(\config_k \rightarrow \config^{\prime}\right)$ is larger than a random number $a \in [0 ,1)$, the new configuration is added to the chain, i.e., $\config_{k+1}=\config^\prime$, else the initial configuration is added to the chain, or, $\config_{k+1}=\config_k$.
    \item Last, the estimators $\mathcal{Q}$ of the desired collection of observables $\{ \hat{O}_i \}$ (see section \ref{sec: observable}), defined in \eqref{eq: expval mc integral}, are evaluated for the added member of the chain $\config_{k+1}$.
    \item Set $k$ to $k+1$.
\end{enumerate}
This process of steps (iii-vii) is then repeated until enough Markov chain entries are collected. In the case of a rejection, the old configuration must be still counted as a member of the chain. Including only accepted configurations leads to wrong results.

\begin{table}[!ht]
    \centering
    \caption{Jargon regarding the Path integral Monte Carlo simulation.}
    \label{tab: jargon}
    \begin{tabular}{m{0.33\textwidth} >{\centering\arraybackslash}m{0.17\textwidth} m{0.40\textwidth}}
    \thickhline\xrowht[()]{15pt}
    Bead & $\mathbf{r}_{k, j}$ & a single coordinate of the $k$-th particle at the $j$-th imaginary time index \\
    \xrowht[()]{20pt}
    Time slice & $\mathbf{R}_{j}$ & a set of beads at the $j$-th imaginary time index \\
    \xrowht[()]{20pt}
    Configuration & $\left\{\mathbf{R}_{0}, \cdots, \mathbf{R}_{J}\right\}$ & the set of all time slices which as a whole defines the configuration\\
    \xrowht[()]{20pt}
    Link & $\rho\left(\mathbf{R}_{j}, \mathbf{R}_{j+1} ;  \tau\right)$ & the propagator connecting two consecutive time slices \\
    \xrowht[()]{20pt}
    Single-particle link & $\rho\left(\mathbf{r}_{k, j}, \mathbf{r}_{k, j+1} ;  \tau\right)$ & the propagator connecting two consecutive beads \\
    \xrowht[()]{20pt}
    Weight function & $\mathcal{W}(\config)$ & the weight of configuration $\config$ \\
    \xrowht[()]{20pt}
    Probability density function & $\pi(\config)$ & the normalized probability to be in configurations $\config$ \\
    \xrowht[()]{20pt}
    Estimator function  & $\mathcal{Q}(\config)$ &  the estimator of observable to be evaluated for each configuration $\config$ in the Markov chain \\
    \xrowht[()]{20pt}
    Transition probability & $\mathcal{T}\left(\config \rightarrow \config^{\prime}\right)$ & the probability a configuration $\config$ transitions to configuration $\config^\prime$ \\
    \xrowht[()]{20pt}
    Proposal probability & $\mathcal{P}\left(\config \rightarrow \config^{\prime}\right)$ & the probability a configuration $\config^\prime$ is proposed when the system is in $\config$ \\
    \xrowht[()]{20pt}
    Acceptance probability & $\mathcal{A}\left(\config \rightarrow \config^{\prime}\right)$ & the probability a configuration $\config^\prime$ is accepted when the system is in $\config$ \\
    \thickhline
\end{tabular}
\end{table}

\subsubsection{Comparison with other methods}
The last part of this section I would like to devote to how \gls{PIMC}\footnote{We use \gls{PIMC} to refer to the core method of the simulation, as \gls{WA} only adds additional updates.} compares to other methods and why it is the method of choice for our purposes. In doing this, \citet{Ceperley1995, YanBlume2017, king_scaling_2021, Pollet2012} were considered. 

First and foremost, \gls{PIMC} is used by a lot of others to simulate these ultracold atoms in optical lattices \cite{Cinti2014,Pascual2021,Carleo2013,Gauter2021}\footnote{Which is a priori, of course, not a strong argument on why to use \gls{PIMC}.}. It is well accepted as the standard in the field for computing the accurate estimate of finite-temperature equilibrium statistics with dimensions higher than one \cite{king_scaling_2021}. That said, the main reason why it has this status is that it is completely unbiased. The only input \gls{PIMC} needs is a microscopic Hamiltonian and does not require any a priori assumptions, i.e., it is an ab-initio simulation. By contrast, comparable methods such as Path Integral Ground State (\gls{PIGS}) \cite{YanBlume2017}, Variational Monte Carlo (\gls{VMC}), Green's-function Monte Carlo (\gls{GFMC}) and Diffusion Monte Carlo (\gls{DMC}) need a trail wavefunction to start from. However, in general, when humans appreciate the outcome of a simulation, they tend to not look deeper into the result. Hence, a systematic inaccuracy is unavoidably introduced in the above-mentioned methods. Moreover, for complicated systems, a good quality trail function has far too many potential parameters, making it difficult to optimize properly. One may refer to \gls{PIMC} as \emph{numerical exact} for boson systems, which means that all the approximations required for the fact that it is a numerical technique can be decreased arbitrarily as long as sufficient computing is performed. Although one has to remember that for fermions, one encounters the Fermi sign problem. Another advantage of \gls{PIMC} is that observables related to particle statistics are easier expressed with comparison to other methods \cite{Ceperley1995}.

The disadvantage of \gls{PIMC} is that it is quite involved compared to other methods such as \gls{VMC}, both in structure and computationally. \gls{VMC} is simple to comprehend and to program. As a result, for a system with for example a lot of symmetry, it can be preferred. In addition, the particle statistics remain difficult for \gls{WAPIMC}. Sampling worldlines with many-particle permutations still scales unfavourably with the number of particles (see section \ref{sec: Worms}). Although \gls{WA} greatly improved the problem, it remains the main obstacle even when one uses \gls{WA} \cite{Pollet2012}. Also, note that \gls{WAPIMC} does not have the ability to measure strictly zero-temperature systems. Nevertheless, in the context of comparing with experimental data, this is no problem.

A similar formulation of \gls{PIMC} exists, where one restricts the spatial location of the particles to the vertices of the lattice. In doing this, one can formulate the algorithm such that there is no time step error. However, in practice, the time step error does not seem to be the main bottleneck of the former, i.e., the more ab-initio continuous space \gls{PIMC} variant.

\section{Small-time propagator} \label{sec: stp}
One obstacle with working in Feynman's path integral formulation is that one has an expression in terms of the many-body density matrix. However, this object is unknown in exact form for any non-trivial system. Hence, one has to find good approximations. Depending on the system at hand, one has to construct clever approximations which are valid in the correct regime. For ultracold atoms in optical lattices, the Hamiltonian is written as follows:
\begin{align}\label{eq: system Hamiltonian}
    \ham = \sum_{i=1}^{N} -\lambda\hat{\nabla_{\mathrm{i}}}^{2}+\sum_{i=1}^{N} V\left(\hat{\mathbf{r}}_{i}\right)+\sum^{N-1}_{i=1}\sum^{N}_{k=i+1} U\left(\hat{\mathbf{r}}_{i}-\hat{\mathbf{r}}_{k}\right),
\end{align}
where we defined $\lambda=\frac{\hbar^2}{2 m}$ for convenience. The first term on the right side of the equality represents the kinetic energy contribution, V the optical landscape or external potential at a given position, and U the interaction between the relative positions of two particles $i$ and $k$, i.e., $\hat{\vb{r}}_{ik} = \hat{\mathbf{r}}_{i}-\hat{\mathbf{r}}_{k}$. As discussed in chapter \ref{ch: Ultracold Atoms In Optical Lattices}, ultracold gases are often diluted gases. As the particles rarely collide, one can consider only the two-body interaction which is done in the last term of the Hamiltonian. This is the first approximation we make, nevertheless, a good experimentally confirmed one \cite{Bloch2008}.

Considering that the interactions only depend on the relative position of the particles, it is favourable to extract the relative Hamiltonian of the reduced single particle from equation \eqref{eq: system Hamiltonian}, i.e., $\ham^{\mathrm{rel}} = \ham_{0\mathrm{,rel}}(\hat{\vb{r}}_{ik})+ U(\hat{\vb{r}}_{ik})$ with the non-interacting reduced Hamiltonian $\ham_{0\mathrm{,rel}}(\hat{\vb{r}}_{ik})=\hbar^2\nabla^2_{ik}/2m^*$ and reduced mass $m^* = m/2$. Splitting the Hamiltonian in this way, we approximate the density matrix using the \emph{pair product approximation} \cite{Ceperley1995,Yan2015,Barker1979} given by
\begin{align}\label{eq: pair product approx}
    \rho\left(\mathbf{R}, \mathbf{R}^{\prime}; \tau\right) \approx \left(\prod_{i=1}^{N} \rho_{1}\left(\mathbf{r}_{i}, \mathbf{r}_{i}^{\prime}, \tau\right)\right)\left( \prod^{N-1}_{i=i}\prod_{k=i+1}^{N} \frac{\rho^{\mathrm{rel}}\left(\mathbf{r}_{i k}, \mathbf{r}_{i k}^{\prime}; \tau\right)}{\rho_{0}^{\mathrm{rel}}\left(\mathbf{r}_{i k}, \mathbf{r}_{i k}^{\prime}; \tau\right)}\right)+\mathcal{O}(\tau^3),
\end{align}
where $\rho_{1}$ is the \emph{single-particle density matrix} of the Hamiltonian $\ham_1$, $\rho^{\mathrm{rel}}$ is the \emph{two-body density matrix} of the interacting system $\ham^{\mathrm{rel}}$, and $\rho_0^{\mathrm{rel}}$ the two-body density matrix of the non-interacting part. The form of the relative contribution comes from the use of the Feynman-Kac formula \cite{FeyKac1949}. The exact derivation of the Feynman-Kac formula and pair product approximation can be found in \citet{militzer2000path}. In the following two subsections, I would like to dissect and compute each part of the above expression.

\subsubsection*{Non-interacting propagator} \label{sec: n-int prop}
The non-interacting single-particle density matrix $\rho_1$ is handled by applying the primitive approximation as done in section \ref{sec: pimc}. However, instead of using \gls{BCH} only ones, you can apply it recursively to generate higher-order corrections, e.g.
\begin{align} \label{eq: second order trotter}
    e^{-\tau(\hat{A}+\hat{B})}=e^{-\frac{1}{2} \tau \hat{B}} e^{-\tau \hat{A}} e^{-\frac{1}{2} \tau \hat{B}}+O\left(\tau^{3}\right).
\end{align}
This scheme of computing higher order is called \emph{Trotter-Suzuki factorization} \cite{suzuki_generalized_1976}. In practice higher orders of Trotter-Suzuki factorization are used, nevertheless, the second order is sufficient to illustrate our purposes.
Applying equation \ref{eq: second order trotter} to the non-interacting single particle density matrix $\rho_1$, i.e. setting $\hat{A} = \ham_0$ and $\hat{B} = V$, we can write
\begin{align} \label{eq: rho_1}
     \rho_{1}\left(\mathbf{R}, \mathbf{R}^{\prime}; \tau\right)=e^{-\frac{\tau}{2}\sum^N_{i=1}\left(V\left(\mathbf{r}_{i}\right)+V\left(\mathbf{r}_{i}^{\prime}\right)\right)}\mel{\mathbf{R}}{e^{-\tau \hat{H}_0}}{\mathbf{R}^{\prime}}.
\end{align}
The only remaining unknown is the free particle propagator $\rho_{0}\left(\mathbf{R}, \mathbf{R}^{\prime}; \tau\right)=\mel{\mathbf{R}}{e^{-\tau \hat{H}_0}}{\mathbf{R}^{\prime}}$. Nevertheless, the free part of the Hamiltonian is $\hat{H}_0$ not diagonal in the spatial basis. This can be resolved by decomposing the position eigenstates in terms of the free particle plane waves or momentum eigenstates. More concretely, we write
\begin{align}
    \rho_{0}\left(\mathbf{R}, \mathbf{R}^{\prime}; \tau\right)
    &= \prod^N_{i=1} \int \frac{\dd{\vb{k}_{i}} }{(2 \pi)^{d}} \int \frac{\dd{\vb{k}_{i}^{\prime}}}{(2 \pi)^{d}} \mathrm{e}^{-i \vb{k}_{i} \cdot \vb{r}_{i}} \mathrm{e}^{i \vb{k}_{i}^{\prime} \cdot \vb{r}_{i}^{\prime}}
    \mel{\vb{k}_{1}, \cdots, \vb{k}_{N}}{\mathrm{e}^{-\tau \lambda \sum_{j=1}^{N} \hat{\nabla}_{j}^{2}}}{\vb{k}_{1}^{\prime}, \cdots, \vb{k}_{N}^{\prime}},  \\ \label{eq: derivation free particle propagator}
    &= \prod^N_{i=1} \int \frac{\dd{\vb{k}_{i}}}{(2 \pi)^{d}} \int \frac{\dd{\vb{k}_{i}^{\prime}}}{(2 \pi)^{d}} \mathrm{e}^{-\lambda \tau\left|\vb{k}_{i}^{\prime}\right|^{2}-i \vb{k}_{i} \cdot \vb{r}_{i}+i \vb{k}_{i}^{\prime} \cdot \vb{r}_{i}^{\prime}}
    \braket{\vb{k}_{1}, \cdots, \vb{k}_{N}}{\vb{k}_{1}^{\prime}, \cdots, \vb{k}_{N}^{\prime}}, \\
    &= \prod^N_{i=1} \int \frac{d \vb{k}_{i}}{(2 \pi)^{d}} \mathrm{e}^{-\lambda \tau\left|\vb{k}_{i}\right|^{2}+i \vb{k}_{i} \cdot\left(\vb{r}_{i}^{\prime}-\vb{r}_{i}\right)}, \\
    &=(4 \pi \lambda \tau)^{-N d / 2}\mathrm{e}^{-\frac{1}{4 \lambda \tau} \sum_{i=1}^{N}\left(\vb{r}_{i}-\vb{r}_{i}^{\prime}\right)^{2}}.
\end{align}
The result is the well-known expression for the free one particle propagator in $d$ dimensions,
\begin{align} \label{eq: free particle propagator}
    \rho_{0}\left(\mathbf{r}_{i}, \mathbf{r}_{i}^{\prime}, \tau\right)
    =\frac{1}{\lambda_{\tau}^{d}} \exp \left(-\frac{\pi}{\lambda_{\tau}^{2}} \sum_{i=1}^{N}\left(\mathbf{r}_{i}-\mathbf{r}^\prime_{i}\right)^{2}\right),
\end{align}
where $\lambda_\tau = \sqrt{4\pi\lambda\tau}$ is the de Broglie wavelength at the temperature set by imaginary time step $\tau$. Substituting this in the non-interacting single particle density matrix \eqref{eq: rho_1}, we find
\begin{align} \label{eq: rho_1 complete}
    \rho_{1}\left(\mathbf{R}, \mathbf{R}^{\prime}; \tau\right)=\frac{1}{\lambda_{\tau}^{dN}} \exp \left(-\frac{\pi}{\lambda_{\tau}^{2}} \sum_{i=1}^{N}\left(\mathbf{r}_{i}-\mathbf{r}^\prime_{i}\right)^{2}-\frac{\tau}{2}\sum^N_{i=1}\left(V\left(\mathbf{r}_{i}^{\prime}\right)+V\left(\mathbf{r}_{i}\right)\right)\right).
\end{align}
Examining the result, we can already infer two findings. On the one hand, smooth and straight paths generally have a higher probability of being generated. On the other hand, paths with high potential energy have a low probability of occurring.

There is a small detail I left out until now. Often, one wants to work with periodic boundary conditions. This is ideal if one wants to study the bulk properties of systems featuring infinite spatial extension. In addition, some physical observables, such as superfluid fractions, are easier to measure if one applies periodic boundary conditions to the simulation box. However, in enforcing this, the size of the box becomes a key element in the numerical approach, necessitating cautious extrapolation of all results. Consequently, to write the momentum eigenstate in equation \ref{eq: derivation free particle propagator} over an integral, one must assume that the wavelength of one imaginary time step $\tau$ is much smaller than the size of the box with volume $V=L^d$, i.e., \cite{Ceperley1995,militzer2000path,Spada2022}
\begin{align}
    \lambda_{\beta} \ll L.
\end{align}
Not respecting this condition will lead to biased results. Recently, a new formulation of \gls{WAPIMC} has been published by \citet{Spada2022} which does not have such a limitation\footnote{Unfortunately, this was published way too late into my thesis, but shows how relevant \gls{WAPIMC} still is.}.

\subsubsection*{Interaction propagator} \label{sec: int prop}
The interaction part is completely described by the second factor in the expression \eqref{eq: pair product approx} where we have that the numerator and denominator, respectively, are given by
\begin{align}
    \rho^{\mathrm{rel}}\left(\mathbf{r}_{i k}, \mathbf{r}_{i k}^{\prime}; \tau\right)=\mel{\mathbf{r}_{i k}^{\prime}}{\mathrm{e}^{-\tau \hat{H}^{\mathrm{rel}}}}{\mathbf{r}_{i k}}
    \qq{and}
    \rho^{\mathrm{rel}}_0\left(\mathbf{r}_{i k}, \mathbf{r}_{i k}^{\prime}; \tau\right)=\mel{\mathbf{r}_{i k}^{\prime}}{\mathrm{e}^{-\tau \hat{H}^{\mathrm{rel}}_0}}{\mathbf{r}_{i k}}.
\end{align}
The latter, or the relative free density matrix $\rho^{\mathrm{rel}}_0$, will be the same as expression \eqref{eq: free particle propagator} but with the replacement $m\to m^*=m/2$. The former is often not that easy to find an expression for. The relative two-body density matrix for a specific interaction potential can be computed by solving the radial Schr\"odinger equation (see appendix \ref{app: rel int prop}). Nevertheless, often one has to perform a partial wave decomposition, which leads to a non-closed analytical expression and thereby must be numerically evaluated\footnote{There exist other numerical methods as well, such as the matrix squaring technique \cite{Ceperley1995}.}. However, in some cases, the relative interaction propagator can be found in a closed analytical form, e.g., for the one-dimensional delta or contact interaction potential $U= g\delta\left(\mathbf{r}_{i j}\right)$ with $g$ an interaction coupling constant \citet{Yan2015} computed
\begin{align}
\rho^{\mathrm{rel}}\left(\mathbf{r}_{ik}, \mathbf{r}_{i k}^{\prime}; \tau\right)=1-\exp \left(-\frac{m^*\left(\mathbf{r}_{i k} \mathbf{r}_{i k}^{\prime}+\left|\mathbf{r}_{i k} \mathbf{r}_{i k}^{\prime}\right|\right)}{\tau \hbar^{2}}\right) \times \sqrt{\frac{\pi m^* \tau}{2}} \frac{g}{\hbar} \operatorname{erfc}(u) \exp (u),
\end{align}
with $ u=m^* \left(r_{i k}+r_{i k}^{\prime}+g \tau\right) / \sqrt{2 m^* \tau \hbar^{2}}$ and $\erfc(.)$ is the complementary error function. Another, noticeable example is by \citet{CaoBerne1992} which found a three-dimensional closed analytical approximation for the hard-sphere potential
\begin{align}
    \frac{\rho^{\mathrm{rel}}\left(\mathbf{r}_{i k}, \mathbf{r}_{i k}^{\prime}; \tau\right)}{\rho^{\mathrm{rel}}_{0}\left(\mathbf{r}_{i k}, \mathbf{r}_{i k}^{\prime}; \tau\right)}=1-\frac{a\left(r_{ik}+r_{ik}^{\prime}\right)-a^{2}}{r_{ik} r_{ik}^{\prime}} e^{-\left[r_{ik} r_{ik}^{\prime}+a^{2}-a\left(r_{ik}+r_{ik}^{\prime}\right)\right](1+\cos \theta) m^* /\left(2 h^{2} \tau\right)},
\end{align}
with $a$ the cut-off (equivalent to the scattering length) of the hard-sphere potential and $\theta$ the angle between $\vb{r}_{ik}$ and $\vb{r}^\prime_{ik}$. To our knowledge, no two-dimensional closed analytical form exist for any non-trivial interaction potential. In section \ref{sec: derivation int prop} we shall derive non-closed approximation which can be numerically evaluated.


\section{Monte Carlo Updates} \label{sec: update}
In this section, we will construct the updates or moves to generate new entries in the Markov chain. This is done by modifying the worldlines in the configuration. One important thing to keep in mind is that we have to guarantee the ergodicity of the updates, i.e., the whole configuration space must be accessible starting from any configuration. In addition, any update must respect the detailed balance equation \eqref{eq: detailed balance equation}.

To maintain the high acceptance rate and, thereby, make the move more efficient, we should suggest new configurations in a way that the ratio in the acceptance probability $\mathcal{A}$, i.e., equation \ref{eq: acceptance probability}, is around unity. One way of doing this is to employ local updates, which means we only change a small portion of the configuration such that the improbable configurations are avoided. It is important to note that the efficiency of an update does not affect the converged distribution, i.e., the obtained physics. It only affects the speed of the convergence.

Let us start from a configuration $\config$. A simple but naive example for an update would be moving a bead of worldline $i$ at time slice $j$ inside a $d$-dimensional Ball $\mathcal{B}(\vb{r}_{i, j},\Delta)$ with radius $\Delta$ to a position $\vb{r}^\prime_{i, j}$.\footnote{Note that the interval must symmetric, as otherwise the detail balance is violated.} The exact distance can be uniformly generated from the cube. However, as discussed in section \ref{sec: n-int prop}, in the case one works with periodic boundary conditions, there can exist ambiguity for the periodic image of the new position if $\Delta>L / 2$ with $L$ being the length of the sides of the simulation box. Hence, one must ensure $\Delta<L / 2$.

The proposal probability $\mathcal{P}\left(\config \rightarrow \config^{\prime}\right)$ to start from configuration $\config$ and generate this new configuration $\config^\prime$ is than equal to starting form $\config^\prime$ and going to $\config$. Hence, the acceptance probability defined in expression \eqref{eq: acceptance probability} is given by
\begin{align}
    A(\mathbf{X} \rightarrow \tilde{\mathbf{X}})
    =\min \left(1, e^{\tau \Delta S(\mathbf{X} \rightarrow \tilde{\mathbf{X}})}
    \frac{\rho_{0}\left(\mathbf{r}_{i, j-1}, \mathbf{r}^\prime_{i, j}; \tau\right)\rho_{0}\left(\mathbf{r}^\prime_{i, j}, \mathbf{r}_{i, j+1}; \tau\right)}{\rho_{0}\left(\mathbf{r}_{i, j-1}, \mathbf{r}_{i, j}; \tau\right)\rho_{0}\left(\mathbf{r}_{i, j}, \mathbf{r}_{i, j+1}; \tau\right)} \right).
\end{align}
Because of the $\beta$-periodicity we assume, without sacrificing generality, throughout this thesis that whenever we write the $j+m$ time slice with $m\in\mathbb{Z}$, the enumeration happens in modular $J$, i.e, $j+m\equiv j+m \mod J$. Furthermore, notice, that the expression \eqref{eq: acceptance probability} is greatly reduced, as only the affected time slices of the update contribute to the probability. Nevertheless, the average acceptance probability will be low as for large distances the kinetic free density matrices will vanish when the distance between $\mathbf{r}_{i, j-1}$ or $\mathbf{r}_{i, j+1}$ and $\mathbf{r}^\prime_{i, j}$ will become too large.

To make the updates more efficient, we make use of the fact that the proposal probability $\mathcal{P}$ can be chosen freely. We can construct an update such that the proposal probability $\mathcal{P}$ cancels factors in the acceptance probability $\mathcal{A}$ to achieve unity for both the update and the inverse. In other words, we perform \emph{importance sampling}, to generate a configuration with a high acceptance probability.

For example, to expand on the above update, we could generate the new position $r^\prime_{i, j}$ out of the kinetic propagators, such that the proposal probability is given by $\mathcal{P}\left(\config \rightarrow \config^{\prime}\right)=\tfrac{1}{JN}\rho_{0}\left(\mathbf{r}_{i, j-1}, \mathbf{r}^\prime_{i, j}; \tau\right)\rho_{0}\left(\mathbf{r}^\prime_{i, j}, \mathbf{r}_{i, j+1}; \tau\right)$ and the acceptance probability becomes\footnote{The reverse proposal probability is given by $\mathcal{P}\left(\config^\prime \rightarrow \config\right)=\tfrac{1}{JN}\rho_{0}\left(\mathbf{r}_{i, j-1}, \mathbf{r}_{i, j}; \tau\right)\rho_{0}\left(\mathbf{r}_{i, j}, \mathbf{r}_{i, j+1}; \tau\right)$.}
\begin{align}
    A(\config \to \config^\prime)
    =\min \left(1, e^{\tau \Delta S(\config \to \config^\prime)} \right).
\end{align}
The $\tfrac{1}{JN}$ comes from selecting a bead to move. This is possible because the kinetic propagator can be seen as a Gaussian distribution. There are only a couple of probability distributions where one can sample directly, and the Gaussian is one of them. The remaining factor in the above expression cannot be removed, as one cannot simply sample a new configuration out of it.

There is one last problem to consider. Single time slice moves are extremely insufficient. Not because the acceptance probability is small, but because the newly subsequent generated configurations are so similar that the configuration space is explored too slowly. \citet{Ceperley1995} found that the amount of computation time required to modify the overall form of a single path scales as $J^3$. This will be addressed in the next subsections, where we formulate the official updates used in our \gls{WAPIMC} code.

\subsection{Canonical updates} \label{sec: Can update}
\subsubsection*{Center-of-mass update} \label{sec: COM}
Instead of moving individual beads, one could just move a whole worldline by a distance $\vb{r}^\prime$. Now, remember that the worldlines are permuted. Hence, one can not simply move one worldline, as the link connecting the last and first bead will become highly improbable, and thereby, the update gets easily rejected in the Metropolis-Hastings question. To solve this one has to move to the whole permutation cycle or polymer. The move is called the \emph{centre-of-mass update} as only the centre-of-mass of the polymer is moved --- the relative position of the bead on the polymer remains the same. This has the consequence that there will be no kinetic component in the acceptance probability. The move is very efficient if the average polymer size is less than the polymer distance or the temperature is higher than the degeneracy temperature. This is because there will be less overlap between the polymers. At low temperatures, it can be more efficient to only move one particle of the polymer.
\begin{figure}[htbp]
    \centering
    \includegraphics[width=0.6\textwidth]{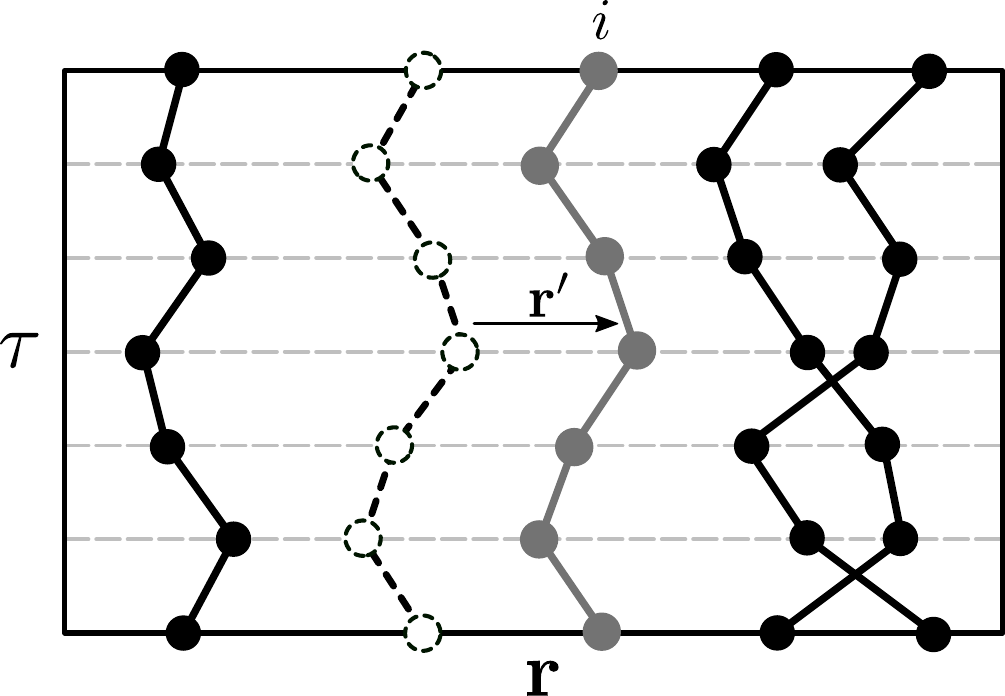}
    \caption{Illustration of the Center-of-mass update. A permutation cycle of worldlines or polymer gets translated by a vector $\vb{r}^\prime$. Notice the links between two consecutive beads are not modified. The unchanged worldlines are portrayed in black, the old beads and links are dashed, and the modified moved closed worldline $i$ is in solid grey.}
    \label{fig: com}
\end{figure}

More concretely, the centre-of-mass update goes as follows:
\begin{enumerate}[i)]
    \item Choose a single worldline $i$ uniformly from the last configuration in the Markov chain.
    \item Sample the polymer displacement $\vb{r}^\prime$ uniformly from the symmetric interval $\left[-\Delta, \Delta\right]^d$. $\Delta$ is a free parameter that can be optimized to have a desirable acceptance ratio of the centre-of-mass update during the simulation.
    \item Move the polymer by adding $\mathbf{r}^\prime$ to the position of all the beads and compute the weights of the old and new configurations.
    \item  Determine if the new configuration is accepted by the Metropolis-Hastings question with the acceptance probability
    $$
    A_{\text {Move }}(\config \rightarrow \config^\prime)=\min \left(1, e^{\tau \Delta S(\config \rightarrow \config^\prime)}\right).
    $$
\end{enumerate}

\subsubsection*{Reshape update} \label{sec: Wiggle}
\begin{figure}[htbp]
    \centering
    \includegraphics[width=0.8\textwidth]{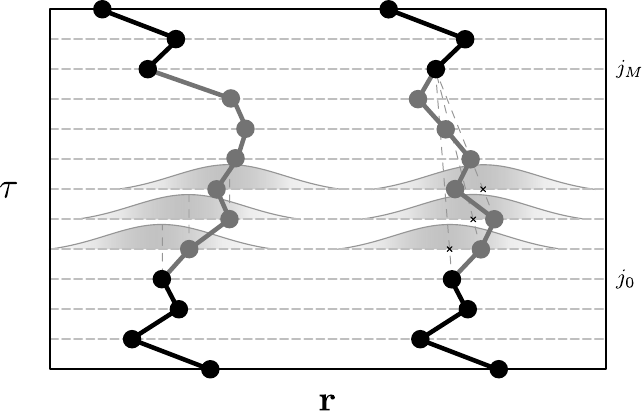}
    \caption{Illustration of the different reshape schemes. The left path is sampled using the proposal naive probability 
    whereas on the right the path is sampled taking the endpoint into account.
    The Gaussian distributions are shown for the first three beads of every Brownian bridge. The new paths are portrayed in solid grey.}
    \label{fig: ReshapeVSbissection}
\end{figure}

Let us go back to modifying individual beads. Instead of moving one individual bead of a polymer by sampling the new position out of the kinetic propagator $\rho_0$, we could modify $M-2$ beads of the polymer starting from a time slice $j_0$ and ending at $j_M$, with $M$ smaller than the total length of the selected polymer. This procedure of generating the path $\left(\mathbf{r}_{j_{0}}, \mathbf{r}_{j_{0}+1}^{\prime}, \ldots, \mathbf{r}_{j_{0}+M-1}^{\prime}, \mathbf{r}_{j_{M}}\right)$ is called generating a \emph{Brownian bridge} \cite{krauth2006} between $j_0$ and $j_M$. The proposal probability of the direct and reverse update is given by
\begin{align} \label{eq: naive proposal prop reshape}
    \mathcal{P}\left(\config \rightarrow \config^{\prime}\right)
    &=
    \frac{1}{JN}\rho_{0}\left(\mathbf{r}_{\mathrm{p}(j_0), j_0}, \mathbf{r}^\prime_{\mathrm{p}(j_0+1), j_0+1}; \tau\right)\cdots\rho_{0}\left(\mathbf{r}^\prime_{\mathrm{p}(j_M-1), j_M-1}, \mathbf{r}_{\mathrm{p}(j_M), j_M}; \tau\right),\\
    \mathcal{P}\left(\config^\prime \rightarrow \config\right)
    &=
    \frac{1}{JN}\rho_{0}\left(\mathbf{r}_{\mathrm{p}(j_0), j_0}, \mathbf{r}_{\mathrm{p}(j_0+1), j_0+1}; \tau\right)\cdots\rho_{0}\left(\mathbf{r}_{\mathrm{p}(j_M-1), j_M-1}, \mathbf{r}_{\mathrm{p}(j_M), j_M}; \tau\right),
\end{align}
with $\mathrm{p}(j)$ the worldline index of the particle in the polymer at time slice $j$. The resulting acceptance probability is then
\begin{align} \label{eq: acc prop reshape}
    A(\config \to \config^\prime)
    =\min \left(1, e^{\tau \Delta S(\config \to \config^\prime)} \right).
\end{align}
In figure \ref{fig: ReshapeVSbissection}, a visualization of such a procedure is shown in the left worldline. Notice the Gaussian distributions at every newly sampled bead centred around the previous bead. This is the Gaussian where the bead is sampled from. However, it only depends on the previous bead and is thus independent of the last bead $\mathbf{r}_{j_{M}}$ of the new path. This has the result that the last generated bead can be far from $\mathbf{r}_{j_{M}}$, which leads to a very small acceptance probability.

A solution would be to generate the Brownian bridge with the Levy construction \cite{krauth2006}, which takes the last bead $\mathbf{r}_{j_{M}}$ of the path into account. This adds up to complementing each factor in the proposal probability \eqref{eq: naive proposal prop reshape} with a kinetic propagator to the last bead $\mathbf{r}_{j_{M}}$, which yields
\begin{multline}
    \mathcal{P}\left(\config \rightarrow \config^{\prime}\right)
    = \frac{1}{JN}\frac{\rho_{0}\left(\mathbf{r}_{j_0}, \mathbf{r}^\prime_{j_0+1}, \tau\right) \rho_{0}\left(\mathbf{r}^\prime_{j_0+1}, \mathbf{r}_{j_M}, (M-1) \tau\right)}{\rho_{0}\left(\mathbf{r}_{j_0}, \mathbf{r}_{j_M},M \tau\right)} \times \\
    \hspace{1cm}\frac{\rho_{0}\left(\mathbf{r}^\prime_{j_0+1}, \mathbf{r}^\prime_{j_0+2}, \tau\right) \rho_{0}\left(\mathbf{r}^\prime_{j_0+2}, \mathbf{r}_{j_M},(M-2) \tau\right)}{\rho_{0}\left(\mathbf{r}^\prime_{j_0+1}, \mathbf{r}_{j_M}, (M-1) \tau\right)} \times \cdots \\
    \cdots \times \frac{\rho_{0}\left(\mathbf{r}^\prime_{j_M-2}, \mathbf{r}^\prime_{j_M-1}, \tau\right) \rho_{0}\left(\mathbf{r}^\prime_{j_M-1}, \mathbf{r}_{j_M}, \tau\right)}{\rho_{0}\left(\mathbf{r}^\prime_{j_M-2}, \mathbf{r}_{j_M}, 2 \tau\right)}.
\end{multline}
Here, we neglected the index notation for the worldlines to make the expression more readable. We can simplify the expression by cancelling some factors such that the direct and reverse proposal probabilities give
\begin{align} \label{eq: proposal prop reshape}
    \mathcal{P}_{\mathrm{Reshape}}\left(\config \rightarrow \config^{\prime}\right)
    &=
    \frac{1}{JN}\frac{\rho_{0}\left(\mathbf{r}_{\mathrm{p}(j_0), j_0}, \mathbf{r}^\prime_{\mathrm{p}(j_0+1), j_0+1}, \tau\right)\cdots\rho_{0}\left(\mathbf{r}^\prime_{\mathrm{p}(j_M-1), j_M-1}, \mathbf{r}^\prime_{\mathrm{p}(j_M), j_M}, \tau\right)}{\rho_{0}\left(\mathbf{r}_{\mathrm{p}(j_0), j_0}, \mathbf{r}_{\mathrm{p}(j_M),j_M},M \tau\right)},
    \\
    \mathcal{P}_{\mathrm{Reshape}}\left(\config^\prime \rightarrow \config\right)
    &=
    \frac{1}{JN}\frac{\rho_{0}\left(\mathbf{r}_{\mathrm{p}(j_0), j_0}, \mathbf{r}_{\mathrm{p}(j_0+1), j_0+1}, \tau\right)\cdots\rho_{0}\left(\mathbf{r}_{\mathrm{p}(j_M-1), j_M-1}, \mathbf{r}_{\mathrm{p}(j_M), j_M}, \tau\right)}{\rho_{0}\left(\mathbf{r}_{\mathrm{p}(j_0), j_0}, \mathbf{r}_{\mathrm{p}(j_M),j_M},M \tau\right)},
\end{align}
and the acceptance probability of equation \ref{eq: acc prop reshape} remains the same. We call the move corresponding to the above proposal probabilities the \emph{reshape update}. In practice, a position $\vb{r}_k$ of a single bead in the path with $k\in[1, M]$ is sampled out of the Gaussian probability distribution per dimension
\begin{align}\label{eq: levy}
    P\left(r_k\right)=\frac{1}{\sqrt{2 \pi} \sigma_{k}} \exp \left(-\frac{\left({r}_k-{\xi}_{k}\right)^{2}}{2 {\sigma}_{k}^{2}}\right),
\end{align}
with the variance and mean, respectively, given by
\begin{align}
    {\sigma}_{k}=\sqrt{\frac{\alpha_{k}}{2 \pi}}\lambda_\tau=\sqrt{2\lambda\alpha_{k}\tau}\qq{and} {\xi}_{k}=\alpha_{k} {r}_{k-1}+\left(1-\alpha_{k}\right) {r}_{M+1}, \qq{where} \alpha_{k}=\frac{M+1-k}{M+2-k}.
\end{align}
In figure \ref{fig: ReshapeVSbissection}, the difference in the sampling is shown in the right worldline. Now, the Gaussian where the beads are sampled from is not centred around the previous bead any more. Instead, the intersection of the line between the previous bead and the last bead of the path with the horizontal line of the time slice serves as the mean of the Gaussian. The advantage being that the last bead of the path is taken into account, and thus large gaps, as in figure \ref{fig: ReshapeVSbissection} on the right, do not occur.

In conclusion, the reshape update is formulated as follows:
\begin{enumerate}[i)]
    \item Choose a single worldline uniformly from the last configuration in the Markov chain.
    \item Select the time slice $j_0\in[1, J]$ and number of beads to be reshaped $M\in[1, \bar{M}]$ uniformly. $\bar{M}$ is a free parameter to be optimized during the \gls{WAPIMC} simulation.
    \item Construct the new path from time slice $j_0$ until $j_M=j_0+M$ with the Levy construction.
    \item Determine if the new configuration is accepted by the Metropolis-Hastings question with the acceptance probability given by equation \ref{eq: acc prop reshape}.
\end{enumerate}


\subsubsection*{Permutation update} \label{sec: Permutation}
Until now, no update has been introduced which can sample configurations where permutations between worldlines are present. However, these permutations represent the quantum statistics of the system, which is a crucial ingredient of the physics of ensembles of indistinguishable particles. It underlies phenomena such as Bose-Einstein condensation and superfluidity --- both conditions of interest for our purposes.

\begin{figure}[htb]
    \centering
    \includegraphics[width=0.8\textwidth]{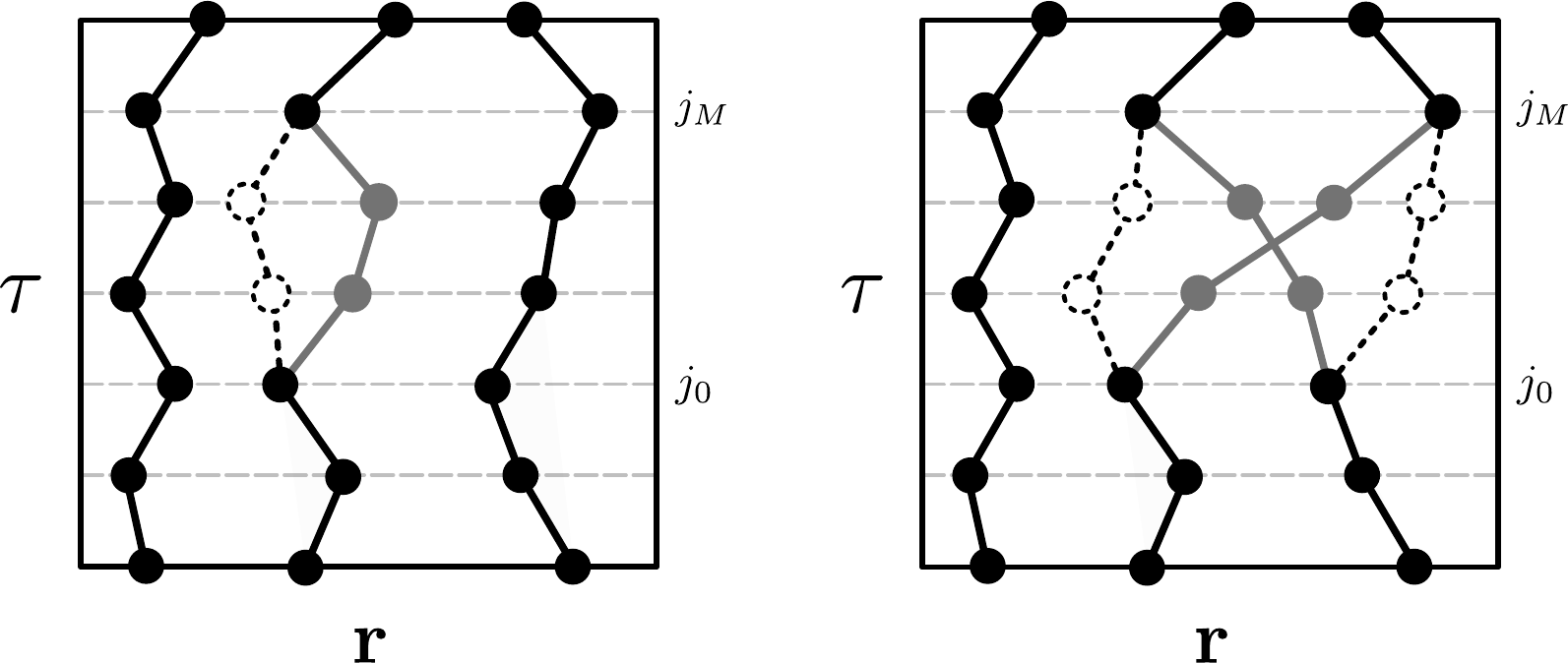}
    \caption{Illustration of the reshape \textcolor{red}{(a)} and permutation \textcolor{red}{(b)} update with the levy construction. The unchanged beads are portrayed in black, the old beads and links are dashed, and the new path or Brownian bridge is depicted in solid grey.}
    \label{fig: reshape and permute}
\end{figure}

With the reshape update in mind, one can easily think of a similar update that can permute the worldlines. Instead of creating a Levy construction between two beads of the same worldlines, one instead creates a bridge between beads of two different worldlines. A simple example and comparison with the reshape update is shown in figure \ref{fig: reshape and permute}. The same proposal probabilities analysis as in the reshape update can be worked out, which results in the same acceptance probability as in equation \ref{eq: acc prop reshape}.

Naturally, the update is likely to be rejected if the to-be connected beads of the corresponding two worldlines are far apart. Hence, one can again do importance sampling. Suppose we have selected a worldline $i_1$. We can sample a worldline $i_2$ in the neighbourhood of $i_1$ with the probability
\begin{align} \label{eq: sample particle permutation}
    p\left(i_{2}\right)=\frac{\rho_{0}\left(\mathbf{r}_{i_1, j_0}, \mathbf{r}_{i_2,j_M};M \tau\right)\rho_{0}\left(\mathbf{r}_{i_2, j_0}, \mathbf{r}_{i_1,j_M};M \tau\right)}{\sum_{i=1}^{N} \rho_{0}\left(\mathbf{r}_{i_1, j_0}, \mathbf{r}_{i,j_M};M \tau\right)\rho_{0}\left(\mathbf{r}_{i, j_0}, \mathbf{r}_{i_1,j_M};M \tau\right)}.
\end{align}
This way we guarantee that the distance $\abs{\mathbf{r}_{i_1, j_0}-\mathbf{r}_{i_2,j_M}}$ is of the order $\sqrt{M \tau}$, consequently, the permutation update is more probable to be accepted, and the sampling process can proceed more swiftly across the configuration space.

In conclusion, the permutation update is formulated as follows:
\begin{enumerate}[i)]
    \item Choose a single worldline uniformly from the last configuration in the Markov chain.
    \item Select the time slice $j_0\in[1, J]$ and number of beads to be reshaped $M\in[1, \bar{M}]$ uniformly. $\bar{M}$ is a free parameter to be optimized during the \gls{WAPIMC} simulation.
    \item Sample a second wordline $i_2$ from the probability distribution \eqref{eq: sample particle permutation}. If $i_1=i_2$ reject the update.
    \item Construct the  new paths from the bead $(i_1, j_0)$ to $(i_2, j_M=j_0+M)$ and bead $(i_2, j_0)$ to $(i_1, j_M=j_0+M)$ with the Levy construction.
    \item Determine if the new configuration is accepted by the Metropolis-Hastings question with the acceptance probability given by equation \ref{eq: acc prop reshape}.
\end{enumerate}

There is still one major problem with this type of permutation sampling. In the presence of repulsive and/or hard-core potential, the permutation update is bound to become inefficient, i.e., the proposed update has a high likelihood of being rejected. This is because when there are a lot of particles, it becomes impossible to sample a path in an unbiased way where no two particles become close to each other. As a result, it is hard to generate long exchange cycles, which are the configurations related to the superfluid fraction, and thereby, lead to large errors\footnote{It will lead to large autocorrelation times, which is translated into an error (see \ref{sec: error}).}. This is an ergodicity problem as the whole configuration space can not be sampled in a reasonable computation time. This worsens exponentially in the function of the number of particles $N$. In general, such a problem is called a \emph{critical slowing down phenomenon}.

\subsection{Worm algorithm} \label{sec: Worms}
\begin{figure}[htb]
    \centering
    \hspace*{0.04\linewidth}
    \begin{subfigure}[b]{.45\linewidth}
        \begin{overpic}[width=0.85\linewidth]{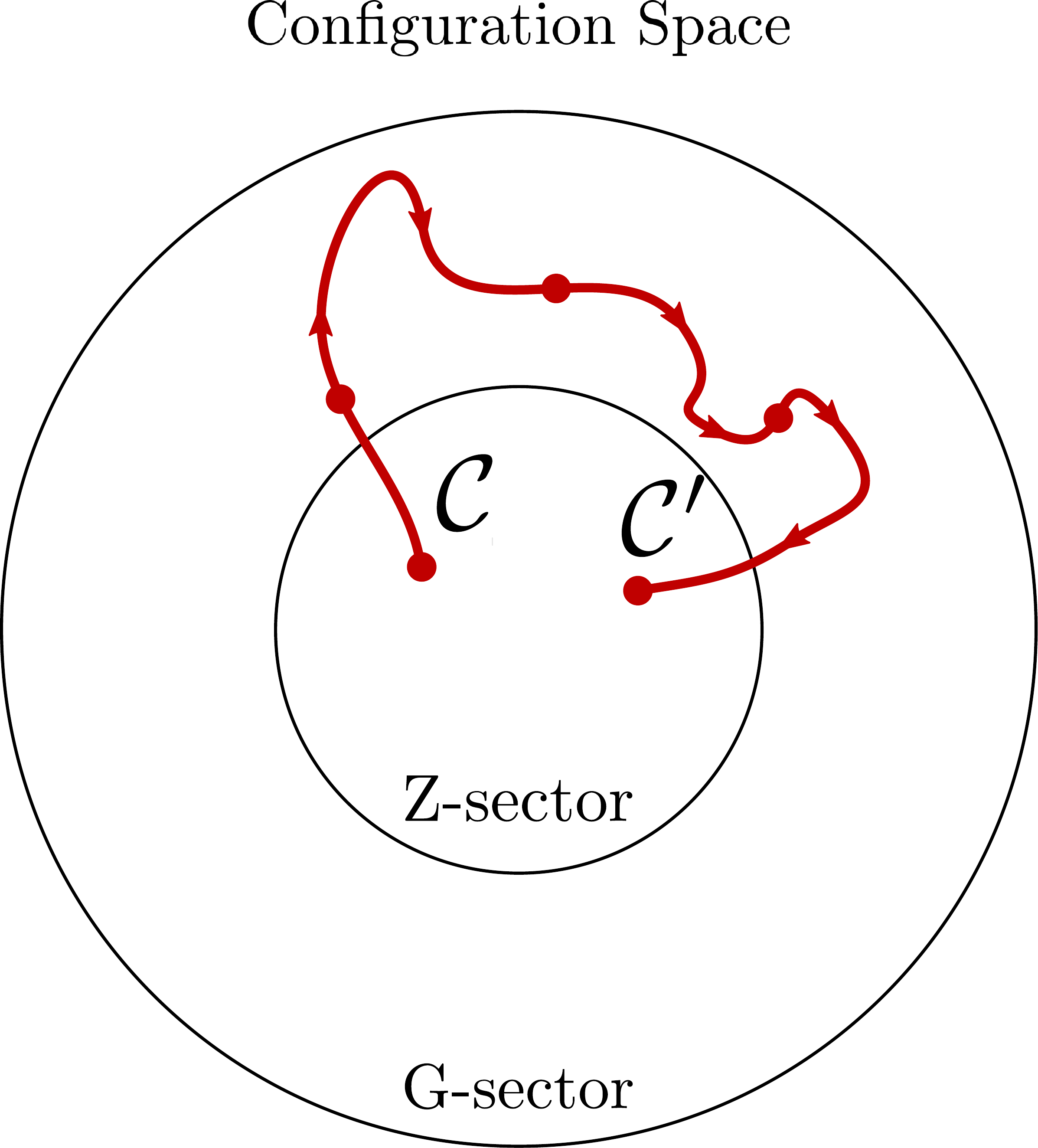}
        \put(43,-10){(a)}
        \end{overpic}
        \phantomsubcaption\label{fig: extendingconfig}
    \end{subfigure}
    \hspace*{0.03\linewidth}
    \begin{subfigure}[b]{.45\linewidth}
        \begin{overpic}[width=0.90\linewidth]{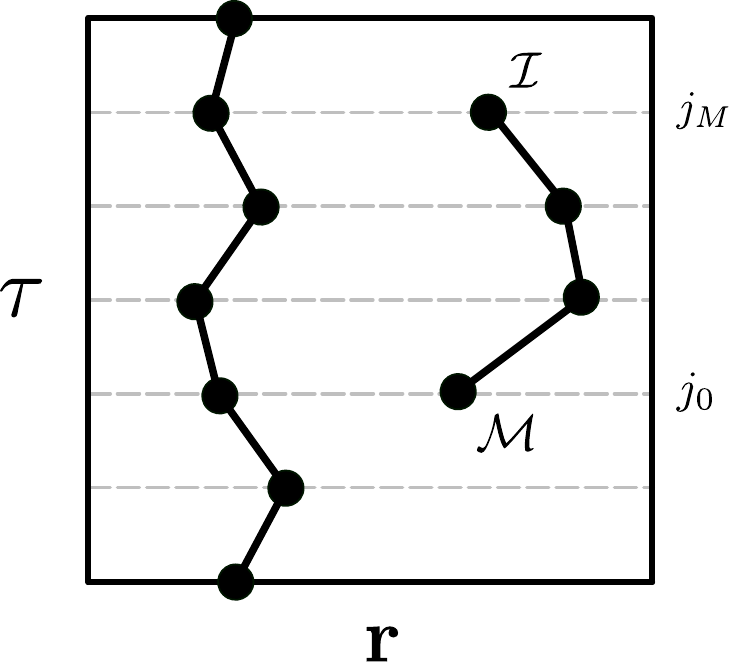}
        \put(48,-11){(b)}
        \end{overpic}
        \phantomsubcaption\label{fig: Gsector config}
    \end{subfigure}
    \vspace*{0.6cm}
    \caption{Figures adjoining the Worm Algorithm. \subref{fig: extendingconfig} Sketch representing a walk in the configuration space from an initial configuration $\config$ to a final configuration $\config^\prime$. The walker takes a detour in the G-sector to reach a point in the Z-sector more easily. \subref{fig: Gsector config} A configuration in the G-sector with one closed worldline and a worm. The head (Ira) and tail (Masha) are labelled by $\ira$ and $\masha$.}
\end{figure}

\tikz[remember picture, overlay] \node[opacity=1,inner sep=0pt] at ($(0,0)-(0cm, 16cm)$){\includegraphics[width=2.5cm,height=2.5cm]{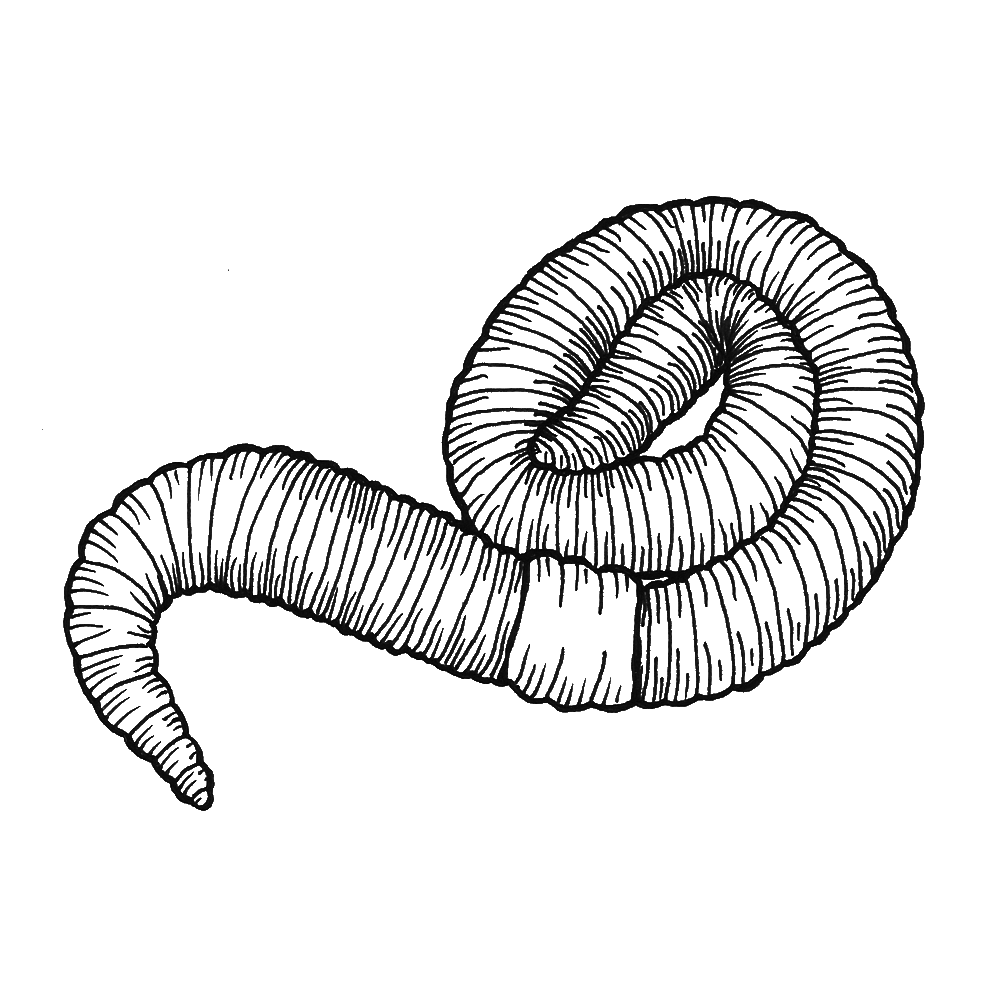}};
To deal with this critical slowing down phenomenon of generating permuted configurations, the Worms Algorithm was created, initially in the discrete spatial variation of \gls{PIMC} \cite{prokofev1998sec} and afterwards in the continuous version \cite{Boninsegn2006PRL, Boninsegn2006}. Figure \ref{fig: extendingconfig} depicts the primary concept. Certain sections of the configuration space are difficult to reach with the help of local updates described above, so, why not extend the configuration space by allowing \emph{open worldlines}? In that way, one could take a detour in this so-called \emph{G-sector}\footnote{We call this subset of the grand-canonical configuration space the G-sector because the configuration is associated with sampling the Matsubara Green's function.}, i.e., the set of all configurations in the ensemble that contain open worldlines or \emph{worms}, to generate configurations in the complement or \textit{Z-sector}. In figure \ref{fig: Gsector config}, an example of a configuration in the G-sector is shown. For historical reasons, the tail of the worm, i.e., the first bead, is called \emph{Masha} $\mathcal{M}$ and the head or last bead is called \emph{Ira} $\mathcal{I}$.

\begin{figure}[htb]
    \centering
    \includegraphics[width=0.7\textwidth]{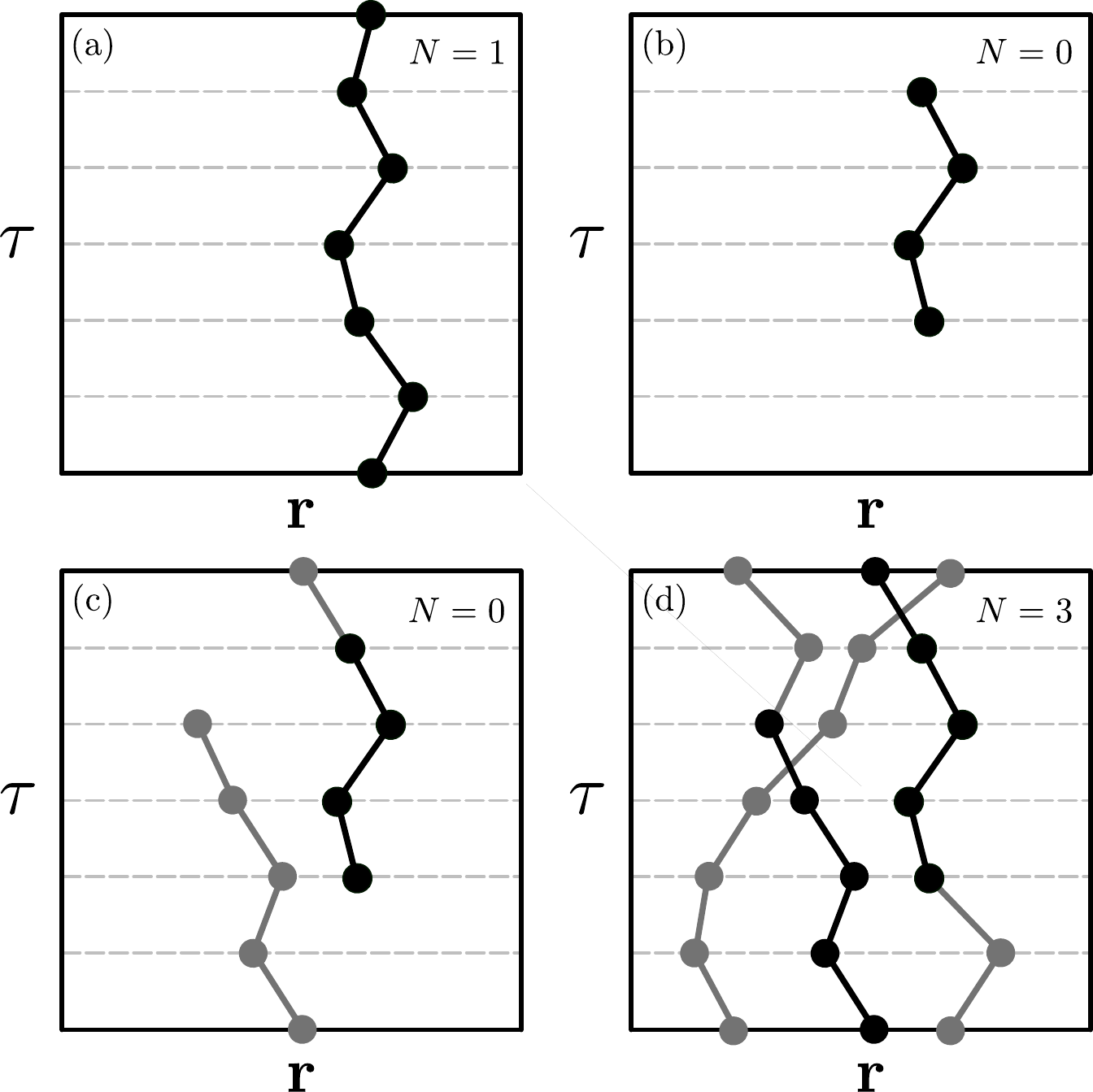}
    \caption{Four worldline configurations demonstrating how the number of closed worldlines $N$ can increase with the help of the Worm Algorithm updates. First the closed worldine is opened to a worm (a~$\to$~b). Next, the head of the worm is advanced (b~$\to$~c). Lastly, the worm is closed (c~$\to$~d). For each state the number of closed worldlines $N$ is given.}
    \label{fig: Example G-sector}
\end{figure}

The worm, or open worldline, of the G-sector configurations break the critical slowing down phenomena by serving as a mediator and assisting the worldlines to generated permutations by using the Monte Carlo update called \emph{swap} (see section \ref{sec: updates WA}). Furthermore, extending the configuration space allows \gls{PIMC} to sample the grand-canonical ensemble. This is best explained with the help of figure \ref{fig: Example G-sector}. One introduces Monte Carlo updates, called \emph{advance} and \emph{recede}, so the worm can, respectively, grow and shrink. By virtue of the $\beta$-periodicity of the worldlines, the particle number or closed wordlines can change if one back closed the worm.

Using the Worm algorithm has other advantages other than solving ergodicity problem for permutated configurations. It provides \gls{PIMC} with additional physical quantities (see section \ref{sec: observable}). In addition to accurately determining the superfluid fraction $f$, one can determine the compressibility $\kappa$, as it relates to the fluctuation of the number of particles. Furthermore, as one also generates configuration in the G-sector, one can measure observables related to the \emph{Matsubara Green's function}, such as correlation functions and the momentum distribution. Lastly, also dynamical properties can be extracted, as the Matsubara Green function is connected to the single-particle excitation spectrum through a Laplace transform \cite{dornheim2014,FilinovBonitz2012}.

\subsubsection{Extending the partition function}\label{sec: worm Z}
In this section, the formalism of the Worms Algorithm is explained. It is represented in the grand canonical ensemble. Nevertheless, the worm algorithm in the canonical ensemble is done analogously by taking the limit $\mu\to0$.

First, let us introduce some jargon. As alluded to earlier, the configuration space of the grand-canonical ensemble is split into two complemented parts or sectors, the G-sector and Z-sector, which refer to the configuration containing a worm or \emph{off-diagonal} configuration and the closed configuration or \emph{diagonal} configurations, respectively.

To add the Worms Algorithm to \gls{PIMC} we must rewrite the partition function in terms of the grand-canonical Hamiltonian $\hat{K} = \ham - \mu\hat{N}$ where $\ham$ is the Hamiltonian of the canonical ensemble, $\hat{N}$ is the particle number operator, and $\mu$ the chemical potential. A direct consequence will be that the particle number $N$ can fluctuate and instead the chemical potential $\mu$ will be the third fixed-parameter besides the volume and temperature. Given the definition of the partition function \eqref{eq: def expval}, we can write
\begin{align}\label{eq: gc partition function}
    Z_{\mathrm{GC}}=\sum_{N=0}^{+\infty} \frac{1}{N !} \sum_{\sigma \in \Pi} \int \dd{\vb{R}}\mel{\sigma \cdot \mathbf{R}}{e^{-\beta \hat{K}}}{\mathbf{R}},
\end{align}
where the trace was expanded in the spatial position basis and all the possible particle numbers. Again, as in section \ref{sec: feynman path integral}, splitting $\beta$ into $J$ imaginary time slices of size $\tau$ yields the weight of a closed configuration with $N$ worldlines or particles
\begin{align}
    \mathcal{W}(\config)=e^{\beta \mu N}\mel{\sigma \cdot \mathbf{R}_{\mathbf{0}}}{e^{-\tau \hat{H}}}{ \mathbf{R}_{\mathbf{J}-\mathbf{1}}}\times \ldots \times\mel{\mathbf{R}_{\mathbf{1}}}{e^{-\tau \hat{H}}}{\mathbf{R}_{\mathbf{0}}}.
\end{align}
Notice the $1/N!$ factor disappears as there are $N!$ variations in labelling the worldlines in the configuration.

As explained above, we do not only want to extend the configuration space to the grand-canonical ensemble with closed worldlines but also add the configuration containing open worldlines. The latter does not contribute to the partition function \eqref{eq: gc partition function}, but rather to an additional term given by
\begin{align}
    Z^\prime = Z_{\mathrm{GC}}\sum_{j_\masha}\sum_{j_\ira}\int \dd{\vb{r}_\masha} \dd{\vb{r}_\ira} G(\vb{r}_\masha, \vb{r}_\ira;\tau(j_\ira-j_\masha)),
\end{align}
where the under scripts $\masha$ and $\ira$ stand for Masha and Ira, respectively, and $G$ is the Matsubara Green's function given by
\begin{align}
    G(\vb{r}_j, \vb{r}_k; \tau)
    =\expval{\hat{\mathcal{T}}\hat{\psi}(\vb{r}_\ira, \tau)\hat{\psi}^\dagger(\vb{r}_\masha, 0)}.
\end{align}
Here is $\hat{\mathcal{T}}$ the time-ordering operator, $\hat{\psi}$ and $\hat{\psi}^\dagger$, respectively, the bosonic annihilation and creation operators in the Matsubara representation. The latter are both operators evolved in imaginary time in the Heisenberg representation \cite{altland_simons_2010}. The Green's function corresponds to the worm as it creates a particle at the bead $\masha$ and annihilates it at $\ira$. The generalized partition function becomes $Z_\mathrm{tot}=Z_\mathrm{GC} +CZ^\prime$ where $C\in\mathbb{R}^+$ is an optimization parameter which remains constant during simulation. The constant parameter determines the ratio of the diagonal and off-diagonal configurations during the \gls{WAPIMC} simulation. More concretely, by labelling the number of diagonal configurations sampled in the Markov chain as $N_\mathrm{Z}$ and off-diagonal configurations as $N_\mathrm{G}$, so that the total number of configurations in the Markov chain $N_\mathrm{MC} = N_\mathrm{Z}+N_\mathrm{G}$, we can write
\begin{align}
    \frac{N_\mathrm{Z}}{N_\mathrm{MC}}=\frac{Z_\mathrm{GC}}{Z_\mathrm{tot}} \qq{and} \frac{N_\mathrm{G}}{N_\mathrm{MC}}=C\frac{Z^\prime}{Z_\mathrm{tot}} \qq{if} N_\mathrm{MC}\to\infty.
\end{align}
Combining both expressions yields
\begin{align}
    \frac{N_\mathrm{G}}{N_\mathrm{Z}}=C\frac{Z_\mathrm{G}}{Z} \qq{if} N_\mathrm{MC}\to\infty.
\end{align}
It follows that $C$ is linearly proportional to the ratio $N_\mathrm{G} / N_\mathrm{Z}$. This relation allows for equilibrated transitions between Z and the G-sector.

\subsubsection{Updates in the Worm Algorithm}\label{sec: updates WA}
The Worms Algorithm proposes an additional set of Monte Carlo Markov chain updates that modify the worms of the configurations, thereby, changing the configuration to the G- or Z-sector. Naturally, they must fulfil the detail balanced \eqref{eq: detailed balance equation} and ergodicity conditions. In the following, we present the set updates for the grand-canonical ensemble originally constructed by \citet{Boninsegn2006PRL}. For each update, the proposal and acceptance probabilities are discussed. Afterwards, the set of updates is adjusted to the sample in the canonical ensemble.
\begin{itemize}[leftmargin=*]
    \item \textbf{Insert} and \textbf{close}: This pair of complementary update ``inserts'' a worm into a diagonal configuration or ``removes'' a worm from an off-diagonal configuration. Hence, a transition from the Z-sector to the G-sector is made or vice versa. Together, the updates respect the detailed-balance condition. An example can be seen in figure \ref{fig: Insert and Remove}. The insertion of a worm goes by the following steps:
    \begin{enumerate}[i)]
        \item Choose the time slice $j_0=j_\masha\in[1, J]$ and spatial coordinate $\vb{r}_\masha$ for the tail $\masha$ uniformly in the volume $V$ of the system.
        \item Sample the number of beads to be created $M\in[1, \bar{M}]$ uniformly. $\bar{M}$ is a free parameter to be optimized during the \gls{WAPIMC} simulation.
        \item Sample the coordinate of the worm's head on time slice $j_\ira=j_M=j_0+M$ from the kinetic distribution $\rho_0(j_\masha, j_\ira; M\tau)$.
        \item Construct the open worldline from the tail $\masha$ to the head $\ira$ with the Levy construction by \eqref{eq: levy}.
    \end{enumerate}
    As a result, we get the proposal probability
    \begin{align}
        \mathcal{P}_{\mathrm{Insert}}(\config\to\config^\prime)
        =\frac{\rho_0(\vb{r}_\masha, \vb{r}_\ira; M\tau)}{VJ\bar{M}}\prod^{j_\ira-1}_{j=j_\masha}\frac{\rho_0(\vb{r}_j, \vb{r}_{j+1}; \tau)}{\rho_0(\vb{r}_\masha, \vb{r}_\ira; M\tau)}
        = \frac{1}{VJ\bar{M}}\prod^{j_\ira-1}_{j=j_\masha}\rho_0(\vb{r}_j,\vb{r}_{j+1}; \tau),
    \end{align}
    where denominator contribution from the levy construction $\rho_0(j_\masha, j_\ira; M\tau)$ cancels because we sample the worm's head from the kinetic distribution $\rho_0(j_\masha, j_\ira; M\tau)$. The removal goes as follows:
    \begin{enumerate}[i)]
        \item Count the number of links $M$ the worm contains. If $M>\bar{M}$ reject the update where $\bar{M}$ is the same optimization parameter as for \textit{insert}.
        \item Remove the worldline.
    \end{enumerate}
    The proposal probability of the \textit{remove} update is given by $\mathcal{P}_{\mathrm{Insert}}(\config^\prime\to\config)=1$. Hence, the acceptance probabilities of the \textit{insert} and \textit{remove} update yield
    \begin{align}
        \mathcal{A}_{\mathrm{Insert}}(\config\to\config^\prime)
        &=\min \left(1, CVJ\bar{M}e^{\tau M \mu}e^{\tau \Delta S(\config \to \config^\prime)} \right),\\
        \mathcal{A}_{\mathrm{Remove}}(\config^\prime\to\config)
        &=\min \left(1, \frac{1}{CVJ\bar{M}} e^{-\tau M \mu} e^{-\tau \Delta S(\config^\prime \to \config)} \right).
    \end{align}
    It is important to again state that the optimization parameter $\bar{M}$ must be the same for both \textit{insert} and \textit{remove}. Otherwise, the detailed-balance condition is not satisfied.
    \begin{figure}[htbp]
        \centering
        \includegraphics[width=0.8\textwidth]{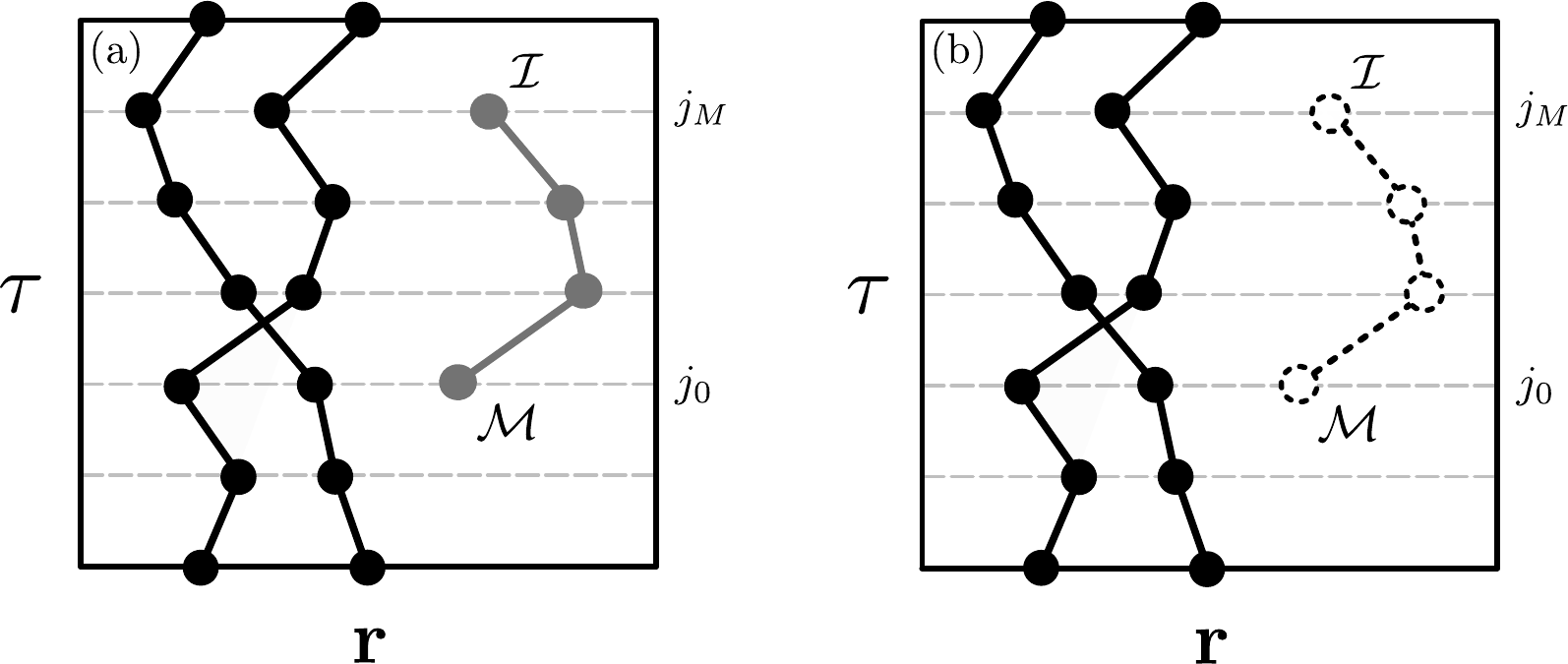}
        \caption{Illustration of the insert \textcolor{red}{(a)} and remove \textcolor{red}{(b)} update. The unchanged beads are portrayed in black, the old beads and links are dashed, and the newly generated path or Brownian bridge in solid grey.}
        \label{fig: Insert and Remove}
    \end{figure}

    \item \textbf{Open} and \textbf{Close}: This pair of complementary update ``opens'' a closed worldline in a diagonal configuration or ``closes'' the worm in an off-diagonal configuration. Hence, a transition from the Z-sector to the G-sector is made or vice versa. An example can be seen in figure \ref{fig: Open and Close}. Together, the updates respect the detailed-balance condition \eqref{eq: detailed balance equation}. The opening of a diagonal worldline goes by the following steps:
    \begin{enumerate}[i)]
        \item Choose uniformly a closed worldline $i$ from all $N$ worldlines and a time slice $j_0=j_\ira\in[1, J]$ for the head $\ira$ of the new worm.
        \item Sample the number of links to be removed $M\in[1, \bar{M}]$ uniformly. $\bar{M}$ is a free parameter to be optimized during the \gls{WAPIMC} simulation.
    \end{enumerate}
    Hence, the proposal probability is therefore given as
    \begin{align}
        \mathcal{P}_{\mathrm{open}}(\config\to\config^\prime)=\frac{1}{JN\bar{M}}.
    \end{align}
    The \textit{close} update is implemented as follows:
    \begin{enumerate}[i)]
        \item Count the number of beads $M$ that must be constructed to close the worm. If $M>\bar{M}$ reject the update where $\bar{M}$ is the same optimization parameter as for \textit{open}.
        \item Close the open worldline from the head $\ira$ to the tail $\masha$ with the Levy construction by \eqref{eq: levy}.
    \end{enumerate}
    which yields a proposal probability
    \begin{align}
        \mathcal{P}_{\mathrm{close}}(\config^\prime\to\config)=\prod^{j_\masha-1}_{j=j_\ira}\frac{\rho_0(j, j+1; \tau)}{\rho_0(j_\masha, j_\ira; M\tau)}.
    \end{align}
    The above proposal probabilities lead to acceptance probabilities
    \begin{align}
        \mathcal{A}_{\text {Close }}(\config \to \config^\prime)
        &=\min \left(1, \frac{\rho\left(\mathbf{r}_{\masha}, \mathbf{r}_{\ira}; M \tau\right) e^{\tilde{M} \tau \mu} e^{\tau \Delta S(\config \to \config^\prime)}}{C JN\bar{M}}\right), \\
        \mathcal{A}_{\text {Open }}(\config^\prime \to \config)
        &=\min \left(1, \frac{CJN\bar{M} e^{-M \tau \mu} e^{-\tau \Delta S(
        \config^\prime \to \config)}}{\rho\left(\mathbf{r}_{\masha}, \mathbf{r}_{\ira}; M \tau \right)}\right).
    \end{align}
    Again, note that the optimisation parameter $\bar{M}$ must be the same for both \textit{open} and \textit{close} when the update is proposed to advance the Markov chain. Furthermore, $N$ is the number of closed worldlines if we open the configuration. For example, if we want to propose to close a configuration in the G-sector which only contains a worm with a bead smaller than $M$, the reverse update would be to open the closed configuration, so $N=1$.
    \begin{figure}[htbp]
        \centering
        \includegraphics[width=0.8\textwidth]{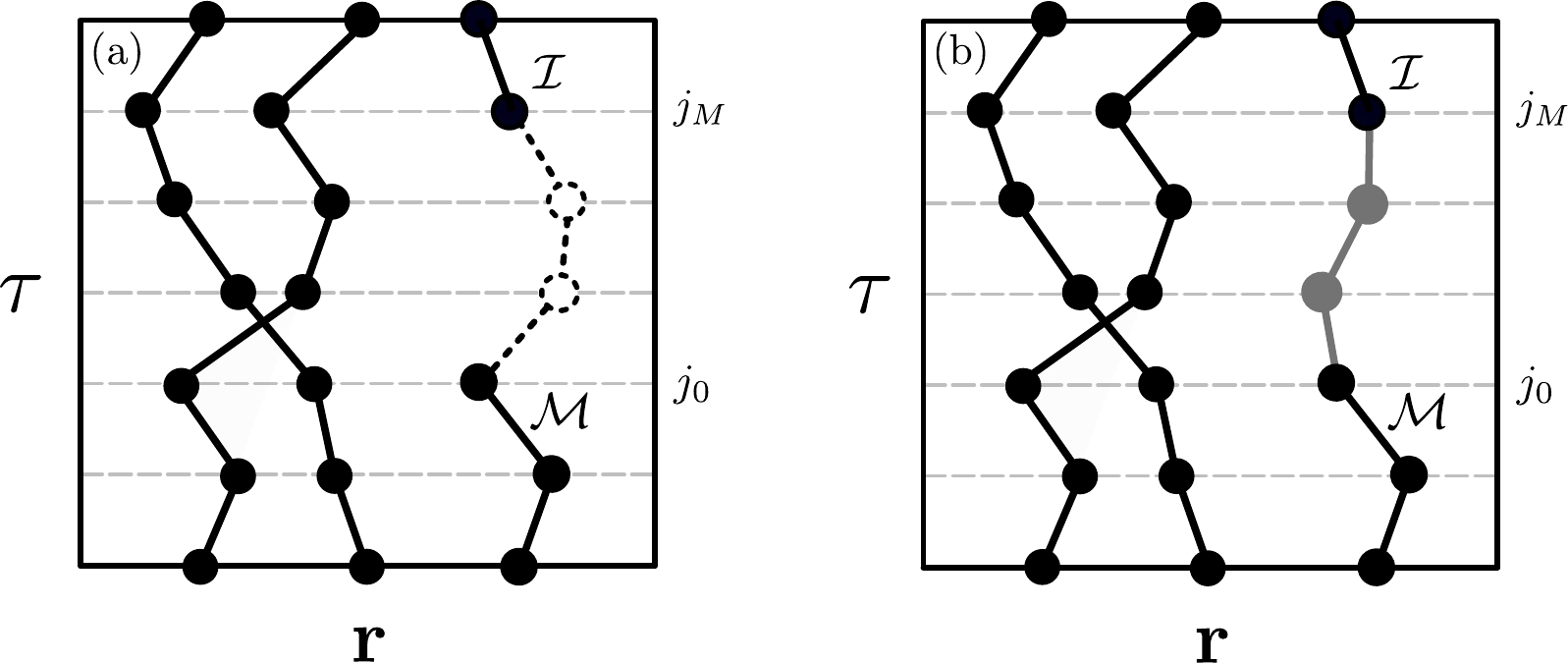}
        \caption{Illustration of the open \textcolor{red}{(a)} and close \textcolor{red}{(b)} update. The unchanged beads are portrayed in black, the old beads and links are dashed, and the newly generated path or Brownian bridge in solid grey.}
        \label{fig: Open and Close}
    \end{figure}

    \item \textbf{Advance} and \textbf{recede}: Of course, only modifying configurations to contain or disclude a worm is not enough. We would also like to modify the beads of the worm. Naturally, the canonical updates \textit{reshape} and \textit{center-of-mass} of section \ref{sec: Can update} can be used to adjust the open worldline. However, we would also like to grow or shrink the worm. This is done with the \textit{advance} and \textit{recede} update, respectively. An example is shown in figure \ref{fig: Advance and Recede}. To advance the worm, we undergo the steps:
    \begin{enumerate}[i)]
        \item Sample the number of beads to be created $M\in[1, \bar{M}]$ uniformly. $\bar{M}$ is a free parameter to be optimised during the \gls{WAPIMC} simulation.
        \item Sample the coordinate of the new worm's head on time slice $j_\ira=j_M=j_0+M$ from the kinetic distribution $\rho_0(\vb{r}_0, \vb{r}_\ira; M\tau)$ where $j_0$ and $\vb{r}_0$ are the time slice and position of the old head, respectively.
        \item Construct the worldline from the old head to the new head with the Levy construction by \eqref{eq: levy}.
    \end{enumerate}
    Conversely, to recede the worm we must:
    \begin{enumerate}[i)]
        \item Sample the number of beads to be removed $M\in[1, \bar{M}]$ uniformly. $\bar{M}$ is the same free parameter as for the \textit{advance} update. If $M$ is greater than the number of beads the worm contains, reject the update.
        \item Remove the beads of the worm on the time slices $j_\ira$, $\ldots$, $j_{\ira-M}$.
    \end{enumerate}
    The above procedures lead to the proposal probabilities
    \begin{align}
        \mathcal{P}_{\mathrm{Advance}}(\config\to\config^\prime)
        &=\frac{\rho_0(\vb{r}_0, \vb{r}_\ira; M\tau)}{\bar{M}}\prod^{j_\ira-1}_{j=j_\masha}\frac{\rho_0(\vb{r}_j, \vb{r}_{j+1}; \tau)}{\rho_0(\vb{r}_0, \vb{r}_\ira; M\tau)},\\
        \mathcal{P}_{\mathrm{Recede}}(\config\to\config^\prime)
        &=\frac{1}{\bar{M}}.
    \end{align}
    As a result, the acceptance probabilities of the \textit{advance} and \textit{recede} update in the Metropolis question yields
    \begin{align}
        \mathcal{A}_{\text {Advance}}(\config \to \config^\prime)
        &=\min \left(1, e^{\tilde{M} \tau \mu} e^{\tau \Delta S(\config \to \config^\prime)}\right), \\
        \mathcal{A}_{\text {Recede}}(\config^\prime \to \config)
        &=\min \left(1, e^{-M \tau \mu} e^{-\tau \Delta S(\config^\prime \to \config)}\right).
    \end{align}
    Once more, the optimisation parameter $\bar{M}$ must be the same for \textit{advance} and \textit{recede} when one of them is proposed.
    \begin{figure}[htbp]
        \centering
        \includegraphics[width=0.8\textwidth]{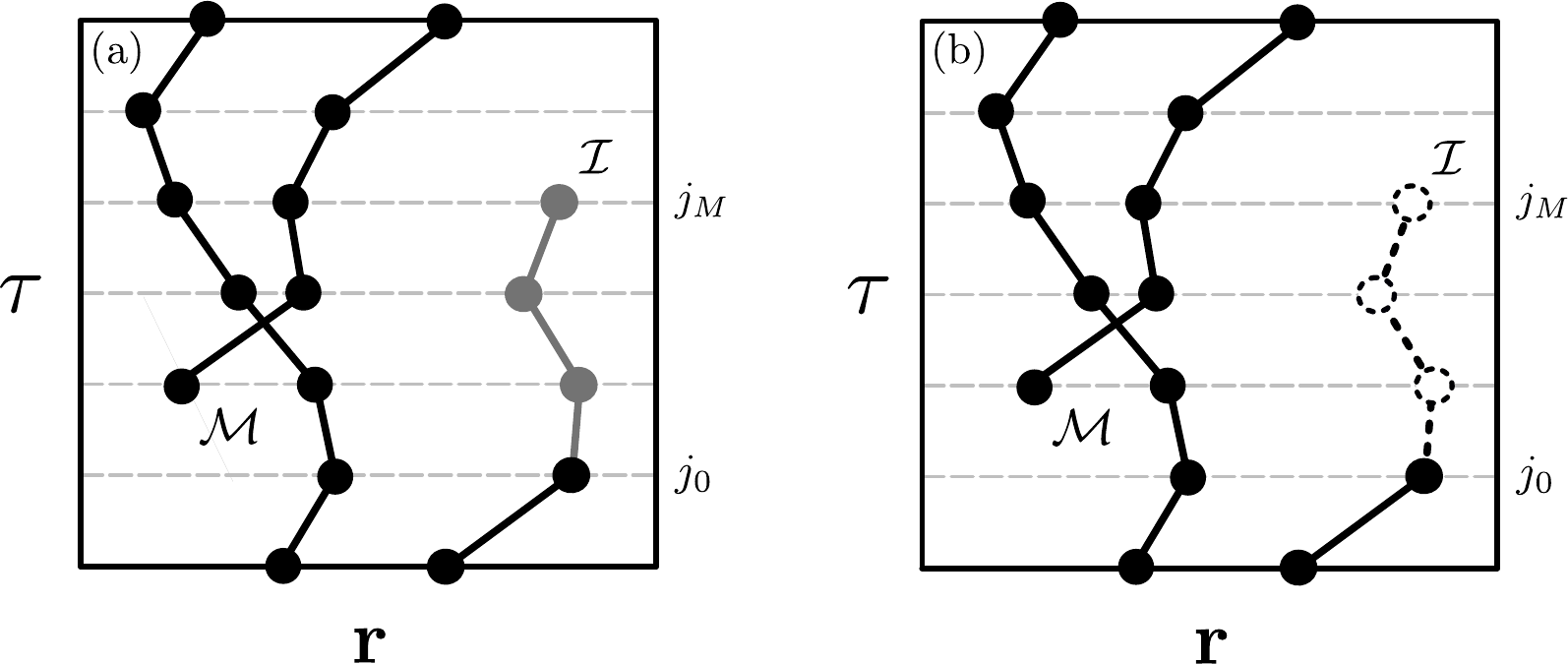}
        \caption{Illustration of the advance \textcolor{red}{(a)} and recede \textcolor{red}{(b)} update. The unchanged beads are portrayed in black, the old beads and links are dashed, and the newly generated path or Brownian bridge in solid grey.}
        \label{fig: Advance and Recede}
    \end{figure}

    \item \textbf{Swap}: Last but not least, we must address the critical slowing down phenomenon where we introduce the Worm algorithm. That is the sampling of permutations for the configuration in the Z-sector, which is done with the \textit{swap} update. The concept is best explained visually which can be found in figure \ref{fig: Swap}. The steps are as follows:
    \begin{enumerate}[i)]
        \item The number of beads to be modified is $\bar{M}$. $\bar{M}$ is a free parameter to be optimised during the \gls{WAPIMC} simulation.
        \item Construct the list $\chi$ of all kinetic propagators $\rho_0(\vb{r}_\ira, \vb{r}_{k, j_{\bar{M}}}, \bar{M}\tau)$ with $k=1, \ldots, N$, $j_{\bar{M}}=j_\ira+\bar{M}$ and $N$ the number of closed worldlines.
        \item Sample a bead $\alpha$ discretely by the weights $w(k)=\chi(k)/\Sigma$ where $\Sigma=\sum^N_{k=1}\chi(k)$.
        \item Locate the bead $\delta$ on the time slice of the head of the worm $\ira$ by backtracking on the polymer from the bead $\alpha$ with $\bar{M}$ beads. Label the path as $(\mathbf{r}_{j_\ira}, \mathbf{r}_{j_\ira+1}, \ldots, \mathbf{r}_{j_\ira+\bar{M}-1},\mathbf{r}_{j_\ira+\bar{M}})$ where $\mathbf{r}_{j_\ira}=\mathbf{r}_\delta$ and $\mathbf{r}_{j_\ira+\bar{M}}=\mathbf{r}_\alpha$.
        \item Construct the list $\chi^\prime$ of all kinetic propagators $\rho_0(\vb{r}_\delta, \vb{r}_{k, j_{\bar{M}}}, \bar{M}\tau)$ with $k=1, \ldots, N$ and compute $\Sigma^\prime=\sum^N_{k=1}\chi^\prime(k)$.
        \item Remove all beads in between $\alpha$ and $\delta$ and generate a new path $(\mathbf{r}^\prime_{j_\ira}, \mathbf{r}^\prime_{j_\ira+1}, \ldots, \mathbf{r}^\prime_{j_\ira+\bar{M}-1},\mathbf{r}^\prime_{j_\ira+\bar{M}})$ from the head of the worm to the bead $\alpha$ with the levy construction where $\mathbf{r}^\prime_{j_\ira}=\mathbf{r}_\ira$ and $\mathbf{r}^\prime_{j_\ira+\bar{M}}=\mathbf{r}_\alpha$.
    \end{enumerate}
    The proposal probabilities by following such a scheme give
    \begin{align}
        \mathcal{P}_{\mathrm{Swap}}(\config\to\config^\prime)
        &=\frac{\rho_0\left(\mathbf{r}_{\ira}, \mathbf{r}_{\alpha}; \bar{M} \tau\right)}{\Sigma_{1}} \frac{\prod_{k=0}^{\bar{M}-1} \rho_0\left(\mathbf{r}_{j_\ira+k}^{\prime}, \mathbf{r}_{j_\ira+k+1}^{\prime}; \tau\right)}{\rho_0\left(\mathbf{r}_{\ira}, \mathbf{r}_{\alpha}; \bar{M} \tau\right)}=\frac{\prod_{k=0}^{\bar{M}-1} \rho_0\left(\mathbf{r}_{j_\ira+k}^{\prime}, \mathbf{r}_{j_\ira+k+1}^{\prime}; \tau\right)}{\Sigma_1},\\
        \mathcal{P}_{\mathrm{Swap}}(\config^\prime\to\config)
        &=\frac{\rho_0\left(\mathbf{r}_{\ira}, \mathbf{r}_{\alpha}; \bar{M} \tau\right)}{\Sigma_{2}} \frac{\prod_{k=0}^{\bar{M}-1} \rho_0\left(\mathbf{r}_{j_\ira+k}, \mathbf{r}_{j_\ira+k+1}; \tau\right)}{\rho_0\left(\mathbf{r}_{\ira}, \mathbf{r}_{\alpha}; \bar{M} \tau\right)}= \frac{\prod_{k=0}^{\bar{M}-1} \rho_0\left(\mathbf{r}_{j_\ira+k}, \mathbf{r}_{j_\ira+k+1}; \tau\right)}{\Sigma_2},
    \end{align}
    which results in the acceptance probabilities
    \begin{align}
        A_{\mathrm{Swap}}(\config \to \config^\prime)
        &=\min \left(1,\frac{\Sigma}{\Sigma^\prime} e^{\tau \Delta S(\config \to \config^\prime)} \right), \\
        A_{\mathrm{Swap}}(\config^\prime \to \config)
        &=\min \left(1, \frac{\Sigma^\prime}{\Sigma}e^{\tau \Delta S(\config^\prime \to \config)} \right).
    \end{align}
    Note, that we again perform importance sampling by choosing a worldline to swap the worm with from the list $\chi$, thereby ensuring that the distance is smaller than the order $\sqrt{\bar{M}\tau}$. This makes the update local, which together with the fact that only a single permutation is made at a time and modifications can be made many times before the configuration is closed, ensures that exchanges or permutations are made if the system wants so.

    \begin{figure}[htbp]
        \centering
        \includegraphics[width=0.4\textwidth]{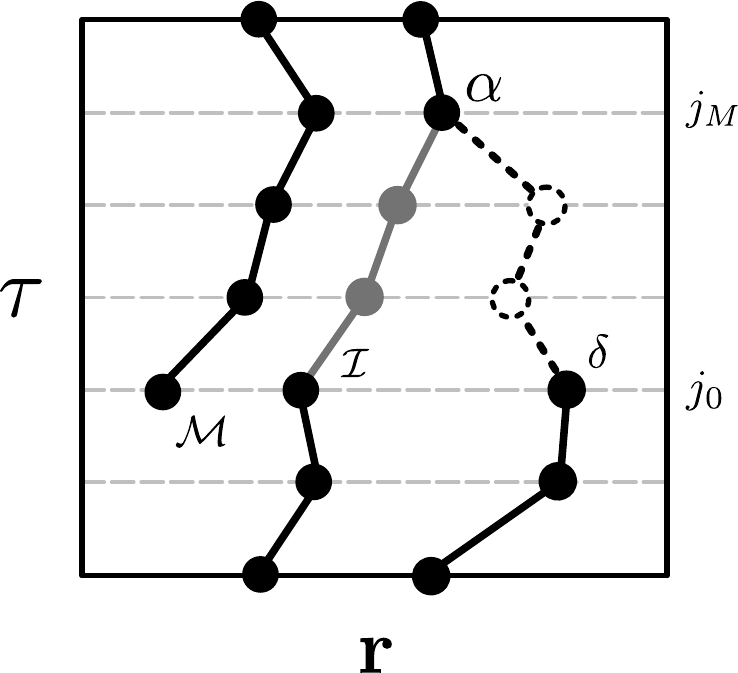}
        \caption{Illustration of the swap update. The unchanged beads are portrayed in black, the old beads and links are dashed, and the newly generated path or Brownian bridge in solid grey.}
        \label{fig: Swap}
    \end{figure}
\end{itemize}
The Markov chain updates, described above, combined with the canonical updates (see section \ref{sec: Can update}), form an ergodisic set of updates where each individual update or complementary pair holds the detailed balance condition. This set allows for conversion between the Z- and G-sector in a equilibriated and local manner.

\subsubsection{Worms algorithm in the Canonical ensemble}\label{sec: Worms canonical}
The set of updates in the Worm algorithm are in a natural way design to sample the grand-canonical ensemble. Nevertheless, often one want to only take measurement of a specific canonical ensemble $\{N, V, T\}$. One way of doing this is choosing a optimal chemical potential $\mu$ such that the number of particles fluctuates around the desired $N$ and only take measurements when the system is in ensemble $\{N, V, T\}$\footnote{This process is described in section \ref{sec: measering sheme}.}. Nevertheless, this can be quite inefficient. Furthermore, this can not be done if want the simulate non-interacting bosons as any non-zero chemical potential will lead to a infinite of particles.

Another alternative is restricting the number of closed worldlines or particle numbers to remain constant in the Z-sector. To enforce this on the Worm algorithm, one often adds restriction on the set of updates. For example, if one:
\begin{itemize}
    \item limits the set of the Worm algorithm to only include $\{$Open, Close, Advance, Recede, Swap$\}$,
    \item impede the time slices of head and tail to advance/recede passed each other,
    \item confine $\bar{M} < J$,
\end{itemize}
one cannot change the particle number $N$ for configuration in the Z-sector. Another option --- which can lead to better performance --- would be to design new set of updates for the Worm algorithm. This is for example done in \citet{RotaPIMC}.

\section{Measure Observables} \label{sec: observable}
In the previous section, we explained how to generate new configurations in the Markov chain. The only thing that remains is to extract physical properties out of the configurations. As shown in section \ref{sec: feynman path integral}, this comes down to determining the estimator $\mathcal{Q}(\config)$ of a thermodynamic observable $\hat{O}$. This is done by writing out the definition \eqref{eq: def expval} explicitly using the small-time propagator you will be using in simulating the system and, afterwards, comparing it to the form of equation \ref{eq: expval mc integral}. Often one has also the option to extract it directly from the partition function by using their thermodynamic definition, see for example section \ref{sec: energy}. If the estimator of an observable is identified, it comes down to taking the average of this estimator for all entries in the Markov chain as in equation \ref{eq: estimator average}.

We will cover the observables interesting for ultracold atoms in optical lattices. However, note that \gls{WAPIMC} is not limited by the observables covered in this manuscript.

\subsection{Energy}\label{sec: energy}
The estimator for the total energy of the system is often computed by using the thermodynamic definition given by
\begin{align}\label{eq: thermo def energy}
    E=\expval{\ham}_Z=-\frac{1}{Z}\pdv{Z}{\beta}.
\end{align}
The subscript $Z$ emphasises the fact that the average is only taken over the Z-sector. How this is explicitly done is described in section \ref{sec: measering sheme}. The form of the estimator is, of course, system dependent and hence relies on the content of the density matrix. Let us compute the total energy estimator $\mathcal{Q}_E$ for a non-interacting Bose gas in an external potential described by the density matrix in \eqref{eq: rho_1 complete}\footnote{Note that we used the second-order trotter formula to derive the density matrix and hence the result may change if higher orders are used.}. The derivation is analogous when adding interaction to the system. The partition function can then be written as
\begin{align}\label{eq: noninteracting partition function}
Z
&=\int \dd{\config}\pi(\config) = \frac{1}{N!}\int \dd{\mathbf{R}_{0}} \ldots \dd{\mathbf{R}_{J-1}}\prod_{i=0}^{J-1}\rho_0\left(\mathbf{R}_{j}, \mathbf{R}_{j+1};\tau\right) \mathrm{e}^{- S\left(\mathbf{R}_{j}, \mathbf{R}_{j+1};\tau\right)} \\
&=\frac{1}{N!}\int \dd{\mathbf{R}_{0}} \ldots \dd{\mathbf{R}_{J-1}} \frac{1}{\lambda_{\beta}^{d NJ}} \prod_{i=0}^{J-1}  \exp \left(-\frac{\pi}{\lambda_{\tau}^{2}} \sum_{k=1}^{N}\left(\mathbf{r}_{k, j}-\mathbf{r}_{k, j+1}\right)^{2}-\frac{\tau}{2}\sum_{k=1}^{N}( V\left(\mathbf{r}_{k, j}\right)+V\left(\mathbf{r}_{k, j+1}\right))\right).
\end{align}
where we remember that $\lambda_{\tau}=\sqrt{4\pi\lambda\tau}$ and $S$ is the action. Substituting this partition function \eqref{eq: noninteracting partition function} in the expression for the expectation value of the total energy \eqref{eq: thermo def energy} yields
\begin{align} \label{eq: subresult energy}
    E=\frac{1}{Z}\int \dd{\config}\pdv{\pi(\config)}{\beta}.
\end{align}
Further, one easily derives the identity
\begin{align}
\partial_{\beta}\left(\frac{1}{\lambda_{\beta}^{d NJ}}\right) =-\frac{d NJ}{2 \beta}\left(\frac{1}{\lambda_{\beta}}\right)^{d NJ},
\end{align}
combined with applying the product rule of derivatives to equation \ref{eq: subresult energy} we find that the expectation value for the total energy of the system is given by
\begin{align}
    E=
    &\frac{1}{Z}\int \dd{\config}\pi(\config)\times\left(\frac{d N}{2 \tau}-\frac{1}{4\lambda\tau^2 J} \sum_{j=0}^{J-1} \sum_{k=1}^{N}\left(\mathbf{r}_{k, j}-\mathbf{r}_{k, j+1}\right)^{2}+ \frac{1}{2J}\sum_{j=0}^{J-1} \sum_{k=1}^{N}( V\left(\mathbf{r}_{k, j}\right)+V\left(\mathbf{r}_{k, j+1}\right))\right).
\end{align}
Comparing this result to the form of equation \ref{eq: expval mc integral}, the total energy estimator yields
\begin{align}
\mathcal{Q}_{E_T}
=\frac{d N}{2 \tau}-\frac{1}{4\lambda\tau^2 J} \sum_{j=0}^{J-1} \sum_{k=1}^{N}\left(\mathbf{r}_{k, j}-\mathbf{r}_{k, j+1}\right)^{2}+ \frac{1}{J}\sum_{j=0}^{J-1} \sum_{k=1}^{N}( V\left(\mathbf{r}_{k, j}\right)+V\left(\mathbf{r}_{k, j+1}\right)).
\end{align}
The result is interpreted very easily. The first term corresponds to the well-known \emph{equipartition theorem} for the kinetic energy of an ideal gas, e.g., $E_\mathrm{kin} = 3Nk_BT/2$ in a three-dimensional system. The second term is a quantum correction to kinetic energy. Indeed, in the classical limit where only $J=1$ the correction term vanishes. The last term corresponds to the contribution of potential energy. Notice that the kinetic energy becomes small when the beads are far away from each other. It shows that worldlines in imaginary time that combine to form a configuration do not represent real physics, i.e., do not correspond to any real-time propagation, and are only is a convenient representation of the quantum mechanical path integral structure.

\begin{figure}[ht]
    \centering
    \includegraphics[width=0.7\linewidth]{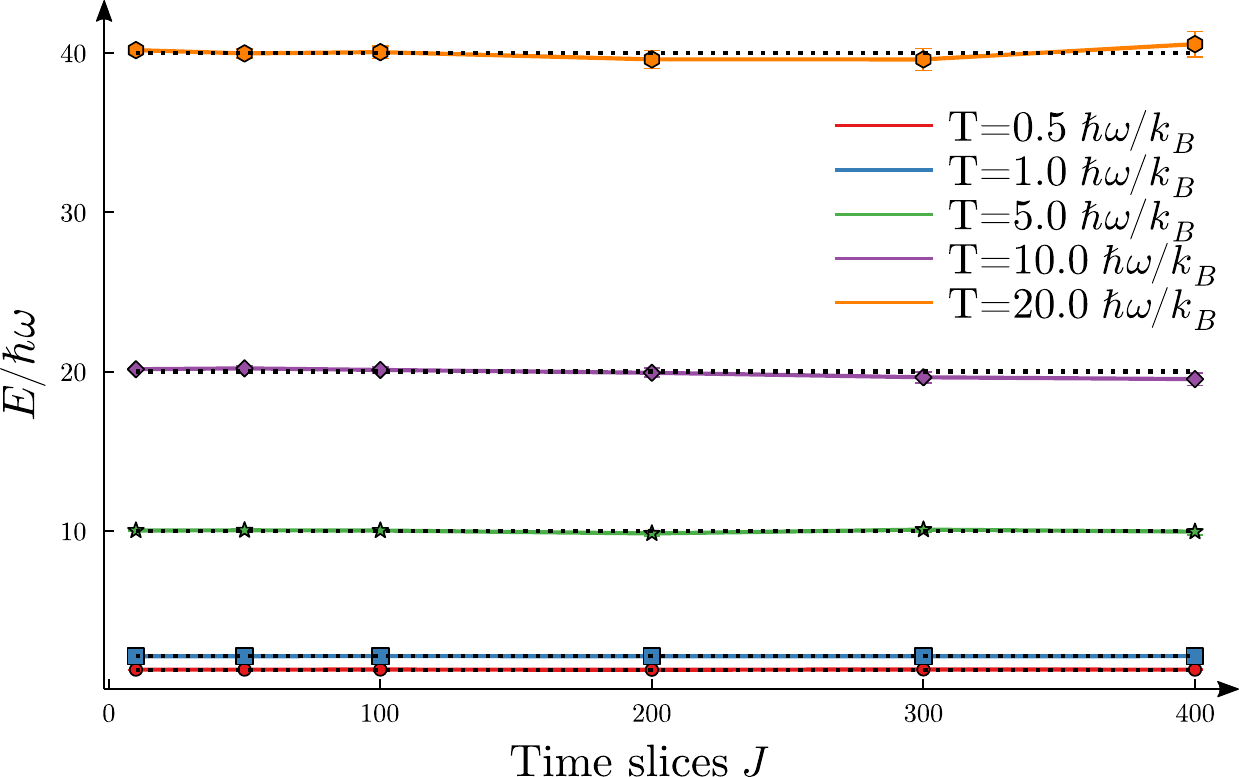}
    \caption{Total energy of a boson in a harmonic trap $V(\vb{r})=\frac{1}{2}\hbar\omega\vb{r}^2$ at different temperatures $T$ and time slices $J$. The black dashed lines represent the exact value. The coloured markers are the simulated values with the help of our implementation of \gls{WAPIMC}. Notice that the error bar becomes larger for increasing $J$.}
    \label{fig: Energy_trapped_boson}
\end{figure}

The above estimator is known as the \emph{thermodynamic estimator} for the total energy\footnote{One could also compute the total energy estimator from the explicit definition in equation \ref{eq: expval mc integral} which will yield a different result for the potential contribution. Nevertheless, both will converge to the true expectation value in the limit $\tau\to0$. A third estimator can also be derived from the virial theorem. The combination of the three estimates can be used to reduce the statistical error for the energy estimation \cite{Ceperley1995}.}. However, it is notorious to have a large error for a large $J$. This is due to the the quantum correction term where in increasing $J$ the variance of the sum diverges linearly \cite{Wolfhard1997}. Indeed, in figure \ref{fig: Energy_trapped_boson}, we observe empirically that the error for the total energy diverges when one uses time slices for a boson in a harmonic potential. To circumvent the problem, we need to propose a better kinetic energy estimator with a lower fluctuation. To this goal, one often uses the \emph{virial theorem}, $\expval{\frac{\hat{p}^{2}}{2 m}}=\frac{1}{2}\expval{\hat{x} V^{\prime}(\hat{x})}$ where we have used the abbreviation $V^{\prime}(x)=d V / d x$. Using the expression, one can derive a ``variance reduced'' or \emph{virial estimator}, given by
\begin{align}\label{eq: virial estimator}
    \mathcal{Q}_{E_V} =
\frac{dN}{2 \beta}
&+\frac{1}{4 \lambda \tau J} \sum_{k=1}^{N}\left(\mathbf{r}_{k, J-1}-\mathbf{r}_{k, J}\right)\left(\mathbf{r}_{k, J}-\mathbf{r}_{k, 0}\right)\\
&+\frac{1}{2 \beta} \sum_{j=1}^{J-1} \sum_{k=1}^{N}\left(\mathbf{r}_{k,j}-\mathbf{r}_{k,0}\right) \frac{\partial}{\partial \mathbf{r}_{k,j}}\left(S\left(\mathbf{r}_{k,j+1}, \mathbf{r}_{k,j} ; \tau\right)+S\left(\mathbf{r}_{k,j}, \mathbf{r}_{k,j-1} ; \tau\right)\right)\\
&+\sum_{j=0}^{J-1}\sum_{k=1}^{N} \frac{\partial S\left(\mathbf{r}_{k,j+1}, \mathbf{r}_{k,j} ; \tau\right)}{\partial \tau},
\end{align}
with $S$ the action or potential contribution to the density matrix and the bead notation is always understood the $\beta$-periodic sense. This expression has been taken from \citet{Spada2022}. Although, different versions are achievable, e.g., in \citet{RotaPIMC}. For a derivation and comparison between different energy estimators, \citet{Ceperley1995} is recommended.

We conclude with two final remarks. First, the second term in the virial estimator \eqref{eq: virial estimator} represents the particle statistics. Indeed, if no permutations between the worldlines are present, we have that $\mathbf{r}_{k, J}=\mathbf{r}_{k, 0}$ and the term vanishes. Next, the difference $\mathbf{r}_{k, J}-\mathbf{r}_{k, 0}$ in the second must be evaluated continuously along the worldline\footnote{Periodic boundary conditions may only apply between two succeeding beads!}. It follows, that the absolute value of this difference can be larger than the size of the simulation box.

\subsection{Spatial structure}
\subsubsection{Density}
The spatial density distribution $n(\mathbf{r})$ of a system often reveals interesting  features of the system. It turns out the density of the system is one of the variables that is easily obtained as it is diagonal in the position representation. This allows us to easily compute the definition of the estimator \eqref{eq: expval mc integral} to be $\mel{\vb{R}_0}{\hat{n}(\vb{r})}{\vb{R}_0}=\sum^N_{i=1}\delta(\vb{r}-\vb{r}_{i, 0})$. However, we can improve the statistics error by noticing that the translation invariance in imaginary time makes that a time slice is not particularly special. As a result, an improved version is found for the local spatial density estimator at a position $\mathbf{r}_{\alpha}$, given by
\begin{align}
    \mathcal{Q}_n(\mathbf{r}_{\alpha}) = \frac{1}{J} \sum_{i=1}^{N} \sum_{j=0}^{J-1} \frac{\delta\left(\vb{r}_{\alpha},\mathbf{r}_{i, j}\right)}{V_{\alpha}}.
\end{align}
Here, the spatial volume is divided in a discrete grid of hyper-cubes in $d$ dimensions where each bin $\alpha$ has a center coordinate $\mathbf{r}_{\alpha}$ with volume $V_\alpha$ such that
\begin{align}\label{eq: implementation of the delta function}
    \delta\left(\mathbf{r}_{j, k}, \mathbf{r}_{\alpha}\right)= \begin{cases}1, & \qq{if} \mathbf{r}_{i, j} \in \alpha, \\ 0, & \text { otherwise }.\end{cases}
\end{align}

\subsubsection{Radial Distribution}
Using the same approach as for the density, other spatial observables can easily be extracted from the configuration, e.g., the radial distribution function $g(r)$, a pair correlation function that explains how the atoms in a system are radially distributed around each other. This is a valuable way of representing the average structure of disordered systems as it can be easily measured experimentally. More concretely, it is proportional to the probability of finding a particle at a distance of $r$ away from a given reference particle, which yields
\begin{align}
\mathcal{Q}_g({r}_{\alpha})
= \frac{V}{N^{2} J}\sum_{j=0}^{J-1} \sum_{i \neq k} \frac{\delta\left({r}_{\alpha}-\left({r}_{i,j}-{r}_{k,j}\right)\right)}{L_\alpha},
\end{align}
evaluating the delta-function by $1$ if ${r}_{\alpha}$ lies in the interval $\alpha=[r_\alpha-\Delta r, r_\alpha+\Delta r]$ with length $L_\alpha=2\Delta r$. In general, one could look many other pair correlation function such as the Static structure factor, etc \cite{RotaPIMC}.

\subsubsection{Compressibility}
The spatial density $n$ in combination with working in the grand canonical ensemble gives us the opportunity to compute the compressibility $\kappa$ of the system. It can be defined as the change in spatial density $n$ as the chemical potential $\mu$ is varied at a constant temperature, i.e.,
\begin{align}
\kappa=\left.\frac{\partial n}{\partial \mu}\right|_{T}.
\end{align}
Substituting the grand canonical spatial density operator $\hat{n}=\operatorname{Tr}\left[\hat{N} e^{-\beta(\hat{H}-\mu \hat{N})}\right] / V Z$ yields that the compressibility $\kappa$ is dependent on the variance of the number of particles, i.e.,\cite{Yao2020}
\begin{align}
\kappa=V\beta\frac{\expval{N^{2}}_{\mathrm{Z}}-\expval{N}_{\mathrm{Z}}^{2}}{\expval{N}_{\mathrm{Z}}}.
\end{align}
Here the subscript $Z$ indicates the average in equation \ref{eq: estimator average} only includes a configuration in the Z-sector. The estimators $\mathcal{Q}_N$ and $\mathcal{Q}_{N^2}$ in \eqref{eq: estimator average} are just the number of closed worldlines and the number of closed worldlines squared in a configuration $\config$.

\subsection{Superfluid density}
As \gls{PIMC} solves the path integral formulation of quantum mechanics also pure quantum mechanical properties can be predicted. An example would be superfluidity. Nevertheless, there is at the moment no complete satisfying coherent description theoretical description of superfluidity\footnote{The second chapter of \citet{Boning2007} contains a nice overview of the challenges.}. It is currently well understood in a phenomenological context, although analytical conclusions obtained from microscopic models remain sporadic \cite{Boning2007}. For our purposes, it is enough to understand superfluidity as the decrease of the system's moment of inertia induced by quantum effects, i.e. non-classical rotational inertia. In what follows, we will derive an estimator both for the global superfluid fraction and local superfluid density with the help of \citet{Ceperley1995, Boning2007, Dornheim2018}.
\subsubsection{Global superfluid fraction}
Consider the system is placed in a bucket and is slowly rotated with an angular frequency $\omega$ around the z-axis. The bucket does not have to be cylindrical symmetric. A classical fluid will rotate together with the walls where the work done is $\frac{1}{2} I_\mathrm{c} \omega^2$ with the $I_\mathrm{c}$ the classical moment of inertia. The cycle duration is thereby proportional to the square root of the moment of inertia and, therefore, to the total density n, i.e.
\begin{align}
    \frac{2 \pi}{\omega} \propto \sqrt{I_{\mathrm{c}}} \propto \sqrt{n}.
\end{align}
Assume that at low temperature, only a portion of the density, dubbed $n_\mathrm{n}$, will participate with the rotation, such that
\begin{align}
    \sqrt{I(T)} \propto \sqrt{n_{\mathrm{n}}}=\sqrt{n-n_{\mathrm{sf}}},
\end{align}
with $n_{\mathrm{sf}}$ the density of the fluid which does not respond to the rotation. What follows is the relation
\begin{align} \label{eq: sf in momenta}
f\equiv\frac{n_{\mathrm{sf}}}{n}=1-\frac{I(T)}{I_{\mathrm{c}}},
\end{align}
where we defined the superfluid fraction $f$ as the ratio of the superfluid and total density. What remains is to find an expression for the momenta of inertia. Following \citet{Sindzingre1989} and \citet{Ceperley1987} one can use the definition
\begin{align}
    I=\left.\dv{F}{\omega^{2}}\right|_{\omega=0}=\left.\dv{\left\langle L_\mathrm{z} \right\rangle}{\omega}\right|_{\omega=0},
\end{align}
where the momenta of inertia is the work done for an infinitesimally small rotation rate. Rewriting the expectation value of the angular momentum operator $L_\mathrm{z}$ in the path integral formulation and decomposition in $J$ time slices yields the following expression \cite{Ceperley1987}:
\begin{align}\label{eq: superfluid fraction}
    \mathcal{Q}_\mathrm{f}=\frac{2 m}{\lambda \beta }\frac{\left\langle A_{z}^{2}\right\rangle}{I_{\mathrm{c}}},
\end{align}
with
\begin{align}\label{eq: additional expressions sf}
    \mathbf{A}=\frac{1}{2} \sum_{i=1}^{N} \sum_{j=0}^{J-1}\left(\mathbf{r}_{i, j} \times \mathbf{r}_{i, j+1}\right)
    \qq{and}
    I_{c}=m\left\langle\sum_{i=1}^{N} \sum_{j=0}^{J-1} \mathbf{r}_{i,  j}^{\perp} \cdot \mathbf{r}_{i, j+1}^{\perp}\right\rangle.
\end{align}
The vector $\vb{A}$ can be seen as the area of the path or ring-polymers. Hence, equation \eqref{eq: superfluid fraction} connects the superfluid fraction to the worldlines mean-squared area normalized by the cross-sectional area of the whole system. This links nicely to the particle exchange picture as the permutated polymers can cover a larger area. Taking a system where the bosons can only move in two spatial dimensions, we can write the superfluid fraction as
\begin{align}
    \mathcal{Q}_\mathrm{f}=\frac{1}{\lambda \beta }\frac{\sum_{i=1}^{N} \sum_{j=0}^{J-1} r^x_{i, j}r^y_{i, j+1}-r^y_{i, j}r^x_{i, j+1}}{\sum_{i=1}^{N} \sum_{j=0}^{J-1} r^x_{i, j}r^x_{i, j+1}+r^y_{i, j}r^y_{i, j+1}},
\end{align}
where the rotation axis has been taken as perpendicular to the two-dimensional plane. Note that the superfluid fraction is indeed dimensionless.

\subsubsection{Local superfluid density}
It is useful to have an estimator that measures the global properties of the system. However, acquiring spatially resolved information about the system of interest is often also necessary. For the superfluid density, this is provided by \citet{Kwon2006}. Although previous efforts were made by \citet{Draeger2003}, \citet{Kwon2006} showed their implementation leads to biased results as the integration does not give the correct moment of inertia $I$.

The estimator works on the basis of the Landau two fluid model, which divides total density into normal $n_\mathrm{n}$ and superfluid contributions of $n_\mathrm{sf}$, i.e., $n=n_{\mathrm{n}}+n_{\mathrm{sf}}$. Therefore, we can write the moment of inertia as
\begin{align}
I(T)=m \int \dd{\mathbf{r}}\left(n(\mathbf{r})-n_{\mathrm{sf}}(\mathbf{r})\right) \mathbf{r}_{\perp}^{2}=I_{\mathrm{cl}} - m \int \dd{\mathbf{r}}n_{\mathrm{sf}}(\mathbf{r})\mathbf{r}_{\perp}^{2},
\end{align}
where we used that the classical moment of inertia $I_{\mathrm{cl}}=m \int \dd{\mathbf{r}} n(\mathbf{r}) \mathbf{r}_{\perp}^{2}$. $\mathbf{r}_{\perp}$ is the vector where we integrate over with the components of the rotation axis (the z-axis) set to zero. Using the relation \eqref{eq: sf in momenta} and the estimator \eqref{eq: superfluid fraction} that we deduced in the previous subsection, we can write
\begin{align}
m \int \dd{\mathbf{r}} n_{\mathrm{sf}}(\mathbf{r}) \mathbf{r}_{\perp}^{2} = \frac{2 m}{\lambda\beta}\left\langle A_{z}\right\rangle^{2}.
\end{align}
with $A_z$ the z-component of the area vector defined in \eqref{eq: additional expressions sf}. Above relation yields
\begin{align}
\mathcal{Q}_{n_{\mathrm{sf}}}(\mathbf{r})=\frac{2}{\lambda\beta}\frac{ A_{z} A_{z, \mathrm{loc}}(\mathbf{r})}{\vb{r}_{\perp}^{2}},
\qq{with}
\mathbf{A}_{\mathrm{loc}}(\mathbf{r})=\frac{1}{2} \sum_{i=1}^{N} \sum_{j=0}^{J-1}\left(\mathbf{r}_{i, j} \times \mathbf{r}_{i, j+1}\right) \delta\left(\mathbf{r}-\mathbf{r}_{i, j}\right).
\end{align}
The ladder constitutes the local contribution of all trajectories, as it is the sum of all area segments travelling through a differential volume enforced by the delta function. The delta function is dealt with in the same manner as the density, i.e., by dividing space into discrete hyper-cubes and using equation \eqref{eq: implementation of the delta function}.

\subsection{Measuring scheme} \label{sec: measering sheme}
\begin{figure}[htbp]
    \centering
    \includegraphics[width=0.9\linewidth]{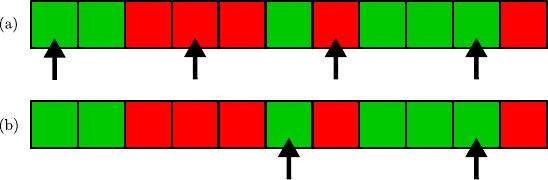}
    \caption{Illustration of two different measuring schemes \textcolor{red}{(a)} and \textcolor{red}{(b)}. Each square in the two series describes an element of the Markov chain which contains diagonal (green) and off-diagonal (red) configurations. The black arrow corresponds to attempting a measurement. The concept of the figure is based on \cite{dornheim2014}.}
    \label{fig: measering shemes}
\end{figure}
As explained in section \ref{sec: pimc}, the canonical expectation value of an observable $\oper$ is given by 
\begin{align} 
\expval*{\oper}\approx \frac{\mathcal{Q}\left(\config_{1}\right)+\ldots+\mathcal{Q}\left(\config_{n_{\mathrm{MC}}}\right)}{n_{\mathrm{MC}}},
\end{align}
where the set $\{\config_k\}=\{\config_{1}, \ldots, \config_{N_{\mathrm{MC}}}\}$ represents the the Markov chain of canonical configurations. However, \gls{WAPIMC} generates configuration in an extended configuration space. Hence, we need a measuring scheme to extract the canonical date from the Markov chain $\{\config_k\}$.

In figure \ref{fig: measering shemes}, two different measuring schemes represented by \textcolor{red}{(a)} and \textcolor{red}{(b)} are shown. In general both schemes are equally valid and which is preferred depends on the system. Measure scheme \textcolor{red}{(a)} goes as follows:
\begin{itemize}
    \item Keep track of the amount of elements in the Markov chain.
    \item Check if the walker is in the z-sector (green square) every $k$ updates (black arrow). If affirmative then perform the canonical meaurements and add them to the corresponding canonical set. If the walker is in G-sector (red square) then do nothing
    \item Reiterate until enough measurements are taken for the desired canonical set of measurements.
\end{itemize}
The second option would be:
\begin{itemize}
    \item Count every new diagonal configuration in the Markov chain with a specific number of particle $N$ stored in $\operatorname{ctr}(N)$.
    \item If one of the counters $\operatorname{ctr}(N)$ is divisible by $k$, then perform the canonical measurements and add them to the corresponding canonical set. Afterwards, reset the counter $\operatorname{ctr}(N)$.
    \item Reiterate until enough measurements are taken for the desired canonical set of measurements.
\end{itemize}
Both meaures schemes are equally valid. Which of the two extraction strategies is more efficient depends on the specific use case. The first strategy is applied in this work.

\subsection{Systematic and Statistical errors in PIMC} \label{sec: error}
As in every simulation, we need error bars in our simulation results to assess if the obtained results are decent.  The following summary of the error analyses of a \gls{PIMC} simulation is based on \citet{Vinay2010,Young2014, Janke2004}. The thesis's \citet{dornheim2014,Yao2020} also provide a section on the topic.

The \gls{PIMC} simulation results include two types of errors that are common. The first is the systematic error caused by the inaccuracy of the density matrix, an error that goes as an increasing function of $\tau$. This finite value has an effect on any result that uses the small-time propagator. The error is often handled by observing the convergence of the system for $\tau\to0$ where the result is extrapolated.

The other type of error is a statistical error by virtue of the limited number of configurations generated during the simulation, contributing to the expectation value of the observable in equation \ref{eq: estimator average}. As stated earlier in section \ref{sec: mc}, if there are a sufficient enough number of uncorrelated configurations, the central limit theorem states that the obtained average $\bar{O}$ of an observable $\hat{O}$ is normally distributed around the exact expectation value with a finite variance given by
\begin{align}\label{eq: def variance}
    \sigma_{\bar{O}}^{2}=\expval{\bar{O}^{2}}- \expval{\bar{O}}^{2}=\frac{\sigma_{O_i}^{2}}{n},
\end{align}
where $n$ is the number of Markov chain elements and $\sigma_{{O}_k}$ the variance of the individual measurement ${O}_i$ with $i\in{1, \ldots, n}$. It implies the notorious Monte Carlo property that the error vanishes as $1/\sqrt{n}$.

\subsubsection{Autocorrelation}
Nevertheless, the result is different if we do not neglect the correlation between the individual configurations. This is easily observed if we substitute the expression for the mean $\bar{O}=\sum^n_{i=1}O_i/n$ into equation \ref{eq: def variance} which yields
\begin{align}
    {\sigma_{\bar{O}}}^{2}
    =\frac{1}{n^{2}} \sum_{i=1}^{n}\left(\expval{O_{i}^{2}}-\expval{O_{i}}^{2}\right)
    +\frac{1}{n^{2}} \sum_{i \neq k}^{n}\left(\expval{O_{i} O_{k}}-
    \expval{O_{i}}\expval{O_{j}}\right).
\end{align}
From the first term, we identify the variance $\sigma_{O_{i}}^{2}$ of an individual Markov chain element. The result can be simplified further by using the summation of the summation in the second term to write $\sum_{i \neq k}^{n}=2\sum_{i =1}^{n}\sum_{k= i+1}^{n}$. Furthermore, using the time translation invariance, we obtain
\begin{align}
    {\sigma_{\bar{O}}}^{2}
    =\frac{1}{n}\left[\sigma_{O_{i}}^{2}
    +2 \sum_{k=1}^{n}\left(\expval{ O_{1} O_{1+k}}-\expval{ O_{1}}\expval{ O_{1+k}}\right)\left(1-\frac{k}{n}\right)\right].
\end{align}
One can factor out $\sigma_{O_{i}}^{2}$ to write the variance of the mean in its usual form, i.e.,
\begin{align}
    {\sigma_{\bar{O}}}^{2}=2\frac{\sigma_{O_{i}}^{2}}{n} \tau_{\mathrm{auto}}
\end{align},
where the integrated autocorrelation time $\tau_{\mathrm{auto}}$ represents a measure for the length between two independent configurations in the Markov chain, given by
\begin{align} \label{eq: def autocorrelation}
    \tau_{\mathrm{auto}}
    =\frac{1}{2}+\sum_{k=1}^{n} O(k)\left(1-\frac{k}{n}\right)
    \qq{with}
    O(k)=\frac{\expval{O_{i} O_{i+k}}-\expval{O_{i}}\expval{O_{i}}}{\expval{O_{i}^{2}}-\expval{O_{i}}\expval{O_{i}}}.
\end{align}
It is clear that correlations between configurations increases the variance.

\chapter{Implementation and Benchmarks}
\label{ch: Implementation}
This chapter gives an overview of the technical information for the \gls{WAPIMC} implementation and discusses the system and software decisions taken. The algorithm is written in Julia, a high-performance, dynamic programming language that is suitable for numerical analysis and computational research due to its use of type-stability through specialization via multiple-dispatch \cite{Julia2012}. The chapter begins with a brief overview of the unit system, interactions, and treatment of the optimization parameters. Afterwards, a discussion of implementation tests, including a comparison to analytical solutions, is given. Divergence of the Worm extension to the \gls{PIMC} algorithm with analytical solutions is observed and discussed. Finally, some preliminary results of density distributions for \gls{SRL}s are presented.

\section{WA-PIMC in practise: Our implementation}
\subsection{Units}
Our system of interest, i.e., ultracold atoms in two-dimensional optical lattices, can be described by the Hamiltonian of the form
\begin{align}
    H&=-\lambda\sum^N_i\grad^2_i+\sum^N_iV(\vb{r}_i)+\sum^N_{i<k} U(\vb{r}_i-\vb{r}_k),
\end{align}
with $V=V_0V^\prime$ the effective potential of the optical lattice and $U$ the pairwise interaction potential and $\lambda=\frac{\hbar^2}{2m}$ the prefactor of the kinetic term. Here we split the optical lattice potential into its depth $V_0$ and the normalized interference pattern. For numerical analyses, it is convenient to work in dimensionless units. From our discussion in chapter \ref{ch: Ultracold Atoms In Optical Lattices}, it is clear that the recoil energy $E_r=\frac{\pi^2\hbar^2}{2ma^2}=\frac{\pi^2\lambda}{a^2} \sim [ML^2T^{-2}\Theta^0]$\footnote{The SI dimension in terms of physical dimensions of time ($T$), length ($L$), mass ($M$), temperature ($\Theta$).} and the lattice spacing $a$ present themselves as natural measures of the energy and length scales of the system. The temperature can then be taken in units of $E_r/k_B$. The resulting dimensionless Hamiltonian is given by
\begin{align}
    \tilde{H}
    &=\frac{-1}{\pi^2}\tilde{\grad}^2 + \tilde{V}_0\sum^N_i \tilde{V}^\prime(\tilde{\vb{r}}_i) +  \sum^N_{i<k} \tilde{U}(\tilde{\vb{r}}_i-\tilde{\vb{r}}_k),
\end{align}
where we use a tilde over the variable $\tilde{x}$ to indicate it is dimensionless. Hence, to work dimensionless in our notation, one must take $\lambda=1/\pi^2$ when evaluating the propagators from chapter \ref{ch: PIMC}. The lattice depth $\tilde{V}_0$ is a tunable parameter that corresponds to varying the laser intensity in an experimental setting. The dimensionless interaction potential  $\tilde{U}$ is handled in the next subsection. When discussing the results in section \ref{sec: WA-PIMC UAOL} the tilde notation is dropped for convenience, and we are assumed to work dimensionless.

\subsection{Interactions}\label{sec: interaction implementation}
In chapter \ref{ch: Ultracold Atoms In Optical Lattices}, we discussed that ultracold atoms interact in the low-energy s-wave scattering regime such that the interaction is fully characterized by the two-dimensional scattering length $a_{2\mathrm{D}}$. Nonetheless, we saw that, in practice, the two-dimensional ultracold atomic gas is realized by confining the gas with a harmonic trap. Hence, we related the two-dimensional scattering length $a_{2\mathrm{D}}$ to the three-dimensional scattering length (see equation \eqref{eq: a_2D}). This allowed us to express the dimensionless coupling strength $\tilde{g}$ in terms of the two-dimensional scattering length $a_{2\mathrm{D}}$:
\begin{align}\label{eq: dimensionless coupling constant}
\tilde{g} \simeq \frac{1}{\tilde{g}_0^{-1}+(4\pi)^{-1}\ln \left(\Lambda E_{\mathrm{r}} / \mu\right)},
\qq{with}
    \tilde{g}_{0}\equiv\frac{2 \pi}{\ln \left(a / a_{2 \mathrm{D}}\right)},
\end{align}
where $\mu$ is the chemical potential, $E_r=\frac{\pi^2\lambda}{a^2}$ the recoil energy, $a$ the lattice spacing, and $\Lambda \approx 0.141$ a numerical constant. The lattice spacing $a$ in the both term can be cancelled with each other such that the coupling strength $\tilde{g}$ only depends on the two-dimensional scattering length $a_{2\mathrm{D}}$ and the chemical potential $\mu$. Indeed, using the recoil energy as units of energy and lattice spacing as units of length, only $\tilde{a}_{2\mathrm{D}}$ and $\tilde{\mu}$ remain. In practice, we will use $\tilde{g}_{0}$ as the parameter to tune the interaction.

\begin{figure}[htbp]
    \centering
    \includegraphics[width=\linewidth]{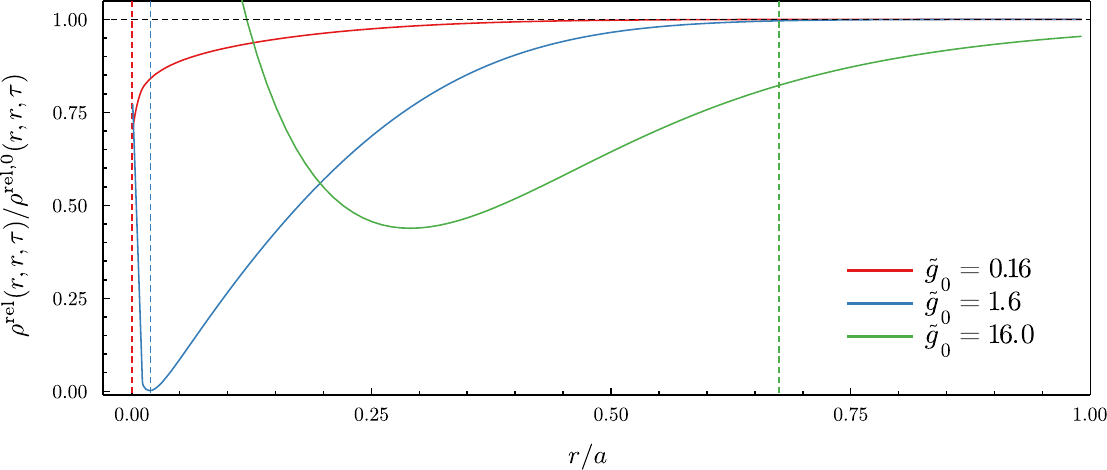}
    \caption{The relative diagonal interaction propagator for an inverse temperature of $\tau=0.1$ three values of the interaction strength $\tilde{g}_0=0.16, 1.6, 16.0$. The dashed black indicates the non-interacting value. The coloured vertical dashed line indicate the two-dimensional scattering length $a_{2\mathrm{D}}$ corresponding to each coupling strength where the propagator is cut off.}
    \label{fig: diagonal_propagator}
\end{figure}

Nevertheless, we still need an expression for the relative interaction propagator $\rho^{\mathrm{rel}}$. For this, we follow the approach from \citet{Gauter2021} by solving for eigenfunctions of the radial Schr\"odinger equation in the repulsive interaction s-wave scattering limit. This is worked out in appendix \ref{app: rel int prop}. We find that the relative interaction propagator yields
\begin{align}
\rho^{\mathrm{rel}}\left(\mathbf{r}, \mathbf{r}^{\prime}, \tau\right)
=\rho_0^\mathrm{rel}\left(\mathbf{r}, \mathbf{r}^{\prime}, \tau\right) &-\frac{1}{2 \pi} \int \dd{k} k \mathrm{e}^{-\tau E_k} \frac{t_{k}^{2}}{1+t_{k}^{2}} J_{0}(k r) J_{0}\left(k r^{\prime}\right) \\
&-\frac{1}{2 \pi} \int \dd{k} k \mathrm{e}^{ -\tau E_k} \frac{t_{k}}{1+t_{k}^{2}}\left[J_{0}(k r) Y_{0}\left(k r^{\prime}\right)+J_{0}\left(k r^{\prime}\right) Y_{0}(k r)\right] \\
&+\frac{1}{2 \pi} \int \dd{k} k \mathrm{e}^{-\tau E_k} \frac{t_{k}^{2}}{1+t_{k}^{2}} Y_{0}(k r) Y_{0}\left(k r^{\prime}\right) ,
\end{align}
where $\rho_0^\mathrm{rel}$ is the non-interacting propagator for a particle with reduced mass $m^*=m/2$, $J_{0}$ the zeroth order Bessel function of the first kind, $Y_{0}$ the zeroth order Bessel function of the second kind, and $t_{k} =\pi / 2 \ln \left(\eta k a_{2 \mathrm{D}}\right)$ with $\eta\approx0.89054$ a numerical constant. Using equation \eqref{eq: dimensionless coupling constant}, we can write the interaction propagator in the relevant coupling constant $\tilde{g_0}$, i.e., we find the relation
\begin{align}
    \frac{1}{t_{k}}+\frac{4}{\tilde{g}_{0}}=\frac{2}{\pi} \ln (\eta k a) .
\end{align}
Substituting this dimensionless identity, the interaction propagator is solved numerically beforehand and used as an interpolation function during the simulation. In figure \ref{fig: diagonal_propagator} the relative diagonal interaction propagator $\rho^{\mathrm{rel}}\left(r, r, \tau\right)$ with $\tau=0.1$ is plotted for different values of the relevant coupling constant $\tilde{g_0}$. It represents the likelihood of encountering the two interacting particles at a distance $r$, whereas the non-diagonal represents the chance of their moving from $r$ to $r^\prime$. The corresponding scattering lengths $\tilde{a}_{2\mathrm{D}}$ is indicated by the dashed coloured vertical line. As required, the diagonal interaction propagator increases with $r$ if $r>a_{2\mathrm{D}}$, eventually converging to the non-interacting regime (horizontal dashed black line). However, because of the non-exact treatment in the short-range region, it diverges for $r \ll a_{2\mathrm{D}}$. To address this issue, we opted to split the propagator for radii less than the scattering length, i.e., $\rho^{\mathrm{rel}}\left(\mathbf{r}, \mathbf{r}^{\prime}, \epsilon\right)=0$ if $r<a_{2 \mathrm{D}}$ or $r^{\prime}<a_{2 \mathrm{D}}$. It enforces the simulation to respect the condition $na_{2\mathrm{D}}<1$ with $n$ the two-dimensional density of the ultracold atomic gas, a condition already set by the s-wave scattering approximation.

\subsection{Nearest-neighbours}
To assure scalability when adding many particles to the system, a nearest-neighbour algorithm is implemented such that the pairwise interaction propagator must only be evaluated for beads within a certain range. Furthermore, it can be used for optimizing the Monte Carlo updates by allowing for importance sampling.

At all time slices, we employ a nearest-neighbour list or table \citet{allen2017computer}. That is, the volume of the system is partitioned into identical bins for each time slice $j=0,1,2, \ldots,(J-1)$, with a list of all the beads that are present in each bin. The size is taken to be the distance where the diagonal interaction propagator becomes unity (see figure \ref{fig: diagonal_propagator}). If one wants to determine which beads should be considered in evaluating the interaction propagator, one collects all particles in the bin of the relevant bead and all neighbouring bins. Next, a ball-tree\footnote{Here we opted for a ball tree instead of a more efficient k-d tree as the latter does not work with periodic boundary conditions.} is created from the collected beads. Finally, a nearest-neighbour search is performed with a radius of the bin size.


\subsection{Optimalization parameters}
As we have seen in chapter \ref{ch: PIMC}, all the Monte Carlo updates discussed contain an optimalisation parameter. These parameters do not affect the result of the Monte Carlo simulation but only determine the efficiency with which the walker traverses configuration space. They are tied with the acceptance ratios of the Monte Carlo updates. For example, for the reshape update, the parameter $\bar{M}$ controls the maximum number of time slices that can be reshaped. On the one hand, if it is too large, the updates will get more rejected and the acceptance ratio will be small. On the other hand, if it is too small, the updates are always accepted. In both cases, the Markov chain will cover the configuration space less effectively. Hence, during the simulation, the optimalization parameters are adjusted such that the acceptance ratio lies between 20 to 80 percent.

\subsection{Error analyses}
In section \ref{sec: error}, we discussed that to compute errors for our obtained calculation, one must take the correlation effects between configurations into account. One often computes the autocorrelation time with a method called the binning analysis, complemented by jackknife resampling. It computes the error and the autocorrelation time by using the value representative of intervals of various bin sizes during the Monte Carlo step to give a reliable errorbar value. The error analysis in our application is done with the code\footnote{The ALPS package \cite{Bauer_2011} in C\texttt{++} is also worth looking into.} in \citet{Nichols2020}.

\section{Benchmarks}
\begin{figure}[htbp]
    \centering
    \centerline{
    \includegraphics[width=1.05\linewidth]{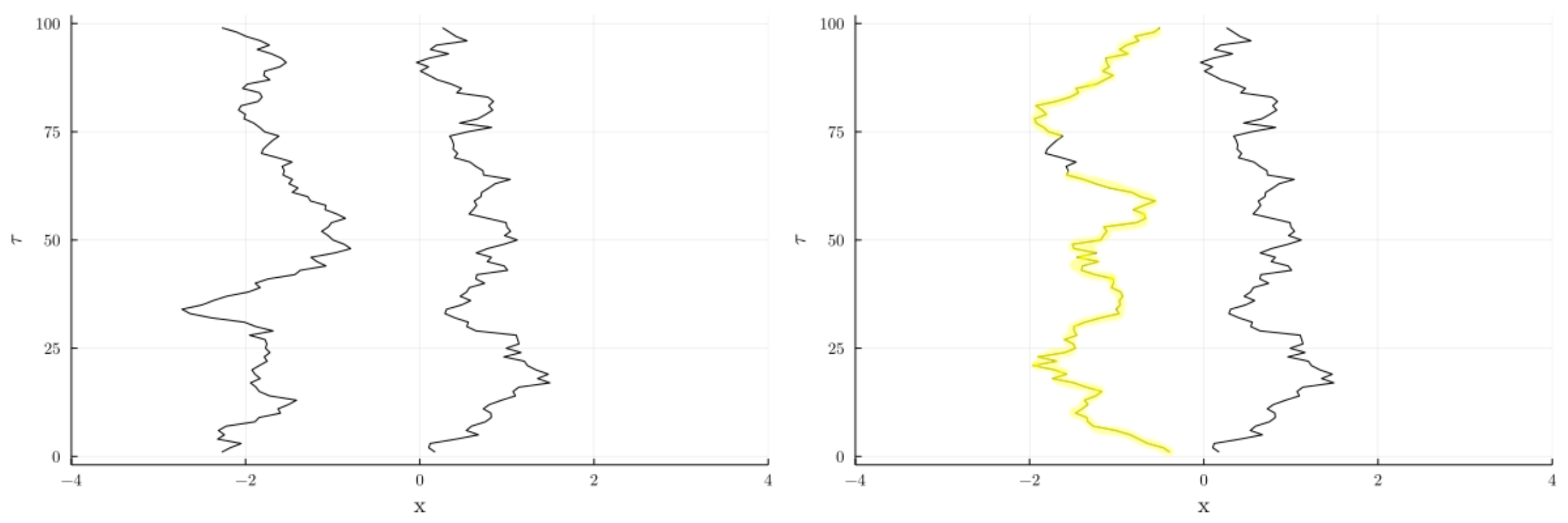}}
    \caption{Snapshots of the $x$ coordinate of the worldlines of $J=100$ beads in imaginary time at a temperature $T=1.0$ in a box with volume $V = 64.0$ with periodic boundary conditions. The snapshots are taken before (left) and after (right) a reshape update. The modified path is highlighted in yellow.}
    \label{fig: reshape_before_after}
\end{figure}

Naturally, in software development, it is important to benchmark and test all components of the code. Unit testing is a common testing approach that involves analyzing individual functions or subroutines of the source code to see if they are acceptable for use. Each module of our algorithm has undergone this procedure, from every bead having a next and previous bead, to correspondence between the nearest neighbour bins and the position of the beads. Other than standard unit testing, \gls{WAPIMC} leans itself to examine the path visually. For example, in figure \ref{fig: reshape_before_after}, a before-after is seen of the reshape update on a worldline with $J=100$ beads where the modified path is highlighted in yellow. The worldlines are plotted in imaginary time, but also the polymer perspective can be observed. Bugs that would be difficult to identify with unit tests can be found easier utilizing these visual inspections.

In what follows, we discuss the validity of the written code by comparing results calculated using our PIMC implementation with analytical solutions. The discussion is split into two parts: a part on the core PIMC implementations and a part on the Worm algorithm extension.

\subsection{Path Integral Monte Carlo}
A reliable way of testing the integration of all software modules and components --- including updates and measurements --- at one is by simulating systems where the results can be computed analytically. We will analyse a non-interacting ideal Bose gas trapped in a harmonic potential. In the literature, the non-interacting Bose gas model is typically addressed in the grand-canonical ensemble and in the thermodynamic limit. However, the $N$-particles partition function $Z_{N}(\beta)$ at inverse temperature $\beta$ can be computed exactly with the help of the recursion formula \cite{Borrmann1993, krauth2006}
\begin{align}\label{eq: recursion formula}
Z_{N}(\beta)=\frac{1}{N} \sum_{k=1}^{N} z(k \beta) Z_{N-k}(\beta),
\end{align}
which includes the $(N-k)$-particles partition function $Z_{N-k}(\beta)$ at the same temperature and the one-particle partition function $Z_1(k \beta)=z(k \beta)$ evaluated at $k$ times the inverse temperature $\beta$.

\subsubsection{Harmonically trapped Bose gas}
In order to simulate the Bose a gas trapped in a two-dimensional harmonic potential, it is convenient to use dimensionless units\footnote{Note that the SI dimensions are indeed correct, i.e., $\hbar\omega \sim [ML^2T^{-2}]\sim [E]$ and $\sqrt{\frac{\hbar}{m \omega}}\sim \left[\sqrt{\frac{ML^2T^{-1}}{MT^{-1}}}\right]=[L]$, such that $\tilde{x}$ and $\tilde{E}$ are indeed dimensionless.} of energy and length. Proposing energy units $\hbar \omega$ and length units $ (\frac{\hbar}{m \omega})^{1/2}$, indeed results in the dimensionless Hamiltonian
\begin{align}\label{eq: Ham QHO}
\ham = \sum^N_{i=1}-\frac{\hbar^{2}}{2 m} \hat{\grad}^2+\frac{1}{2} m \omega^{2} \hat{\vb{r}}^{2}
\ \to \
\tilde{H} = \sum^N_{i=1}-\frac{1}{2}  \tilde{\grad}^2+\frac{1}{2} \tilde{\vb{r}}^{2},
\end{align}
with $\omega$ the angular frequency of the oscillator and the tilde over the operators indicates it is dimensionless. By taking in mind our energy units we use the temperature units of $\frac{\hbar\omega}{k_B}$ such that $\tau=\tilde{\tau} \frac{1}{\hbar \omega}$. From now one we drop the tilde notation and assume we work dimensionless.

The energy eigenvalues for a two-dimensional quantum harmonic oscillator corresponding to the dimensionless Hamiltonian \eqref{eq: Ham QHO} are well-known to be $\varepsilon\left(n_{x}, n_{y}\right)=(n_{x}+\frac{1}{2})+(n_{y}+\frac{1}{2})$ \cite{sakurai2017} which results in the one-particle partition function
\begin{align}
z(\beta)
&=\sum_{n_x,n_y=0}^{\infty} e^{-\beta\varepsilon\left(n_{x}, n_{y}\right)}
=\left(\sum_{n=0}^{\infty} e^{-\beta\left(n+\frac{1}{2}\right)} \right)^2
=e^{-\beta }\left( \sum_{n=0}^{\infty}\left[e^{-\beta }\right]^{n}\right)^2
= \left(\frac{e^{-\frac{1}{2}\beta }}{1-e^{-\beta }}\right)^2.
\end{align}
Because the partition function is known analytically, the total energy expectation value for a boson trapped in a harmonic potential is simply computed from the thermodynamic definition
\begin{align}
E_{1}
=-\frac{\partial}{\partial \beta} \ln Z_{1}(\beta)
=-\frac{\partial}{\partial \beta} \ln z(\beta)
=\coth\left(\frac{\beta}{2}\right).
\end{align}
For $N>1$ bosons one needs to implement Bose-Einstein statistics into the partition function. For non-interacting bosons, this is reflected in the recursion formula given in equation \eqref{eq: recursion formula}. For example, the two-bosons and three-bosons partition functions are given by 
\begin{align}
\begin{aligned}
Z_{2}(\beta)=\frac{1}{2}\left[z(\beta)^{2}+z(2 \beta)\right] \qq{and}
Z_{3}(\beta)=\frac{1}{3}\left[z(\beta)^{3}+z(\beta)^{2} z(2 \beta)+z(3 \beta)\right] .
\end{aligned}
\end{align}
If one sets the particle exchange terms $z(k \beta)$ with $k>1$ to zero and gets rids of the normalization factor, the partition function represents so-called Bolzmannons. An example of Bolzmannons gas would be $N$ boson atoms of different species.\footnote{Bolzmannons are still quantum mechanical.} The total energy for $N=2, 3$ is then again computed by taking the derivative to the inverse temperature, $\beta$. For higher $N$, one can easily write a solver which computes the energy in terms of the inverse temperature $\beta$ symbolically. Or even better, one could compute the exact value numerically by using methods such as \emph{automatic differentiation} \cite{Hoffmann2016}.
\begin{figure}[htbp]
    \centering
    \includegraphics[width=0.75\linewidth]{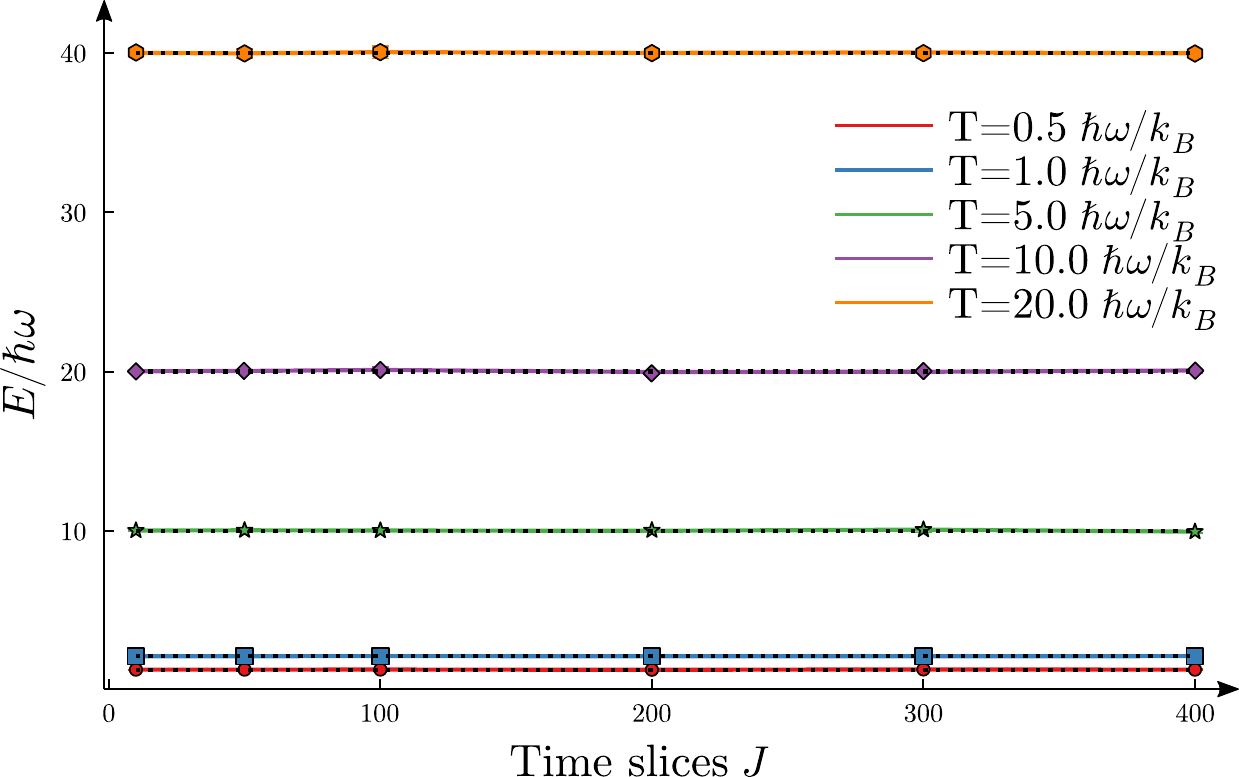}
    \caption{Total energy for a boson trapped in a harmonic potential for different temperatures $T$ and time slices $J$. The errors are smaller than the symbol size. The dashed black lines are the exact analytically computed values.}
    \label{fig: Energy_trapped_boson_alt}
\end{figure}

\begin{figure}[htbp]
    \centering
    \includegraphics[width=0.85\linewidth]{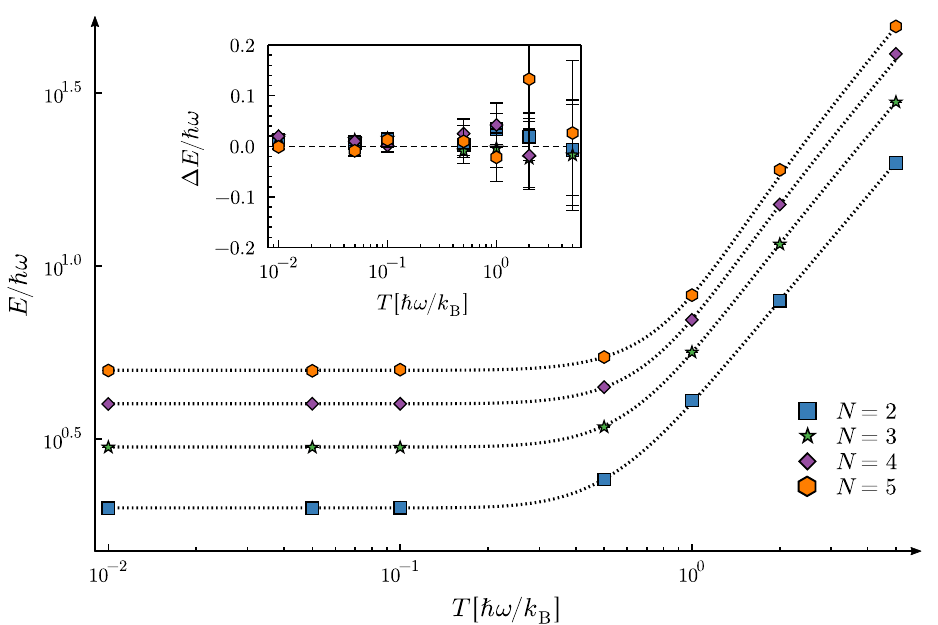}
    \caption{Total energy for a non-interacting Bose gas trapped in a harmonic potential for different number of particles $N$ and temperatures $T$. Most of the errors are smaller than the symbol size. The inset contains the difference between the thermodynamic and virial estimator.}
    \label{fig: Energy_trapped_boson_varying_T}
\end{figure}

First, we analyse the situation where $N = 1$. In this case, there are no permutation updates in the \gls{PIMC} simulation, but sets a rigorous benchmark for all the other updates. Moreover, it is a simple test for the core modules of \gls{PIMC}. In figure \ref{fig: Energy_trapped_boson_varying_T}, the total energy computed for one boson is plotted against different values of temperature. The energies are obtained using the virial estimator \eqref{eq: virial estimator}. We ran the experiment by altering the total number of beads $J$ while leaving the other simulation parameters the same. The black dashed lines indicate the exact values of energy. The simulation with $10^7$ number of measurements yielded statistical error on the fourth digit and the exact value within single error bars. Hence, a complete agreement with numerical results is observed for all different combinations of time slices and temperatures.

The energy of a harmonically trapped boson seems trivial. However, it is a key test for our computational technique. Indeed, to extract a scalar from the configuration with a very low error requires sampling the stationary distribution with great precision. Having a good fit of simulated data to the analytical values is an important indicator of the correctness of the sampling technique.

We continue with the harmonically trapped non-interacting Bose gas, i.e., $N > 1$ system where the particle statistics and therefore the permutation update become relevant. In figure \ref{fig: Energy_trapped_boson_varying_T}, the simulated total energy for two to five particles at different values of the temperature can be seen. Again, the virial estimator \eqref{eq: virial estimator} was used and the number of time slices are taken such $\tau\approx0.01$. The black dashed line corresponds to the exact computed energies from the partition function \eqref{eq: recursion formula}. The simulated results lie within one standard deviation of the theoretical values.

In table \ref{tab: accaptance ratios}, one observes the acceptance ratios of the Markov chain updates for a harmonically trapped Bose gas of $N=5$ particles at temperature $T=1.0$ and with $J=100$ beads. All updates have a medium to high probability, indicating that there is no ergodicity problem and just minor autocorrelations in the Markov chain.
\begin{table}[H]
    \centering
        \caption{Acceptance ratios for a harmonically trapped Bose gas of $N=5$ particles at temperature $T=1.0$ and with $J=100$ beads.}
    \begin{tabular}{lll}
        \toprule
        \textbf{Center-Of-Mass} & \textbf{Reshape} & \textbf{Permutation}\\
        \midrule[\heavyrulewidth]
        0.636975 & 0.73971 & 0.34014 \\
        \bottomrule
    \end{tabular}
    \label{tab: accaptance ratios}
\end{table}

As discussed in section \ref{sec: observable}, assuming the virial theorem, one can obtain two different estimators for the energy. Hence, it provides a good test to observe if the virial theorem is respected by our \gls{PIMC} algorithm. Indeed, the inset in figure \ref{fig: Energy_trapped_boson_varying_T} shows the difference between the thermodynamic and virial estimator. The expected value always lies within the error bar, although the variance of the thermodynamic estimator becomes apparent for high temperatures.

\begin{figure}[htbp]
    \centering
    \includegraphics[width=0.87\linewidth]{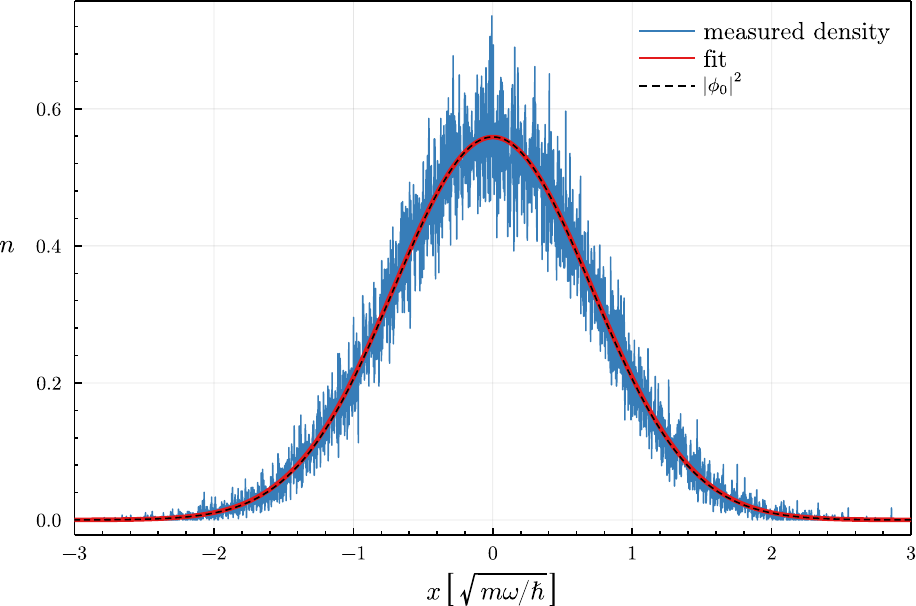}
    \caption{Density of a non-interacting harmonically trapped Bose gas at very low temperature. The blue solid line corresponds to the measured density with a PIMC simulation of $10^5$ measurements of $N=5$ bosons at $T=0.03 \ \hbar\omega/k_\mathrm{B}$ of $J=500$ beads. The red line is a fitted Gaussian through the measured density. The black dotted line corresponds to the analytically exact ground state density.}
    \label{fig: ground_state_density}
\end{figure}

Not only the energy can be computed analytically, but also the eigenfunctions. Hence, in the dimensionless units defined earlier, the ground state wave function for one boson is given by
\begin{align}
\phi(\vb{r})=\pi^{-1 / 4} e^{-\frac{1}{2} \vb{r}^{2}}.
\end{align}
Furthermore, given that the density operator is $\hat{n}(\vb{r})=\delta(\vb{r}-\vb{r}^\prime)$, the ground state density can be written as
\begin{align*}
n(\vb{r})=\frac{\langle\phi|\hat{n}(\vb{r})| \phi\rangle}{\langle\phi \mid \phi\rangle}=\int \dd{\vb{r}^\prime} \phi(\vb{r}^\prime) \delta(\vb{r}^\prime-\vb{r}) \psi(\vb{r}^\prime)=\phi(\vb{r})^{2}=a e^{-b \vb{r}^{2}},
\end{align*}
with coefficients $a=\pi^{-1 / 2}$ and $b=1$, as expected. At low temperatures $T$, our \gls{PIMC} algorithm should get very close to the ground state density as all bosons will be at the ground state energy level. Indeed, taking a quick sample of $10^5$ measurements of $N=5$ bosons at temperatures $T=0.03 \ \hbar\omega/k_\mathrm{B}$ and fitting a Gaussian to the measured density (see figure \ref{fig: ground_state_density}), we find the coefficients $a=0.5662\pm 0.0031\approx\pi^{-1 / 2}$ $b= 1.0043\pm 0.0073\approx1$ with a 95\% confidence level.

At this point, one may infer that the \gls{PIMC} code works correctly for a non-interacting Bose gas. Unfortunately, there are no analytical results for interacting or correlated many-body systems. Nevertheless, in this case, one can compare different regimes of the system.

\subsection{Worms algorithm}
\begin{figure}[htbp]
    \centering
    \includegraphics[width=\linewidth]{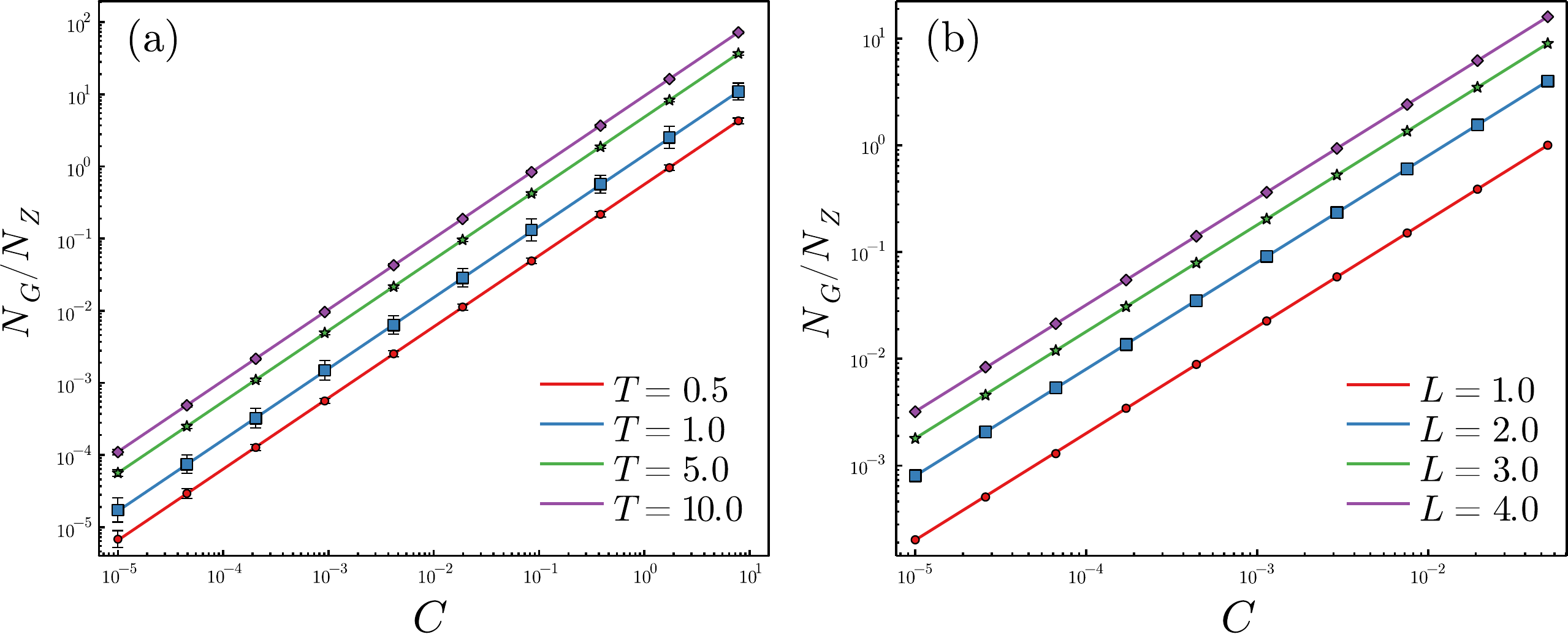}
    \caption{Log–log plot of the ratio of the number of configuration in the G- or Z-sector in the Markov chain in function of the optimization parameter $C$ in the total partition function $Z_{\mathrm{tot}} = Z+CZ^\prime$ with logarithmic error bars. The solid lines are one-parameter fits.\textcolor{red}{(a)} The open and close Markov chain updates for different temperatures $T$ each indicated by a different marker and colour. \textcolor{red}{(b)} The insert and remove Markov chain updates for different box sides $L$ each indicated by a different marker and colour.}
    \label{fig: log_Nz-Ng_plot_box}
\end{figure}

Now, that we established that \gls{PIMC} indeed samples the right stationary distribution, we move on to \gls{WAPIMC}. A typical first test for the worm algorithm is the relation 
\begin{align}\label{eq: ratio Nz-Ng}
    \frac{N_\mathrm{G}}{N_\mathrm{Z}}=C\frac{Z_\mathrm{G}}{Z} \qq{if} N_\mathrm{MC}\to\infty.
\end{align}
In other words, the optimization parameter $C$ is linearly proportional to the ratio of $N_\mathrm{G} / N_\mathrm{Z}$, i.e., the number of configurations in the Markov chain that are in the G-sector versus the Z-sector. Indeed, in figure \ref{fig: log_Nz-Ng_plot_box}, we present the ratios found at various $C$ values, together with one-parameter fits. In \figa~the temperature is varied, which influences the metropolis question of the open and close update by the free propagator. In \figb, the volume of the simulation box --- with periodic boundary conditions --- is changed, which influences the metropolis question of the insert and remove update.

\begin{figure}[htbp]
    \centering
    \includegraphics[width=1.0\linewidth]{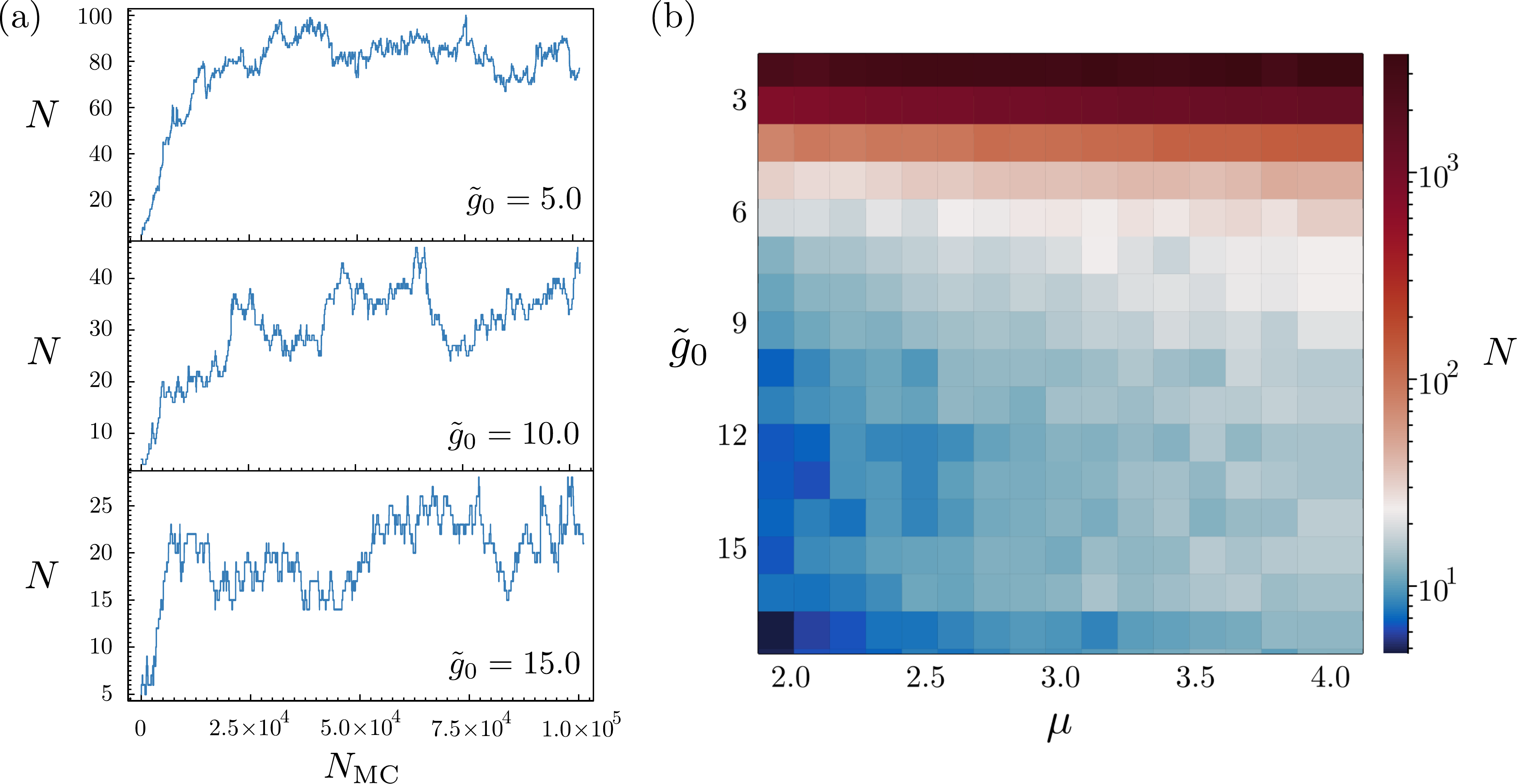}
    \caption{Dependence of the number of particles on the interaction strength $\tilde{g}_0$ and chemical potential $\mu$ of a weakly-interacting Bose gas in a box with periodic boundary conditions. \figa~The number of particles fluctuating in the Markov chain for interaction strengths $\tilde{g}_0=5.0, 10.0, 15.0$. \figb~Heatmap of the average number of particles in the Markov chain with respect to the chemical potential and the interaction strength.}
    \label{fig: Nheatmap}
\end{figure}

A feature of the Worm algorithm is that it allows simulating in the grand canonical ensemble instead of the canonical, with the effect being that the number of particles fluctuates around average depending on the size of the simulation box, chemical potential, and interaction strength. In figure \ref{fig: Nheatmap}\figa, you can observe the number of closed worldlines fluctuating in the Markov chain for different values of the interaction strength $\tilde{g}_0$. Notice that it converges to an average depending on the interaction strength. Figure \ref{fig: Nheatmap}\figb~shows a heatmap of the average number of particles in the Markov chain for a free (no external potential) weakly-interacting Bose gas, varying the chemical potential and the interaction strength.
\begin{table}[htbp] \centering
    \caption{Mean and standard deviation for $10^6$ energy measurements of one boson in a harmonic trap at $T=1.0\hbar\omega/k_\mathrm{B}$ with $J=100$. The worm updates open and close (OC) and/or insert and remove (IR) are added to the set of the Z-sector. Variation with open, close, advance, recede, and/or swap is present. An asterisk indicates a restriction has been added such that the particle number cannot change.}
    \begin{footnotesize}
    \begin{tabular}{llllll}
    \toprule
    \textbf{Energy} & \textbf{OC} & \textbf{IR} & \textbf{OC*+IR*} & \textbf{OC+AR*} & \textbf{OC+AR*+SW}\\
    \midrule[\heavyrulewidth]
    Exact energy & 2.164 & 2.164 & 2.164 & 2.164 & 2.164 \\ 
    Thermodynamic energy & 2.158 $\pm$ 0.041 & 2.167 $\pm$ 0.058 & 2.162 $\pm$ 0.032 & \color{red} 2.006 $\pm$ \color{red}  3.332 &  \color{red} 2.011 $\pm$ 3.329 \\ 
    Virial energy & 2.161 $\pm$ 0.035 & 2.164 $\pm$ 0.053 & 2.166 $\pm$  0.033 & \color{red} 2.153 $\pm$ 1.979 & \color{red} 2.159 $\pm$ 1.994 \\ 
    \bottomrule
    \end{tabular}
    \end{footnotesize}
    \label{tab: boson energy worms}
    \end{table}

Like the updates of the Z-sector, the individual updates of the worm algorithm or G-sector can be checked by performing visual and unit tests. Comparable analytical results for a Bose gas have only been found for non-interacting systems. However, as discussed in section \ref{sec: Worms canonical}, if the interactions are switched off, one cannot work in a grand-canonical ensemble, as the particle number will diverge due to a non-zero chemical potential. We explained that one has to restrict the updates of the worm algorithm (see section \ref{sec: Worms canonical}). In table \ref{tab: boson energy worms}, we report the result of simulating one boson in a harmonic trap at $T=1.0 \ \hbar\omega/k_\mathrm{B}$ with $J=100$ by step-by-step adding the complementary G-sector updates to the set of the Z-sector. Only adding open and close (OC) or insert and remove (IR) yields results where both the thermodynamic and virial estimator are within one sigma of the expected result. Both adding the OC and IR updates but restricting them such that the number of closed worldlines can not change, also gives energies in agreement with within one sigma. Nevertheless, adding advance and recede (AR) together with OC updates leads to a very large error. Doing the same for two bosons such that the swap (SW) update becomes relevant, gives the same results (see table \ref{tab: two boson energy worms}).

\begin{table}[htbp] \centering
\caption{Mean and standard deviation for $10^6$ energy measurements of two bosons in a harmonic trap at $T=1.0 \ \hbar\omega/k_\mathrm{B}$ with $J=100$. The worm updates open and close (OC) and/or insert and remove (IR) are added to the set of the Z-sector. Variation with open, close, advance, recede, and/or swap is present. An asterisk indicates a restriction has been added such that the particle number cannot change.}
\begin{footnotesize}
\begin{tabular}{llllll}
\toprule
\textbf{Energy} & \textbf{OC} & \textbf{IR} & \textbf{OC*+IR*} & \textbf{OC+AR*} & \textbf{OC+AR*+SW}\\
\midrule[\heavyrulewidth]
Exact energy & 4.328 & 4.328 & 4.328 & 4.328 & 4.328 \\ 
Thermodynamic energy & 4.340 $\pm$ 0.053 & 4.445 $\pm$ 0.082 & 4.325 $\pm$ 0.032 & \color{red} 4.023 $\pm$ 6.589 &\color{red} 3.351 $\pm$ 5.018\\ 
Virial energy & 4.334 $\pm$ 0.036 & 4.334 $\pm$ 0.064 & 4.316 $\pm$  0.033 &  \color{red}4.383 $\pm$ 2.803 &\color{red} 3.761 $\pm$ 2.613\\ 
\bottomrule
\end{tabular}
\end{footnotesize}
\label{tab: two boson energy worms}
\end{table}

We postulate two possible reasons for the error. First, there could be a bug in the modules of the advance and recede update which leads to a wrong sampling of the position of the beads. For example, when advancing the worm and sampling a new position $\mathbf{r}_{\mathcal{I}}$ from the position head $\mathbf{r}_{0}$ with the normal distribution $ \rho_{0}\left(\mathbf{r}_{0}, \mathbf{r}_{\mathcal{I}} ; M \tau\right)$ the variance could be off. Closing the worm after an advance update yields a bias in the average distribution of the position of the beads during the simulation. If the other updates are functioning properly, the bias should only result in a small mismatch of the measured data since they sample the correct stationary distribution. Such bugs are hard to spot as they could just be a wrong factor of two in an implemented expression. Second, we restricted the advance and recede updates by prohibiting the time slices of the head and tail of the worm to advance or recede passed each other. This constraint does what it is supposed to, but could be implemented incorrectly. For example, a wrong implementation could result that the advance and recede updates are not complementary to each other any more. Accordingly, the detailed balance condition required in section \ref{eq: detailed balance equation} would not be fulfilled, i.e, the use of an update would not be reversible. As a result, the Metropolis walker could not produce a stationary distribution. Visual and unit tests do not reveal any insight in either of the open cases.
\begin{figure}[htpb]
    \centering
    \includegraphics[width=0.85\textwidth]{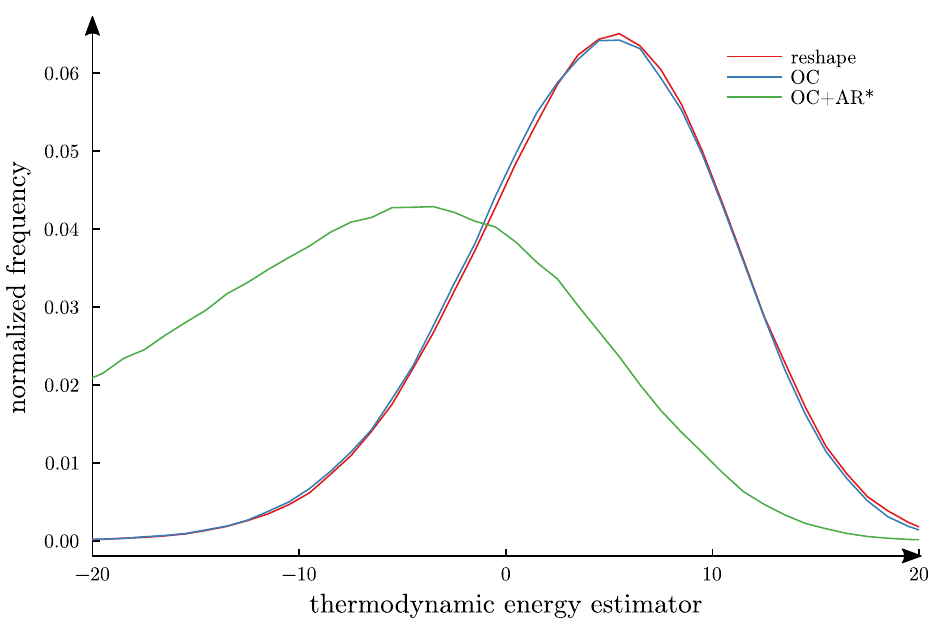}
    \caption{Distribution of the thermodynamic energy estimator for a ``free'' worldline of $J=100$ beads continuously update with the reshape update (red), open and close (OC, blue), and the open and close with restricted advance and recede updates (OC+AR*, green).}
    \label{fig: energy_histogram_better}
\end{figure}

We investigated this further by noticing the discrepancy between the thermodynamic and viral energy estimators. The thermodynamic estimator is consistently lower than the exact value and the virial estimator. The difference between the two estimators is that the thermodynamic estimator depends on the relative distance between two subsequent beads, whereas virial estimator only depends on the position of the individual beads (see section \ref{sec: observable}). Hence, it could be interesting to observe the distribution of thermodynamic energy estimator for one worldline in free space. This is plotted in figure \ref{fig: energy_histogram_better} for only using the reshape update, the open and close updates and open and close with restricted advance and recede updates. Only using reshape or open and close yield the same distribution, whereas also adding advance and recede, replaces and skews the distribution to negative energies. The modification is fairly large and is thought of as the main culprit for the large error observed in table \ref{tab: boson energy worms} and \ref{tab: two boson energy worms}. As no potential is present, the anomaly comes from the relative distance between subsequent beads, i.e., it is too large. In other words, the polymers modified by using the open, close, advance and recede updates are too large in diameter. This could happen by sampling the free propagator with the wrong parameters. However, on first sight this does not seem the case. A closer analyses is needed.



A way to further investigate the matter could be to design simple tests for advance and recede updates. For example, taking a worm of two links long in a specific arrangement. One can advance or recede the worm only one time slice repeatedly from the same arrangement and plot the distribution of the new position. This must correspond to the analytical expression of the free propagator where the new head is sampled from. Also, tests to check detailed balance could be designed. For example, when an update is suggested, one may immediately propose the inverse update and see if the acceptance probability for the two updates fulfils equation \eqref{eq: acceptance probability} from section \ref{sec: mcmc}.

\section{PIMC for Superfluid Ring Lattices} \label{sec: WA-PIMC UAOL}
In this section, we would like to discuss some preliminary results for the SRL lattices shown in chapter \ref{ch: ILM}. We found that the integer lattice $\mathbb{Z}[\zeta_4]$ with field norms $n=25$ and $n=65$ both yield yield square optical lattices with a ring structure in the unit cell. A first interesting observable to quantify, would be the density distribution $n$ of ultracold atoms in these lattices. To construct superfluid rings, one needs that the atoms are trapped/confined in the ring structures of the ILM optical lattices. Therefore, the wavelength of the optical potential is chosen such that the atoms are attracted to the intensity maxima. Whether the atoms are trapped in the rings, will partly depend on the potential depth $V_0$ of the optical lattice and the interaction strength between the atoms $\tilde{g}_0$. The atom's interaction strength is determined by the choice of atomic gas, and the potential depth can be controlled by tuning the power of the laser.

\begin{figure}[htpb]
    \centering
    \includegraphics[width=1\textwidth]{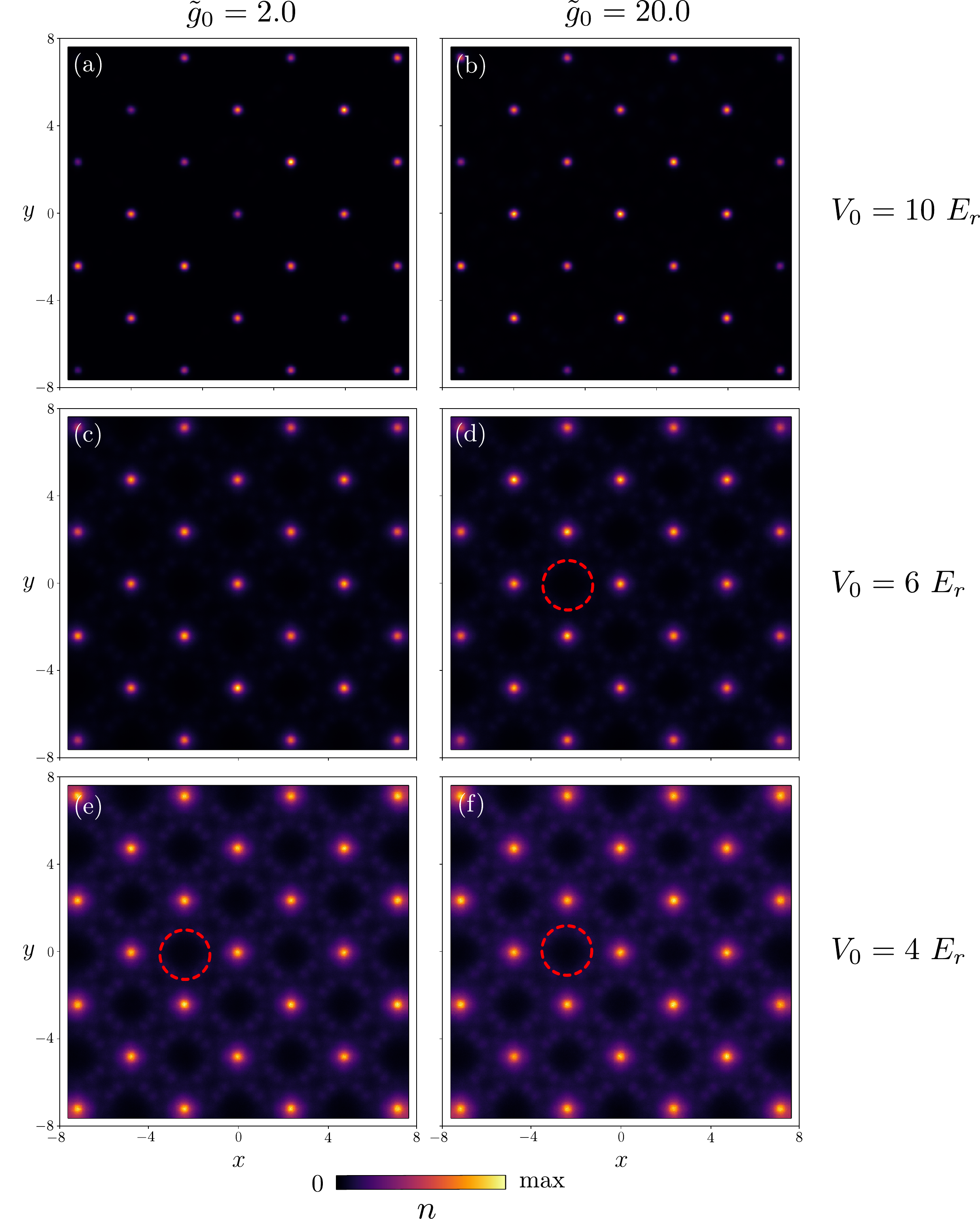}
    \caption{Simulated density for $N=20$ ultracold atoms in the SRL optical lattice with $n=25$ in the integer lattice $\mathbb{Z}[\zeta_4]$ at temperatures $T=0.025 \ E_r/k_B$ with $J=200$ beads in box with sides $L=8 \ a$ and periodic boundary conditions of $10^7$ measurements. The simulation where done varied for different two different interaction strengths $\tilde{g}_0=2.0$ for \textcolor{red}{(a)}, \textcolor{red}{(b)} and \textcolor{red}{(c)} and $\tilde{g}_0=20.0$ for \textcolor{red}{(b)}, \textcolor{red}{(d)} and \textcolor{red}{(f)}. Also, three different potential depths can be seen at $V_0=4.0 \ E_r$ for \textcolor{red}{(a-b)}, $V_0=6.0 \ E_r$ for \textcolor{red}{(c-d)} and $V_0=10.0 \ E_r$ for \textcolor{red}{(e-f)}. The dashed red lines indicates the ring-shaped density profiles if present.}
    \label{fig: densityl25}
\end{figure}

\begin{figure}[htpb]
    \centering
    \includegraphics[width=1\textwidth]{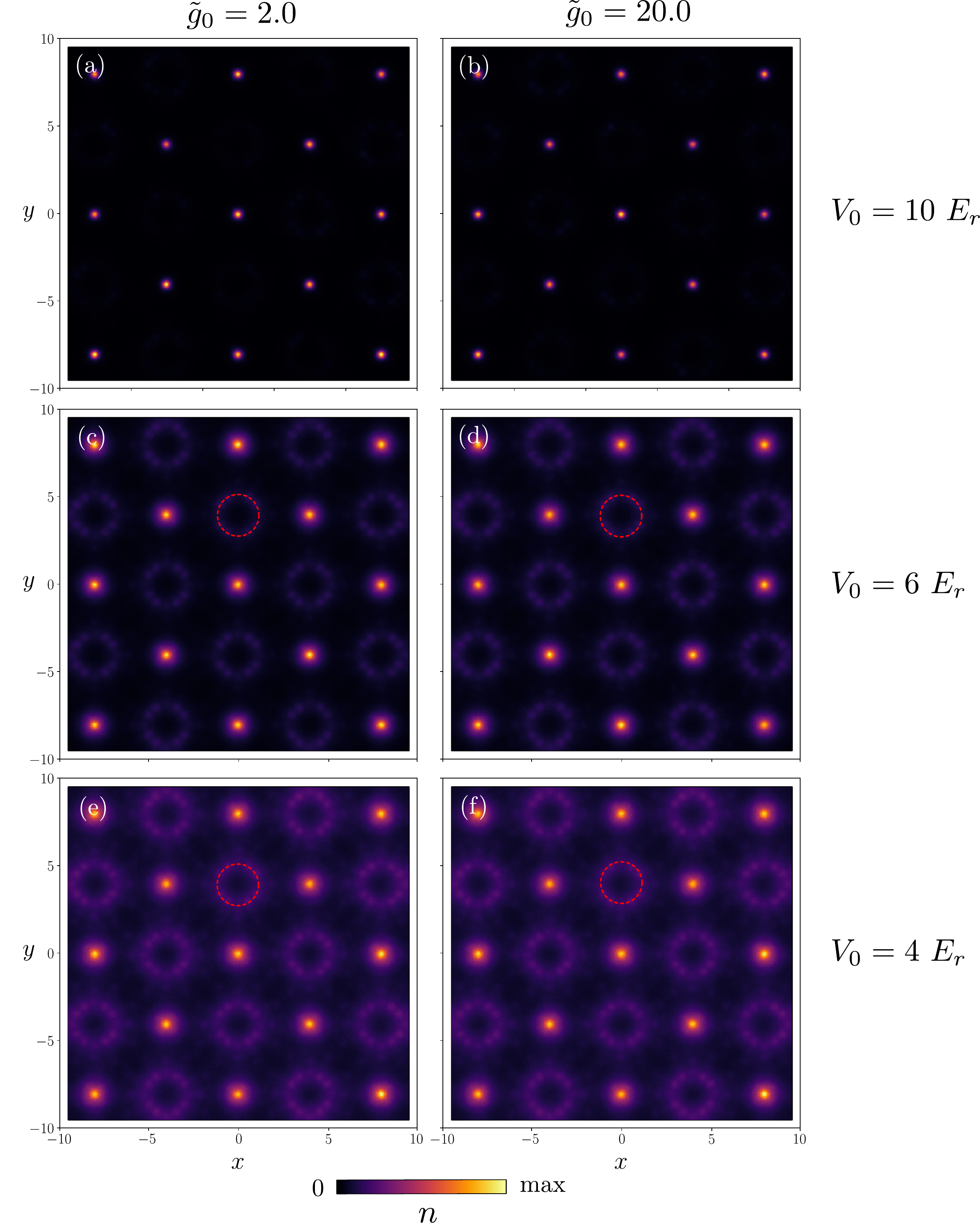}
    \caption{Simulated density for $N=20$ ultracold atoms in the SRL optical lattice with $n=65$ in the integer lattice $\mathbb{Z}[\zeta_4]$ at temperatures $T=0.025 \ E_r/k_B$ with $J=200$ beads in box with sides $L=10 \ a$ and a periodic boundary conditions of $10^7$ measurements. The simulation where done varied for different two different interaction strengths $\tilde{g}_0=2.0$ for \textcolor{red}{(a)}, \textcolor{red}{(b)} and \textcolor{red}{(c)} and $\tilde{g}_0=20.0$ for \textcolor{red}{(b)}, \textcolor{red}{(d)} and \textcolor{red}{(f)}. Also, three different potential depths can be seen at $V_0=4.0 \ E_r$ for \textcolor{red}{(a-b)}, $V_0=6.0 \ E_r$ for \textcolor{red}{(c-d)} and $V_0=10.0 \ E_r$ for \textcolor{red}{(e-f)}. The dashed red lines indicates the ring-shaped density profiles if present.}
    \label{fig: densityl65}
\end{figure}

The \gls{WA} extension for \gls{PIMC} only enhances the sampling of permutation at low temperatures. Using this with the fact that the density only depends on the position of the worldlines beads, we can infer that \gls{WA} is not needed to measure the density distribution $n$ of the optical lattices. In figure \ref{fig: densityl25} and \ref{fig: densityl65} the measured density for $N=20$ bosons of both field norms $n=25$ and $n=65$ of the integer lattice $\mathbb{Z}[\zeta_4]$ using \gls{PIMC} are shown. The measurements were done with $J=200$ beads at a temperature of $T=0.025 \ E_r/k_B$ in box with periodic boundary conditions. For each optical lattice different interaction strengths $\tilde{g}_0=2.0, 20.0$ and potential depths $V_0=4.0, 6.0, 10.0 \  E_r$ were used.

As expected, for high potential depths, e.g., $V_0=10 \ E_r$, the atoms are completely trapped in the interstitial peaks for both lattices independent of the interaction strengths, seen both in figure \ref{fig: densityl25}\textcolor{red}{(a-b)} and \ref{fig: densityl65}\textcolor{red}{(a-b)}. For low potential depths, e.g., $V_0=4 \ _r$), both lattices exhibit circular profiles in their density distributions. Nevertheless, ``leakage'' is seen between the interstitial peaks and the rings. For lack of a better term, ``leakage'' is employed here to explain this property of the equilibrium density distribution irrespective of time domain. It indicates the non-zero density between the intensity peaks and the rings. The effect is stronger for the optical lattice of field norm $n=25$ as the interstitial peaks and rings are closer to each other. Making the interaction larger increases the leakage as fewer atoms can be packed into the interstitial peaks. Finally, taking the potential depth of $V_0=6 \ E_r$ indeed yields a well separated density distribution in the rings and interstitial peaks for the lattice with a field norm of $n=65$, observed in figure \ref{fig: densityl65}\textcolor{red}{(c-d)}. Again, increasing the interaction slightly, increases the atom's density in the rings. Looking closely at the density distribution of field norm $n=25$ with interaction strength $\tilde{g}_0=20.0$, i.e., figure \ref{fig: densityl25}\textcolor{red}{(d)}, the rings can be faintly observed. However, also the leakage is already present. This is probably because the interstitial peaks and rings are too close together. Nevertheless, a closer inspection is needed.

This preliminary result indeed suggests that for the optical lattice with field norm $n=65$ of the integer lattice $\mathbb{Z}[\zeta_4]$ the intensity difference between the interstitial interference and the ring interference pattern is high enough such that the atoms can be trapped in the rings. Despite that, examinations with more ultracold atoms, i.e., higher total densities, are needed. In addition, the local and global superfluidity estimates can give more insight into this query.

\chapter{Conclusion and outlook}
\section{Summary}
As quantum mechanics revolutionized our understanding of nature in the 20th century, it will revolutionize the technology of the 21st century. The most sought-after quantum engineered devices are quantum circuits, which will provide a way to perform a new set of computational algorithms. The main challenges regarding its realization lie in the scalability, reconfigurability, and stability of the many proposed qubit systems. In this thesis, an idea was outlined to use the Integer Lattice Method (\gls{ILM}) to pursue optical lattices which could facilitate superfluid ring qubits in a strict two-dimensional system, dubbed superfluid ring lattices. It may pave the way for system miniaturization and, eventually, on-chip realization.

This work lays the foundation for investigating the possibilities of constructing a multi-qubit system consisting of ultracold atoms loaded in optical lattices, called a superfluid ring lattice. This is done by implementing an algorithm for simulating correlated ultracold atoms undergoing an external potential at a finite temperature. The technique of choice was the Path Integral Monte Carlo (\gls{PIMC}) method, which enables the computation of a quasi-exact result in thermodynamic equilibrium. The fundamental concept behind the method is to describe the partition function as an integral over all feasible paths in imaginary time and sample the stationary distribution with the help of the Metropolis walker. In addition to \gls{PIMC}, an extension to the algorithm was pursued, called Worm Algorithm (\gls{WA}), which allows for a more efficient sampling of the particle statistics needed for low temperatures by including open worldlines or worms.

Several tests have been performed to ensure the validity of the developed software. In addition to visual and unit tests, the \gls{PIMC} algorithm has been benchmarked against an exact analytical result of the non-interacting harmonically trapped Bose gas. In addition, the virial theorem was verified for the system. Also, \gls{WAPIMC} has passed several tests, including the correct response of the average number of particles in a parameter sweep and the controlled equilibration between the G and Z-sector. Nevertheless, when benchmarking the \gls{WAPIMC} simulation to analytical results, a large error arose for the computed energy when adding the advance and recede updates. In addition, a consistent deviation of the mean from the exact analytically computed energy was observed. Further analyses leads to the conclusion that the subsequent beads are sampled too far from each other when using the combination of the open, close, advance and recede updates. Two potential culprits were identified: a bug in the advance and recede updates and/or a wrong implementation of the restriction to work in the canonical ensemble on the G-sector updates. The former could take form by a wrong parameter use in the sampling of the free propagator. The latter could break the detailed balance condition such that the walker could not produce a stationary distribution.

Finally, the density distributions were simulated for the optical lattices --- found using \gls{ILM} --- with field norms $n=25$ and $n=65$ in the integer lattice $\mathbb{Z}[\zeta_4]$. Both contain circular interference patterns in which potentially Superfluid Ring Qubits (\gls{SRQ}s) can be constructed. The lattice with field norm $n=25$ suffered from leakage of the atoms to the interstitial fringes. The lattice with field norm $n=65$, indeed, exhibited the favourable property of compact confinement of the ultracold atoms in the rings regardless of the interaction strength. We have found a first indication that the implementation of \gls{SRQ}s, in a two-dimensional optical lattice constructed by \gls{ILM}, is feasible.

\section{Future Perspectives}
This thesis provides a stepping stone to studying the possibility of constructing superfluid ring lattices with the help of \gls{ILM}. The main goal was to develop the necessary tools in order to accomplish this future aim. Although, the complete objective has not been obtained, the core module of the Monte Carlo algorithm has been validated. Naturally, a first next step, will be to implement a \gls{WA} extension to the \gls{PIMC} module which can pass our designed tests. The unusually large spacing of the beads is a first clue that requires further investigation. This can be further inspected by designing simple tests which for example could examine the sampling of the position of the beads.

After a complete benchmarked \gls{WAPIMC} is acquired, it can be used to further investigate the obtained \gls{ILM} lattices. A first step, would be to show that ring-like structures in the optical lattices can facilitate superfluid currents. For this, the interplay between the global and local superfluid estimators can be used. Making the lattice potential attractive and gradually ramping up the lattice depth will trap/localize the atoms in the focal maxima. At critical potential depth, the superfluid has completely localized into a Mott insulator. The aim would be to tune the potential depth before the critical point, such that the superfluid density is mostly present in the circular structures. This behaviour can be quantified by observing the local superfluid density at the ring intensity patterns and normalizing it with the global superfluid estimate. Almost all preparation for this step have been developed and can be carried as soon a complete benchmarked \gls{WAPIMC} has been obtained.

If we want quantitative results comparable to experiments, we need to simulate at high total densities, i.e., computational measurements for a high number of particles are needed. Nevertheless, the computational time increases rapidly with the number of closed worldlines. Hence, some optimization of the \gls{WAPIMC} will be needed. Other than performance increases by making programming language dependent choses, one can add an artificial potential between the head and tail of the worm. This prevents the head and tail to appear very far away from each other. A situation that arises due to the high probability of permutation at low temperatures. If no artificial potential is present, one must wait until the head and tail are brought close to each other to go back to the Z-sector, increasing computation time. Furthermore, transitions between the G and Z-sectors will be rare, resulting in higher autocorrelations. The solution would be to implement an artificial potential which affect the relative distribution of the distance between the head and tail.

\chapter*{Contributions}
\addcontentsline{toc}{chapter}{Contributions}%
This thesis has been written as a partial fulfilment of the requirements for the degree Master of Science in Physics. The candidate has implemented the algorithm in chapter 4 of this thesis. The numerical calculations were performed using HPC resources from Imec. All the data obtained from the algorithm were analysed and processed by the candidate. The original research question was formalized by Dmitry Kouznetsov. The candidate thanks Niels Verellen and Dmitry Kouznetsov for their insightful comments in writing the thesis.\cleardoublepage
\addcontentsline{toc}{chapter}{Acknowledgement}
\chapter*{Acknowledgement}

First and foremost, I'd want to express my gratitude to supervisors: Professor Pol van Dorpe, Niels Verellen and Dmitry Kouznetsov. They provided me with the opportunity to learn about numerous topics regarding this thesis, e.g., computational physics, optics, quantum engineering, etc. A lot of students get a ready-made research assignment where they are weekly corrected by their supervisor. Instead, at Imec, I was dropped onto an open research project where it was not clear on what to do next. This allowed my to guide the project and to conduct the research together with my daily supervisor, Dmitry. I am very grateful for this. I want to thank Dmitry personally for all the effort he put in this thesis. He was always available to give research and mental support. I want to thank both Dmitry and Niels for helping me to write my motivation letter and preparing my interviews which eventually led into catching a PhD position.

Finally, my appreciation to Pilar, William, Lucas, Niels, Ann and Johan for being great friends and comfort.\cleardoublepage

\begin{appendices}
    \addtocontents{toc}{\protect\setcounter{tocdepth}{1}}
    

\chapter{Relative Interaction Propagator}\label{app: rel int prop}

\subsubsection{Interaction propagator}\label{sec: derivation int prop}
In a general form the Hamiltonian of interests is given by
\begin{align} \label{eq: General Hamiltonian}
    \ham = \ham_1 + \ham_{\mathrm{int}} 
    = \sum_{i=1}^{N} -\lambda\hat{\nabla_{\mathrm{i}}}^{2}+\sum_{i=1}^{N} V\left(\hat{\mathbf{r}}_{i}\right)+\sum^{N-1}_{i=1}\sum^{N}_{k=i+1} U\left(\hat{\mathbf{r}}_{i}-\hat{\mathbf{r}}_{k}\right),
\end{align}
where $\ham_1$ is the sum of the single-particle Hamiltonian of N particles with mass $m$ and $\ham_\mathrm{int}$ the sum of all two-body interactions between the particles determined by their relative positions $\mathbf{r}_{i k}=\mathbf{r}_{i}-\mathbf{r}_{k}$. As discussed in section \ref{sec: stp} the propagator of the Hamiltonian \eqref{eq: General Hamiltonian} can be approximated by the pair-product approximation which with the help the interacting relative Hamiltonian \cite{Yan2015}
\begin{align}
\ham^{\mathrm{rel}}\left(\hat{\mathbf{r}}_{jk}\right) = \ham_{0}^{\mathrm{rel}} \left(\hat{\mathbf{r}}_{j k}\right) + U \left(\hat{\mathbf{r}}_{j k}\right)
\end{align}
with $H_{0}^{\mathrm{rel}}\left(\hat{\mathbf{r}}_{j k}\right)= \lambda_*\hat{\nabla}_{\mathbf{r}_{j k}}^{2}$ the relative non-interacting Hamiltonian with $\lambda_*=\hbar^2/2m_*$ and reduced mass $m_*=m/2$, can be written as
\begin{align}
\rho\left(\mathbf{R}, \mathbf{R}^{\prime}, \tau\right) \approx \prod_{j=1}^{N} \rho_{1}\left(\mathbf{r}_{j}, \mathbf{r}_{j}^{\prime}, \tau\right) \prod_{j<k}^{N} \frac{\rho^{\mathrm{rel}}\left(\mathbf{r}_{j k}, \mathbf{r}_{j k}^{\prime}, \tau\right)}{\rho_{0}^{\mathrm{rel}}\left(\mathbf{r}_{j k}, \mathbf{r}_{j k}^{\prime}, \tau\right)}.
\end{align}
Here is $\rho_1$ the propagator of the non-interacting Hamiltonian $\ham_1$, $\rho^{\mathrm{rel}}$ the propagator of the relative Hamiltonian $\ham^{\mathrm{rel}}$ and $\rho^{\mathrm{rel}}_0$ the relative non-interacting Hamiltonian $\ham_{0}^{\mathrm{rel}}$.

\subsubsection{Relative interaction propagator}
The task remains to derive a proper expression for the relative propagators in two dimensions. However, the problem has reduced to the two-dimensional time independent Schr\"odinger equation of a particle with mass $m_*$ and position $\mathbf{r}$ in a potential $U$. Furthermore, because the potential $U$ is spherically symmetric, i.e. $U(\mathbf{r})=U(r)$, the Schr\"odinger equation may be solved by separating variables in polar coordinates. Hence, for every eigenvalue $E_{k}=\lambda_* k^{2}= \hbar^2k^{2} / 2m_*$, i.e., in the center-of-mass frame, propose a separable eigenstate of the form \cite{Zaslow1967}
\begin{align}
    \psi_{k l}(r, \theta)=R_{k l}(r) \Theta_{l}(\theta)
\end{align} with $\Theta_{l}(\theta) = \mathrm{e}^{i l \theta} / \sqrt{2 \pi}$ the eigenstates of the angular momentum operator normal to the 2D plane and $l \in \mathbb{Z}$ the corresponding quantum number. Above ansatz is inspired by noticing that the angular momentum operators $\hat{L}^2$ and $\hat{L}$ commute with the Hamiltonian $\ham^{\mathrm{rel}}$. The propagator can than be written in its eigenbasis decomposition given by
\begin{align}\label{eq: relative propagator}
\rho^{\mathrm{rel}}\left(\mathbf{r}, \mathbf{r}^{\prime}, \tau\right)=\sum_{l} \int \dd{k} \psi_{k l}^{*}\left(\mathbf{r}^{\prime}\right) \mathrm{e}^{-\tau E_{k}} \psi_{k l}(\mathbf{r})
\end{align}
Substituting the eigenstates $\psi_{k l}$ in the two-dimensional Schr\"odinger equation yields the following radial equation
\begin{align} \label{eq: radial 2D TISE}
-\lambda_*\left(\frac{1}{r} \dv[2]{}{r}\left(rR_{kl}\right)\right)+\lambda_*\frac{l^2}{r^{2}} R_{k l}(r)+U(r) R_{k l}(r)
=E_{k} R_{k l}(r)
\end{align}
We can gain some insight on the influence of angular momentum on the solution by performing the substitution $R_{k l} = u_{k l}/r$ in equation \eqref{eq: radial 2D TISE} which yields
\begin{align} \label{eq: eff TISE scattering}
-\lambda_* \dv[2]{}{r} u_{k l}+\left(U(r)+\lambda_*\frac{l^2 }{ r^{2}}\right) u_{k l}(r)=E_k u_{k l}(r)
\end{align}
Here $u_{E l}(r)$ can be interpreted as a wave function of a particle in one dimension for a particle moving in an effective potential $V_\mathrm{eff} = U(r)+\lambda_*\frac{l^2 }{ r^{2}}$. It shows the particle experiences a ``angular momentum barrier'' or ``centrifugal barrier'' if $l\neq0$ and the probability of the r being near the origin becomes small. Notice s-states do not experience this effect.

Equation \eqref{eq: radial 2D TISE} can be rewritten in the form
\begin{align}
\frac{\mathrm{d}^{2} R_{k l}}{\mathrm{~d} r^{2}}+\frac{1}{r} \frac{\mathrm{d} R_{k l}}{\mathrm{~d} r}+\left(k^{2}-\frac{l^{2}}{r^{2}}\right) R_{k l}=\frac{1 }{\lambda_*}U(r) R_{k l}.
\end{align}
which if there are no interactions, i.e. $U(r)=0$, becomes the Bessel's differential equation. The general solutions can be written as \cite{Khuri2009}
\begin{align}\label{eq: radial solution}
R_{k l}(r)=C_{k l}\left[\cos \left(\delta_{k l}\right) J_{l}(k r)-\sin \left(\delta_{k l}\right)  Y_{l}(k r)\right].
\end{align}
where $J_{l}$ and $Y_{l}$ the Bessel functions of the first and second kind respectively, $\delta_{k l}$ a to-be determined phase shift set by the boundary conditions, and, $C_{k l}$ a normalization constant.

We argue that the phase shift $\delta_{k l}$ has to be zero if no interaction are present \cite{sakurai2017}. Let us look at the radial component of the probability flux given by
\begin{align}
j_{r} &=\hat{\mathbf{r}} \cdot \mathbf{j}=\frac{\hbar}{m_*} \operatorname{Im}\left(\psi^{*} \frac{\partial}{\partial r} \psi\right) =\frac{\hbar}{m_*} R_{k l}(r) \frac{d}{d r} R_{k l}(r).
\end{align}
Notice the fact that for the Bessel functions we have that $J_{l}(k r)\propto r^l$ and $Y_{l}(k r)\propto r^{-(l+1)}$ as $r\to0$. Consequently, if $\delta_{k l}\neq0$, than will $R_{k l}(r)\propto r^{-(l+1)}$ as $r\to0$ and thus $j_{r}\propto (l+1)r^{-2l-3}$. However, the probability escaping of a small sphere is $4\pi r^2 j_{r}\propto (l+1)r^{-2l-1}\to \infty$ for $r\to0$. This violates the probabilistic interpretation of quantum mechanics. We conclude that $\delta_{k l}=0$.

Now using the normalization and completeness relations for the eigenstates $\psi_{k l}$, one finds that $C_{k l}=\sqrt{k}$, such that
\begin{align} \label{eq: eigenstates}
\psi_{k l}(r, \theta)=\sqrt{\frac{k}{2 \pi}} J_{l}(k r) \mathrm{e}^{i l \theta}.
\end{align}
Rest us to compute the propagator \eqref{eq: relative propagator}. In doing this the following identities of Bessel functions are useful \cite{Gauter2021}:
\begin{align} \label{eq: Bessel identity 1}
\int k \mathrm{e}^{-\tau E_k} J_{l}(k r) J_{l}\left(k r^{\prime}\right) \dd{k}&=\frac{1}{2\lambda_* \tau} \exp \left(-\frac{r^{2}+r^{\prime 2}}{4\lambda_* \tau}\right) I_{l}\left(\frac{r r^{\prime}}{2\lambda_* \tau}\right)\\
\sum_{l} I_{l}(x) t^{l}&=\exp \left(\frac{1}{2} x\left(t+\frac{1}{t}\right)\right)\label{eq: Bessel identity 2}
\end{align}
Finally, as expected, we find equation \eqref{eq: relative propagator} with eigenstates \eqref{eq: eigenstates} to give
\begin{align}
\rho^{\mathrm{rel}, 0}\left(\mathbf{r}, \mathbf{r}^{\prime}, \tau\right)=\frac{1}{4 \pi \lambda_* \tau} \mathrm{e}^{-\frac{\left(\mathbf{r}-\mathbf{r}^{\prime}\right)^{2}}{4\lambda_* \tau}}=\frac{m_*}{2 \pi \hbar^2 \tau} \mathrm{e}^{-\frac{m_*\left(\mathbf{r}-\mathbf{r}^{\prime}\right)^{2}}{2\hbar^2 \tau}} .
\end{align}
i.e., the free particle propagator of a particle with reduced mass $m_*=m/2$.

Let us now do the same when interaction is present. In the context of cold atoms trapped in optical lattices the dilute nature of the gas and low temperatures (under the sub-milikelvin regime \cite{Bloch2008}) allow us to treat the interactions between the charge neutral atoms as primarily two-body and of zero-range. Considering above partial wave expansion (see equation \eqref{eq: eff TISE scattering}), i.e. decomposing each incoming wave into its constituent angular-momentum $l=0,1,2,\ldots$ components and solving separately, this amounts to only using $l=0$. Furthermore, we can assume that outside the interaction range, the eigenstates have the same shape as in the non-interacting case \eqref{eq: radial solution} with $l=0$. The complete solution is than found by requiring the 2D zero-range potential (\gls{ZRP}) boundary condition\footnote{The boundary condition originated from \citet{Bethe1935} were it was employed the theory of deutron's. Fermi utilized a simplified version of ZRPM to calculate the scattering of neutrons (1934). Hence, in literature the boundary condition is also often referred to as Bethe-Peierls, Fermi-Huang boundary condition\cite{fermi1936, Huang1957, Zbigniew2006}. An accessible review and derivation of the boundary condition in \emph{three dimensions} can be found in \citet{Drukarev1978} and a more modern approach in \citet{PopoviczSeidel2019}. Also, an overview of the physical motivation is given.} \cite{Whitehead2016, Averbuch1986, Lim1980, Wang2016, Dalibard2021}
\begin{align}
\left(r \dv{r}-\frac{1}{\ln (r / a_{2D})}\right) R_{k0}(r)=0 \qq{for} r\to 0.
\end{align}
with $a_{2D}$ the 2D scattering length {of the particles}. The boundary condition can be derived by exactly solving the time independent Schr\"odinger equation for a spherical potential well with a potential range $R$ and imposing continuity the first derivative at $r=R$ of the scattering wavefunction. Assuming that the potential range $R$ vanishes, i.e. $R\to0$, yields above condition. The latter is motivated from the observation that when the energies of the interacting is particles the potential range $R$ is substantially smaller than de Broglie wavelength $\lambda$. As a result, the potential range $R$ can be disregarded and taking to vanish. Furthermore, notice that the scattering length $a_{2D}$ {of the Bose gas} will become a parameter of the interaction propagator. An explicit derivation can be found in appendix {...}.

So instead of solving the radial equation in \eqref{eq: radial 2D TISE}, we replace the problem with a boundary condition imposed on the non-interacting case. The result yields a phase shift \cite{Farrell2010, Whitehead2016,Khuri2009}
\begin{align}
t_k \equiv\tan \delta_{k 0}=\pi / 2 \ln \left(\eta k a_{2 \mathrm{D}}\right).
\end{align}
Using everything in equation \eqref{eq: relative propagator} gives the final result
\begin{align}
\rho^{\mathrm{rel}}\left(\mathbf{r}, \mathbf{r}^{\prime}, \tau\right) \label{eq: 2 dimensional computed relative propagator}
=\rho^{\mathrm{rel}, 0}\left(\mathbf{r}, \mathbf{r}^{\prime}, \tau\right) &-\frac{1}{2 \pi} \int \dd{k} k \mathrm{e}^{-\tau E_k} \frac{t_{k}^{2}}{1+t_{k}^{2}} J_{0}(k r) J_{0}\left(k r^{\prime}\right) \\
&-\frac{1}{2 \pi} \int \dd{k} k \mathrm{e}^{ -\tau E_k} \frac{t_{k}}{1+t_{k}^{2}}\left[J_{0}(k r) Y_{0}\left(k r^{\prime}\right)+J_{0}\left(k r^{\prime}\right) Y_{0}(k r)\right] \\
&+\frac{1}{2 \pi} \int \dd{k} k \mathrm{e}^{-\tau E_k} \frac{t_{k}^{2}}{1+t_{k}^{2}} Y_{0}(k r) Y_{0}\left(k r^{\prime}\right) .
\end{align}

\end{appendices}

\addcontentsline{toc}{chapter}{Bibliography}
\printbibliography

@article{Dmitry22,
  author    = {Dmitry Kouznetsov and Ongun Arisev and Pol Van Dorpe and Niels Verellen},
  year      = {2022},
  title     = {Inverse design assisted coherent optical lattices},
  doi       = {10.1364/OE.455466},
  journal   = {Opt. Express},
  month     = {3},
  number    = {7},
  pages     = {11384--11393},
  publisher = {Optica Publishing Group},
  url       = {http://opg.optica.org/oe/abstract.cfm?URI=oe-30-7-11384},
  volume    = {30}
}

@article{DmitryBEC2022,
  author    = {Kouznetsov, Dmitry and Van Dorpe, Pol and Verellen, Niels},
  year      = {2022},
  title     = {Sieve of Eratosthenes for Bose-Einstein condensates in optical moir\'e lattices},
  doi       = {10.1103/PhysRevA.105.L021304},
  issue     = {2},
  journal   = {Phys. Rev. A},
  month     = {2},
  numpages  = {5},
  pages     = {L021304},
  publisher = {American Physical Society},
  url       = {https://link.aps.org/doi/10.1103/PhysRevA.105.L021304},
  volume    = {105}
}

@article{Spada2022,
  author  = {Spada, Gabriele and Giorgini, Stefano and Pilati, Sebastiano},
  year    = {2022},
  title   = {Path-Integral Monte Carlo Worm Algorithm for Bose Systems with Periodic Boundary Conditions},
  doi     = {10.3390/condmat7020030},
  issn    = {2410-3896},
  journal = {MPDI Condensed Matter},
  number  = {2},
  pages   = {30},
  url     = {https://www.mdpi.com/2410-3896/7/2/30},
  volume  = {7}
}

@article{amico2021,
  author  = {Amico,L.  and Boshier,M.  and Birkl,G.  and Minguzzi,A.  and Miniatura,C.  and Kwek,L.-C.  and Aghamalyan,D.  and Ahufinger,V.  and Anderson,D.  and Andrei,N.  and Arnold,A. S.  and Baker,M.  and Bell,T. A.  and Bland,T.  and Brantut,J. P.  and Cassettari,D.  and Chetcuti,W. J.  and Chevy,F.  and Citro,R.  and De Palo,S.  and Dumke,R.  and Edwards,M.  and Folman,R.  and Fortagh,J.  and Gardiner,S. A.  and Garraway,B. M.  and Gauthier,G.  and Günther,A.  and Haug,T.  and Hufnagel,C.  and Keil,M.  and Ireland,P.  and Lebrat,M.  and Li,W.  and Longchambon,L.  and Mompart,J.  and Morsch,O.  and Naldesi,P.  and Neely,T. W.  and Olshanii,M.  and Orignac,E.  and Pandey,S.  and Pérez-Obiol,A.  and Perrin,H.  and Piroli,L.  and Polo,J.  and Pritchard,A. L.  and Proukakis,N. P.  and Rylands,C.  and Rubinsztein-Dunlop,H.  and Scazza,F.  and Stringari,S.  and Tosto,F.  and Trombettoni,A.  and Victorin,N.  and Klitzing,W. von  and Wilkowski,D.  and Xhani,K.  and Yakimenko,A. },
  year    = {2021},
  title   = {Roadmap on Atomtronics: State of the art and perspective},
  doi     = {10.1116/5.0026178},
  journal = {AVS Quantum Science},
  number  = {3},
  pages   = {039201},
  volume  = {3}
}

@phdthesis{Dalibard2021,
  author = {Dalibard, Jean},
  year   = {2021},
  title  = {{Interactions between Atoms in Quantum Gases: From van der Waals universality to Fano–Feshbach resonances}},
  school = {{Coll\`ege de France}},
  type   = {Lecture notes},
  url    = {https://pro.college-de-france.fr/jean.dalibard/}
}

@article{Dmitriev2021,
  author  = {Dmitriev,A. Yu.  and Astafiev,O. V. },
  year    = {2021},
  title   = {A perspective on superconducting flux qubits},
  doi     = {10.1063/5.0047690},
  journal = {Appl. Phys. Lett.},
  number  = {8},
  pages   = {080501},
  volume  = {119}
}

@article{Gauter2021,
  author    = {Gautier, Ronan and Yao, Hepeng and Sanchez-Palencia, Laurent},
  year      = {2021},
  title     = {Strongly Interacting Bosons in a Two-Dimensional Quasicrystal Lattice},
  doi       = {10.1103/PhysRevLett.126.110401},
  issue     = {11},
  journal   = {Phys. Rev. Lett.},
  month     = {3},
  numpages  = {7},
  pages     = {110401},
  publisher = {American Physical Society},
  url       = {https://link.aps.org/doi/10.1103/PhysRevLett.126.110401},
  volume    = {126}
}

@article{king_scaling_2021,
  author  = {King, Andrew D. and Raymond, Jack and Lanting, Trevor and Isakov, Sergei V. and Mohseni, Masoud and Poulin-Lamarre, Gabriel and Ejtemaee, Sara and Bernoudy, William and Ozfidan, Isil and Smirnov, Anatoly Yu. and Reis, Mauricio and Altomare, Fabio and Babcock, Michael and Baron, Catia and Berkley, Andrew J. and Boothby, Kelly and Bunyk, Paul I. and Christiani, Holly and Enderud, Colin and Evert, Bram and Harris, Richard and Hoskinson, Emile and Huang, Shuiyuan and Jooya, Kais and Khodabandelou, Ali and Ladizinsky, Nicolas and Li, Ryan and Lott, P. Aaron and {MacDonald}, Allison J. R. and Marsden, Danica and Marsden, Gaelen and Medina, Teresa and Molavi, Reza and Neufeld, Richard and Norouzpour, Mana and Oh, Travis and Pavlov, Igor and Perminov, Ilya and Prescott, Thomas and Rich, Chris and Sato, Yuki and Sheldan, Benjamin and Sterling, George and Swenson, Loren J. and Tsai, Nicholas and Volkmann, Mark H. and Whittaker, Jed D. and Wilkinson, Warren and Yao, Jason and Neven, Hartmut and Hilton, Jeremy P. and Ladizinsky, Eric and Johnson, Mark W. and Amin, Mohammad H.},
  year    = {2021},
  title   = {Scaling advantage over path-integral Monte Carlo in quantum simulation of geometrically frustrated magnets},
  doi     = {10.1038/s41467-021-20901-5},
  issn    = {2041-1723},
  journal = {Nat. Commun.},
  month   = {2},
  number  = {1},
  pages   = {1113},
  url     = {https://doi.org/10.1038/s41467-021-20901-5},
  volume  = {12}
}

@article{Pascual2021,
  author    = {Pascual, Gerard and Boronat, Jordi},
  year      = {2021},
  title     = {Quasiparticle Nature of the Bose Polaron at Finite Temperature},
  doi       = {10.1103/PhysRevLett.127.205301},
  issue     = {20},
  journal   = {Phys. Rev. Lett.},
  month     = {11},
  numpages  = {6},
  pages     = {205301},
  publisher = {American Physical Society},
  url       = {https://link.aps.org/doi/10.1103/PhysRevLett.127.205301},
  volume    = {127}
}

@article{Grant2020,
  author   = {Erica K. Grant and Travis S. Humble},
  year     = {2020},
  title    = {Adiabatic Quantum Computing and Quantum Annealing},
  doi      = {10.1093/acrefore/9780190871994.013.32},
  journal  = {Oxford University Press},
  location = {Oxford},
  month    = {7}
}

@article{Kjaergaard2020,
  author  = {Kjaergaard, Morten and Schwartz, Mollie E. and Braumüller, Jochen and Krantz, Philip and Wang, Joel I.-J. and Gustavsson, Simon and Oliver, William D.},
  year    = {2020},
  title   = {Superconducting Qubits: Current State of Play},
  doi     = {10.1146/annurev-conmatphys-031119-050605},
  journal = {Annu. Rev. Condens. Matter Phys.},
  number  = {1},
  pages   = {369-395},
  volume  = {11}
}

@article{DmitryILM2020,
  author    = {Kouznetsov, Dmitry and Deng, Qingzhong and Van Dorpe, Pol and Verellen, Niels},
  year      = {2020},
  title     = {Revival and Expansion of the Theory of Coherent Lattices},
  doi       = {10.1103/PhysRevLett.125.184101},
  issue     = {18},
  journal   = {Phys. Rev. Lett.},
  month     = {10},
  numpages  = {6},
  pages     = {184101},
  publisher = {American Physical Society},
  url       = {https://link.aps.org/doi/10.1103/PhysRevLett.125.184101},
  volume    = {125}
}

@thesis{Nichols2020,
  author    = {Nathan S. Nichols},
  year      = {2020},
  title     = {Error Analysis of quantum Monte Carlo data},
  publisher = {GitHub},
  type      = {GitHub repository},
  url       = {https://github.com/nscottnichols/erroranalysis-py}
}

@article{ryu_quantum_2020,
  author       = {Ryu, C. and Samson, E. C. and Boshier, M. G.},
  year         = {2020},
  title        = {Quantum interference of currents in an atomtronic {SQUID}},
  doi          = {10.1038/s41467-020-17185-6},
  issn         = {2041-1723},
  journaltitle = {Nat. Commun.},
  number       = {1},
  pages        = {3338},
  url          = {https://doi.org/10.1038/s41467-020-17185-6},
  volume       = {11}
}

@phdthesis{Yao2020,
  author      = {Yao, Hepeng},
  year        = {2020},
  title       = {{Strongly-correlated one-dimensional bosons in continuous and quasiperiodic potentials}},
  hal_id      = {tel-03065015},
  hal_version = {v1},
  month       = {10},
  number      = {2020IPPAX057},
  school      = {{Institut Polytechnique de Paris}},
  type        = {Ph.D. dissertation},
  url         = {https://tel.archives-ouvertes.fr/tel-03065015}
}

@article{Yao2020LiebLiniger,
  author    = {Yao, Hepeng and Giamarchi, Thierry and Sanchez-Palencia, Laurent},
  year      = {2020},
  title     = {Lieb-Liniger Bosons in a Shallow Quasiperiodic Potential: Bose Glass Phase and Fractal Mott Lobes},
  doi       = {10.1103/PhysRevLett.125.060401},
  issue     = {6},
  journal   = {Phys. Rev. Lett.},
  month     = {8},
  numpages  = {7},
  pages     = {060401},
  publisher = {American Physical Society},
  url       = {https://link.aps.org/doi/10.1103/PhysRevLett.125.060401},
  volume    = {125}
}

@thesis{Dmitrythesis,
  author      = {Kouznetsov, Dmitry},
  year        = {2019},
  title       = {Optical Lattice Microscopy: Design of Optical Interference Patterns for On-Chip Fluorescence Microscopy},
  institution = {{KU} Leuven},
  location    = {Leuven},
  month       = {6},
  type        = {Ph.D. dissertation},
  url         = {https://repository-teneo-libis-be.kuleuven.e-bronnen.be/delivery/DeliveryManagerServlet?dps_pid=IE12713515}
}

@article{PopoviczSeidel2019,
  author  = {Popovicz Seidel,Eliton  and Arretche,Felipe },
  year    = {2019},
  title   = {Zero range potential approximation in quantum scattering problems},
  doi     = {10.1119/1.5125110},
  journal = {Am. J. Phys},
  number  = {10},
  pages   = {796-801},
  url     = {https://doi.org/10.1119/1.5125110},
  volume  = {87}
}

@article{Antonio2018,
  author    = {Antonio Ac{\'i}n and Immanuel Bloch and Harry Buhrman and Tommaso Calarco and Christopher Eichler and Jens Eisert and Daniel Esteve and Nicolas Gisin and Steffen J Glaser and Fedor Jelezko and Stefan Kuhr and Maciej Lewenstein and Max F Riedel and Piet O Schmidt and Rob Thew and Andreas Wallraff and Ian Walmsley and Frank K Wilhelm},
  year      = {2018},
  title     = {The quantum technologies roadmap: a European community view},
  doi       = {10.1088/1367-2630/aad1ea},
  journal   = {New J. Phys.},
  month     = {8},
  number    = {8},
  pages     = {080201},
  publisher = {{IOP} Publishing},
  volume    = {20}
}

@book{Benov2018,
  author    = {Benov, Dobriyan and Mazhdrakov, Metodi and Valkanov, Nikolai},
  year      = {2018},
  title     = {The Monte Carlo Method: Engineering Applications},
  month     = {11},
  publisher = {ACMO Academic Press}
}

@article{Dornheim2018,
  author    = {Dornheim, T. and Groth, S. and Vorberger, J. and Bonitz, M.},
  year      = {2018},
  title     = {Ab initio Path Integral Monte Carlo Results for the Dynamic Structure Factor of Correlated Electrons: From the Electron Liquid to Warm Dense Matter},
  doi       = {10.1103/PhysRevLett.121.255001},
  issue     = {25},
  journal   = {Phys. Rev. Lett.},
  month     = {12},
  numpages  = {8},
  pages     = {255001},
  publisher = {American Physical Society},
  url       = {https://link.aps.org/doi/10.1103/PhysRevLett.121.255001},
  volume    = {121}
}

@phdthesis{Gerbier2018,
  author = {Gerbier, Fabrice},
  year   = {2018},
  title  = {Quantum gases in optical lattices},
  school = {{Institut de Physique du Coll\`ege de France}},
  type   = {Lecture notes},
  url    = {http://www.lkb.upmc.fr/boseeinsteincondensates/gerbier/}
}

@article{Haug2018,
  author    = {Haug, Tobias and Tan, Joel and Theng, Mark and Dumke, Rainer and Kwek, Leong-Chuan and Amico, Luigi},
  year      = {2018},
  title     = {Readout of the atomtronic quantum interference device},
  doi       = {10.1103/PhysRevA.97.013633},
  issue     = {1},
  journal   = {Phys. Rev. A},
  month     = {1},
  numpages  = {12},
  pages     = {013633},
  publisher = {American Physical Society},
  volume    = {97}
}

@article{Agostini2018,
  author  = {M. Agostini and K. Altenmüller and S. Appel and V. Atroshchenko and Z. Bagdasarian and D. Basilico and G. Bellini and J. Benziger and D. Bick and G. Bonfini and L. Borodikhina and D. Bravo and B. Caccianiga and F. Calaprice and A. Caminata and M. Canepa and S. Caprioli and M. Carlini and P. Cavalcante and A. Chepurnov and K. Choi and D. D'Angelo and S. Davini and A. Derbin and X.F. Ding and L. {Di Noto} and I. Drachnev and K. Fomenko and A. Formozov and D. Franco and F. Froborg and F. Gabriele and C. Galbiati and C. Ghiano and M. Giammarchi and M. Goeger-Neff and A. Goretti and M. Gromov and C. Hagner and T. Houdy and E. Hungerford and Aldo Ianni and Andrea Ianni and A. Jany and D. Jeschke and V. Kobychev and D. Korablev and G. Korga and D. Kryn and M. Laubenstein and E. Litvinovich and F. Lombardi and P. Lombardi and L. Ludhova and G. Lukyanchenko and I. Machulin and M. Magnozzi and G. Manuzio and S. Marcocci and J. Martyn and E. Meroni and M. Meyer and L. Miramonti and M. Misiaszek and V. Muratova and B. Neumair and L. Oberauer and B. Opitz and F. Ortica and M. Pallavicini and L. Papp and A. Pocar and G. Ranucci and A. Razeto and A. Re and A. Romani and R. Roncin and N. Rossi and S. Schönert and D. Semenov and P. Shakina and M. Skorokhvatov and O. Smirnov and A. Sotnikov and L.F.F. Stokes and Y. Suvorov and R. Tartaglia and G. Testera and J. Thurn and M. Toropova and E. Unzhakov and A. Vishneva and R.B. Vogelaar and F. {von Feilitzsch} and H. Wang and S. Weinz and M. Wojcik and M. Wurm and Z. Yokley and O. Zaimidoroga and S. Zavatarelli and K. Zuber and G. Zuzel},
  year    = {2018},
  title   = {The Monte Carlo simulation of the Borexino detector},
  doi     = {https://doi.org/10.1016/j.astropartphys.2017.10.003},
  issn    = {0927-6505},
  journal = {Astropart. Phys.},
  pages   = {136-159},
  url     = {https://www.sciencedirect.com/science/article/pii/S0927650517301330},
  volume  = {97}
}

@book{allen2017computer,
  author    = {Allen, Michael P and Tildesley, Dominic J},
  year      = {2017},
  title     = {Computer simulation of liquids},
  publisher = {Oxford university press}
}

@article{Cinti2017,
  author    = {Cinti, Fabio and Cappellaro, Alberto and Salasnich, Luca and Macr{\`i}, Tommaso},
  year      = {2017},
  title     = {Superfluid Filaments of Dipolar Bosons in Free Space},
  doi       = {10.1103/PhysRevLett.119.215302},
  issue     = {21},
  journal   = {Phys. Rev. Lett.},
  month     = {11},
  numpages  = {6},
  pages     = {215302},
  publisher = {American Physical Society},
  url       = {https://link.aps.org/doi/10.1103/PhysRevLett.119.215302},
  volume    = {119}
}

@book{sakurai2017,
  author    = {Sakurai, J. J. and Napolitano, Jim},
  year      = {2017},
  title     = {Modern Quantum Mechanics},
  doi       = {10.1017/9781108499996},
  edition   = {3},
  place     = {Cambridge},
  publisher = {Cambridge University Press}
}

@article{YanBlume2017,
  author    = {Yangqian Yan and D Blume},
  year      = {2017},
  title     = {Path integral Monte Carlo ground state approach: formalism, implementation, and applications},
  doi       = {10.1088/1361-6455/aa8d7f},
  journal   = {J. Phys. B: At. Mol. Opt. Phys.},
  month     = {10},
  number    = {22},
  pages     = {223001},
  publisher = {{IOP} Publishing},
  url       = {https://doi.org/10.1088/1361-6455/aa8d7f},
  volume    = {50}
}

@article{aghamalyanAtomtronicFluxQubit2016,
  author       = {Aghamalyan, D. and Nguyen, N. T. and Auksztol, F. and Gan, K. S. and Valado, M. Martinez and Condylis, P. C. and Kwek, L.-C. and Dumke, R. and Amico, L.},
  year         = {2016},
  title        = {An atomtronic flux qubit: a ring lattice of Bose–Einstein condensates interrupted by three weak links},
  doi          = {10.1088/1367-2630/18/7/075013},
  issn         = {1367-2630},
  journaltitle = {New J. Phys.},
  month        = {6},
  number       = {7},
  pages        = {075013},
  shorttitle   = {An atomtronic flux qubit},
  url          = {https://doi.org/10.1088/1367-2630/18/7/075013},
  volume       = {18}
}

@article{Aharonovich2016,
  author  = {Aharonovich, Igor
             and Englund, Dirk
             and Toth, Milos},
  year    = {2016},
  title   = {Solid-state single-photon emitters},
  day     = {01},
  doi     = {10.1038/nphoton.2016.186},
  issn    = {1749-4893},
  journal = {Nat. Photonics},
  month   = {10},
  number  = {10},
  pages   = {631-641},
  url     = {https://doi.org/10.1038/nphoton.2016.186},
  volume  = {10}
}

@article{Goldman2016,
  author  = {Goldman, N.
             and Budich, J. C.
             and Zoller, P.},
  year    = {2016},
  title   = {Topological quantum matter with ultracold gases in optical lattices},
  day     = {01},
  doi     = {10.1038/nphys3803},
  journal = {Nat. Phys.},
  month   = {7},
  number  = {7},
  pages   = {639-645},
  url     = {https://doi.org/10.1038/nphys3803},
  volume  = {12}
}

@article{Hoffmann2016,
  author  = {Hoffmann, Philipp H. W.},
  year    = {2016},
  title   = {A Hitchhiker's Guide to Automatic Differentiation},
  day     = {01},
  doi     = {10.1007/s11075-015-0067-6},
  issn    = {1572-9265},
  journal = {Numerical Algorithms},
  month   = {7},
  number  = {3},
  pages   = {775-811},
  url     = {https://doi.org/10.1007/s11075-015-0067-6},
  volume  = {72}
}

@article{Wang2016,
  author    = {Jing-Kun Wang,Wei Yi,Wei Zhang},
  year      = {2016},
  title     = {Two-body physics in quasi-low-dimensional atomic gases under spin-orbit coupling},
  doi       = {10.1007/s11467-015-0529-2},
  journal   = {Front. Phys.},
  number    = {3},
  pages     = {118102},
  publisher = {Front. Phys.},
  url       = {https://journal.hep.com.cn/fop/EN/abstract/article_16515.shtml},
  volume    = {11}
}

@article{Whitehead2016,
  author    = {Whitehead, T. M. and Schonenberg, L. M. and Kongsuwan, N. and Needs, R. J. and Conduit, G. J.},
  year      = {2016},
  title     = {Pseudopotential for the two-dimensional contact interaction},
  doi       = {10.1103/PhysRevA.93.042702},
  issue     = {4},
  journal   = {Phys. Rev. A},
  month     = {4},
  numpages  = {10},
  pages     = {042702},
  publisher = {American Physical Society},
  url       = {https://link.aps.org/doi/10.1103/PhysRevA.93.042702},
  volume    = {93}
}

@thesis{aghamalyan2015,
  author = {Aghamalyan, Davit},
  year   = {2015},
  title  = {Atomtronics: {Q}unatum {T}echnology with {C}old {A}toms in {R}ing {S}haped {O}ptical {L}attices},
  month  = {6},
  school = {National University of Singapore},
  type   = {Ph.D. dissertation},
  url    = {https://scholarbank.nus.edu.sg/handle/10635/121371}
}

@article{amicoSuperfluidQubitSystems2015,
  author       = {Amico, Luigi and Aghamalyan, Davit and Auksztol, Filip and Crepaz, Herbert and Dumke, Rainer and Kwek, Leong Chuan},
  year         = {2015},
  title        = {Superfluid qubit systems with ring shaped optical lattices},
  doi          = {10.1038/srep04298},
  issn         = {2045-2322},
  month        = {5},
  number       = {1},
  pages        = {4298},
  shortjournal = {Sci. Rep.},
  url          = {http://www.nature.com/articles/srep04298},
  volume       = {4}
}

@article{Corcoles2015,
  author  = {C{\'o}rcoles, A. D.
             and Magesan, Easwar
             and Srinivasan, Srikanth J.
             and Cross, Andrew W.
             and Steffen, M.
             and Gambetta, Jay M.
             and Chow, Jerry M.},
  year    = {2015},
  title   = {Demonstration of a quantum error detection code using a square lattice of four superconducting qubits},
  day     = {29},
  doi     = {10.1038/ncomms7979},
  issn    = {2041-1723},
  journal = {Nat. Commun.},
  month   = {4},
  number  = {1},
  pages   = {6979},
  url     = {https://doi.org/10.1038/ncomms7979},
  volume  = {6}
}

@article{Aghamalyan_2015,
  author    = {Davit Aghamalyan and Marco Cominotti and Matteo Rizzi and Davide Rossini and Frank Hekking and Anna Minguzzi and Leong-Chuan Kwek and Luigi Amico},
  year      = 2015,
  title     = {Coherent superposition of current flows in an atomtronic quantum interference device},
  doi       = {10.1088/1367-2630/17/4/045023},
  journal   = {New J. Phys.},
  month     = {4},
  number    = {4},
  pages     = {045023},
  publisher = {{IOP} Publishing},
  url       = {https://doi.org/10.1088/1367-2630/17/4/045023},
  volume    = {17}
}

@phdthesis{piatecki2005,
  author      = {Piatecki, Swann},
  year        = {2015},
  title       = {{The Bose gas at large scattering lengths}},
  hal_id      = {tel-01158248},
  hal_version = {v1},
  month       = {3},
  school      = {{ENS Paris - Ecole Normale Sup{\'e}rieure de Paris}},
  type        = {Ph.D. dissertation},
  url         = {https://tel.archives-ouvertes.fr/tel-01158248}
}

@article{Yan2015,
  author    = {Yan, Yangqian and Blume, D.},
  year      = {2015},
  title     = {Incorporating exact two-body propagators for zero-range interactions into $N$-body Monte Carlo simulations},
  doi       = {10.1103/PhysRevA.91.043607},
  issue     = {4},
  journal   = {Phys. Rev. A},
  month     = {4},
  numpages  = {10},
  pages     = {043607},
  publisher = {American Physical Society},
  url       = {https://link.aps.org/doi/10.1103/PhysRevA.91.043607},
  volume    = {91}
}

@article{Amico2014,
  author  = {Amico, Luigi
             and Aghamalyan, Davit
             and Auksztol, Filip
             and Crepaz, Herbert
             and Dumke, Rainer
             and Kwek, Leong Chuan},
  year    = {2014},
  title   = {Superfluid qubit systems with ring shaped optical lattices},
  doi     = {10.1038/srep04298},
  issn    = {2045-2322},
  journal = {Scientific Reports},
  month   = {3},
  number  = {1},
  pages   = {4298},
  volume  = {4}
}

@article{Betzig2014,
  author  = {Bi-Chang Chen  and Wesley R. Legant  and Kai Wang  and Lin Shao  and Daniel E. Milkie  and Michael W. Davidson  and Chris Janetopoulos  and Xufeng S. Wu  and John A. Hammer  and Zhe Liu  and Brian P. English  and Yuko Mimori-Kiyosue  and Daniel P. Romero  and Alex T. Ritter  and Jennifer Lippincott-Schwartz  and Lillian Fritz-Laylin  and R. Dyche Mullins  and Diana M. Mitchell  and Joshua N. Bembenek  and Anne-Cecile Reymann  and Ralph Böhme  and Stephan W. Grill  and Jennifer T. Wang  and Geraldine Seydoux  and U. Serdar Tulu  and Daniel P. Kiehart  and Eric Betzig },
  year    = {2014},
  title   = {Lattice light-sheet microscopy: Imaging molecules to embryos at high spatiotemporal resolution},
  doi     = {10.1126/science.1257998},
  journal = {Science},
  number  = {6208},
  pages   = {1257998},
  url     = {https://www.science.org/doi/abs/10.1126/science.1257998},
  volume  = {346}
}

@article{Cinti2014,
  author  = {Cinti, F.
             and Macr{\`i}, T.
             and Lechner, W.
             and Pupillo, G.
             and Pohl, T.},
  year    = {2014},
  title   = {Defect-induced supersolidity with soft-core bosons},
  day     = {04},
  doi     = {10.1038/ncomms4235},
  issn    = {2041-1723},
  journal = {Nat. Commun.},
  month   = {2},
  number  = {1},
  pages   = {3235},
  url     = {https://doi.org/10.1038/ncomms4235},
  volume  = {5}
}

@mastersthesis{dornheim2014,
  author      = {Dornheim, Tobias},
  year        = {2014},
  title       = {Path Integral Monte Carlo Simulations of Correlated Bosons and Fermions in Traps},
  institution = {University of Kiel},
  langid      = {english},
  location    = {Kiel},
  month       = {10},
  pagetotal   = {128},
  url         = {http://www.theo-physik.uni-kiel.de/bonitz/theses/dornheim_14.pdf}
}

@article{Jotzu2014,
  author  = {Jotzu, Gregor
             and Messer, Michael
             and Desbuquois, R{\'e}mi
             and Lebrat, Martin
             and Uehlinger, Thomas
             and Greif, Daniel
             and Esslinger, Tilman},
  year    = {2014},
  title   = {Experimental realization of the topological Haldane model with ultracold fermions},
  day     = {01},
  doi     = {10.1038/nature13915},
  issn    = {1476-4687},
  journal = {Nature},
  month   = {11},
  number  = {7526},
  pages   = {237-240},
  url     = {https://doi.org/10.1038/nature13915},
  volume  = {515}
}

@article{Martinis2014,
  author  = {Martinis, John M. and Megrant, A.},
  year    = {2014},
  title   = {{UCSB} final report for the {CSQ} program: Review of decoherence and materials physics for superconducting qubits},
  doi     = {10.48550/ARXIV.1410.5793},
  journal = {{CSQ} program of IARPA, ar{X}iv},
  url     = {https://arxiv.org/abs/1410.5793}
}

@article{Simon2014,
  author  = {Simon, Jonathan},
  year    = {2014},
  title   = {Magnetic fields without magnetic fields},
  day     = {01},
  doi     = {10.1038/515202a},
  issn    = {1476-4687},
  journal = {Nature},
  month   = {11},
  number  = {7526},
  pages   = {202-203},
  url     = {https://doi.org/10.1038/515202a},
  volume  = {515}
}

@article{Young2014,
  author    = {Young, Peter},
  year      = {2014},
  title     = {Everything you wanted to know about Data Analysis and Fitting but were afraid to ask},
  doi       = {10.48550/ARXIV.1210.3781},
  journal   = {arXiv, Physics.data-an},
  month     = {10},
  publisher = {arXiv},
  url       = {https://arxiv.org/abs/1210.3781}
}

@book{Stone2013,
  author    = {A. DOUGLAS STONE},
  year      = {2013},
  title     = {Einstein and the Quantum: The Quest of the Valiant Swabian},
  publisher = {Princeton University Press},
  url       = {http://www.jstor.org/stable/j.ctt3fgxvv}
}

@article{Carleo2013,
  author    = {Carleo, Giuseppe and Bo\'eris, Guilhem and Holzmann, Markus and Sanchez-Palencia, Laurent},
  year      = {2013},
  title     = {Universal Superfluid Transition and Transport Properties of Two-Dimensional Dirty Bosons},
  doi       = {10.1103/PhysRevLett.111.050406},
  issue     = {5},
  journal   = {Phys. Rev. Lett.},
  month     = {10},
  numpages  = {5},
  pages     = {050406},
  publisher = {American Physical Society},
  url       = {https://link.aps.org/doi/10.1103/PhysRevLett.111.050406},
  volume    = {111}
}

@book{cox2011primes,
  author    = {Cox, David A},
  year      = {2013},
  title     = {{P}rimes of the form $x^2+ ny^2$: {F}ermat, {C}lass {F}ield {T}heory, and {C}omplex {M}ultiplication},
  edition   = {2},
  location  = {New York},
  number    = {119},
  publisher = {Wiley},
  series    = {Pure and Applied Mathematics: A Wiley Series of Texts, Monographs and Tracts}
}

@article{Edwards2013,
  author   = {Edwards, Mark},
  year     = {2013},
  title    = {Atom SQUID},
  day      = {01},
  doi      = {10.1038/nphys2546},
  issn     = {1745-2481},
  journal  = {Nat. Phys.},
  month    = {2},
  number   = {2},
  pages    = {68-69},
  url      = {https://doi.org/10.1038/nphys2546},
  volume   = {9}
}

@book{landau2013quantum,
  author    = {Landau, Lev Davidovich and Lifshitz, Evgenii Mikhailovich},
  year      = {2013},
  title     = {Quantum mechanics: non-relativistic theory},
  doi       = {https://doi.org/10.1016/B978-0-08-017801-1.50001-4},
  publisher = {Elsevier},
  url       = {https://www.sciencedirect.com/science/article/pii/B9780080178011500014},
  volume    = {3}
}

@thesis{Julia2012,
  author = {Bezanson, Jeff, and Karpinski, Stefan and Shah, Viral B. and Edelman, Alan},
  year   = {2012},
  title  = {Why We Created Julia},
  type   = {Blog post, Julialang.Org},
  url    = {https://julialang.org/blog/2012/02/why-we-created-julia/}
}

@article{FilinovBonitz2012,
  author    = {Filinov, A. and Bonitz, M.},
  year      = {2012},
  title     = {Collective and single-particle excitations in two-dimensional dipolar Bose gases},
  doi       = {10.1103/PhysRevA.86.043628},
  issue     = {4},
  journal   = {Phys. Rev. A},
  month     = {10},
  numpages  = {26},
  pages     = {043628},
  publisher = {American Physical Society},
  url       = {https://link.aps.org/doi/10.1103/PhysRevA.86.043628},
  volume    = {86}
}

@article{Fowler2012,
  author    = {Fowler, Austin G. and Mariantoni, Matteo and Martinis, John M. and Cleland, Andrew N.},
  year      = {2012},
  title     = {Surface codes: Towards practical large-scale quantum computation},
  doi       = {10.1103/PhysRevA.86.032324},
  issue     = {3},
  journal   = {Phys. Rev. A},
  month     = {9},
  numpages  = {48},
  pages     = {032324},
  publisher = {American Physical Society},
  url       = {https://link.aps.org/doi/10.1103/PhysRevA.86.032324},
  volume    = {86}
}

@book{Lewenstein2012,
  author    = {Lewenstein, Maciej and Sanpera, Anna and Ahufinger, Ver{\`o}nica},
  year      = {2012},
  title     = {Ultracold Atoms in Optical Lattices: Simulating quantum many-body systems},
  address   = {Oxford},
  doi       = {10.1093/acprof:oso/9780199573127.001.0001},
  publisher = {Oxford University Press},
  url       = {https://oxford.universitypressscholarship.com/10.1093/acprof:oso/9780199573127.001.0001/acprof-9780199573127}
}

@article{Pollet2012,
  author    = {Lode Pollet},
  year      = {2012},
  title     = {Recent developments in quantum Monte Carlo simulations with applications for cold gases},
  doi       = {10.1088/0034-4885/75/9/094501},
  journal   = {Rep. Prog. Phys.},
  month     = {8},
  number    = {9},
  pages     = {094501},
  publisher = {{IOP} Publishing},
  url       = {https://doi.org/10.1088/0034-4885/75/9/094501},
  volume    = {75}
}

@article{Moulder2012,
  author    = {Moulder, Stuart and Beattie, Scott and Smith, Robert P. and Tammuz, Naaman and Hadzibabic, Zoran},
  year      = {2012},
  title     = {Quantized supercurrent decay in an annular Bose-Einstein condensate},
  doi       = {10.1103/PhysRevA.86.013629},
  issue     = {1},
  journal   = {Phys. Rev. A},
  month     = {7},
  numpages  = {7},
  pages     = {013629},
  publisher = {American Physical Society},
  url       = {https://link.aps.org/doi/10.1103/PhysRevA.86.013629},
  volume    = {86}
}

@article{Bauer_2011,
  author    = {B Bauer and L D Carr and H G Evertz and A Feiguin and J Freire and S Fuchs and L Gamper and J Gukelberger and E Gull and S Guertler and A Hehn and R Igarashi and S V Isakov and D Koop and P N Ma and P Mates and H Matsuo and O Parcollet and G Paw{\l}owski and J D Picon and L Pollet and E Santos and V W Scarola and U Schollwöck and C Silva and B Surer and S Todo and S Trebst and M Troyer and M L Wall and P Werner and S Wessel},
  year      = {2011},
  title     = {The {ALPS} project release 2.0: open source software for strongly correlated systems},
  doi       = {10.1088/1742-5468/2011/05/p05001},
  journal   = {J. Stat. Mech. Theory Exp.},
  month     = {5},
  number    = {05},
  pages     = {P05001},
  publisher = {{IOP} Publishing},
  url       = {https://doi.org/10.1088/1742-5468/2011/05/p05001},
  volume    = {2011}
}

@article{Dalibard2011,
  author    = {Dalibard, Jean and Gerbier, Fabrice and Juzeli\ifmmode \bar{u}\else \={u}\fi{}nas, Gediminas and \"Ohberg, Patrik},
  year      = {2011},
  title     = {Colloquium: Artificial gauge potentials for neutral atoms},
  doi       = {10.1103/RevModPhys.83.1523},
  issue     = {4},
  journal   = {Rev. Mod. Phys.},
  month     = {11},
  numpages  = {0},
  pages     = {1523--1543},
  publisher = {American Physical Society},
  url       = {https://link.aps.org/doi/10.1103/RevModPhys.83.1523},
  volume    = {83}
}

@article{DelMaestro2011,
  author    = {Del Maestro, Adrian and Boninsegni, Massimo and Affleck, Ian},
  year      = {2011},
  title     = {$^{4}\mathrm{He}$ Luttinger Liquid in Nanopores},
  doi       = {10.1103/PhysRevLett.106.105303},
  issue     = {10},
  journal   = {Phys. Rev. Lett.},
  month     = {3},
  numpages  = {4},
  pages     = {105303},
  publisher = {American Physical Society},
  url       = {https://link.aps.org/doi/10.1103/PhysRevLett.106.105303},
  volume    = {106}
}

@phdthesis{RotaPIMC,
  author = {Rota, Riccardo},
  year   = {2011},
  title  = {Path Integral Monte Carlo and Bose-Einstein condensation in quantum fluids and solids},
  school = {Polytechnic University of Catalonia},
  type   = {Ph.D. dissertation},
  url    = {http://hdl.handle.net/2117/94513}
}

@book{altland_simons_2010,
  author    = {Altland, Alexander and Simons, Ben D.},
  year      = {2010},
  title     = {Condensed Matter Field Theory},
  doi       = {10.1017/CBO9780511789984},
  edition   = {2},
  place     = {Cambridge},
  publisher = {Cambridge University Press}
}

@article{Vinay2010,
  author  = {Ambegaokar,Vinay  and Troyer,Matthias },
  year    = {2010},
  title   = {Estimating errors reliably in Monte Carlo simulations of the Ehrenfest model},
  doi     = {10.1119/1.3247985},
  journal = {American Journal of Physics},
  number  = {2},
  pages   = {150-157},
  url     = {https://doi.org/10.1119/1.3247985},
  volume  = {78}
}

@article{Chin2010,
  author    = {Chin, Cheng and Grimm, Rudolf and Julienne, Paul and Tiesinga, Eite},
  year      = {2010},
  title     = {Feshbach resonances in ultracold gases},
  doi       = {10.1103/RevModPhys.82.1225},
  issue     = {2},
  journal   = {Rev. Mod. Phys.},
  month     = {4},
  numpages  = {0},
  pages     = {1225--1286},
  publisher = {American Physical Society},
  url       = {https://link.aps.org/doi/10.1103/RevModPhys.82.1225},
  volume    = {82}
}

@article{Farrell2010,
  author  = {Farrell, Aaron and van Zyl, Brandon P.},
  year    = {2010},
  title   = {s-wave scattering and the zero-range limit of the finite square well in arbitrary dimensions},
  doi     = {10.1139/P10-061},
  journal = {Can. J. Phys.},
  number  = {11},
  pages   = {817-824},
  url     = {https://doi.org/10.1139/P10-061},
  volume  = {88}
}

@article{Trotzky2010,
  author  = {Trotzky, S.
             and Pollet, L.
             and Gerbier, F.
             and Schnorrberger, U.
             and Bloch, I.
             and Prokof'ev, N. V.
             and Svistunov, B.
             and Troyer, M.},
  year    = {2010},
  title   = {Suppression of the critical temperature for superfluidity near the Mott transition},
  day     = {01},
  doi     = {10.1038/nphys1799},
  issn    = {1745-2481},
  journal = {Nat. Phys.},
  month   = {12},
  number  = {12},
  pages   = {998-1004},
  url     = {https://doi.org/10.1038/nphys1799},
  volume  = {6}
}

@book{amidror2009theory,
  author    = {Amidror, Isaac},
  year      = {2009},
  title     = {The Theory of the Moir{\'e} Phenomenon: Volume I: Periodic Layers},
  doi       = {https://doi.org/10.1007/978-1-84882-181-1},
  edition   = {2},
  langid    = {english},
  location  = {London},
  number    = {38},
  publisher = {Springer},
  publisher = {Springer},
  series    = {Computational Imaging and Vision},
  url       = {https://doi.org/10.1007/978-1-84882-181-1},
  volume    = {38}
}

@article{Bernard2009,
  author    = {Bernard, Etienne P. and Krauth, Werner and Wilson, David B.},
  year      = {2009},
  title     = {Event-chain Monte Carlo algorithms for hard-sphere systems},
  doi       = {10.1103/PhysRevE.80.056704},
  issue     = {5},
  journal   = {Phys. Rev. E},
  month     = {11},
  numpages  = {5},
  pages     = {056704},
  publisher = {American Physical Society},
  url       = {https://link.aps.org/doi/10.1103/PhysRevE.80.056704},
  volume    = {80}
}

@article{Boninsegni2009,
  author    = {Boninsegni, Massimo},
  year      = {2009},
  title     = {Quantum statistics and the momentum distribution of liquid parahydrogen},
  doi       = {10.1103/PhysRevB.79.174203},
  issue     = {17},
  journal   = {Phys. Rev. B},
  month     = {5},
  numpages  = {7},
  pages     = {174203},
  publisher = {American Physical Society},
  url       = {https://link.aps.org/doi/10.1103/PhysRevB.79.174203},
  volume    = {79}
}

@article{Fetter2009,
  author    = {Fetter, Alexander L.},
  year      = {2009},
  title     = {Rotating trapped Bose-Einstein condensates},
  doi       = {10.1103/RevModPhys.81.647},
  issue     = {2},
  journal   = {Rev. Mod. Phys.},
  month     = {3},
  numpages  = {0},
  pages     = {647--691},
  publisher = {American Physical Society},
  url       = {https://link.aps.org/doi/10.1103/RevModPhys.81.647},
  volume    = {81}
}

@article{Buluta2009,
  author  = {Iulia Buluta  and Franco Nori },
  year    = {2009},
  title   = {Quantum Simulators},
  doi     = {10.1126/science.1177838},
  journal = {Science},
  number  = {5949},
  pages   = {108-111},
  volume  = {326}
}

@article{Khuri2009,
  author  = {Khuri,N. N.  and Martin,Andr'  and Richard,J.-M.  and Wu,Tai Tsun },
  year    = {2009},
  title   = {Low-energy potential scattering in two and three dimensions},
  doi     = {10.1063/1.3167803},
  journal = {J. Math. Phys.},
  number  = {7},
  pages   = {072105},
  url     = {https://doi.org/10.1063/1.3167803},
  volume  = {50}
}

@book{senechal2009quasicrystals,
  author    = {Senechal, Marjorie},
  year      = {2009},
  title     = {Quasicrystals and geometry},
  location  = {Cambridge},
  publisher = {Cambridge University Press},
  url       = {https://www.cambridge.org/academic/subjects/mathematics/mathematical-physics/quasicrystals-and-geometry}
}

@article{Bloch2008,
  author    = {Bloch, Immanuel and Dalibard, Jean and Zwerger, Wilhelm},
  year      = {2008},
  title     = {Many-body physics with ultracold gases},
  doi       = {10.1103/RevModPhys.80.885},
  issue     = {3},
  journal   = {Rev. Mod. Phys.},
  month     = {7},
  numpages  = {79},
  pages     = {885--964},
  publisher = {American Physical Society},
  url       = {https://link.aps.org/doi/10.1103/RevModPhys.80.885},
  volume    = {80}
}

@article{Clarke2008,
  author  = {Clarke, John
             and Wilhelm, Frank K.},
  year    = {2008},
  title   = {Superconducting quantum bits},
  doi     = {10.1038/nature07128},
  journal = {Nature},
  month   = {6},
  number  = {7198},
  pages   = {1031-1042},
  url     = {https://doi.org/10.1038/nature07128},
  volume  = {453}
}

@article{Corboz2008,
  author    = {Corboz, Philippe and Boninsegni, Massimo and Pollet, Lode and Troyer, Matthias},
  year      = {2008},
  title     = {Phase diagram of $^{4}\text{H}\text{e}$ adsorbed on graphite},
  doi       = {10.1103/PhysRevB.78.245414},
  issue     = {24},
  journal   = {Phys. Rev. B},
  month     = {12},
  numpages  = {6},
  pages     = {245414},
  publisher = {American Physical Society},
  url       = {https://link.aps.org/doi/10.1103/PhysRevB.78.245414},
  volume    = {78}
}

@book{hardy2008introduction,
  author    = {Hardy, Godfrey Harold and Wright, Edward Maitland and others},
  year      = {2008},
  title     = {An introduction to the theory of numbers},
  doi       = {https://global.oup.com/academic/product/an-introduction-to-the-theory-of-numbers-9780199219865},
  edition   = {6},
  publisher = {Oxford university press},
  url       = {https://global.oup.com/academic/product/an-introduction-to-the-theory-of-numbers-9780199219865}
}

@thesis{Boning2007,
  author = {B\"{o}ning, J.},
  year   = {2007},
  title  = {Superfluidity in mesoscopic systems of charged bosons},
  month  = {1},
  school = {University of Kiel},
  type   = {Ph.D. dissertation},
  url    = {http://www.itap.uni-kiel.de/theo-physik/bonitz/theses/boening_07.pdf}
}

@article{Pricoupenko2007,
  author    = {Ludovic Pricoupenko and Maxim Olshanii},
  year      = {2007},
  title     = {Stability of two-dimensional Bose gases in the resonant regime},
  doi       = {10.1088/0953-4075/40/11/009},
  journal   = {J. Phys. B: At. Mol. Opt. Phys.},
  month     = {5},
  number    = {11},
  pages     = {2065--2074},
  publisher = {{IOP} Publishing},
  url       = {https://doi.org/10.1088/0953-4075/40/11/009},
  volume    = {40}
}

@article{Prokofev2007,
  author    = {Prokof'ev, Nikolay and Svistunov, Boris},
  year      = {2007},
  title     = {Bold Diagrammatic Monte Carlo Technique: When the Sign Problem Is Welcome},
  doi       = {10.1103/PhysRevLett.99.250201},
  issue     = {25},
  journal   = {Phys. Rev. Lett.},
  month     = {12},
  numpages  = {4},
  pages     = {250201},
  publisher = {American Physical Society},
  url       = {https://link.aps.org/doi/10.1103/PhysRevLett.99.250201},
  volume    = {99}
}

@article{Boninsegn2006,
  author    = {Boninsegni, M. and Prokof'ev, N. V. and Svistunov, B. V.},
  year      = {2006},
  title     = {Worm algorithm and diagrammatic Monte Carlo: A new approach to continuous-space path integral Monte Carlo simulations},
  doi       = {10.1103/PhysRevE.74.036701},
  issue     = {3},
  journal   = {Phys. Rev. E},
  month     = {9},
  numpages  = {16},
  pages     = {036701},
  publisher = {American Physical Society},
  url       = {https://link.aps.org/doi/10.1103/PhysRevE.74.036701},
  volume    = {74}
}

@article{Boninsegn2006PRL,
  author    = {Boninsegni, Massimo and Prokof'ev, Nikolay and Svistunov, Boris},
  year      = {2006},
  title     = {Worm Algorithm for Continuous-Space Path Integral Monte Carlo Simulations},
  doi       = {10.1103/PhysRevLett.96.070601},
  issue     = {7},
  journal   = {Phys. Rev. Lett.},
  month     = {2},
  numpages  = {4},
  pages     = {070601},
  publisher = {American Physical Society},
  url       = {https://link.aps.org/doi/10.1103/PhysRevLett.96.070601},
  volume    = {96}
}

@article{Zbigniew2006,
  author    = {Idziaszek, Zbigniew and Calarco, Tommaso},
  year      = {2006},
  title     = {Pseudopotential Method for Higher Partial Wave Scattering},
  doi       = {10.1103/PhysRevLett.96.013201},
  issue     = {1},
  journal   = {Phys. Rev. Lett.},
  month     = {1},
  numpages  = {4},
  pages     = {013201},
  publisher = {American Physical Society},
  url       = {https://link.aps.org/doi/10.1103/PhysRevLett.96.013201},
  volume    = {96}
}

@book{krauth2006,
  author     = {Krauth, Werner},
  year       = {2006},
  title      = {Statistical mechanics: algorithms and computations},
  langid     = {english},
  location   = {Oxford},
  number     = {13},
  publisher  = {Oxford University Press},
  series     = {Oxford master series in physics},
  shorttitle = {Statistical mechanics}
}

@article{Kwon2006,
  author    = {Kwon, Yongkyung and Paesani, Francesco and Whaley, K. Birgitta},
  year      = {2006},
  title     = {Local superfluidity in inhomogeneous quantum fluids},
  doi       = {10.1103/PhysRevB.74.174522},
  issue     = {17},
  journal   = {Phys. Rev. B},
  month     = {11},
  numpages  = {10},
  pages     = {174522},
  publisher = {American Physical Society},
  url       = {https://link.aps.org/doi/10.1103/PhysRevB.74.174522},
  volume    = {74}
}

@article{Betzig2005,
  author    = {Betzig, Eric},
  year      = {2005},
  title     = {Sparse and composite coherent lattices},
  doi       = {10.1103/PhysRevA.71.063406},
  issue     = {6},
  journal   = {Phys. Rev. A},
  month     = {6},
  numpages  = {5},
  pages     = {063406},
  publisher = {American Physical Society},
  url       = {https://link.aps.org/doi/10.1103/PhysRevA.71.063406},
  volume    = {71}
}

@article{Troyer2005,
  author    = {Troyer, Matthias and Wiese, Uwe-Jens},
  year      = {2005},
  title     = {Computational Complexity and Fundamental Limitations to Fermionic Quantum Monte Carlo Simulations},
  doi       = {10.1103/PhysRevLett.94.170201},
  issue     = {17},
  journal   = {Phys. Rev. Lett.},
  month     = {5},
  numpages  = {4},
  pages     = {170201},
  publisher = {American Physical Society},
  url       = {https://link.aps.org/doi/10.1103/PhysRevLett.94.170201},
  volume    = {94}
}

@book{Bell2004,
  author    = {Bell, John S.},
  year      = {2004},
  title     = {Speakable and Unspeakable in Quantum Mechanics: Collected Papers on Quantum Philosophy},
  doi       = {10.1017/CBO9780511815676},
  edition   = {2},
  publisher = {Cambridge University Press}
}

@article{Dowling2003,
  author  = {Dowling, Jonathan P.  and Milburn, Gerard J. },
  year    = {2003},
  title   = {Quantum technology: the second quantum revolution},
  doi     = {10.1098/rsta.2003.1227},
  journal = {Philosophical Transactions of the Royal Society of London. Series A: Mathematical, Physical and Engineering Sciences},
  number  = {1809},
  pages   = {1655-1674},
  volume  = {361}
}

@article{Draeger2003,
  author    = {Draeger, E. W. and Ceperley, D. M.},
  year      = {2003},
  title     = {Superfluidity in a Doped Helium Droplet},
  doi       = {10.1103/PhysRevLett.90.065301},
  issue     = {6},
  journal   = {Phys. Rev. Lett.},
  month     = {2},
  numpages  = {4},
  pages     = {065301},
  publisher = {American Physical Society},
  url       = {https://link.aps.org/doi/10.1103/PhysRevLett.90.065301},
  volume    = {90}
}

@article{Carlon2002,
  author    = {Carlon, Enrico and Orlandini, Enzo and Stella, Attilio L.},
  year      = {2002},
  title     = {Roles of Stiffness and Excluded Volume in DNA Denaturation},
  doi       = {10.1103/PhysRevLett.88.198101},
  issue     = {19},
  journal   = {Phys. Rev. Lett.},
  month     = {4},
  numpages  = {4},
  pages     = {198101},
  publisher = {American Physical Society},
  url       = {https://link.aps.org/doi/10.1103/PhysRevLett.88.198101},
  volume    = {88}
}

@article{greiner2002,
  author    = {Greiner, Markus and Mandel, Olaf and Esslinger, Tilman and H{\"a}nsch, Theodor W and Bloch, Immanuel},
  year      = {2002},
  title     = {Quantum phase transition from a superfluid to a Mott insulator in a gas of ultracold atoms},
  journal   = {nature},
  number    = {6867},
  pages     = {39--44},
  publisher = {Nature Publishing Group},
  volume    = {415}
}

@incollection{Janke2004,
  author    = {Janke, Wolfhard},
  year      = {2002},
  title     = {Statistical Analysis of Simulations: Data Correlations and Error Estimation},
  booktitle = {Quantum simulations of complex many-body systems: from theory to algorithms},
  editor    = {Grotendorst, Johannes and Marx, Dominik and Muramatsu, Alejandro},
  pages     = {423--445},
  publisher = {John von Neumann Institute for Computing},
  volume    = {10}
}

@article{Truscott2001,
  author  = {Andrew G. Truscott  and Kevin E. Strecker  and William I. McAlexander  and Guthrie B. Partridge  and Randall G. Hulet},
  year    = {2001},
  title   = {Observation of Fermi Pressure in a Gas of Trapped Atoms},
  doi     = {10.1126/science.1059318},
  journal = {Science},
  number  = {5513},
  pages   = {2570-2572},
  volume  = {291}
}

@article{Petrov2001,
  author    = {Petrov, D. S. and Shlyapnikov, G. V.},
  year      = {2001},
  title     = {Interatomic collisions in a tightly confined Bose gas},
  doi       = {10.1103/PhysRevA.64.012706},
  issue     = {1},
  journal   = {Phys. Rev. A},
  month     = {6},
  numpages  = {14},
  pages     = {012706},
  publisher = {American Physical Society},
  url       = {https://link.aps.org/doi/10.1103/PhysRevA.64.012706},
  volume    = {64}
}

@article{Schreck2001,
  author    = {Schreck, F. and Khaykovich, L. and Corwin, K. L. and Ferrari, G. and Bourdel, T. and Cubizolles, J. and Salomon, C.},
  year      = {2001},
  title     = {Quasipure Bose-Einstein Condensate Immersed in a Fermi Sea},
  doi       = {10.1103/PhysRevLett.87.080403},
  issue     = {8},
  journal   = {Phys. Rev. Lett.},
  month     = {8},
  numpages  = {4},
  pages     = {080403},
  publisher = {American Physical Society},
  url       = {https://link.aps.org/doi/10.1103/PhysRevLett.87.080403},
  volume    = {87}
}

@article{Cornish2000,
  author    = {Cornish, S. L. and Claussen, N. R. and Roberts, J. L. and Cornell, E. A. and Wieman, C. E.},
  year      = {2000},
  title     = {Stable ${}^{85}\mathrm{Rb}$ Bose-Einstein Condensates with Widely Tunable Interactions},
  doi       = {10.1103/PhysRevLett.85.1795},
  issue     = {9},
  journal   = {Phys. Rev. Lett.},
  month     = {8},
  numpages  = {0},
  pages     = {1795--1798},
  publisher = {American Physical Society},
  url       = {https://link.aps.org/doi/10.1103/PhysRevLett.85.1795},
  volume    = {85}
}

@article{Grimm2000,
  author  = {Grimm, Rudolf and Weidemüller, Matthias and Ovchinnikov, Yurii B.},
  year    = {2000},
  title   = {Optical dipole traps for neutral atoms},
  doi     = {https://doi.org/10.1016/S1049-250X(08)60186-X},
  journal = {Adv. At., Mol., Opt. Phys},
  pages   = {95--170},
  url     = {https://www.sciencedirect.com/science/article/pii/S1049250X0860186X},
  volume  = {42}
}

@phdthesis{militzer2000path,
  author = {Militzer, Burkhard},
  year   = {2000},
  title  = {Path integral Monte Carlo simulations of hot dense hydrogen},
  school = {University of Illinois at Urbana-Champaign},
  type   = {Ph.D. dissertation},
  url    = {http://militzer.berkeley.edu/diss/diss_militzer.pdf}
}

@article{Petrov2000,
  author    = {Petrov, D. S. and Holzmann, M. and Shlyapnikov, G. V.},
  year      = {2000},
  title     = {Bose-Einstein Condensation in Quasi-2D Trapped Gases},
  doi       = {10.1103/PhysRevLett.84.2551},
  issue     = {12},
  journal   = {Phys. Rev. Lett.},
  month     = {3},
  numpages  = {0},
  pages     = {2551--2555},
  publisher = {American Physical Society},
  url       = {https://link.aps.org/doi/10.1103/PhysRevLett.84.2551},
  volume    = {84}
}

@article{DeMarco1999,
  author  = {B. DeMarco  and D. S. Jin},
  year    = {1999},
  title   = {Onset of Fermi Degeneracy in a Trapped Atomic Gas},
  doi     = {10.1126/science.285.5434.1703},
  journal = {Science},
  number  = {5434},
  pages   = {1703-1706},
  volume  = {285}
}

@book{conway1998book,
  author    = {Conway, John H and Guy, Richard},
  year      = {1998},
  title     = {The book of numbers},
  doi       = {https://doi.org/10.1007/978-1-4612-4072-3},
  langid    = {english},
  location  = {New York},
  publisher = {Springer},
  publisher = {Springer},
  url       = {https://doi.org/10.1007/978-1-4612-4072-3}
}

@article{Courteille1998,
  author    = {Courteille, Ph. and Freeland, R. S. and Heinzen, D. J. and van Abeelen, F. A. and Verhaar, B. J.},
  year      = {1998},
  title     = {Observation of a Feshbach Resonance in Cold Atom Scattering},
  doi       = {10.1103/PhysRevLett.81.69},
  issue     = {1},
  journal   = {Phys. Rev. Lett.},
  month     = {6},
  numpages  = {0},
  pages     = {69--72},
  publisher = {American Physical Society},
  url       = {https://link.aps.org/doi/10.1103/PhysRevLett.81.69},
  volume    = {81}
}

@article{Inouye1998,
  author  = {Inouye, S.
             and Andrews, M. R.
             and Stenger, J.
             and Miesner, H.-J.
             and Stamper-Kurn, D. M.
             and Ketterle, W.},
  year    = {1998},
  title   = {Observation of Feshbach resonances in a Bose--Einstein condensate},
  day     = {01},
  doi     = {10.1038/32354},
  issn    = {1476-4687},
  journal = {Nature},
  month   = {3},
  number  = {6672},
  pages   = {151-154},
  url     = {https://doi.org/10.1038/32354},
  volume  = {392}
}

@article{Prokofev1998,
  author  = {N.V Prokof'ev and B.V Svistunov and I.S Tupitsyn},
  year    = {1998},
  title   = {“Worm” algorithm in quantum Monte Carlo simulations},
  doi     = {https://doi.org/10.1016/S0375-9601(97)00957-2},
  issn    = {0375-9601},
  journal = {Phys. Lett. A},
  number  = {4},
  pages   = {253-257},
  url     = {https://www.sciencedirect.com/science/article/pii/S0375960197009572},
  volume  = {238}
}

@article{prokofev1998sec,
  author  = {Prokof'ev, N. V. and Svistunov, B. V. and Tupitsyn, I. S.},
  year    = {1998},
  title   = {Exact, complete, and universal continuous-time worldline Monte Carlo approach to the statistics of discrete quantum systems},
  doi     = {10.1134/1.558661},
  issn    = {1090-6509},
  journal = {Journal of Experimental and Theoretical Physics},
  month   = {8},
  number  = {2},
  pages   = {310--321},
  url     = {https://doi.org/10.1134/1.558661},
  volume  = {87}
}

@book{Baker1997,
  author    = {Baker, R. C. and Harman, G. and Pintz, J. and Greaves, G. R. H. and Harman, G. and Huxley, M. N.},
  year      = {1997},
  title     = {Sieve Methods, Exponential Sums, and their Applications in Number Theory},
  doi       = {10.1017/CBO9780511526091},
  place     = {Cambridge},
  publisher = {Cambridge University Press},
  series    = {London Mathematical Society Lecture Note Series}
}

@article{Wolfhard1997,
  author  = {Janke,Wolfhard  and Sauer,Tilman },
  year    = {1997},
  title   = {Optimal energy estimation in path-integral Monte Carlo simulations},
  doi     = {10.1063/1.474309},
  journal = {J. Chem. Phys.},
  number  = {15},
  pages   = {5821-5839},
  url     = {https://doi.org/10.1063/1.474309},
  volume  = {107}
}

@online{Nobelprize1997,
  author       = {{Nobel prize}},
  year         = {1997},
  title        = {Press release: The 1997 Nobel Prize in Physics},
  note         = {Accessed on 05/18/2022},
  organization = {Nobel prize outreach AB},
  url          = {https://www.nobelprize.org/prizes/physics/1997/press-release/}
}

@book{washington1997introduction,
  author    = {Washington, Lawrence C},
  year      = {1997},
  title     = {Introduction to cyclotomic fields},
  doi       = {https://doi.org/10.1007/978-1-4612-1934-7},
  edition   = {2},
  langid    = {english},
  location  = {New York},
  number    = {83},
  publisher = {Springer},
  series    = {Graduate Texts in Mathematics},
  url       = {https://doi.org/10.1007/978-1-4612-1934-7}
}

@article{Bradley1995,
  author    = {Bradley, C. C. and Sackett, C. A. and Tollett, J. J. and Hulet, R. G.},
  year      = {1995},
  title     = {Evidence of Bose-Einstein Condensation in an Atomic Gas with Attractive Interactions},
  doi       = {10.1103/PhysRevLett.75.1687},
  issue     = {9},
  journal   = {Phys. Rev. Lett.},
  month     = {8},
  numpages  = {0},
  pages     = {1687--1690},
  publisher = {American Physical Society},
  volume    = {75}
}

@article{Ceperley1995,
  author    = {Ceperley, D. M.},
  year      = {1995},
  title     = {Path integrals in the theory of condensed helium},
  doi       = {10.1103/RevModPhys.67.279},
  issue     = {2},
  journal   = {Rev. Mod. Phys.},
  month     = {4},
  pages     = {279--355},
  publisher = {American Physical Society},
  url       = {https://link.aps.org/doi/10.1103/RevModPhys.67.279},
  volume    = {67}
}

@article{Davis1995,
  author    = {Davis, K. B. and Mewes, M. -O. and Andrews, M. R. and van Druten, N. J. and Durfee, D. S. and Kurn, D. M. and Ketterle, W.},
  year      = {1995},
  title     = {Bose-Einstein Condensation in a Gas of Sodium Atoms},
  doi       = {10.1103/PhysRevLett.75.3969},
  issue     = {22},
  journal   = {Phys. Rev. Lett.},
  month     = {11},
  numpages  = {0},
  pages     = {3969--3973},
  publisher = {American Physical Society},
  volume    = {75}
}

@article{Anderson1995,
  author    = {M. H. Anderson and J. R. Ensher and M. R. Matthews and C. E. Wieman and E. A. Cornell},
  year      = {1995},
  title     = {Observation of Bose-Einstein Condensation in a Dilute Atomic Vapor},
  doi       = {10.1126/science.269.5221.198},
  journal   = {Science},
  number    = {5221},
  pages     = {198--201},
  publisher = {American Association for the Advancement of Science},
  volume    = {269}
}

@article{Petsas1994,
  author    = {Petsas, K. I. and Coates, A. B. and Grynberg, G.},
  year      = {1994},
  title     = {Crystallography of optical lattices},
  doi       = {10.1103/PhysRevA.50.5173},
  issue     = {6},
  journal   = {Phys. Rev. A},
  month     = {12},
  numpages  = {0},
  pages     = {5173--5189},
  publisher = {American Physical Society},
  url       = {https://link.aps.org/doi/10.1103/PhysRevA.50.5173},
  volume    = {50}
}

@article{Borrmann1993,
  author  = {Borrmann, Peter and Franke, Gert},
  year    = {1993},
  title   = {Recursion formulas for quantum statistical partition functions},
  doi     = {10.1063/1.464180},
  issn    = {0021-9606},
  journal = {J. Chem. Phys.},
  number  = {3},
  pages   = {2484--2485},
  url     = {http://aip.scitation.org/doi/10.1063/1.464180},
  urldate = {2022-04-22},
  volume  = {98}
}

@article{CaoBerne1992,
  author  = {Cao,J.  and Berne,B. J. },
  year    = {1992},
  title   = {A new quantum propagator for hard sphere and cavity systems},
  doi     = {10.1063/1.463076},
  journal = {J. Chem. Phys.},
  number  = {4},
  pages   = {2382-2385},
  url     = {https://doi.org/10.1063/1.463076},
  volume  = {97}
}

@article{tannoudji1992atom,
  author    = {Tannoudji, Claude Cohen and Grynberg, Gilbert and Dupont-Roe, J},
  year      = {1992},
  title     = {Atom-photon interactions},
  doi       = {10.1002/9783527617197},
  location  = {New York},
  publisher = {John Wiley and Sons Inc.}
}

@article{Reynolds1990,
  author  = {Reynolds,Peter J.  and Tobochnik,Jan  and Gould,Harvey },
  year    = {1990},
  title   = {Diffusion Quantum Monte Carlo},
  doi     = {10.1063/1.4822960},
  journal = {Comput. Phys.},
  number  = {6},
  pages   = {662-668},
  url     = {https://aip.scitation.org/doi/abs/10.1063/1.4822960},
  volume  = {4}
}

@book{fowles1989introduction,
  author    = {Fowles, Grant R.},
  year      = {1989},
  title     = {Introduction to modern optics},
  doi       = {https://store.doverpublications.com/0486659577.html},
  edition   = {2},
  location  = {New York},
  publisher = {Dover Publications},
  url       = {https://store.doverpublications.com/0486659577.html}
}

@article{Dalibard1989,
  author    = {J. Dalibard and C. Cohen-Tannoudji},
  year      = {1989},
  title     = {Laser cooling below the Doppler limit by polarization gradients: simple theoretical models},
  doi       = {10.1364/JOSAB.6.002023},
  journal   = {J. Opt. Soc. Am. B},
  month     = {11},
  number    = {11},
  pages     = {2023--2045},
  publisher = {Optica Publishing Group},
  url       = {http://opg.optica.org/josab/abstract.cfm?URI=josab-6-11-2023},
  volume    = {6}
}

@article{Ungar1989,
  author    = {P. J. Ungar and D. S. Weiss and E. Riis and Steven Chu},
  year      = {1989},
  title     = {Optical molasses and multilevel atoms: theory},
  doi       = {10.1364/JOSAB.6.002058},
  journal   = {J. Opt. Soc. Am. B},
  month     = {11},
  number    = {11},
  pages     = {2058--2071},
  publisher = {Optica Publishing Group},
  url       = {http://opg.optica.org/josab/abstract.cfm?URI=josab-6-11-2058},
  volume    = {6}
}

@article{Sindzingre1989,
  author    = {Sindzingre, Philippe and Klein, Michael L. and Ceperley, David M.},
  year      = {1989},
  title     = {Path-integral Monte Carlo study of low-temperature $^{4}\mathrm{He}$ clusters},
  doi       = {10.1103/PhysRevLett.63.1601},
  issue     = {15},
  journal   = {Phys. Rev. Lett.},
  month     = {10},
  numpages  = {0},
  pages     = {1601--1604},
  publisher = {American Physical Society},
  url       = {https://link.aps.org/doi/10.1103/PhysRevLett.63.1601},
  volume    = {63}
}

@article{Metropolis1987,
  author  = {Metropolis, N.},
  year    = {1987},
  title   = {The Beginning of the Monte Carlo Method},
  doi     = {10.2172/1054744},
  journal = {Los Alamos Science},
  month   = {1},
  url     = {https://www.osti.gov/biblio/1054744},
  volume  = {15}
}

@article{Ceperley1987,
  author    = {Pollock, E. L. and Ceperley, D. M.},
  year      = {1987},
  title     = {Path-integral computation of superfluid densities},
  doi       = {10.1103/PhysRevB.36.8343},
  issue     = {16},
  journal   = {Phys. Rev. B},
  month     = {12},
  numpages  = {0},
  pages     = {8343--8352},
  publisher = {American Physical Society},
  url       = {https://link.aps.org/doi/10.1103/PhysRevB.36.8343},
  volume    = {36}
}

@article{Averbuch1986,
  author    = {P G Averbuch},
  year      = {1986},
  title     = {Zero energy divergence of scattering cross sections in two dimensions},
  doi       = {10.1088/0305-4470/19/12/018},
  journal   = {J. Phys. A: Math. Gen.},
  month     = {8},
  number    = {12},
  pages     = {2325--2335},
  publisher = {{IOP} Publishing},
  url       = {https://doi.org/10.1088/0305-4470/19/12/018},
  volume    = {19}
}

@article{Ceperley1984,
  author    = {Pollock, E. L. and Ceperley, D. M.},
  year      = {1984},
  title     = {Simulation of quantum many-body systems by path-integral methods},
  doi       = {10.1103/PhysRevB.30.2555},
  issue     = {5},
  journal   = {Phys. Rev. B},
  month     = {9},
  numpages  = {0},
  pages     = {2555--2568},
  publisher = {American Physical Society},
  url       = {https://link.aps.org/doi/10.1103/PhysRevB.30.2555},
  volume    = {30}
}

@article{MacGillivray1982,
  author  = {MacGillivray, H. T.
             and Dodd, R. J.},
  year    = {1982},
  title   = {Monte-Carlo simulations of galaxy systems},
  doi     = {10.1007/BF00683346},
  issn    = {1572-946X},
  journal = {Astrophys. Space Sci.},
  month   = {9},
  number  = {2},
  pages   = {419-435},
  url     = {https://doi.org/10.1007/BF00683346},
  volume  = {86}
}

@article{ChandlerWolynes1981,
  author  = {Chandler,David  and Wolynes,Peter G. },
  year    = {1981},
  title   = {Exploiting the isomorphism between quantum theory and classical statistical mechanics of polyatomic fluids},
  doi     = {10.1063/1.441588},
  journal = {J. Chem. Phys.},
  number  = {7},
  pages   = {4078-4095},
  url     = {https://doi.org/10.1063/1.441588},
  volume  = {74}
}

@article{Lim1980,
  author    = {Lim, T. K. and Maurone, P. A.},
  year      = {1980},
  title     = {Nonexistence of the Efimov effect in two dimensions},
  doi       = {10.1103/PhysRevB.22.1467},
  issue     = {3},
  journal   = {Phys. Rev. B},
  month     = {8},
  numpages  = {0},
  pages     = {1467--1469},
  publisher = {American Physical Society},
  url       = {https://link.aps.org/doi/10.1103/PhysRevB.22.1467},
  volume    = {22}
}

@article{Barker1979,
  author  = {Barker,J. A. },
  year    = {1979},
  title   = {A quantum‐statistical Monte Carlo method; path integrals with boundary conditions},
  doi     = {10.1063/1.437829},
  journal = {J. Chem. Phys.},
  number  = {6},
  pages   = {2914-2918},
  url     = {https://doi.org/10.1063/1.437829},
  volume  = {70}
}

@article{Drukarev1978,
  author    = {G. Drukarev},
  year      = {1978},
  title     = {The Zero-Range Potential Model and Its Application in Atomic and Molecular Physics},
  doi       = {10.1016/S0065-3276(08)60239-7},
  issn      = {0065-3276},
  journal   = {Adv. Quantum Chem.},
  pages     = {251-274},
  publisher = {Academic Press},
  series    = {Adv. Quantum Chem.},
  url       = {https://www.sciencedirect.com/science/article/pii/S0065327608602397},
  volume    = {11}
}

@article{suzuki_generalized_1976,
  author       = {Suzuki, Masuo},
  year         = {1976},
  title        = {Generalized Trotter's formula and systematic approximants of exponential operators and inner derivations with applications to many-body problems},
  doi          = {10.1007/BF01609348},
  issn         = {1432-0916},
  journaltitle = {Comm. Math. Phys.},
  month        = {6},
  number       = {2},
  pages        = {183--190},
  url          = {https://doi.org/10.1007/BF01609348},
  volume       = {51}
}

@book{feynman1972,
  author    = {Feynman, R.P.},
  year      = {1972},
  title     = {Statistical Mechanics},
  publisher = {Addison-Wesley},
  url       = {http://www.stat.phys.kyushu-u.ac.jp/B4/Papers/Feynman-StatMech-Chap-1-3-Org.pdf}
}

@article{Zaslow1967,
  author  = {Zaslow, B. and Zandler, Melvin E.},
  year    = {1967},
  title   = {Two-Dimensional Analog to the Hydrogen Atom},
  doi     = {10.1119/1.1973790},
  journal = {Am. J. Phys},
  number  = {12},
  pages   = {1118-1119},
  url     = {https://doi.org/10.1119/1.1973790},
  volume  = {35}
}

@book{feynman1965,
  author    = {Feynman, RP and Hibbs, AR},
  year      = {1965},
  title     = {{Q}uantum {M}echanics and {P}ath {I}ntegrals},
  address   = {New-York},
  publisher = {McGraw-Hill},
  url       = {http://www-f1.ijs.si/~ramsak/km1/FeynmanHibbs.pdf}
}

@article{Huang1957,
  author    = {Huang, Kerson and Yang, C. N.},
  year      = {1957},
  title     = {Quantum-Mechanical Many-Body Problem with Hard-Sphere Interaction},
  doi       = {10.1103/PhysRev.105.767},
  issue     = {3},
  journal   = {Phys. Rev.},
  month     = {2},
  numpages  = {0},
  pages     = {767--775},
  publisher = {American Physical Society},
  url       = {https://link.aps.org/doi/10.1103/PhysRev.105.767},
  volume    = {105}
}

@article{Fermi1957,
  author    = {Vajda, S.},
  year      = {1957},
  title     = {Symposium on Monte Carlo Methods},
  doi       = {10.2307/3610170},
  journal   = {The Mathematical Gazette},
  number    = {338},
  pages     = {318–318},
  publisher = {Cambridge University Press},
  volume    = {41}
}

@article{Feynman1953,
  author    = {Feynman, R. P.},
  year      = {1953},
  title     = {Atomic Theory of Liquid Helium Near Absolute Zero},
  doi       = {10.1103/PhysRev.91.1301},
  issue     = {6},
  journal   = {Phys. Rev.},
  month     = {9},
  numpages  = {0},
  pages     = {1301--1308},
  publisher = {American Physical Society},
  url       = {https://link.aps.org/doi/10.1103/PhysRev.91.1301},
  volume    = {91}
}

@article{Ginzburg1950,
  author  = {Ginzburg, V. L. and Landau, L. D.},
  year    = {1950},
  title   = {{On the Theory of superconductivity}},
  doi     = {https://doi.org/10.1016/B978-0-08-010586-4.50078-X},
  journal = {Zh. Eksp. Teor. Fiz.},
  pages   = {1064--1082},
  volume  = {20}
}

@article{FeyKac1949,
  author   = {Kac, M.},
  year     = {1949},
  title    = {On distributions of certain Wiener functionals},
  doi      = {10.2307/1990512},
  journal  = {Trans. Amer. Math. Soc.},
  month    = {1},
  numpages = {13},
  pages    = {1-13},
  url      = {https://doi.org/10.2307/1990512},
  volume   = {65}
}

@article{Metropolis1949,
  author    = {Nicholas Metropolis and  S. Ulam},
  year      = {1949},
  title     = {The Monte Carlo Method},
  doi       = {10.1080/01621459.1949.10483310},
  journal   = {J. Am. Stat. Assoc.},
  number    = {247},
  pages     = {335-341},
  publisher = {Taylor & Francis},
  url       = {https://www.tandfonline.com/doi/abs/10.1080/01621459.1949.10483310},
  volume    = {44}
}

@article{fermi1936,
  author  = {Fermi, E},
  year    = {1936},
  journal = {Ricerca sci.},
  pages   = {13},
  volume  = {7}
}

@article{Bethe1935,
  author  = {Bethe, H. and Peierls, R. and Hartree, Douglas Rayner },
  year    = {1935},
  title   = {Quantum theory of the diplon},
  doi     = {10.1098/rspa.1935.0010},
  journal = {Proc. R. Soc. Lond. A},
  pages   = {146-156},
  url     = {https://royalsocietypublishing.org/doi/abs/10.1098/rspa.1935.0010},
  volume  = {148}
}

@article{Campbell1897,
  author  = {Campbell, J. E.},
  year    = {1897},
  title   = {On a Law of Combination of Operators},
  doi     = {https://doi.org/10.1112/plms/s1-29.1.14},
  journal = {Proc. Math. Soc.},
  pages   = {391--90},
  volume  = {s1-28}
}
\cleardoublepage

\cleartoleftpage{} 
\thispagestyle{empty}
\sffamily
\begin{textblock}{191}(113,-11)
{\color{blueline}\rule{160pt}{5.5pt}}
\end{textblock}
\begin{textblock}{191}(168,-11)
{\color{blueline}\rule{5.5pt}{59pt}}
\end{textblock}
\begin{textblock}{183}(-24,-11)
\textblockcolour{}
\flushright
\fontsize{7}{7.5}\selectfont
\textbf{Institute for Theoretical Physics}\\
Celestijnenlaan 200D box 2415\\
3000 LEUVEN, BELGI\"{E}\\
tel. +32 16 32 72 32\\
https://fys.kuleuven.be/itf/contact\\
\end{textblock}
\begin{textblock}{191}(154,-7)
\textblockcolour{}
\includegraphics*[height=16.5truemm]{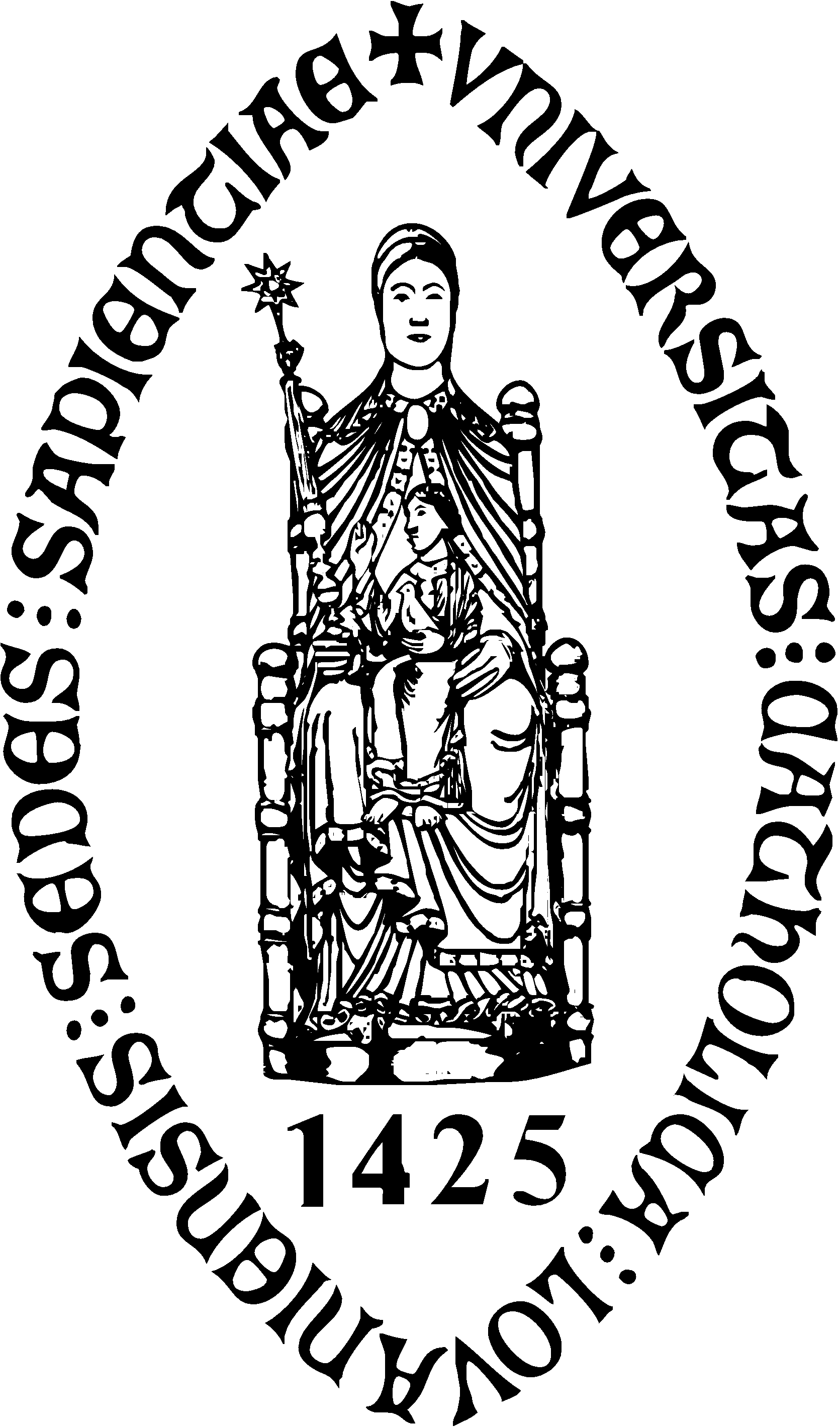}
\end{textblock}
\begin{textblock}{191}(-20,235)
{\color{bluetitle}\rule{544pt}{55pt}}
\end{textblock}

\end{document}